\documentclass[times,twocolumn]{aastex63}

\usepackage{tipa}
\usepackage{CJKutf8}

\submitjournal{AJ}

\shorttitle{Tidal evolution of Gonggong--Xiangliu}
\shortauthors{Arakawa et al.}
\graphicspath{{./}{figures/}}

\begin{document}

\title{Tidal evolution of the eccentric moon around dwarf planet (225088) Gonggong}

\correspondingauthor{Sota Arakawa}
\email{sota.arakawa@nao.ac.jp}

\author[0000-0003-0947-9962]{Sota Arakawa}
\affiliation{Division of Science, National Astronomical Observatory of Japan \\
2-21-1 Osawa, Mitaka, Tokyo, 181-8588, Japan.}

\author[0000-0003-4590-0988]{Ryuki Hyodo}
\affiliation{Department of Solar System Sciences, Institute of Space and Astronautical Science, Japan Aerospace Exploration Agency \\
3-1-1 Yoshinodai, Chuo, Sagamihara, Kanagawa, 252-5210, Japan.}

\author[0000-0001-6423-0698]{Daigo Shoji}
\affiliation{Department of Solar System Sciences, Institute of Space and Astronautical Science, Japan Aerospace Exploration Agency \\
3-1-1 Yoshinodai, Chuo, Sagamihara, Kanagawa, 252-5210, Japan.}

\author[0000-0001-6702-0872]{Hidenori Genda}
\affiliation{Earth-Life Science Institute, Tokyo Institute of Technology \\
2-12-1 Ookayama, Meguro, Tokyo, 152-8550, Japan.}



\begin{abstract}
Recent astronomical observations revealed that (225088) Gonggong, a 1000-km-sized trans-Neptunian dwarf planet, hosts an eccentric satellite, Xiangliu, with an eccentricity of approximately 0.3.
As the majority of known satellite systems around trans-Neptunian dwarf planets have circular orbits, the observed eccentricity of Gonggong--Xiangliu system may reflect the singular properties of the system.
In this study, we assumed that Gonggong--Xiangliu system formed via a giant impact and investigated the following secular tidal evolution of Gonggong--Xiangliu system under the simplifying assumption of homogeneous bodies and of zero orbital inclination.
We conducted coupled thermal--orbital evolution simulations using the Andrade viscoelastic model and included higher-order eccentricity functions.
The distribution of the final eccentricity from a large number of simulations with different initial conditions revealed that the radius of Xiangliu is not larger than 100 km.
We also derived the analytical solution of the semilatus rectum evolution, a function of the radius of Xiangliu.
From the point of view of the final semilatus rectum, the radius of Xiangliu was estimated to be close to 100 km.
Together with the results of the Hubble Space Telescope observations, our findings suggest Gonggong and Xiangliu have similar albedos.
\end{abstract}



\section{Introduction}

There are six (or seven) known trans-Neptunian objects (TNOs) with diameters larger than 1000 km, and (almost) all of them host one or more satellites orbiting the primary \citep[e.g.,][]{Brown+2006,Parker+2016,Kiss+2017}.\footnote{
The diameter of (90377) Sedna is $995 \pm 80\ {\rm km}$ \citep{Pal+2012} and it has no known satellites.
The others, namely, (134340) Pluto, (136199) Eris, (136108) Haumea, (136472) Makemake, (225088) Gonggong, and (50000) Quaoar, have one or more satellites.}
The discovery of satellites around large TNOs provided us with a key to understanding the early history of the outer solar system.

The orbital period of a satellite system allows the determination of the total mass of the satellite system.
We can obtain the density of the primary body when the size is known from radiometry and/or occultation measurements and estimate the bulk composition \citep[e.g., the ice-to-rock ratio;][]{Brown2012}.
Other orbital elements, such as the eccentricity and inclination, are also important to understand how satellites form and evolve.
The majority of moons around 1000-km-sized TNOs have circular orbits.
For example, the orbit of Pluto--Charon system is nearly circular \citep[$e \sim 5 \times 10^{-5}$, where $e$ is the eccentricity;][]{Brozovic+2015}.
Based on the spatially resolved observations from the Atacama Large Millimeter Array by \citet{Brown+2018}, the eccentricity of Eris--Dysnomia system is also small ($e < 4 \times 10^{-3}$).
Their circular orbits are thought to be the result of tidal evolution, as satellite systems are initially eccentric if they formed via giant impacts \citep[e.g.,][]{Canup2005,Arakawa+2019}.

However, the two known exceptions to this trend are Quaoar--Weywot and Gonggong--Xiangliu systems.
Using the adaptive optics facility of the Keck 2 telescope, \citet{Fraser+2013} revealed that the orbit of Quaoar--Weywot system is moderately eccentric ($e = 0.13$--$0.16$).
\citet{Kiss+2019} also reported an eccentricity of $e \simeq 0.3$ for the orbit of Gonggong--Xiangliu system.
If the tidal dissipation primarily occurs inside the satellite rather than inside the host dwarf planets, the eccentricity decreases \citep[e.g.,][]{Ward+2006,Cheng+2014,Arakawa+2019}.
Therefore, the eccentricity evolution strongly depends on the rheological properties of the satellites and their hosts.
The variation in the observed eccentricity among satellite systems of large TNOs may reflect differences in the rheological properties, which should be related to their formation process and thermal histories.

The primary, (225088) Gonggong, was discovered by \citet{Schwamb+2009}.
The satellite of (225088) Gonggong, Xiangliu, was found on archival images of the WFC3 camera of the Hubble Space Telescope obtained in 2009 and 2010 by \citet{Kiss+2017}.
This initial discovery was based only on observations at two epochs, and the orbit of the satellite could not be derived.
\citet{Kiss+2019} conducted recovery observations of Xiangliu taken with the WFC3 camera in 2017 and determined the orbital elements of Gonggong--Xiangliu system.

Table \ref{Table1} shows the orbital elements and fundamental parameters of Gonggong--Xiangliu system.
The radius of Gonggong, $R_{\rm G}$, was also estimated by \citet{Kiss+2019} using thermophysical modeling.
The spin period of Gonggong was obtained from the light curve analysis by \citet{Pal+2016}: $P_{\rm G, obs} = 22.4\ {\rm h}$ or $44.8\ {\rm h}$.
The radius and spin period of Xiangliu were not yet determined.
Assuming equal albedos with the primary, \citet{Kiss+2017} obtained the radius of Xiangliu as $R_{\rm X} = 120\ {\rm km}$.
\citet{Kiss+2019} also discussed the radius of Xiangliu, and used simple estimates on the eccentricity damping timescale to propose that $18\ {\rm km} < R_{\rm X} < 50\ {\rm km}$ might be suitable to explain the observed non-zero eccentricity of the system.
The estimated $R_{\rm X}$ corresponds to a geometric albedo of 1.0--0.2, which is higher than the estimated albedo for the primary \citep[$0.15 \pm 0.02$;][]{Kiss+2019}.

\begin{deluxetable*}{clccl}
\label{Table1}
\tablecaption{
Orbital elements and fundamental properties of Gonggong--Xiangliu \added{assumed in this study}.
Parameters for standard runs are indicated by {\bf boldface}.
\citet{Kiss+2019} reported two orbital solutions.
One is called prograde solution with the orbital period of $P_{\rm orb} = 25.22073 \pm 0.000357\ {\rm day}$, semimajor axis of $a_{\rm obs} = 24021 \pm 202\ {\rm km}$, and the eccentricity of $e_{\rm obs} = 0.2908 \pm 0.0070$.
The other is called retrograde solution with the orbital period of $P_{\rm orb} = 25.22385 \pm 0.000362\ {\rm day}$, semimajor axis of $a_{\rm obs} = 24274 \pm 193\ {\rm km}$, and the eccentricity of $e_{\rm obs} = 0.2828 \pm 0.0063$ \citep[see Table 2 of][]{Kiss+2019}.
Assuming that co-planar primary equator and satellite orbit and spherical shape of the primary, the effective diameter of Gonggong is $D_{\rm eff} \equiv 2 R_{\rm G} =  1224 \pm 55\ {\rm km}$ and $1238 \pm 50\ {\rm km}$ for the prograde solution for the spin period of $P_{\rm G, obs} = 44.8\ {\rm h}$ and $22.4\ {\rm h}$, respectively, and $D_{\rm eff} = 1227 \pm 56\ {\rm km}$ and $1241 \pm 50\ {\rm km}$ for the retrograde solutions with $P_{\rm G, obs} = 44.8\ {\rm h}$ and $22.4\ {\rm h}$ \citep[see Table 3 of][]{Kiss+2019}.
Then we set $a_{\rm obs} = 24000\ {\rm km}$, $e_{\rm obs} = 0.3$, and $R_{\rm G} = 600\ {\rm km}$ for simplicity.
}
\tablewidth{0pt}
\tablehead{
\colhead{Symbol} & \colhead{Property} & \colhead{Value} & \colhead{Unit} & \colhead{Reference}
}
\startdata
$a_{\rm obs}$ & Observed semimajor axis & $2.4 \times 10^{4}$ & ${\rm km}$ & \citet{Kiss+2019} \\
$e_{\rm obs}$ & Observed eccentricity & $0.3$ & --- & \citet{Kiss+2019} \\
$P_{\rm G, obs}$ & Observed spin period of Gonggong & {\boldmath $22.4$} or $44.8$ & ${\rm h}$ & \citet{Pal+2016} \\
$P_{\rm X, obs}$ & Observed spin period of Xiangliu & --- & ${\rm h}$ & --- \\
$M_{\rm tot}$ & Total mass of the system & $1.75 \times 10^{21}$ & ${\rm kg}$ & \citet{Kiss+2019} \\
$R_{\rm G}$ & Radius of Gonggong & $600$ & ${\rm km}$ & \citet{Kiss+2019} \\
$R_{\rm X}$ & Radius of Xiangliu & --- & ${\rm km}$ & --- \\
\enddata
\end{deluxetable*}

Multiple studies address the formation and tidal evolution of Pluto--Charon system \citep[e.g.,][]{Canup2005,Cheng+2014,Barr+2015,Desch2015,Kenyon+2019,Rozner+2020,Renaud+2021}.
The formation and tidal evolution of other systems have also been discussed in several studies \citep[e.g.,][]{Greenberg+2008,Ortiz+2012}.
\citet{Arakawa+2019} found that the general trend of the final eccentricity of satellites around 1000-km-sized TNOs strongly depends on their thermal history, based on higher-order tidal evolution calculations.

\citet{Renaud+2021} studied the tidal evolution of Pluto--Charon system and found that considering higher-order eccentricity terms was necessary to calculate the tidal evolution of eccentric system (considering terms beyond $e^{20}$ would be needed for systems with eccentricities greater than $0.8$).
\citet{Renaud+2021} also found that higher-order spin--orbit resonances exist not only for highly eccentric systems but also for systems with small eccentricity with $e \simeq 0.1$.
In most prior studies \citep[e.g.,][]{Cheng+2014,Arakawa+2019}, the strength of tidal dissipation was estimated by using the traditional constant phase lag model \citep{Kaula1964}.
However, the strength of tidal dissipation and the stability of the higher-order spin--orbit resonances strongly depend on the rheological properties \citep[e.g.,][]{Walterova+2020}.
Thus, we need to use a more realistic rheology model and simultaneously calculate the orbital and thermal evolution.

In this study, we assumed that Gonggong--Xiangliu system formed via a giant impact as an intact fragment and calculated the following secular tidal evolution of the system under the simplifying assumption of homogeneous bodies \citep[e.g.,][]{Ojakangas+1986,Shoji+2013}.
Our study is the first attempt to simulate the secular tidal evolution of Gonggong--Xiangliu system in detail.
We investigated the coupled thermal--orbital evolution using a realistic rheology model and included higher-order eccentricity terms.

The structure of this paper is as follows:
In Section \ref{sec.initial}, we discuss the initial condition of Gonggong--Xiangliu system formed via the giant impact.
In Section \ref{sec.tidal}, we briefly introduce the orbital and thermal evolution models.
In Section \ref{sec.typical}, we present the statistics of the final state of the satellite systems.
We show the distributions of the final spin/orbital properties of the system in Sections \ref{sec.distribution} and \ref{sec.discussion} and summarize them in Section \ref{sec.summary}.

\section{Initial condition of the system}
\label{sec.initial}

In \replaced{Section \ref{sec.initial}}{this study}, we assume that Gonggong--Xiangliu system was formed by a giant impact and discuss their orbital properties immediately after the impact. 
We re-evaluated the results of the giant impact simulations by \citet{Arakawa+2019} in Section \ref{sec.GI} and discuss the initial condition from the point of view of the conservation of the angular momentum in Section \ref{sec.iniGX}.

\subsection{Initial orbit and spin of satellites formed via giant impacts}
\label{sec.GI}

We define $\gamma_{\rm X, G} \equiv M_{\rm X} / M_{\rm G}$ as the secondary-to-primary mass ratio of the system formed after giant impacts: satellite systems with $10^{-3} \lesssim \gamma_{\rm X, G} \lesssim 10^{-1}$ can be directly formed via giant impacts as intact fragments \citep{Canup2005,Arakawa+2019}.
We also define $\gamma_{\rm i, t} \equiv M_{\rm imp} / M_{\rm tar}$ as the impactor-to-target mass ratio, where $M_{\rm tar}$ and $M_{\rm imp}$ are the masses of the target and impactor, respectively.
\citet{Arakawa+2019} performed giant impact simulations with three settings for the targets and impactors: $\gamma_{\rm i, t} = 0.5$ with an ice composition (Set A), $\gamma_{\rm i, t} = 0.2$ with an ice composition (Set B), and $\gamma_{\rm i, t} = 0.5$ with a basalt composition (Set C).
The total mass of the target and the impactor is $M_{\rm tar} + M_{\rm imp} = 6 \times 10^{21}\ {\rm kg}$ in the simulations.

\begin{figure*}
\centering
\includegraphics[width = 0.45\textwidth]{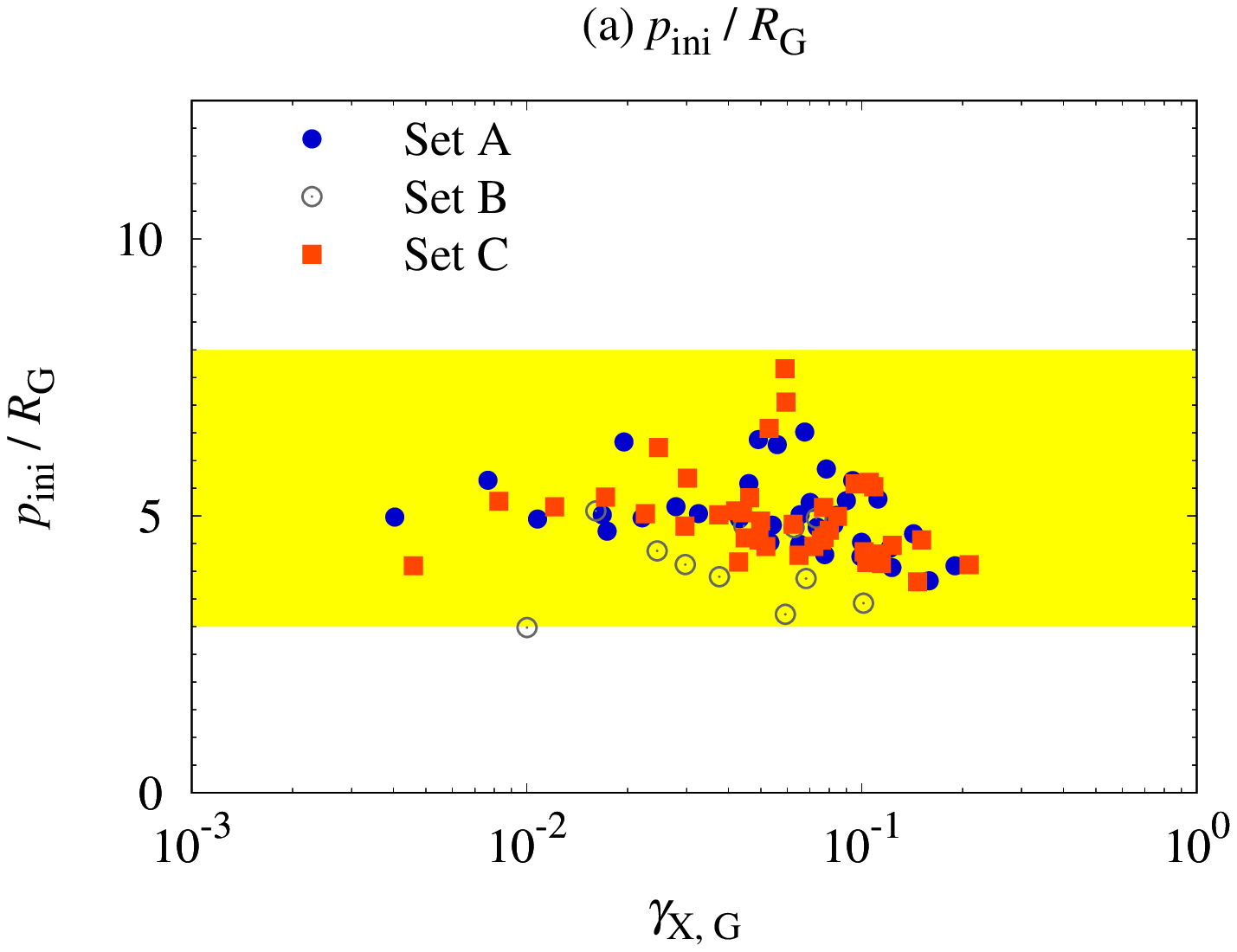}
\includegraphics[width = 0.45\textwidth]{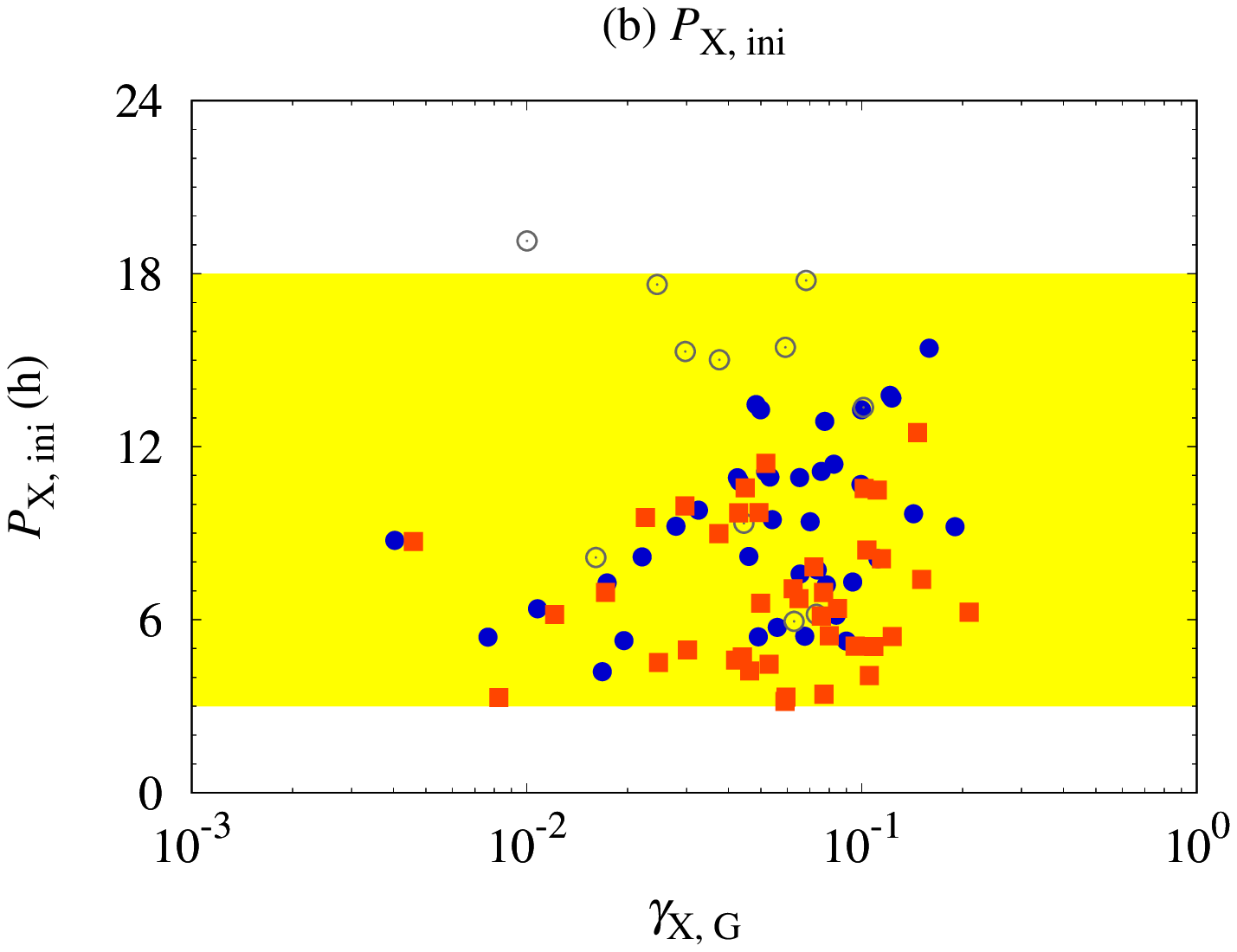}
\caption{
(a) Distribution of the initial semilatus rectum, $p_{\rm ini}$, and the secondary-to-primary mass ratio of the system, $\gamma_{\rm X, G}$, formed after giant impact simulations \citep{Arakawa+2019}.
The yellow region indicates the range of $p_{\rm ini} / R_{\rm G}$ (see relation (\ref{eqpGI})).
(b) Initial spin periods of secondaries, $P_{\rm X, ini}$, formed after the giant impact simulations.
The yellow region indicates the range of $P_{\rm X, ini}$ from the giant impact simulations.
}
\label{figPini}
\end{figure*}

Figure \ref{figPini}(a) shows the distribution of the semilatus rectum from the results of the giant impact simulations \citep{Arakawa+2019}.
The semilatus rectum, \replaced{$p$}{$p_{\rm orb}$}, and the periapsis distance, \replaced{$q$}{$q_{\rm orb}$}, are defined as follows \citep{Murray+1999}:
\begin{eqnarray}
p_{\rm orb} & \equiv & a {\left( 1 - e^{2} \right)} = q_{\rm orb} {\left( 1 + e \right)}, \\
q_{\rm orb} & \equiv & a {\left( 1 - e \right)},
\end{eqnarray}
where $a$ is the semimajor axis and $e$ is the eccentricity.
\citet{Arakawa+2019} performed more than 400 runs of giant impact simulations of 1000-km-sized TNOs, and revealed that the initial periapsis distance of satellites formed via giant impacts, $q_{\rm ini}$, is in the range of
\begin{equation}
3 \lesssim \frac{q_{\rm ini}}{R_{\rm G}} \lesssim 4,
\end{equation}
where $R_{\rm G}$ is the radius of the primary.
As the initial eccentricity of satellites formed via giant impacts, $e_{\rm ini}$, is widely distributed in the range of 0 to 1 \citep{Arakawa+2019}, the initial semilatus rectum, $p_{\rm ini}$, should be in the range of
\begin{equation}
\label{eqpGI}
3 \lesssim \frac{p_{\rm ini}}{R_{\rm G}} \lesssim 8.
\end{equation}
We confirmed that $p_{\rm ini} / R_{\rm G}$ is in the range of 3--8 for all outcomes resulting in satellite formation, as predicted in Equation (\ref{eqpGI}).

Figure \ref{figPini}(b) shows the distribution of the initial spin period of secondaries after giant impacts, $P_{\rm X, ini} = 2 \pi / \dot{\theta}_{\rm X, ini}$, where $\dot{\theta}_{\rm X, ini}$ is the initial spin angular velocity of secondaries.
Although it is difficult to analytically estimate $P_{\rm X, ini}$, the numerical results show that the initial spin period of the secondaries is in the range of $3\ {\rm h} \lesssim P_{\rm X, ini} \lesssim 18\ {\rm h}$.
The lower limit of $P_{\rm X, ini}$ is given by $P_{\rm X, ini} > 2 \pi / \Omega_{\rm cr}$, where
\begin{equation}
\Omega_{\rm cr} \equiv \sqrt{\frac{4 \pi {\mathcal G} \rho}{3}},
\end{equation}
is the critical spin angular velocity for rotational instability \citep[e.g.,][]{Kokubo+2010}; where ${\mathcal G}$ is the gravitational constant, and $\rho$ is the bulk density.

\subsection{Initial spins and orbit of Gonggong--Xiangliu system}
\label{sec.iniGX}

In Section \ref{sec.iniGX}, we discuss the initial spins and orbit of Gonggong--Xiangliu system from the perspective of the angular momentum conservation.
The total angular momentum of the system, $L_{\rm tot}$, is
\begin{equation}
L_{\rm tot} = I_{\rm G} \dot{\theta}_{\rm G} + I_{\rm X} \dot{\theta}_{\rm X} + \frac{M_{\rm G} M_{\rm X}}{M_{\rm tot}} \sqrt{{\mathcal G} M_{\rm tot} p_{\rm orb}},
\end{equation}
and $L_{\rm tot}$ should be constant over time during the tidal evolution of the system.
Here, $M_{\rm tot} \equiv M_{\rm G} + M_{\rm X}$ is the total mass of the satellite system, and $I_{i}$ and $\dot{\theta}_{i}$ are the moment of inertia and the spin angular velocity, respectively.
\replaced{Hereafter, the subscript $i$ denotes the primary ($i = {\rm p}$) or the secondary ($i = {\rm s}$).}{Hereafter, the subscript $i$ denotes the primary ($i = {\rm G}$) or the secondary ($i = {\rm X}$).}
For the case of undifferentiated bodies, the density distribution of the planetary interior is homogeneous; and the moment of inertia is approximately given by
\begin{equation}
I_{i} = \frac{2}{5} M_{i} {R_{i}}^{2}.
\end{equation}

\begin{figure*}
\centering
\includegraphics[width = 0.45\textwidth]{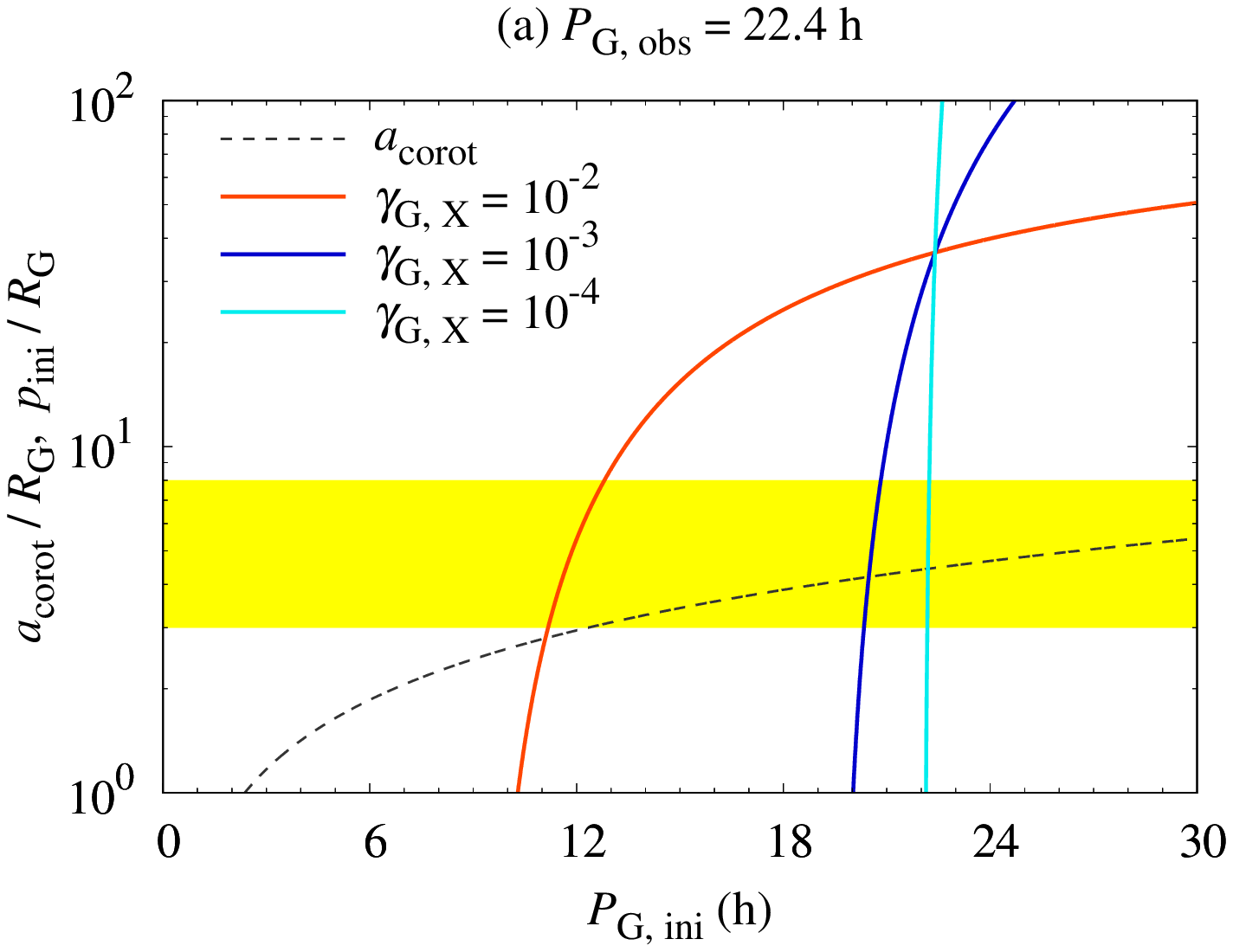}
\includegraphics[width = 0.45\textwidth]{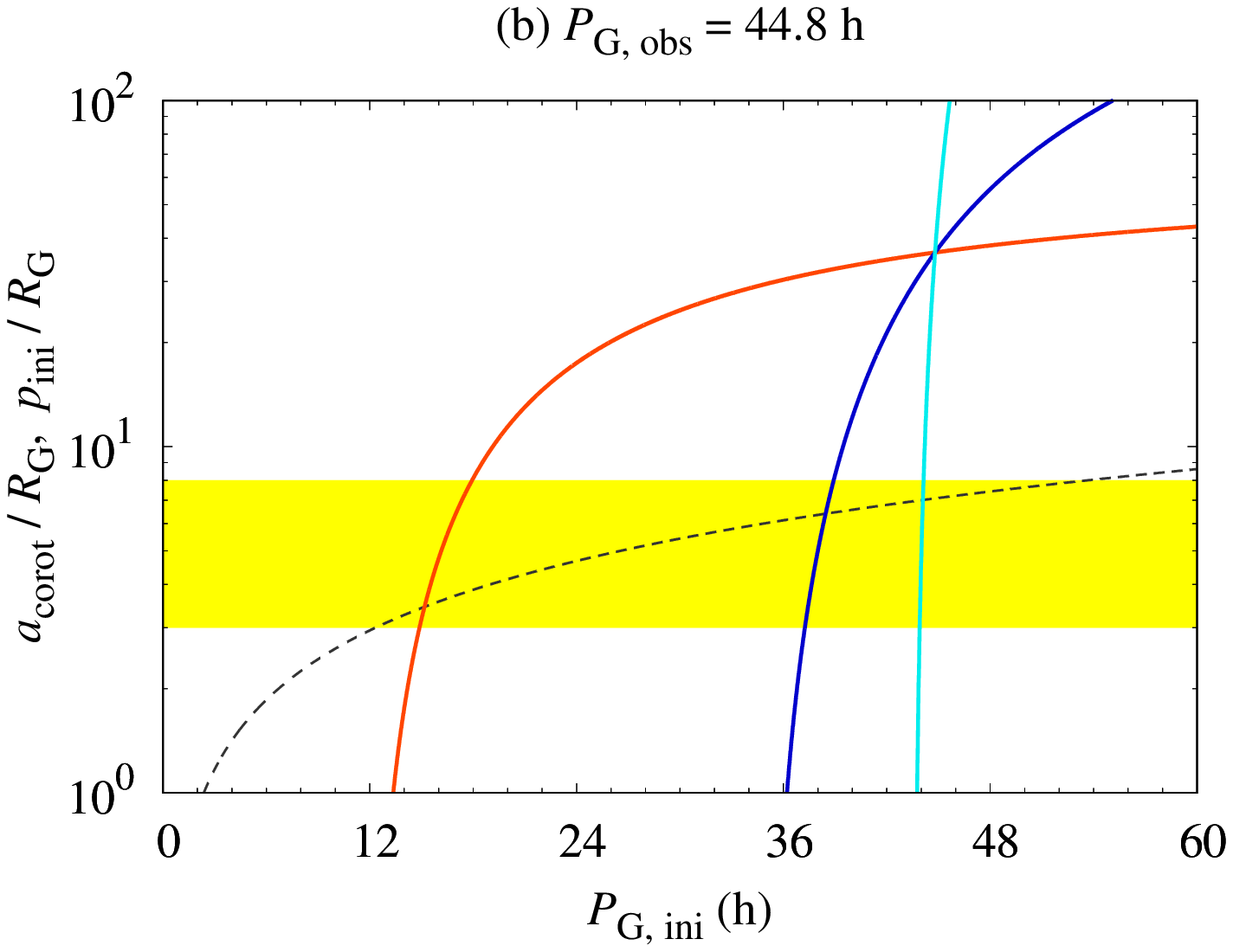}
\caption{
Initial semilatus rectum, $p_{\rm ini}$, as a function of the initial spin period of Gonggong, $P_{\rm G, ini}$, the observed spin period of Gonggong, $P_{\rm G, obs}$, and the secondary-to-primary mass ratio, $\gamma_{\rm X, G}$.
The dashed lines indicate the initial location of the corotation radius, $a_{\rm corot}$.
The yellow regions indicate the range of $p_{\rm ini} / R_{\rm G}$ inferred from the giant impact simulations (Figure \ref{figPini}(a)).
}
\label{figiniGX}
\end{figure*}

As the spin angular momentum of the secondary, Xiangliu, is negligibly small, we obtain the following equation:
\begin{eqnarray}
\label{eqLpLorb}
&& I_{\rm G} \dot{\theta}_{\rm G, ini} + \frac{M_{\rm G} M_{\rm X}}{M_{\rm tot}} \sqrt{{\mathcal G} M_{\rm tot} p_{\rm ini}} \nonumber \\ 
&& = I_{\rm G} \dot{\theta}_{\rm G, obs} + \frac{M_{\rm G} M_{\rm X}}{M_{\rm tot}} \sqrt{{\mathcal G} M_{\rm tot} p_{\rm obs}},
\end{eqnarray}
where $\dot{\theta}_{\rm G, ini}$ and $\dot{\theta}_{\rm G, obs}$ are the initial and observed (current) spin angular velocities of the primary, Gonggong; and $p_{\rm ini}$ and $p_{\rm obs}$ are the initial and observed semilatus recta.
Figure \ref{figiniGX} shows the initial semilatus rectum, $p_{\rm ini}$, as the function of the initial spin period of Gonggong, $P_{\rm G, ini}$, the observed spin period of Gonggong, $P_{\rm G, obs}$, and the secondary-to-primary mass ratio, $\gamma_{\rm X, G}$.

We note that the current spin period of Gonggong has not yet been uniquely determined; the observation by \citet{Pal+2016} suggested $P_{\rm G, obs} = 22.4\ {\rm h}$ or $44.8\ {\rm h}$, although the most likely light-curve solution is the double-peaked solution with a slight asymmetry.
This correspond to the spin period of $P_{\rm G, obs} = 44.8\ {\rm h}$.
In contrast, \citet{Kiss+2019} suggested that the most plausible solution for the system would be a single-peaked visible range light curve (i.e., $P_{\rm G, obs} = 22.4\ {\rm h}$) caused by surface features rather than by a distorted shape, and the primary's equator and the secondary's orbit are co-planar.

We also calculated the initial location of the corotation radius, $a_{\rm corot}$, which is given by
\begin{equation}
a_{\rm corot} = {\left( \frac{\sqrt{{\mathcal G} M_{\rm tot}}}{2 \pi} P_{\rm G, ini} \right)}^{2/3}.
\end{equation}
For an initial eccentricity of $e_{\rm ini} = 0$, the tidal evolution of the satellite results in inward migration when the initial semimajor axis is $a_{\rm ini} < a_{\rm corot}$.
Then, we can expect that the initial semilatus rectum is expected to be $p_{\rm ini} \gtrsim a_{\rm corot}$ to explain the current semimajor axis of the system, $a_{\rm obs}$, which is an order of magnitude larger than $a_{\rm corot}$.

Figure \ref{figiniGX} indicates that the tidal evolution of the satellite results in outward migration when the secondary-to-primary mass ratio is approximately $\gamma_{\rm X, G} = 10^{-2}$ or larger.
In contrast, the tidal evolution of the satellite would result in inward migration when $\gamma_{\rm X, G} \ll 10^{-2}$ and $p_{\rm ini} / R_{\rm G} \lesssim 4$ (i.e., $e_{\rm ini} \simeq 0$).
Therefore, for the case of $\gamma_{\rm X, G} \ll 10^{-2}$, the initial orbit of Gonggong--Xiangliu system formed via a giant impact must have a finite eccentricity; otherwise, the satellite would migrate inwards and become tidally disrupted.
The minimum value of $e_{\rm ini}$ (and $p_{\rm ini}$) is expected to depend on $\gamma_{\rm X, G}$ and $P_{\rm G, ini}$; we confirm this prediction in Sections \ref{sec.distribution} and \ref{sec.discussion}.

\subsection{Initial conditions for tidal evolution simulation}

We summarize the initial parameters for the tidal evolution calculations in Table \ref{tableini}.
In Section \ref{sec.distribution}, we parameterize the radius of Xiangliu, $R_{\rm X}$, the initial eccentricity, $e_{\rm ini}$, and the initial temperature, $T_{\rm ini}$.
For simplicity, we set $\rho_{\rm G} = \rho_{\rm X}$, where $\rho_{\rm G}$ and $\rho_{\rm X}$ are the bulk densities of Gonggong and Xiangliu.
We assume that the initial temperature of Xiangliu is equal to that of Gonggong (i.e., $T_{\rm G, ini} = T_{\rm X, ini} = T_{\rm ini}$) in Section \ref{sec.distribution}.
We also investigated the dependence of tidal evolution outcomes on the current spin period of Gonggong, $P_{\rm G, obs}$, the initial spin periods of Xiangliu, $P_{\rm X, ini}$, the different settings of the initial temperature of Xiangliu (we set $T_{\rm X, ini} = 120\ {\rm K}$ regardless of $T_{\rm G, ini}$), and the value of the reference viscosity (see Table \ref{tablemat}), $\eta_{\rm ref}$, in Section \ref{sec.discussion}.

\begin{deluxetable*}{clccl}
\label{tableini}
\tablecaption{
Initial condition for tidal evolution calculations.
Parameters for standard runs are indicated by {\bf boldface}.
}
\tablewidth{0pt}
\tablehead{
\colhead{Symbol} & \colhead{Parameter} & \colhead{Value} & \colhead{Unit} & \colhead{Comment}
}
\startdata
$q_{\rm ini}$ & Initial periapsis distance & $2.1 \times 10^{3}$ & {\rm km} & $3 \lesssim q_{\rm ini} / R_{\rm G} \lesssim 4$ \citep{Arakawa+2019}\\
$e_{\rm ini}$ & Initial eccentricity & $0.1$, $0.2$, ..., $0.8$ & --- & $0 < e_{\rm ini} < 1$ \citep{Arakawa+2019} \\
$P_{\rm G, ini}$ & Initial spin period of Gonggong & --- & {\rm h} & $P_{\rm G, ini} = 2 \pi / \dot{\theta}_{\rm G, ini}$ is given by Equation (\ref{eqLpLorb}) \\
$P_{\rm X, ini}$ & Initial spin period of Xiangliu & {\boldmath $12$} or $P_{\rm orb, ini}$ & {\rm h} & $3\ {\rm h} < P_{\rm X, ini} < 18\ {\rm h}$ (see Figure \ref{figPini}(b)) \\
$R_{\rm G}$ & Radius of Gonggong & $600$ & ${\rm km}$ & \citet{Kiss+2019} \\
$R_{\rm X}$ & Radius of Xiangliu & $20$, $40$, ..., $120$ & ${\rm km}$ & $R_{\rm X} \gtrsim 18\ {\rm km}$ \citep[see][]{Kiss+2019} \\
$T_{\rm G, ini}$ & Initial temperature of Gonggong & $120$, $140$, ..., $240$ & ${\rm K}$ & $T_{\rm G, ini}$ is below the melting point of ice \\
$T_{\rm X, ini}$ & Initial temperature of Xiangliu & {\boldmath $T_{\rm G, ini}$} or $120$ & ${\rm K}$ & See Section \ref{sec.coldstart} \\
\enddata
\end{deluxetable*}

\added{In this study, we assume that the primary's equator and the secondary's orbit are co-planar, and the orbit of the satellite system is prograde (i.e., the spin--orbit angle is close to zero).
This assumption is also consistent with the numerical simulations of \citet{Arakawa+2019}; however, their simulations do not consider the pre-impact spin of the target and impactor.
Observations by \citet{Kiss+2019} are consistent with both prograde and retrograde orbit.
If Gonggong--Xiangliu system is in retrograde orbit, the semimajor axis will decrease with increasing the time.
Thus the determination of the spin--orbit angle by future observations could provide critical constraints on the origin of the satellite system.
}

\added{We also note that the observed co-planar orbit is for the present-day system, and this does not indicate that the system was born in a co-planar orbit.
Although the giant impact simulations preferred a co-planar orbit, we should discuss the effect of non-zero initial inclination/obliquity on the tidal evolution in future studies.
Several studies \citep[e.g.,][]{Correia2020,Renaud+2021} pointed out that obliquity evolution can affect how bodies fall into and out of higher-order spin--orbit resonances.
}

\section{Tidal Evolution Model}
\label{sec.tidal}

In Section \ref{sec.tidal}, we briefly introduce the orbital evolution models (Section \ref{sec.orb}) and thermal evolution models (Section \ref{sec.thermal}).
We perform 4.5 Gyr tidal evolution calculations with various initial conditions.
We stopped numerical integration when the eccentricity reached $e = 0.9$ or the periapsis distance reached \replaced{$p = R_{\rm p} + R_{\rm s}$}{$q_{\rm orb} = R_{\rm G} + R_{\rm X}$}.

\subsection{Orbital evolution}
\label{sec.orb}

Tides are raised on both the primary and secondary.
The tidal lag caused by friction leads to angular momentum exchange, which also leads to spin and orbital evolution.
In this study, we use the tidal evolution equations following \citet{Boue+2019}.
The orbit-averaged variation of the spin rate of the primary and secondary, $\dot{\theta}_{\rm G}$ and $\dot{\theta}_{\rm X}$, semimajor axis $a$, and the eccentricity $e$ are given by
\begin{widetext}
\begin{eqnarray}
\frac{1}{n} \frac{{\rm d} \dot{\theta}_{\rm G}}{{\rm d} t} & = & - \frac{3 n}{2} \frac{M_{\rm G} M_{\rm X}}{M_{\rm tot}} \frac{M_{\rm X}}{M_{\rm G}} \frac{{R_{\rm G}}^{2}}{I_{\rm G}} {\left( \frac{R_{\rm G}}{a} \right)}^{3} \sum_{q} \mathcal{A}_{{\rm G}, q}, \label{eqSpinp}\\
\frac{1}{n} \frac{{\rm d} \dot{\theta}_{\rm X}}{{\rm d} t} & = & - \frac{3 n}{2} \frac{M_{\rm G} M_{\rm X}}{M_{\rm tot}} \frac{M_{\rm G}}{M_{\rm X}} \frac{{R_{\rm X}}^{2}}{I_{\rm X}} {\left( \frac{R_{\rm X}}{a} \right)}^{3} \sum_{q} \mathcal{A}_{{\rm X}, q}, \\
\frac{1}{a} \frac{{\rm d} a}{{\rm d} t} & = & \frac{1}{a} {\left[ {\left( \frac{{\rm d} a}{{\rm d} t} \right)}_{\rm G} + {\left( \frac{{\rm d} a}{{\rm d} t} \right)}_{\rm X} \right]} \label{eqa}, \\
\frac{1}{a} {\left( \frac{{\rm d} a}{{\rm d} t} \right)}_{\rm G} & = & 2 n \frac{M_{\rm X}}{M_{\rm G}} {\left( \frac{R_{\rm G}}{a} \right)}^{5} \sum_{q} {\left[ \frac{3 {( 2 + q )}}{4} \mathcal{A}_{{\rm G}, q} + \frac{q}{4} \mathcal{B}_{{\rm G}, q} \right]}, \\ 
\frac{1}{a} {\left( \frac{{\rm d} a}{{\rm d} t} \right)}_{\rm X} & = & 2 n \frac{M_{\rm G}}{M_{\rm X}} {\left( \frac{R_{\rm X}}{a} \right)}^{5} \sum_{q} {\left[ \frac{3 {( 2 + q )}}{4} \mathcal{A}_{{\rm X}, q} + \frac{q}{4} \mathcal{B}_{{\rm X}, q} \right]}, \\ 
\frac{1}{e} \frac{{\rm d} e}{{\rm d} t} & = & \frac{1}{e} {\left[ {\left( \frac{{\rm d} e}{{\rm d} t} \right)}_{\rm G} + {\left( \frac{{\rm d} e}{{\rm d} t} \right)}_{\rm X} \right]} \label{eqe}, \\
\frac{1}{e} {\left(  \frac{{\rm d} e}{{\rm d} t} \right)}_{\rm G} & = & \frac{n}{e^{2}} \frac{M_{\rm X}}{M_{\rm G}} {\left( \frac{R_{\rm G}}{a} \right)}^{5} \sum_{q} {\left[ \frac{3 {( 2 + q )} {( 1 - e^{2} )} - 6 \sqrt{1 - e^{2}}}{4} \mathcal{A}_{{\rm G}, q} + \frac{q}{4} {( 1 - e^{2} )} \mathcal{B}_{{\rm G}, q} \right]}, \\
\frac{1}{e} {\left(  \frac{{\rm d} e}{{\rm d} t} \right)}_{\rm X} & = & \frac{n}{e^{2}} \frac{M_{\rm G}}{M_{\rm X}} {\left( \frac{R_{\rm X}}{a} \right)}^{5} \sum_{q} {\left[ \frac{3 {( 2 + q )} {( 1 - e^{2} )} - 6 \sqrt{1 - e^{2}}}{4} \mathcal{A}_{{\rm X}, q} + \frac{q}{4} {( 1 - e^{2} )} \mathcal{B}_{{\rm X}, q} \right]} \label{eqEccs},
\end{eqnarray}
\end{widetext}
where $n = \sqrt{{\mathcal G} M_{\rm tot} / a^{3}}$ denotes the mean motion.
The initial orbital period, $P_{\rm orb, ini}$, is $P_{\rm orb, ini} = 2 \pi / \sqrt{{\mathcal G} M_{\rm tot} / {a_{\rm ini}}^{3}}$.
Here, we assume that the orbital inclinations on both the primary's and secondary's equators are negligibly small.
We note that \citet{Kiss+2019} suggested that the orbit of the secondary would be co-planar with the equator of the primary.
The coefficients $\mathcal{A}_{{\rm G}, q}$, $\mathcal{A}_{{\rm X}, q}$, $\mathcal{B}_{{\rm G}, q}$, $\mathcal{B}_{{\rm X}, q}$, are given by
\begin{eqnarray}
\mathcal{A}_{{\rm G}, q} & = & {\left[ G_{2, 0, q} {( e )} \right]}^{2}\ {\rm Im}{\left[ \tilde{k}_{2, {\rm G}} {\left( {( 2 + q )} n - 2 \dot{\theta}_{\rm G} \right)} \right]}, \\
\mathcal{A}_{{\rm X}, q} & = & {\left[ G_{2, 0, q} {( e )} \right]}^{2}\ {\rm Im}{\left[ \tilde{k}_{2, {\rm X}} {\left( {( 2 + q )} n - 2 \dot{\theta}_{\rm X} \right)} \right]}, \\
\mathcal{B}_{{\rm G}, q} & = & {\left[ G_{2, 1, q} {( e )} \right]}^{2}\ {\rm Im}{\left[ \tilde{k}_{2, {\rm G}} {\left( q n \right)} \right]}, \\
\mathcal{B}_{{\rm X}, q} & = & {\left[ G_{2, 1, q} {( e )} \right]}^{2}\ {\rm Im}{\left[ \tilde{k}_{2, {\rm X}} {\left( q n \right)} \right]}, 
\end{eqnarray}
where $G_{2, 0, q}$ and $G_{2, 1, q}$ are the eccentricity functions (see Appendix \ref{app.G2pq}), and $\tilde{k}_{2, {\rm G}}$ and $\tilde{k}_{2, {\rm X}}$ are the Love numbers of Gonggong and Xiangliu, which depend on the tidal frequency (see Appendix \ref{app.rheology}).
We consider the terms $G_{2, p, q}$ ($p = 0$ or $1$) when the following two conditions, ${| q |} \le 200$ and ${[ G_{2, p, q} ]}^{2} \ge 10^{-20}$, are satisfied.

In our numerical integration, we calculate $\dot{\theta}_{\rm G} / n$ and $\dot{\theta}_{\rm X} / n$ instead of $\dot{\theta}_{\rm G}$ and $\dot{\theta}_{\rm X}$ \citep[e.g.,][]{Cheng+2014}.
The relation between ${{\rm d} n} / {{\rm d} t}$ and ${{\rm d} a} / {{\rm d} t}$ is
\begin{equation}
\frac{1}{n} \frac{{\rm d} n}{{\rm d} t} = - \frac{3}{2} \frac{1}{a} \frac{{\rm d} a}{{\rm d} t},
\end{equation}
and we obtain the following equations:
\begin{eqnarray}
\frac{{\rm d}}{{\rm d} t} {\left( \frac{\dot{\theta}_{\rm G}}{n} \right)} & = & \frac{1}{n} \frac{{\rm d} \dot{\theta}_{\rm G}}{{\rm d} t} + \frac{3 \dot{\theta}_{\rm G}}{2 n} \frac{1}{a} \frac{{\rm d} a}{{\rm d} t}, \\
\frac{{\rm d}}{{\rm d} t} {\left( \frac{\dot{\theta}_{\rm X}}{n} \right)} & = & \frac{1}{n} \frac{{\rm d} \dot{\theta}_{\rm X}}{{\rm d} t} + \frac{3 \dot{\theta}_{\rm X}}{2 n} \frac{1}{a} \frac{{\rm d} a}{{\rm d} t}.
\end{eqnarray}

\subsection{Thermal evolution}
\label{sec.thermal}

In this study, we use the simplifying assumption of \replaced{isothermal}{undifferentiated} homogeneous bodies.
We do not consider the internal temperature structure but calculate the effective temperature of the \deleted{(isothermal)} interior for simplicity\footnote{
Strictly speaking, the simplifying assumption of \deleted{isothermal} homogeneous bodies is a good approximation only for the convective case \added{and for bodies without liquid layers \citep[e.g.,][]{Bolmont+2020}}.
However, in this study, we apply Equation (\ref{eqT}) for both convective and conductive cases.
We should consider the internal temperature structure in future studies, although it must take a high numerical cost.} \citep[e.g.,][]{Ojakangas+1986,Shoji+2013}.
Then, the temperature evolution of the body $i$ is given by the following equation:
\begin{equation}
\label{eqT}
\frac{{\rm d} T_{i}}{{\rm d} t} = \frac{Q_{{\rm tot}, i}}{M_{i} c_{i}}, 
\end{equation}
where $c_{i}$ is the specific heat capacity of the primary and secondary.
The total heat generation rate within bodies, $Q_{{\rm tot}, i}$, is given by
\begin{equation}
Q_{{\rm tot}, i} = Q_{{\rm tide}, i} + Q_{{\rm con}, i} + Q_{{\rm dec}, i}, 
\end{equation}
where $Q_{{\rm tide}, i}$, $Q_{{\rm con}, i}$, and $Q_{{\rm dec}, i}$ are the tidal heating, the conduction/convection cooling, and the decay heating terms, respectively.
\added{We note that the main heat source is not the tidal heating but the decay heating.
We discuss the balance between the conduction/convection cooling and the decay heating for the primary in Appendix \ref{app.coolingrate}.}

\subsubsection{Tidal heating}

The tidal heating rate, $Q_{{\rm tide}, i}$, is given by the sum of two terms:
\begin{eqnarray}
Q_{{\rm tide}, i} & = & Q_{{\rm spin}, i} + Q_{{\rm orb}, i}, \\
Q_{{\rm spin}, i} & = & - I_{i} \dot{\theta}_{i} \frac{{\rm d} \dot{\theta}_{i}}{{\rm d} t}, \\
Q_{{\rm orb}, i} & = & - \frac{G M_{\rm p} M_{\rm s}}{2 a^{2}} {\left( \frac{{\rm d} a}{{\rm d} t} \right)}_{i}. 
\end{eqnarray}
However, that the contribution of the tidal heating is negligibly small when we consider the tidal evolution of 100-km-sized satellites around 1000-km-sized dwarf planets \citep[e.g.,][]{Arakawa+2019}.

\subsubsection{Conduction/convection cooling}

The cooling term by conduction/convection, $Q_{{\rm con}, i}$, is given by
\begin{eqnarray}
Q_{{\rm con}, i}  & = & \min{\left( Q_{{\rm cond}, i}, Q_{{\rm conv}, i} \right)}, \\
Q_{{\rm cond}, i} & = & - 4 \pi {R_{i}}^2 \rho_{i} c_{i} \kappa_{i} \frac{T_{i} - T_{\rm surf}}{R_{i}}, \\
Q_{{\rm conv}, i} & = & {\rm Nu}_{i} Q_{{\rm cond}, i},
\end{eqnarray}
where $\kappa_{i}$ is the thermal diffusivity, which is given by
\begin{equation}
\kappa_{i} = \frac{k_{{\rm th}, i}}{\rho_{i} c_{i}},
\end{equation}
and $k_{{\rm th}, i}$ is the thermal conductivity.
\citet{Reese+1999b} proposed that the Nusselt number, ${\rm Nu}_{i}$, is given by the following scaling equation \citep[see also][]{Solomatov+2000}:
\begin{eqnarray}
{\rm Nu}_{i} & = & 2.51 {\Theta_{i}}^{- 1.2} {{\rm Ra}_{i}}^{0.2}, \\
\Theta_{i}   & = & \frac{E_{\rm a} {\left( T_{i} - T_{\rm surf} \right)}}{R_{\rm gas} {T_{i}}^{2}}, \\ 
{\rm Ra}_{i} & = & \frac{\alpha_{\rm exp} \rho_{i} g_{i} {\left( T_{i} - T_{\rm surf} \right)}}{\kappa_{i} \eta_{i}} {R_{i}}^{3}, 
\end{eqnarray}
where $\Theta_{i}$ and ${\rm Ra}_{i}$ are the Frank-Kamenetskii parameter and the Rayleigh number, respectively.
The surface gravity, $g_{i}$, is given by $g_{i} = {\mathcal G} M_{i} / {R_{i}}^{2}$, and $E_{\rm a}$ is the activation energy, $R_{\rm gas}$ is the gas constant, $\alpha_{\rm exp}$ is the thermal expansion coefficient, and $\eta_{i}$ is the viscosity.
The temperature dependence of the viscosity is given as follows \citep[e.g.,][]{Goldsby+2001}:
\begin{equation}
\eta_{i} = \eta_{\rm ref} \exp{\left[ \frac{E_{\rm a}}{R_{\rm gas} T_{\rm ref}} {\left( \frac{T_{\rm ref}}{T_{i}} - 1 \right)} \right]},
\end{equation}
where $\eta_{\rm ref}$ is the reference viscosity, and $T_{\rm ref}$ is the reference temperature (Table \ref{tablemat}).
We set the surface temperature $T_{\rm surf} = 40\ {\rm K}$ as assumed in previous studies on the thermal evolution of dwarf planets \citep[e.g.,][]{Robuchon+2011,Kamata+2019}.
The material parameters used in this study are summarized in Table \ref{tablemat}.

\begin{deluxetable*}{clccl}
\label{tablemat}
\tablecaption{
Material properties used in this study.
Parameters for standard runs are indicated by {\bf boldface}.
}
\tablewidth{0pt}
\tablehead{
\colhead{Symbol} & \colhead{Property} & \colhead{Value} & \colhead{Unit} & \colhead{Reference}
}
\startdata
$c$ & Specific heat capacity & $8.8 {\left( T / {\rm K} \right)}$ & ${\rm J}\ {\rm kg}^{-1}\ {\rm K}^{-1}$ & \citet{Hammond+2016} \\
$k_{\rm th}$ & Thermal conductivity & $0.48 + 488 {\left( T / {\rm K} \right)}^{-1}$ & ${\rm W}\ {\rm m}^{-1}\ {\rm K}^{-1}$ & \citet{Hammond+2016} \\
$\alpha_{\rm exp}$ & Thermal expansion coefficient & $1 \times 10^{-4}$ & ${\rm K}^{-1}$ & \citet{Kamata+2019} \\
$E_{\rm a}$ & Activation energy & $60$ & ${\rm kJ}\ {\rm mol}^{-1}$ & \citet{Kamata+2019} \\
$T_{\rm ref}$ & Reference temperature & $273$ & ${\rm K}$ & \citet{Kamata+2019} \\
$\eta_{\rm ref}$ & Reference viscosity at $T_{\rm ref}$ & {\boldmath $10^{14}$} or $10^{10}$ & ${\rm Pa}\ {\rm s}$ & See Section \ref{sec.softice} \\
$\mu$ & Shear modulus & $3.33$ & ${\rm GPa}$ & \citet{Robuchon+2011} \\
$\alpha$ & Andrade exponent & $0.33$ & --- & \citet{Rambaux+2010} \\
\enddata
\end{deluxetable*}

We stress that the temperature evolution of bodies significantly impacts their spin/orbital evolution.
As the viscosity is a function of the temperature, the complex Love number also depends on the temperature (see Appendix \ref{app.rheology}).
Then the complex Love number influences the spin/orbital evolution, which is calculated from Equations (\ref{eqSpinp})--(\ref{eqEccs}).

\added{Although it is beyond the scope of this study, ices in undifferentiated icy bodies would not be a pure water ice and the partial melting of ice mixtures may have an great impact on the effective viscosity \citep[e.g.,][]{Henning+2009}.
For the case of ice mixtures containing 1\% ammonia with respect to water, the effective viscosity is orders of magnitude lower than that of pure water ice when the temperature is above the ammonia--water eutectic temperature \citep[$176\ {\rm K}$; e.g.,][]{Arakawa+1994,Neveu+2017}.
We could mimic this effect by changing the reference viscosity, which is similar to the impact of crystal grain size on the reference viscosity (see Section \ref{sec.softice}).
}

\subsubsection{Decay heating}

For 100--1000-km-sized planetary bodies, the radioactive decay of long-lived isotopes is the principal heat source \citep[e.g.,][]{Robuchon+2011}.
We assume that the elemental abundances of long-lived isotopes of the rocky parts of TNOs are equal to those of carbonaceous chondrites \citep{Lodders2003,Robuchon+2011}.
We consider four species as heat sources, namely, $^{238}$U, $^{235}$U, $^{232}$Th, and $^{40}$K.
The half-life, $t_{{\rm HL}, j}$, and the initial heat generation rate per unit mass of rock, $H_{0, j}$, of the element $j$ are shown in Table \ref{tableRD}.

\begin{deluxetable}{ccc}
\label{tableRD}
\tablecaption{Radioactive species and decay data \citep{Robuchon+2011}.}
\tablewidth{0pt}
\tablehead{
\colhead{Element $j$} & \colhead{Half life $t_{{\rm HL}, j}$ (Myr)} & \colhead{$H_{0, j}$ (${\rm W}\ {\rm kg}^{-1}$ of rock)} 
}
\startdata
$^{238}$U  & $4468$   & $1.88 \times 10^{-12}$ \\
$^{235}$U  & $703.81$ & $3.07 \times 10^{-12}$ \\
$^{232}$Th & $14030$  & $1.02 \times 10^{-12}$ \\
$^{40}$K   & $1277$   & $2.15 \times 10^{-11}$ \\
\enddata
\end{deluxetable}

Then the decay heating term is given by
\begin{equation}
Q_{{\rm dec}, i} = f_{\rm rock} M_{i} \sum_{j} 2^{- t / t_{{\rm HL}, j}} H_{0, j}, \\
\end{equation}
where $f_{\rm rock}$ is the mass fraction of rock in bodies.
We set $f_{\rm rock} = 0.4$ in this study.
\citet{Kiss+2019} determined that the bulk density of Gonggong is $1.75 \times 10^{3}\ {\rm kg}\ {\rm m}^{-3}$; our assumption of $f_{\rm rock} = 0.4$ seems within an acceptable range if TNOs are made of a mixture of ices, rocks, and organic materials.
In future studies, we will investigate the effect of the difference in $f_{\rm rock}$ on the thermal and orbital evolution.

\section{Classification of tidal evolution pathways}
\label{sec.typical}

In Section \ref{sec.typical}, we present the statistics of the final state of the satellite systems.
We classify the outcome of tidal evolution into four types, namely, 1:1 spin--orbit resonance (Type A), higher-order spin--orbit resonance (Type B), non-resonance (Type C), and collision with the primary (Type Z).
\added{No systems reached a dual-synchronous state unlike Pluto--Charon system.
}
We introduce typical tidal evolution pathways of satellite systems in Appendix \ref{app.typical}.

\subsection{Summary of the final state}
\label{sec.summary-of-fs}

\begin{figure*}
\centering
\includegraphics[width = 0.45\textwidth]{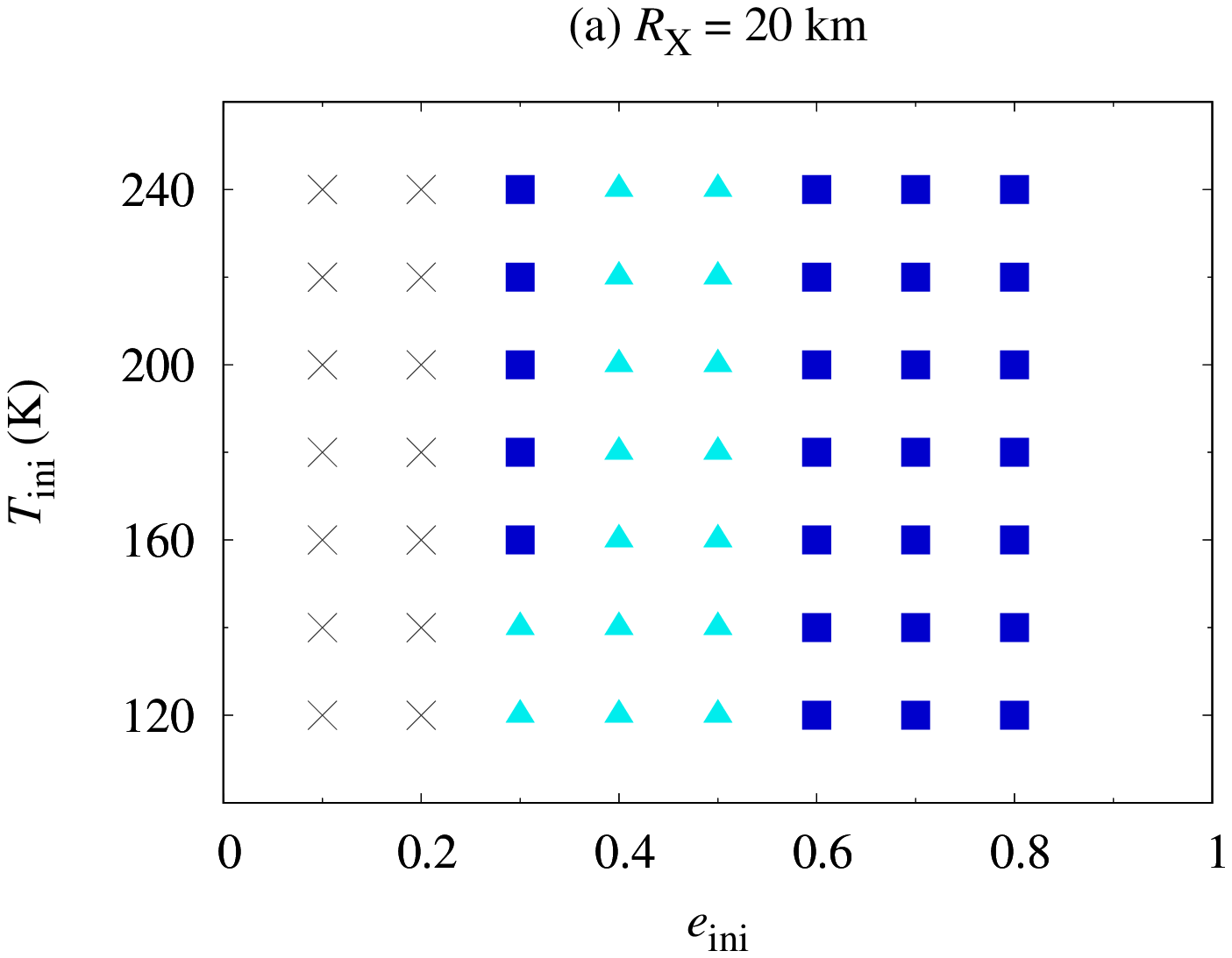}
\includegraphics[width = 0.45\textwidth]{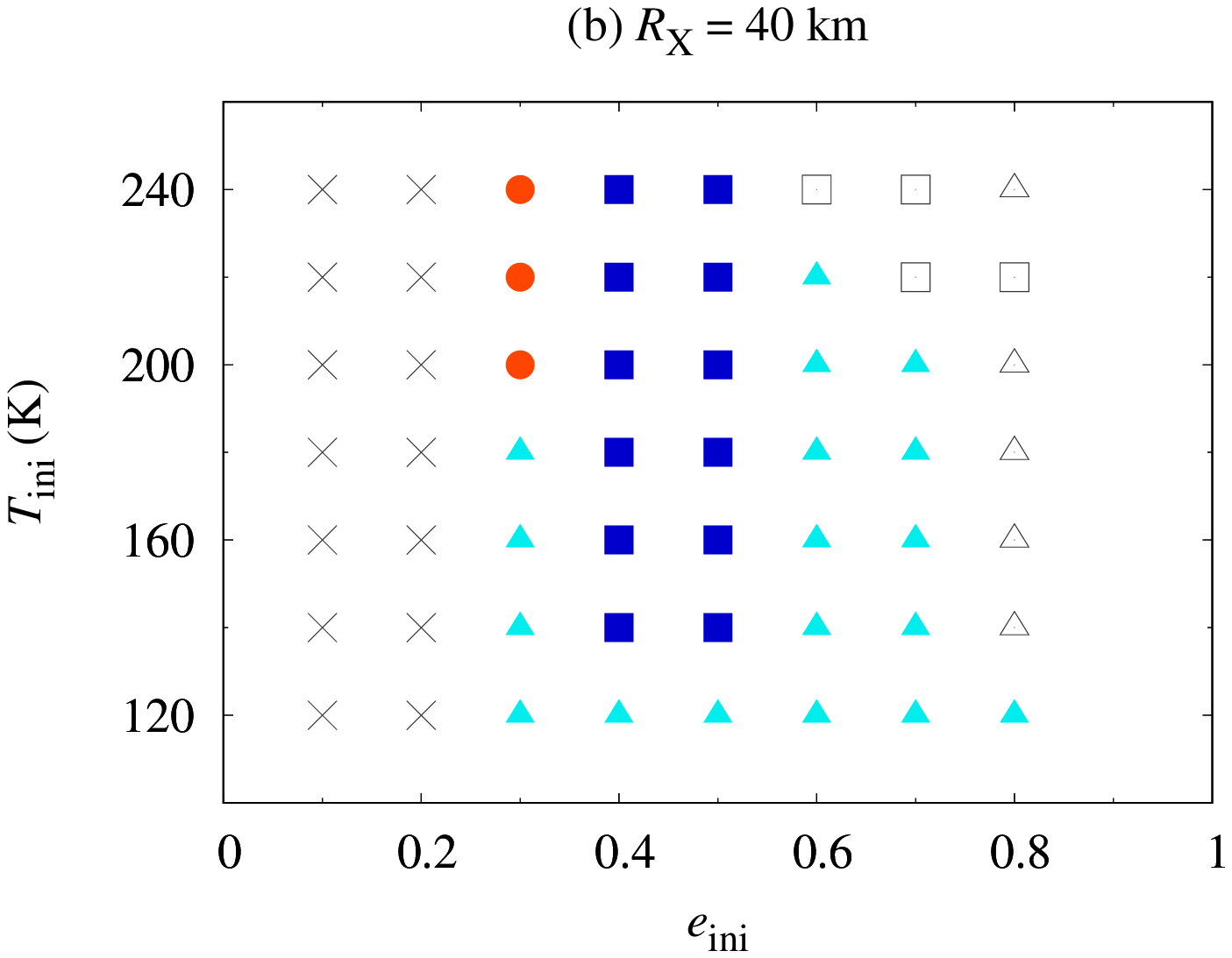}
\includegraphics[width = 0.45\textwidth]{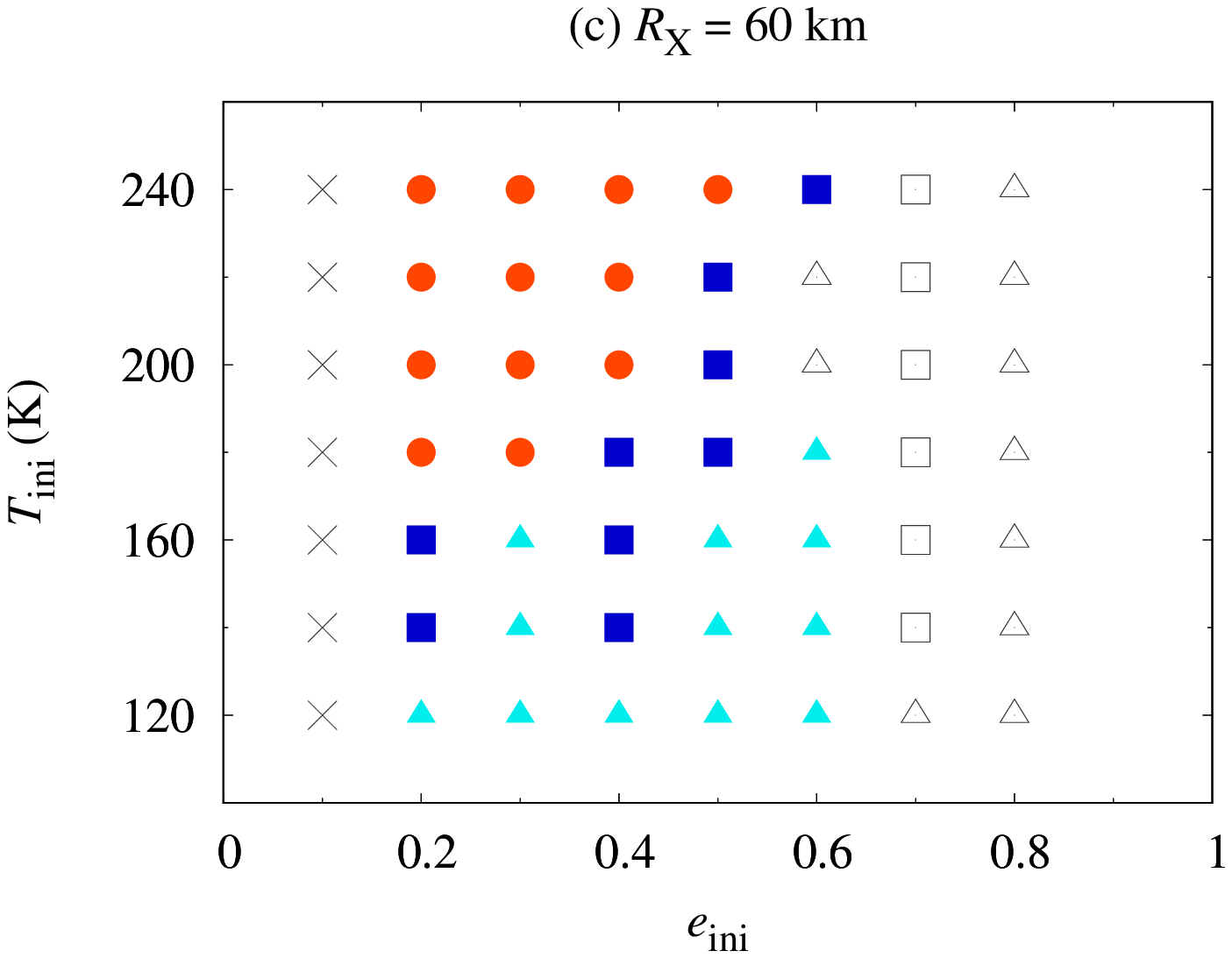}
\includegraphics[width = 0.45\textwidth]{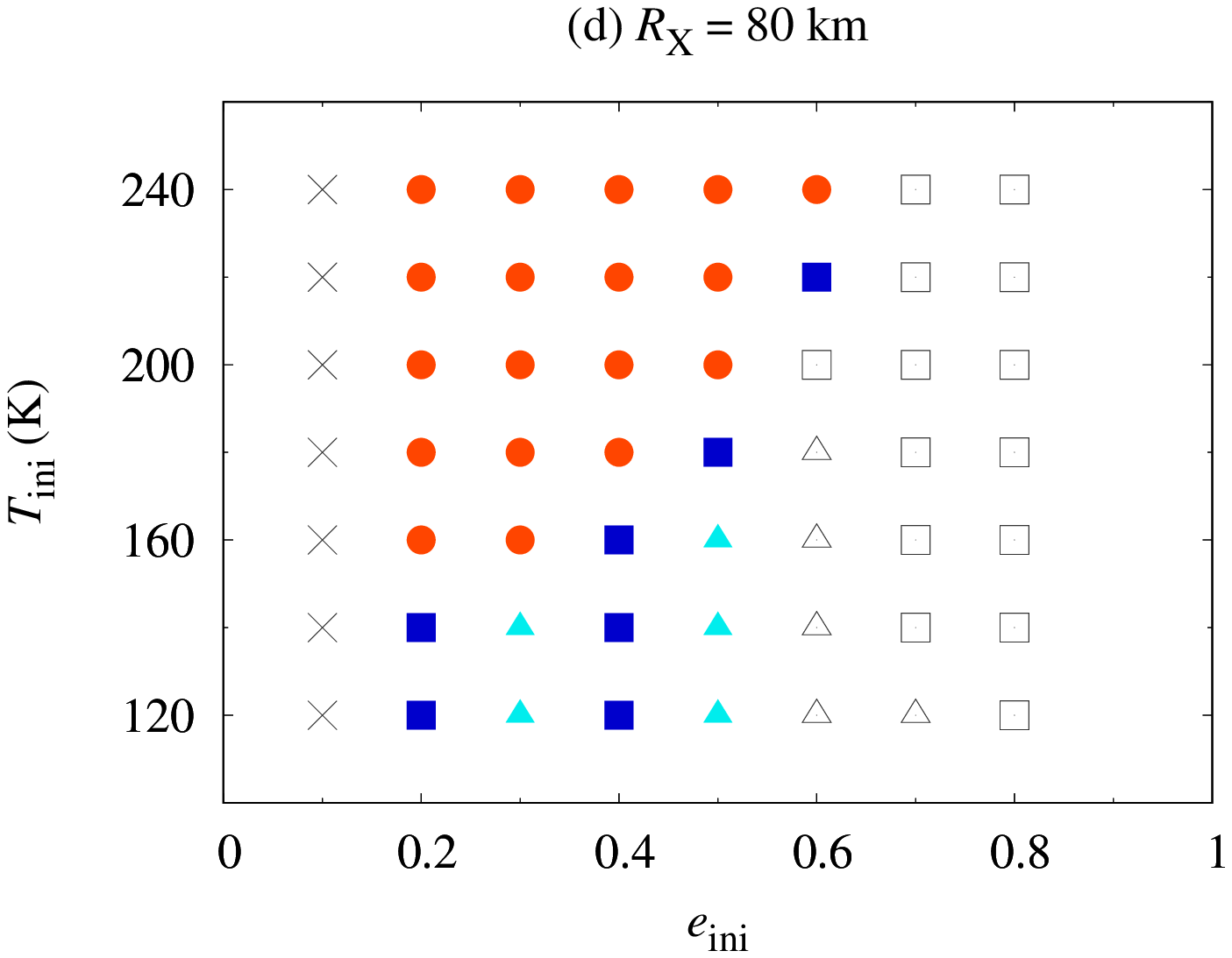}
\includegraphics[width = 0.45\textwidth]{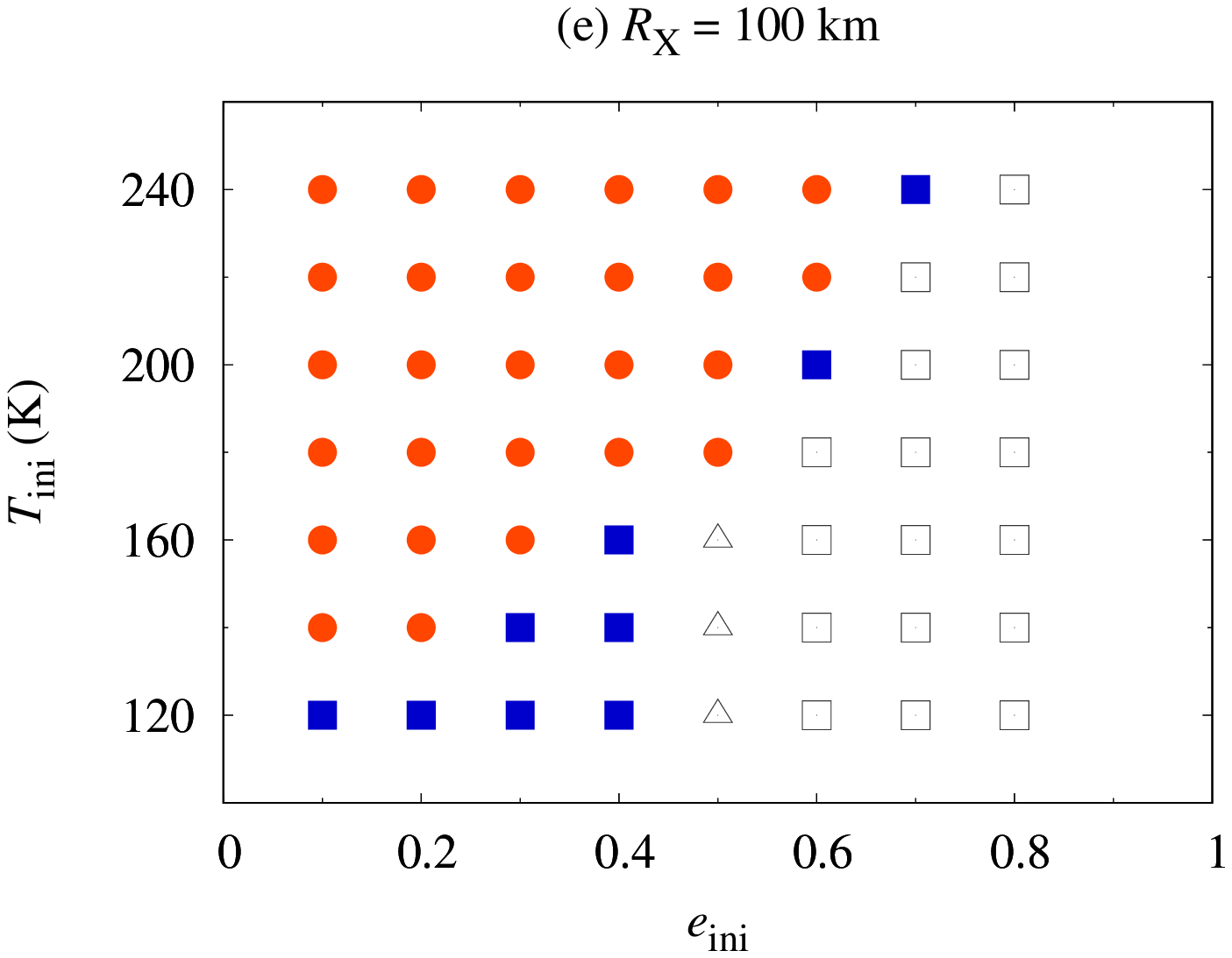}
\includegraphics[width = 0.45\textwidth]{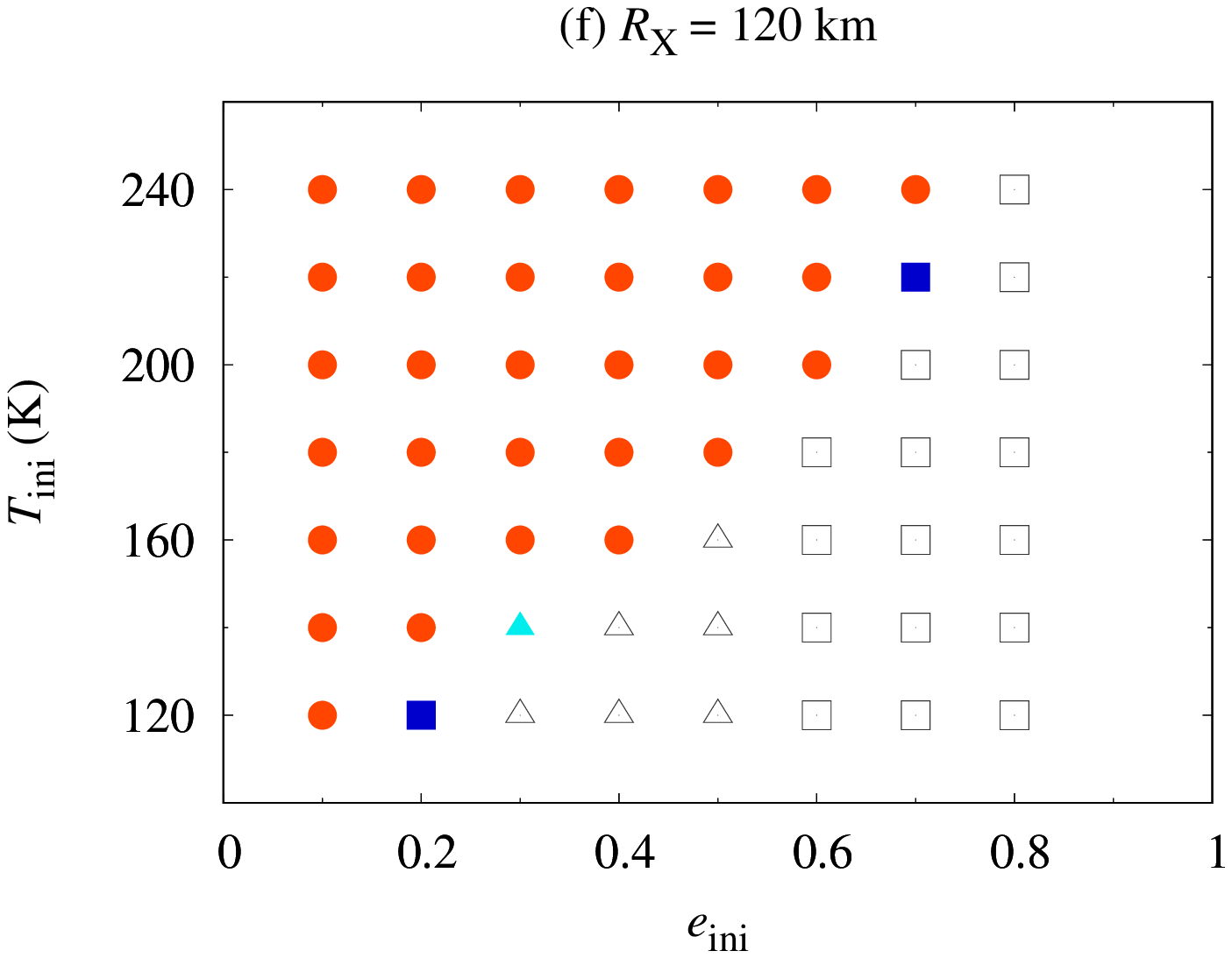}
\includegraphics[width = 0.45\textwidth]{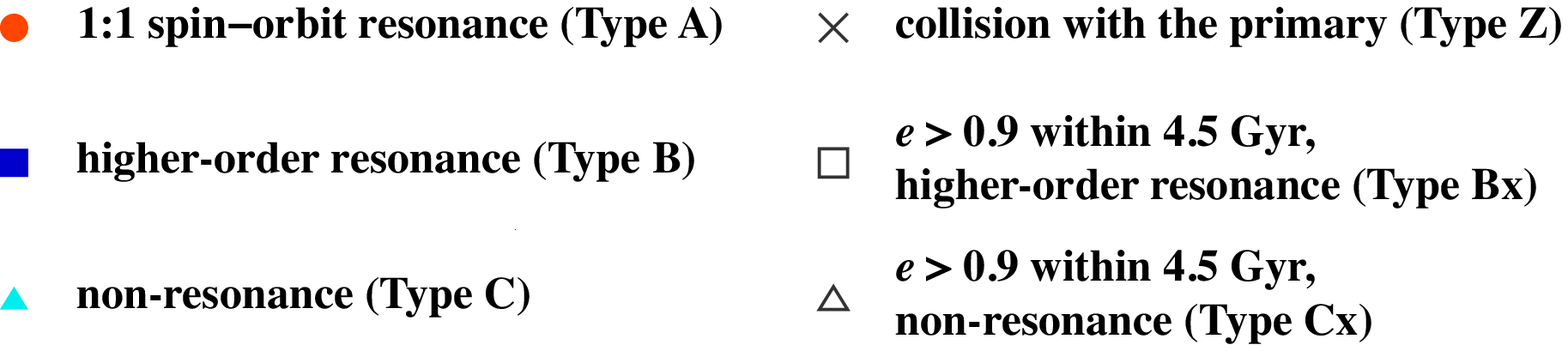}
\caption{
Summary of the final state of the standard runs of our simulation (i.e., $P_{\rm G, obs} = 22.4\ {\rm h}$, $P_{\rm X, ini} = 12\ {\rm h}$, $T_{\rm X, ini} = T_{\rm G, ini} = T_{\rm ini}$, and $\eta_{\rm ref} = 10^{14}\ {\rm Pa}\ {\rm s}$).
We changed the radius of Xiangliu, $R_{\rm X}$, the initial eccentricity, $e_{\rm ini}$, and the initial temperature, $T_{\rm ini}$, as parameters.
The different markers represent the different final states of the tidal evolution classified according to the spin state of Xiangliu, semimajor axis, and eccentricity.
}
\label{figType}
\end{figure*}

Figure \ref{figType} shows the summary of the final state for standard runs of our simulation.
In our standard runs, we set $P_{\rm G, obs} = 22.4\ {\rm h}$, $P_{\rm X, ini} = 12\ {\rm h}$, $T_{\rm X, ini} = T_{\rm G, ini} = T_{\rm ini}$, and $\eta_{\rm ref} = 10^{14}\ {\rm Pa}\ {\rm s}$.
We change the radius of Xiangliu, $R_{\rm X}$, the initial eccentricity, $e_{\rm ini}$, and the initial temperature, $T_{\rm ini}$.
The different markers represent the different final states of the satellite system classified according to the spin state of Xiangliu, semimajor axis, and eccentricity.

Figure \ref{figType}(a) shows the outcomes of the tidal evolution for the cases of $R_{\rm X} = 20\ {\rm km}$.
In this setting, Xiangliu migrates inward and finally collides with Gonggong when the initial eccentricity is $e_{\rm ini} \le 0.2$ (Type Z).
The critical value of $e_{\rm ini}$ required for the outward migration of Xiangliu is clearly dependent on the radius of Xiangliu; smaller satellites have a larger critical value of $e_{\rm ini}$.
This trend is consistent with our analytic arguments in Section \ref{sec.iniGX}.

Figure \ref{figType}(b) shows the outcomes of the tidal evolution for the cases of $R_{\rm X} = 40\ {\rm km}$.
In this setting, some runs with large $e_{\rm ini}$ and high $T_{\rm ini}$ result in extremely eccentric systems with $e > 0.9$ within 4.5 Gyrs (Types Bx/Cx).
The parameter space resulting in Types Bx/Cx is wider for larger values of $R_{\rm X}$.
In addition, the runs with $e_{\rm ini} \gtrsim 0.7$ tend to become extremely eccentric when $R_{\rm X} \ge 40\ {\rm km}$.

Figure \ref{figType}(f) shows the outcomes of the tidal evolution for the cases of $R_{\rm X} = 120\ {\rm km}$.
With this setting, runs with small $e_{\rm ini}$ and high $T_{\rm ini}$ result in Type A, i.e., Xiangliu is trapped in 1:1 spin--orbit resonance at $t = 4.5\ {\rm Gyr}$.
In general, runs with small $e_{\rm ini}$ and high $T_{\rm ini}$ tend to turn into Type A after tidal evolution unless their initial eccentricity is too small to satisfy the condition for the outward migration of Xiangliu (see Figures \ref{figType}(c) and \ref{figType}(d)).

We did not find any clear trends for the boundary between Type B and C.
In other words, it is difficult to predict whether the secondary is trapped in a higher-order spin--orbit resonance or not trapped in resonances.
The fractions of Type B and Type C were 65/336 = 19{\%} and 52/336 = 15{\%}, respectively, according to all our runs with the standard setting (see Table \ref{tableStat}).
We note that we did not consider the realistic probability distributions of $R_{\rm X}$, $e_{\rm ini}$, and $T_{\rm ini}$ for satellite systems around TNOs.

Table \ref{tableStat} shows the statistics of the final state for standard runs of our simulation.
Figure \ref{figType} shows that the fraction of each type is clearly dependent on the radius of the secondary, $R_{\rm X}$.
The fractions of Type A, Type Bx, and Type Cx increase with increasing $R_{\rm X}$.
In contrast, the fractions of Type B, Type C, and Type Z decrease with increasing $R_{\rm X}$.

\begin{deluxetable*}{cccccccc}
\label{tableStat}
\tablecaption{Statistics of the final state for standard runs of our simulation (see also Figure \ref{figType}).}
\tablewidth{0pt}
\tablehead{
\colhead{} & \colhead{(a) $R_{\rm X} = $ 20 km} & \colhead{(b) 40 km} & \colhead{(c) 60 km} & \colhead{(d) 80 km} & \colhead{(e) 100 km} & \colhead{(f) 120 km} & \colhead{Total}
}
\startdata
Type A  &  0 (0\%)  &  3 (5\%)  & 12 (21\%) & 18 (32\%) & 27 (48\%) & 31 (55\%) & {\bf 91 (27\%)} \\
Type B  & 26 (46\%) & 12 (21\%) &  9 (16\%) &  7 (13\%) &  9 (16\%) &  2 (4\%)  & {\bf 65 (19\%)} \\
Type C  & 16 (29\%) & 18 (32\%) & 12 (21\%) &  5 (9\%)  &  0 (0\%)  &  1 (2\%)  & {\bf 52 (15\%)} \\
Type Bx &  0 (0\%)  &  4 (7\%)  &  6 (11\%) & 14 (25\%) & 17 (30\%) & 16 (29\%) & {\bf 57 (17\%)} \\
Type Cx &  0 (0\%)  &  5 (9\%)  & 10 (18\%) &  5 (9\%)  &  3 (5\%)  &  6 (11\%) & {\bf 29 (9\%)}  \\
Type Z  & 14 (25\%) & 14 (25\%) &  7 (13\%) &  7 (13\%) &  0 (0\%)  &  0 (0\%)  & {\bf 42 (13\%)} \\
\enddata
\end{deluxetable*}

Our results also indicate that other satellites around 1000-km-sized TNOs may not be in 1:1 spin--orbit resonance.
In Haumea--Hi'iaka system, the spin period of the secondary, Hi'iaka, is approximately 120 times faster than its orbital period \citep{Hastings+2016}.
Although the fast spin of Hi'iaka could be explained by a scenario in which the current satellite system was formed via the catastrophic disruption of the first-generation moon and subsequent re-accretion of fragments \citep[e.g.,][]{Schlichting+2009,Cuk+2013}, we could also explain the fast spin of Hi'iaka without the disruption event.
We note, however, that the radius of Hi'iaka is 150 km \citep{Ragozzine+2009}, and the fraction of the non-resonant case (Type C) is small in our standard simulations for Gonggong--Xiangliu system with $R_{\rm X} \gtrsim 100\ {\rm km}$.
Moreover, the spin period of the primary, Haumea, is $P_{\rm G, obs} = 3.9\ {\rm h}$ \citep{Rabinowitz+2006}, which is an order of magnitude shorter than that of Gonggong.
In addition, the non-spherical and highly elongated shape of Haumea \citep[e.g.,][]{Ortiz+2017} may also play an important role in their tidal evolution.
\replaced{Thus, we need to apply the tidal evolution model for Haumea--Hi'iaka system for further future discussion.}{We will apply our tidal evolution model for Haumea--Hi'iaka system to unveil their spin--orbit evolution in future studies.}

\subsection{Condition for spin--orbit coupling}
\label{sec.analyticalspin}

Section \ref{sec.summary-of-fs} shows some cases in which the secondary is not in spin--orbit resonances after a 4.5 Gyr orbital evolution (i.e., Type C).
Here we discuss the condition for captured into spin--orbit resonances.

Appendix \ref{app.rheology} shows that the imaginary part of the Love number, ${\rm Im}{[ \tilde{k}_{2, {\rm X}} {( \omega_{{\rm X}, q} )} ]}$, takes the minimum/maximum when $\omega_{{\rm X}, q} \simeq \pm {\left( \mu_{\rm eff, X} \tau_{\rm A, X} \right)}^{-1}$, where $\omega_{{\rm X}, q} = {( 2 + q )} n - 2 \dot{\theta}_{\rm X}$ is the tidal frequency.
When the secondary is in the spin--orbit resonance of order $q$, the tidal frequency satisfies $- {\left( \mu_{\rm eff, X} \tau_{\rm A, X} \right)}^{-1} < \omega_{{\rm X}, q} < {\left( \mu_{\rm eff, X} \tau_{\rm A, X} \right)}^{-1}$, and the spin angular velocity of the secondary, $\dot{\theta}_{\rm X}$, is given by
\begin{equation}
\dot{\theta}_{\rm X} = \frac{2 + q}{2} n + \frac{x}{2 \mu_{\rm eff, X} \tau_{\rm A, X}},
\end{equation}
where ${| x |} < 1$ is the dimensionless parameter.
For ${| x |} \ll 1$, the imaginary part of the Love number is given by
\begin{equation}
{\rm Im}{\left[ \tilde{k}_{2, {\rm X}} {\left( {( 2 + q )} n - 2 \dot{\theta}_{\rm X} \right)} \right]} \simeq \frac{3}{2} x.
\end{equation}

Here, we assume that the following relations are satisfied for integers $q' \neq q$ (see Section \ref{sec.orb} for the definitions of $\mathcal{A}_{{\rm X}, q}$ and $\mathcal{B}_{{\rm X}, q}$):
\begin{equation}
\mathcal{A}_{{\rm X}, q} \gg \mathcal{A}_{{\rm X}, q'}, \mathcal{B}_{{\rm X}, q'}, \mathcal{B}_{{\rm X}, q}.
\end{equation}
Under this assumption, ${{\rm d} \dot{\theta}_{\rm X}}/{{\rm d} t}$, ${( {{\rm d} a}/{{\rm d} t} )}_{\rm X}$, and ${{\rm d}} {( {\dot{\theta}_{\rm X}} / {n} )} / {{\rm d} t}$ near the spin--orbit resonance are approximately given by
\begin{widetext}
\begin{eqnarray}
\frac{1}{n} \frac{{\rm d} \dot{\theta}_{\rm X}}{{\rm d} t} & \simeq & - \frac{9}{4} \frac{M_{\rm G} M_{\rm X}}{M_{\rm tot}} \frac{M_{\rm G}}{M_{\rm X}} \frac{{R_{\rm X}}^{2}}{I_{\rm X}} {\left( \frac{R_{\rm X}}{a} \right)}^{3} {\left[ G_{2, 0, q} {( e )} \right]}^{2} x n \simeq \frac{45}{8} \frac{M_{\rm G}}{M_{\rm X}} {\left( \frac{R_{\rm X}}{a} \right)}^{3} {\left[ G_{2, 0, q} {( e )} \right]}^{2} x n, \\
\frac{1}{a} {\left( \frac{{\rm d} a}{{\rm d} t} \right)}_{\rm X} & \simeq & \frac{9 {( 2 + q )}}{4} \frac{M_{\rm G}}{M_{\rm X}} {\left( \frac{R_{\rm X}}{a} \right)}^{5} {\left[ G_{2, 0, q} {( e )} \right]}^{2} x n, \\
\frac{{\rm d}}{{\rm d} t} {\left( \frac{\dot{\theta}_{\rm X}}{n} \right)} & \simeq & - \frac{45}{8} {\left[ {\left( \frac{a}{R_{\rm X}} \right)}^{2} - \frac{6}{5} {\left( \frac{2 + q}{2} \right)}^{2} \right]} \frac{M_{\rm G}}{M_{\rm X}} {\left( \frac{R_{\rm X}}{a} \right)}^{5} {\left[ G_{2, 0, q} {( e )} \right]}^{2} x n + \frac{3 {( 2 + q )}}{4} \frac{1}{a} {\left( \frac{{\rm d} a}{{\rm d} t} \right)}_{\rm G}.
\end{eqnarray}
\end{widetext}
Assuming that ${( {{\rm d} a} / {{\rm d} t} )}_{\rm G} > 0$, the required conditions for the stable equilibrium around $\dot{\theta}_{\rm X} \simeq {(2 + q)} n / 2 + x / {(\mu_{\rm eff, X} \tau_{\rm A, X})}$ with ${| x |} \ll 1$ are 
\begin{eqnarray}
{\left( \frac{a}{R_{\rm X}} \right)}^{2} & \gg & \frac{6}{5} {\left( \frac{2 + q}{2} \right)}^{2}, \\
{\left[ G_{2, 0, q} {( e )} \right]}^{2} & \gg & \frac{2 {(2 + q)}}{15 n} {\left( \frac{a}{R_{\rm G}} \right)}^{3} \frac{1}{a} {\left( \frac{{\rm d} a}{{\rm d} t} \right)}_{\rm G}.
\end{eqnarray}
On the other hand, for the case of ${\left( {{\rm d} a} / {{\rm d} t} \right)}_{\rm G} < 0$, the required conditions for a stable equilibrium are
\begin{eqnarray}
{\left( \frac{a}{R_{\rm X}} \right)}^{2} & \ll & \frac{6}{5} {\left( \frac{2 + q}{2} \right)}^{2}, \\
{\left[ G_{2, 0, q} {( e )} \right]}^{2} & \gg & - \frac{4}{9 {(2 + q)} n} \frac{M_{\rm X}}{M_{\rm G}} {\left( \frac{a}{R_{\rm X}} \right)}^{5} \frac{1}{a} {\left( \frac{{\rm d} a}{{\rm d} t} \right)}_{\rm G}.
\end{eqnarray}
Except for the case of $q = 0$, ${\left[ G_{2, 0, q} {( e )} \right]}^{2} \to 0$ when $e \to 0$ (see Appendix \ref{app.G2pq}).
Thus, a finite eccentricity is needed to be captured into higher-order spin--orbit resonances with $q \ne 0$.


\section{Distributions of final spin/orbital properties of the system}
\label{sec.distribution}

In Section \ref{sec.distribution}, we present the distributions of the final spin/orbital properties of the system.

\subsection{Semimajor axis and eccentricity}

\begin{figure*}
\centering
\includegraphics[width = 0.45\textwidth]{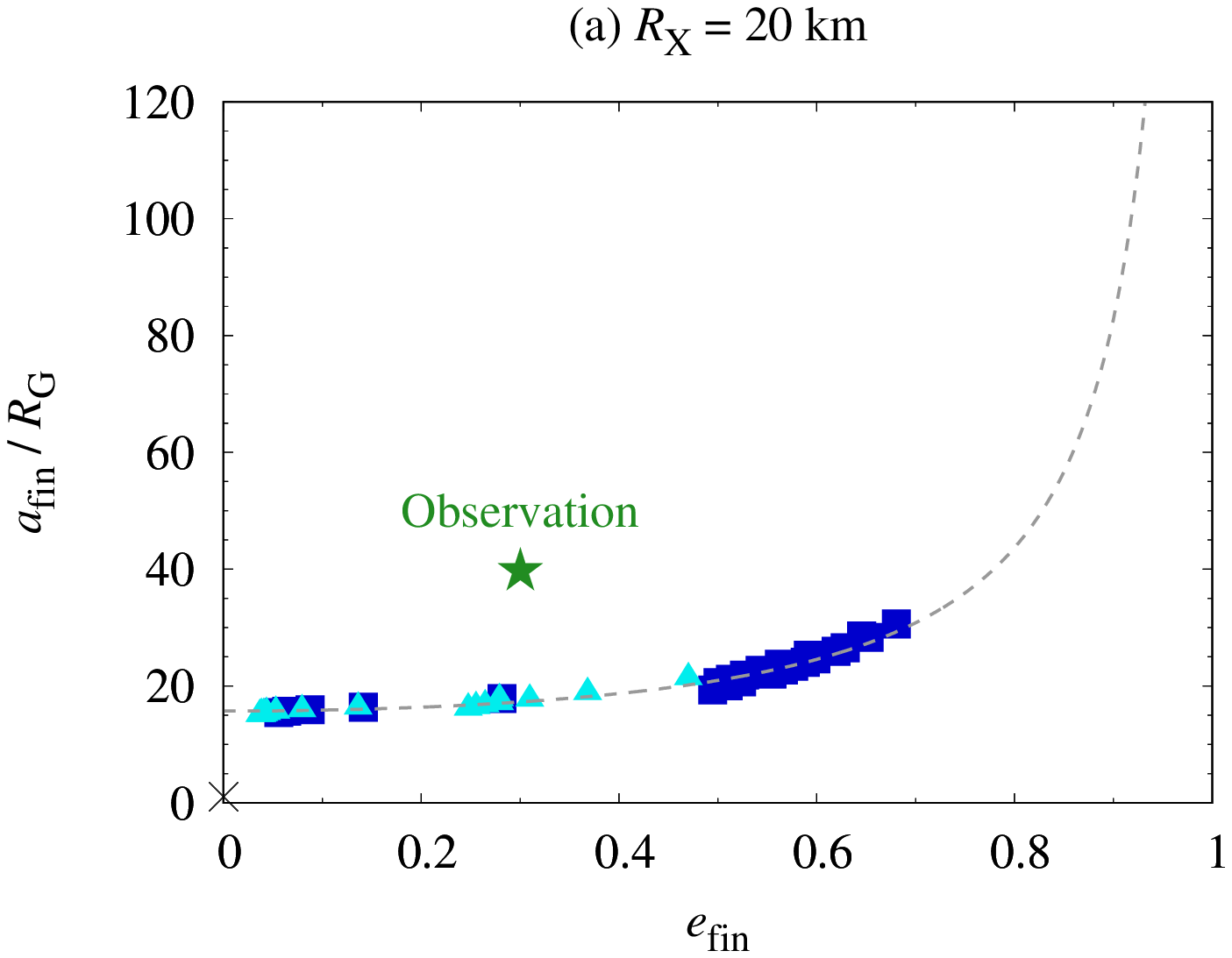}
\includegraphics[width = 0.45\textwidth]{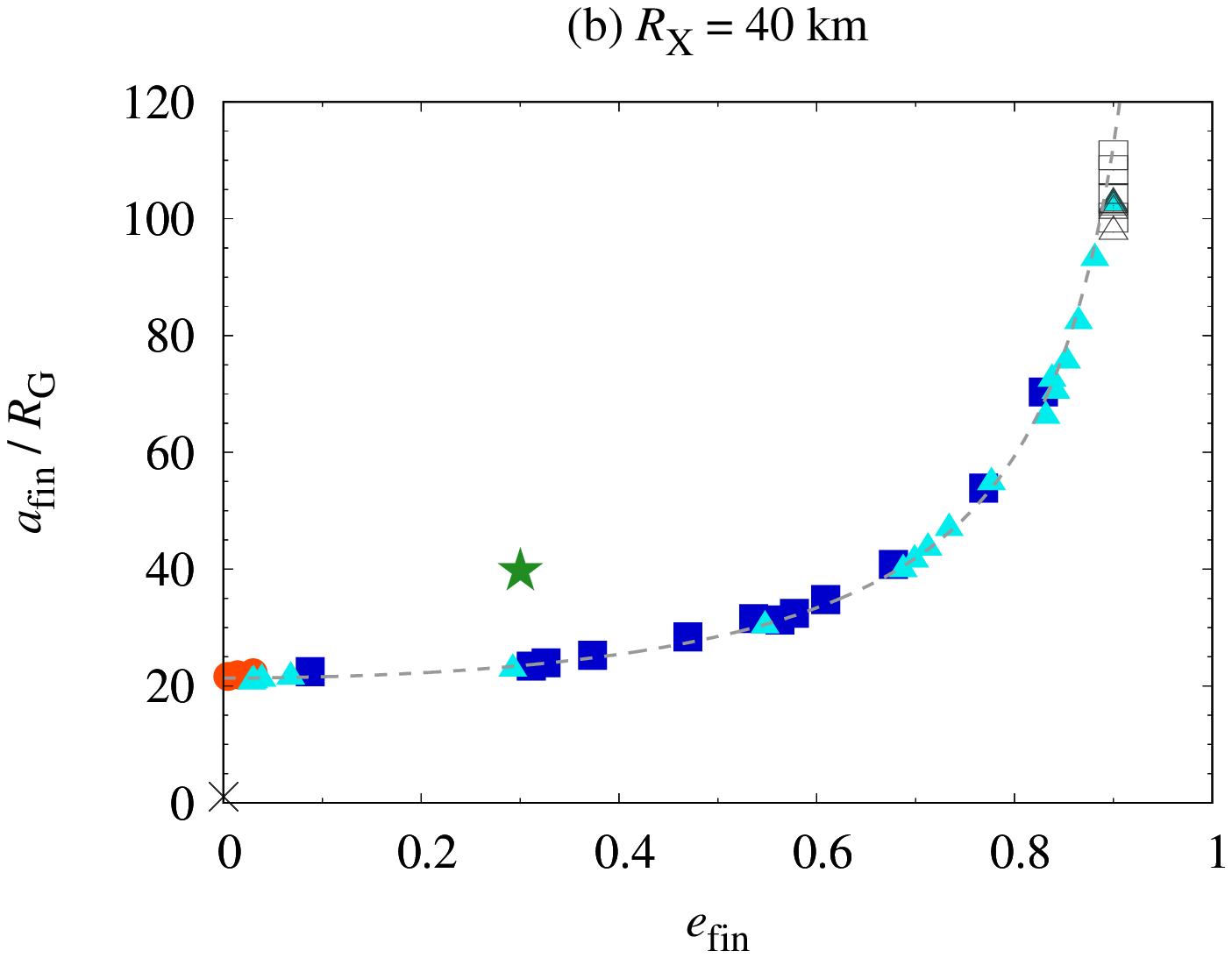}
\includegraphics[width = 0.45\textwidth]{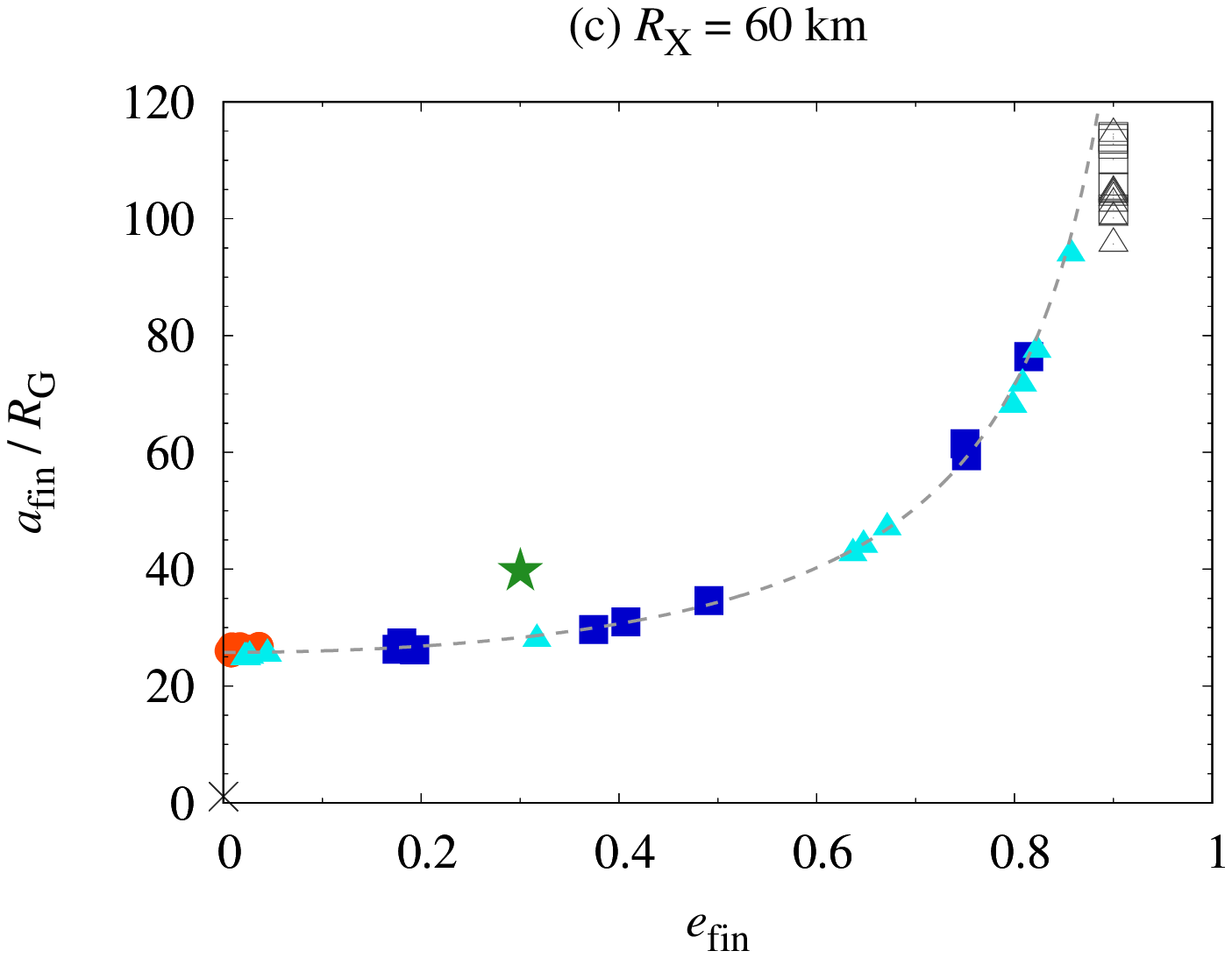}
\includegraphics[width = 0.45\textwidth]{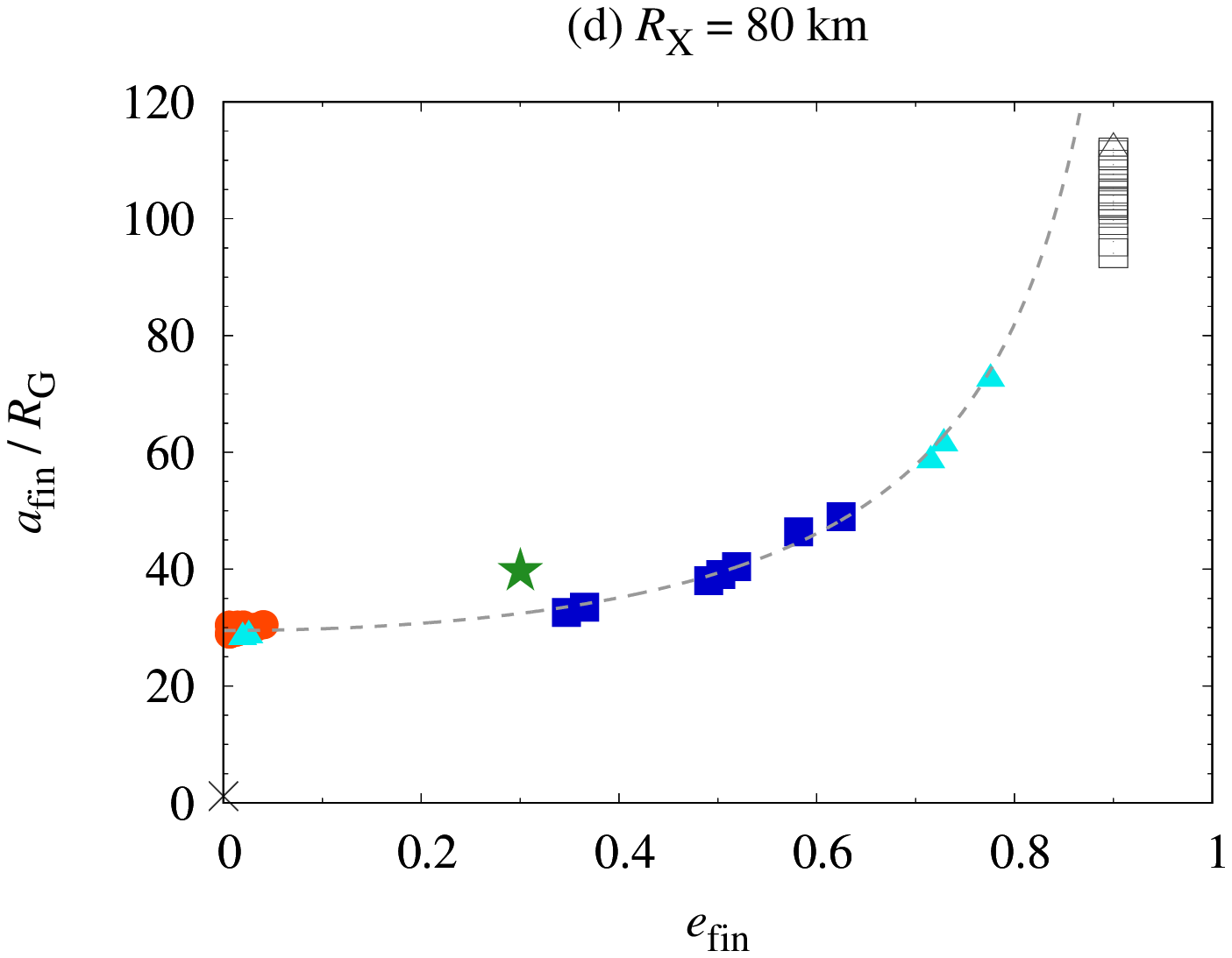}
\includegraphics[width = 0.45\textwidth]{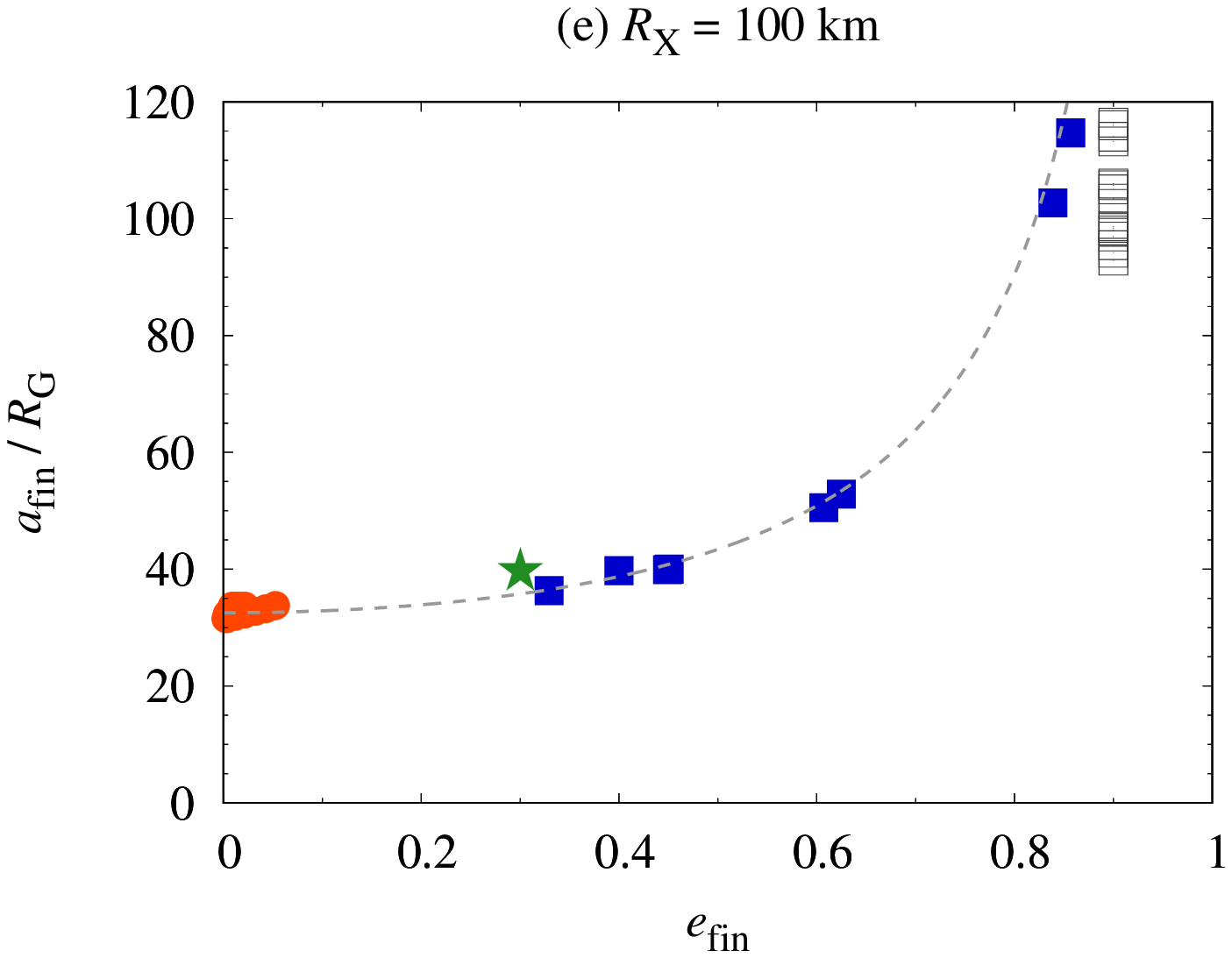}
\includegraphics[width = 0.45\textwidth]{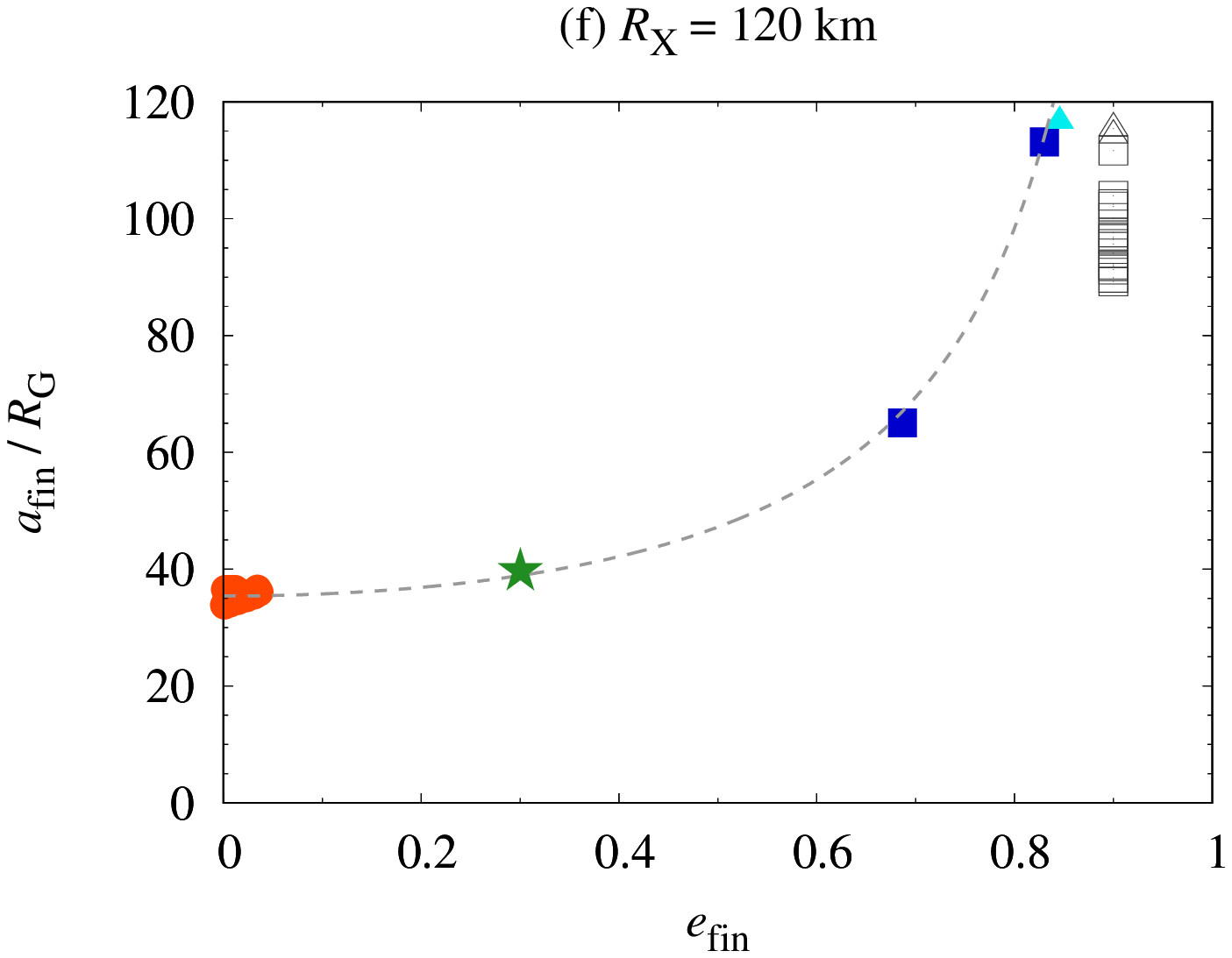}
\caption{
Distribution of the final eccentricity and semimajor axis, $e_{\rm fin}$ and $a_{\rm fin}$, for standard runs of our simulation.
The different markers represent the different final states of the tidal evolution classified according to the spin state of Xiangliu and the final semimajor axis and eccentricity.
For the cases of Type Bx/Cx/Z, we stopped numerical integrations when the eccentricity reached $e = 0.9$ or when the periapsis distance reached $q_{\rm orb} = R_{\rm G} + R_{\rm X}$.
}
\label{figa_ecc}
\end{figure*}

Figure \ref{figa_ecc} shows the distribution of the final eccentricity and semimajor axis, $e_{\rm fin}$ and $a_{\rm fin}$.
The observed values for Gonggong--Xiangliu system are $e_{\rm fin} = 0.3$ and $a_{\rm fin} / R_{\rm G} = 24000\ {\rm km} / 600\ {\rm km} = 40$; the green stars indicate the observed values.

We found that the final eccentricity is $e_{\rm fin} \lesssim 0.05$ for Type A.
Xiangliu has a moderate eccentricity of $e_{\rm obs} = 0.3$ \citep{Kiss+2019} and, therefore, it would not be in 1:1 spin--orbit resonance.
The final eccentricity of satellite systems classified as Type B and Type C is in the wide range \added{($0 < e_{\rm fin} < 0.9$)}.
We note, however, that the fraction of the system whose $e_{\rm fin}$ is comparable to $e_{\rm obs}$ strongly depends on the radius of Xiangliu.
The fraction of the system whose final eccentricity is in the range of $0.2 \le e_{\rm fin} \le 0.5$ is as follows: 10/56 = 18{\%} for the case of $R_{\rm X} = 20\ {\rm km}$, 5/56 = 8.9{\%} for $40\ {\rm km}$, 4/56 = 7.1{\%} for $60\ {\rm km}$, 3/56 = 5.4{\%} for $80\ {\rm km}$, 4/56 = 7.1{\%} for $100\ {\rm km}$, and 0/56 = 0{\%} for $120\ {\rm km}$.
Thus, the radius of Xiangliu is likely not larger than $100\ {\rm km}$.

We also found a clear relation between $a_{\rm fin}$ and $e_{\rm fin}$.
The dashed lines in Figure \ref{figa_ecc} are given by
\begin{equation}
\label{eqpconst}
p_{\rm fin} = a_{\rm fin} {\left( 1 - {e_{\rm fin}}^{2} \right)} = {\rm const.}
\end{equation}
for each $R_{\rm s}$. 
We explain the reason why $a_{\rm fin}$ and $e_{\rm fin}$ are given by Equation (\ref{eqpconst}) in Section \ref{sec.pfin}.
Figure \ref{figa_ecc} shows that the radius of Xiangliu might be close to $100\ {\rm km}$ from the point of view of $p_{\rm fin}$.
We note that we should consider the dependence of $p_{\rm fin}$ on many properties of the system, such as the viscoelastic properties of icy bodies and their thermal histories (see Sections \ref{sec.coldstart} and \ref{sec.softice}).

Our results suggest that not only Gonggong--Xiangliu but also Quaoar--Weywot system would not be in 1:1 spin--orbit resonance.
\citet{Fraser+2013} reported that the eccentricity of Quaoar--Weywot system is $e = 0.13$--$0.16$.
This eccentricity indicates that the system might be classified as Type B or Type C, as systems classified as Type A have a small eccentricity of $e \lesssim 0.05$ in general.

\subsection{Analytic solution of the semilatus rectum}
\label{sec.pfin}

Figure \ref{figa_ecc} shows that the final semilatus rectum, $p_{\rm fin}$, hardly depends on the final eccentricity, $e_{\rm fin}$.
In Section \ref{sec.pfin}, we derive an analytic solution of the evolution of the semilatus rectum.

The time derivative of $p_{\rm orb}$ is given by the following equation:
{\footnotesize
\begin{eqnarray}
\frac{1}{p_{\rm orb}} \frac{{\rm d} p_{\rm orb}}{{\rm d} t} & = & \frac{1}{a} \frac{{\rm d} a}{{\rm d} t} - \frac{2 e^{2}}{1 - e^{2}} \frac{1}{e} \frac{{\rm d} e}{{\rm d} t}, \nonumber \\
& = & 3 n \frac{M_{\rm X}}{M_{\rm G}} {\left( \frac{R_{\rm G}}{a} \right)}^{5} {\left( 1 - e^{2} \right)}^{- 1/2} \sum_{q = - \infty}^{+ \infty} \mathcal{A}_{{\rm G}, q} \nonumber \\
&& + 3 n \frac{M_{\rm G}}{M_{\rm X}} {\left( \frac{R_{\rm X}}{a} \right)}^{5} {\left( 1 - e^{2} \right)}^{- 1/2} \sum_{q = - \infty}^{+ \infty} \mathcal{A}_{{\rm X}, q}.
\end{eqnarray}
}Here, we assume that the contribution of the tidal torques caused by the secondary is negligible:
{\small
\begin{equation}
\frac{1}{p_{\rm orb}} \frac{{\rm d} p_{\rm orb}}{{\rm d} t} \simeq 3 n \frac{M_{\rm X}}{M_{\rm G}} {\left( \frac{R_{\rm G}}{a} \right)}^{5} {\left( 1 - e^{2} \right)}^{- 1/2} \sum_{q = - \infty}^{+ \infty} \mathcal{A}_{{\rm G}, q}.
\end{equation}
}

Assuming that the spin period of the primary is sufficiently shorter than the orbital period (i.e., $\dot{\theta}_{\rm G} \gg n$), we can rewrite the term $\sum_{q} \mathcal{A}_{{\rm G}, q}$ as follows:
\begin{equation}
\sum_{q = - \infty}^{+ \infty} \mathcal{A}_{{\rm G}, q} \simeq {\rm Im}{\left[ \tilde{k}_{2, {\rm G}} {\left( - 2 \dot{\theta}_{\rm G} \right)} \right]} \sum_{q = - \infty}^{+ \infty} {\left[ G_{2, 0, q} {( e )} \right]}^{2}.
\end{equation}
We found that $\sum_{q} {\left[ G_{2, 0, q} {( e )} \right]}^{2}$ is approximately given by
\begin{equation}
\sum_{q = - \infty}^{+ \infty} {\left[ G_{2, 0, q} {( e )} \right]}^{2} \simeq {\left( 1 - e^{2} \right)}^{-6},
\end{equation}
in the range of $0 \le e \le 0.8$ (see Figure \ref{figG_ecc}).
Therefore, the time evolution of $p_{\rm orb}$ is approximately given by the following equation: 
{\footnotesize
\begin{equation}
\frac{{\rm d} p_{\rm orb}}{{\rm d} t} \simeq 3 \sqrt{{\mathcal G} M_{\rm tot}} \frac{M_{\rm X}}{M_{\rm G}} {R_{\rm G}}^{5} {p_{\rm orb}}^{- 11/2} \cdot {\rm Im}{\left[ \tilde{k}_{2, {\rm G}} {\left( - 2 \dot{\theta}_{\rm G} \right)} \right]},
\end{equation}
}and we obtain $p_{\rm orb} = p_{\rm orb} {(t)}$ as
\begin{widetext}
\begin{equation}
p_{\rm orb} {(t)} \simeq {\left[ \frac{39}{2} \sqrt{{\mathcal G} M_{\rm tot}} {R_{\rm G}}^{2} \cdot \int_{0}^{t} {\rm d}t'\ {\rm Im}{\left[ \tilde{k}_{2, {\rm G}} {\left( - 2 \dot{\theta}_{\rm G} \right)} \right]} \right]}^{2/13} {R_{\rm X}}^{6/13},
\label{eqp}
\end{equation}
\end{widetext}
which does not include the eccentricity, $e$, in the equation.

\begin{figure}
\centering
\includegraphics[width = 0.45\textwidth]{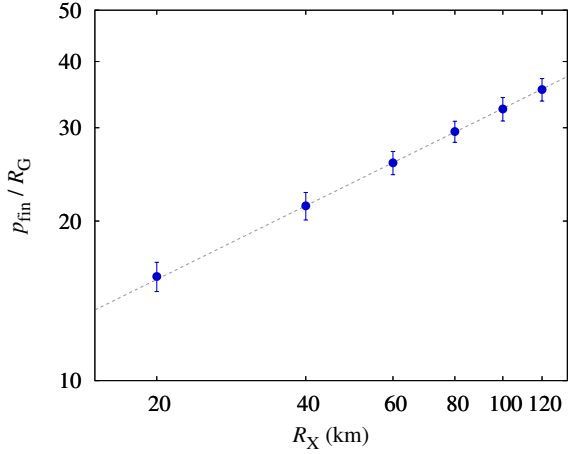}
\caption{
Final semilatus rectum, $p_{\rm fin}$, as a function of the radius of the secondary, $R_{\rm X}$.
The dashed line represents the least squares fit, $p_{\rm fin} / R_{\rm G} = 3.9 {\left( R_{\rm X} / {\rm km} \right)}^{6/13}$; the error bars are the twice the standard deviation for each $R_{\rm X}$.
}
\label{figaprime}
\end{figure}

Figure \ref{figaprime} shows the final semilatus rectum, $p_{\rm fin}$, as a function of the radius of the secondary, $R_{\rm X}$.
We only used the data classified as Type A/B/C in the figure.
We plotted the average value and twice the standard error of $p_{\rm fin}$ for each $R_{\rm X}$.
Equation (\ref{eqp}) shows that the final semilatus rectum is proportional to ${R_{\rm X}}^{6/13}$; our numerical results are consistent with the theoretical prediction.

\subsection{Spin period of the secondary}

\begin{figure*}
\centering
\includegraphics[width = 0.45\textwidth]{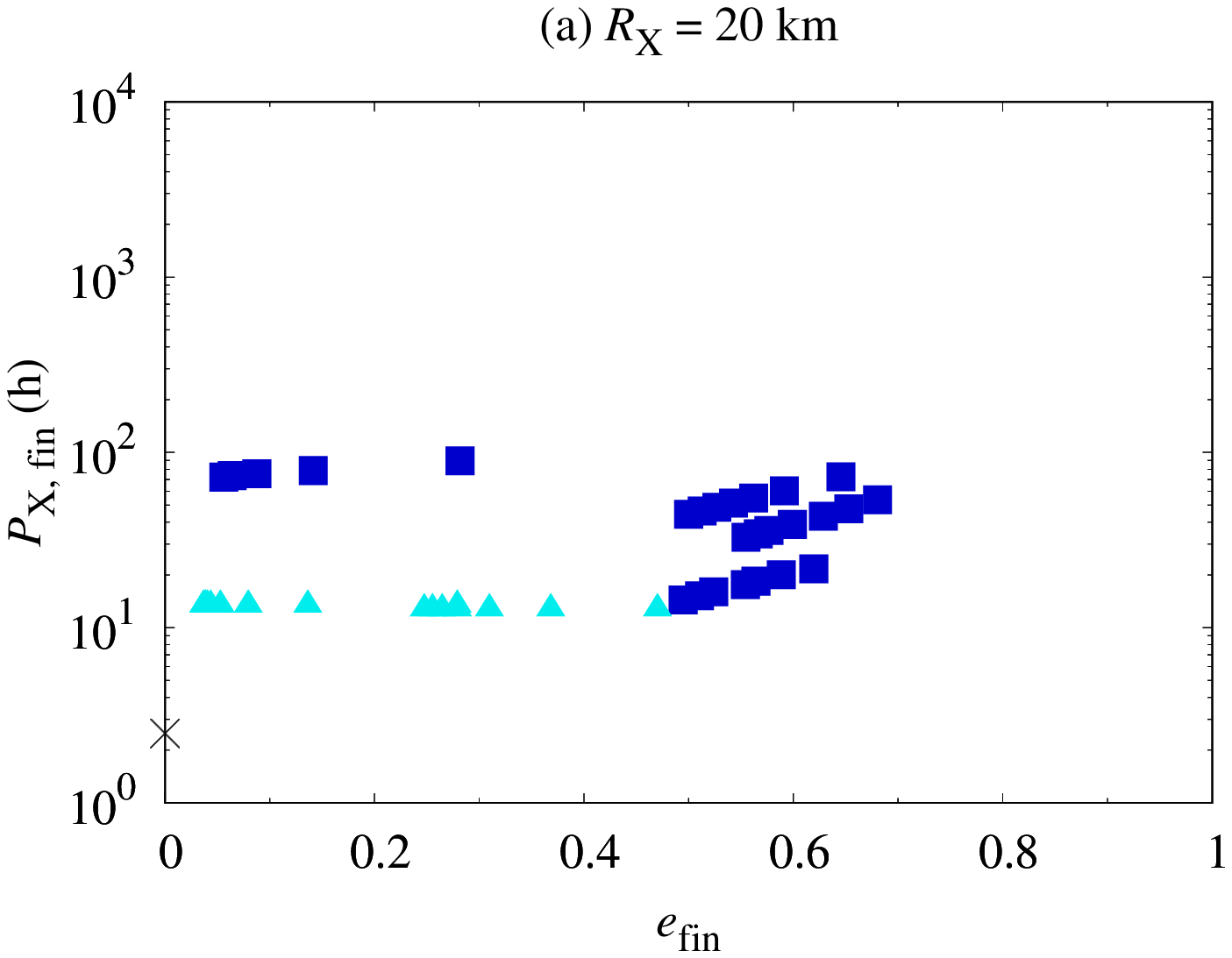}
\includegraphics[width = 0.45\textwidth]{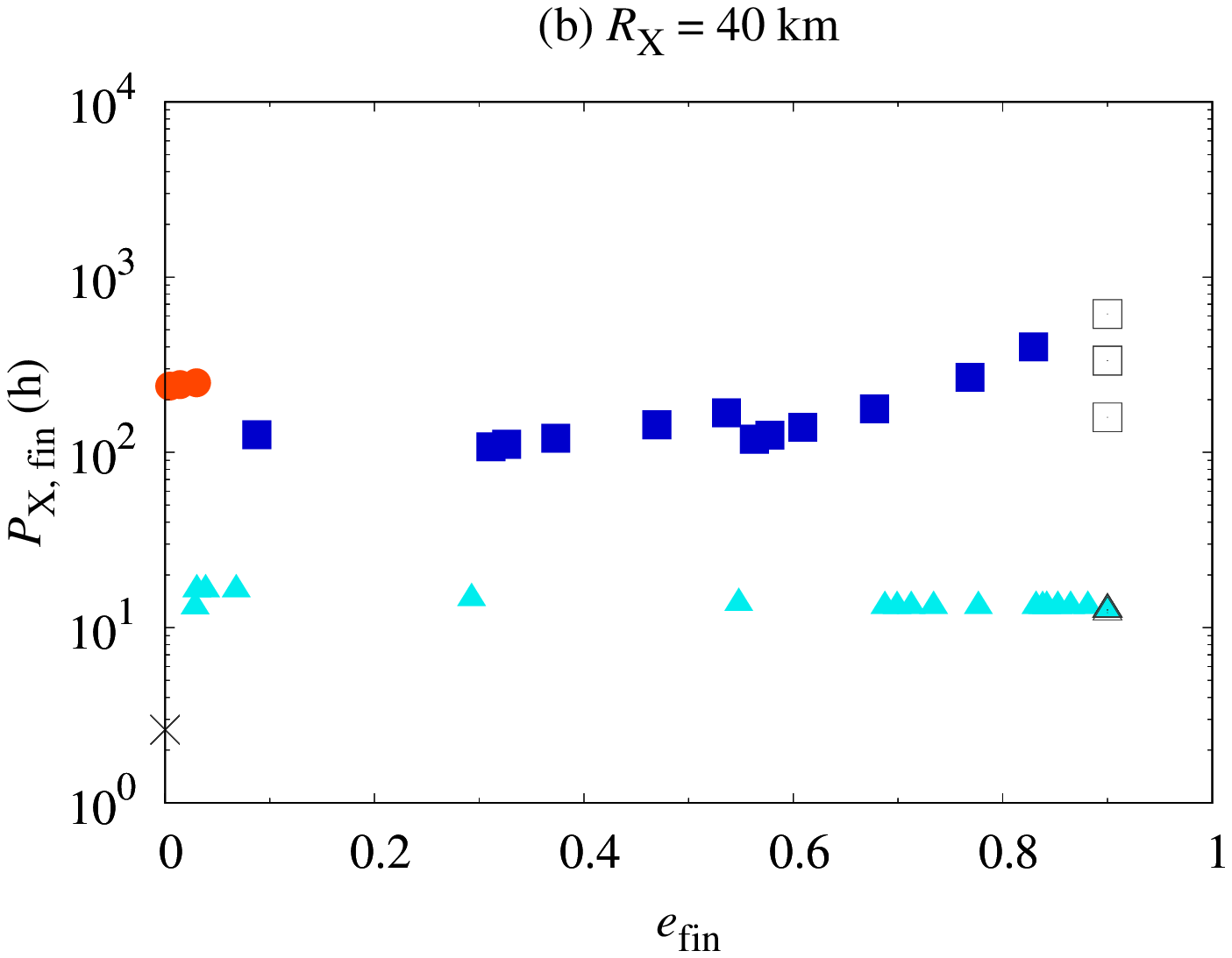}
\includegraphics[width = 0.45\textwidth]{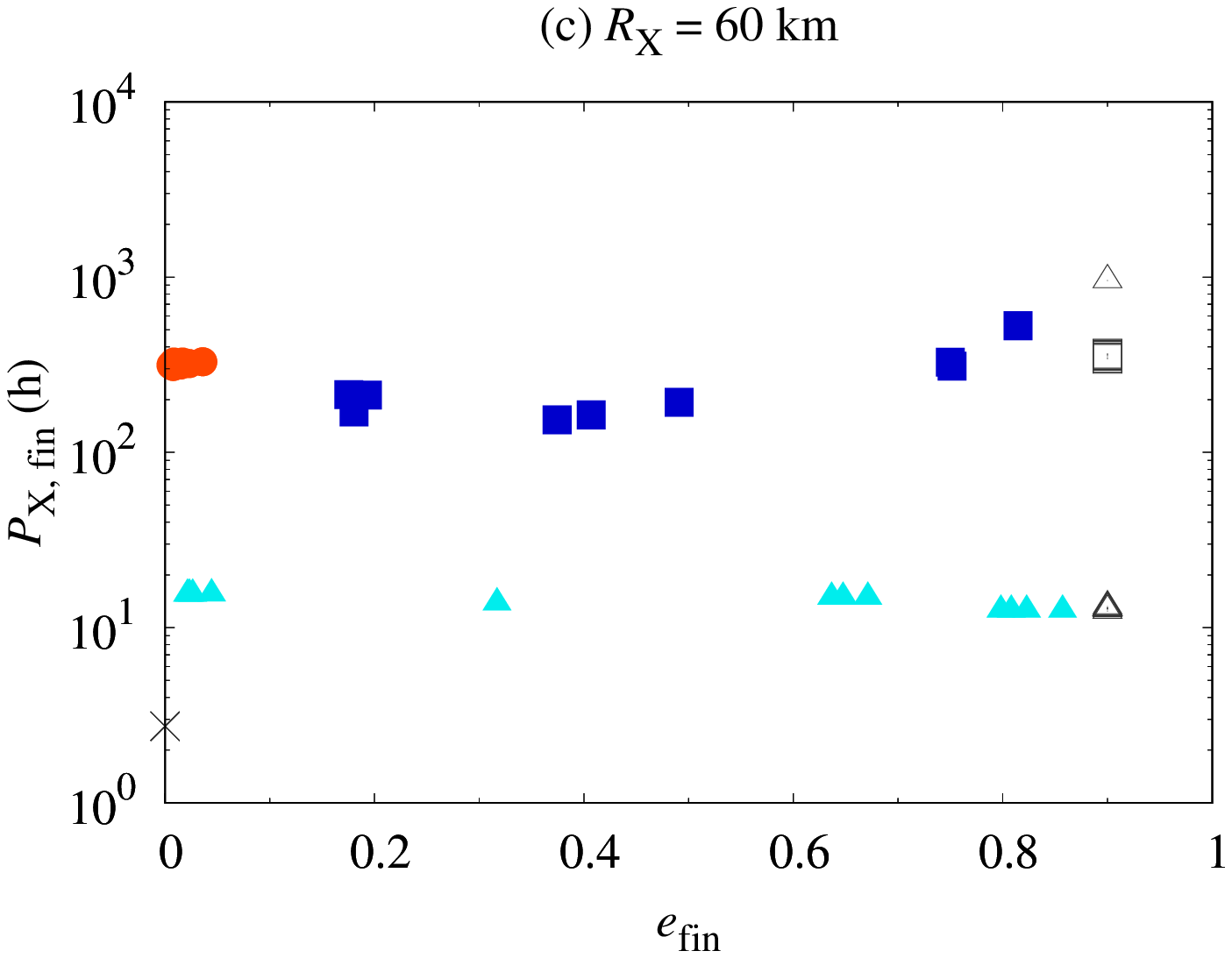}
\includegraphics[width = 0.45\textwidth]{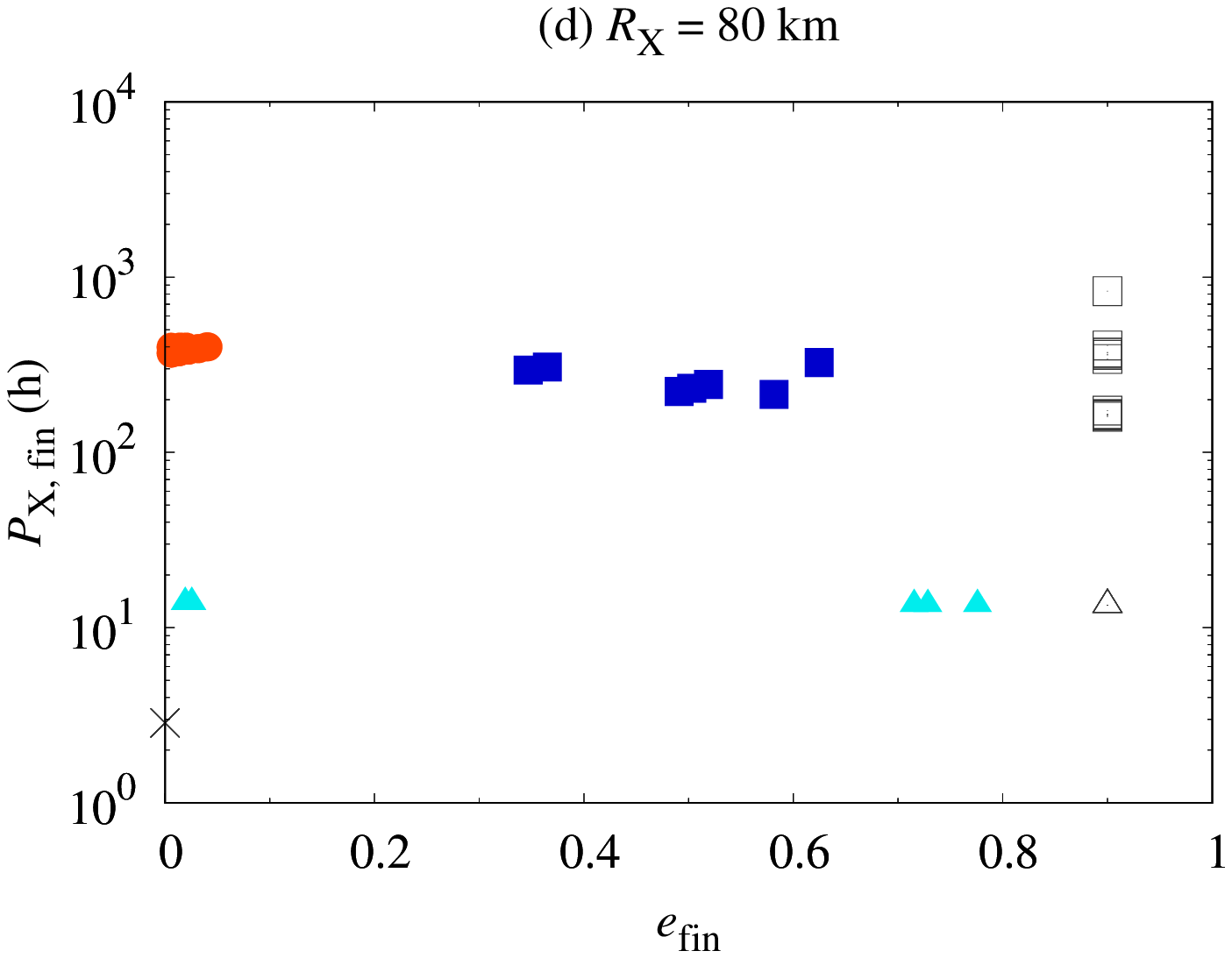}
\includegraphics[width = 0.45\textwidth]{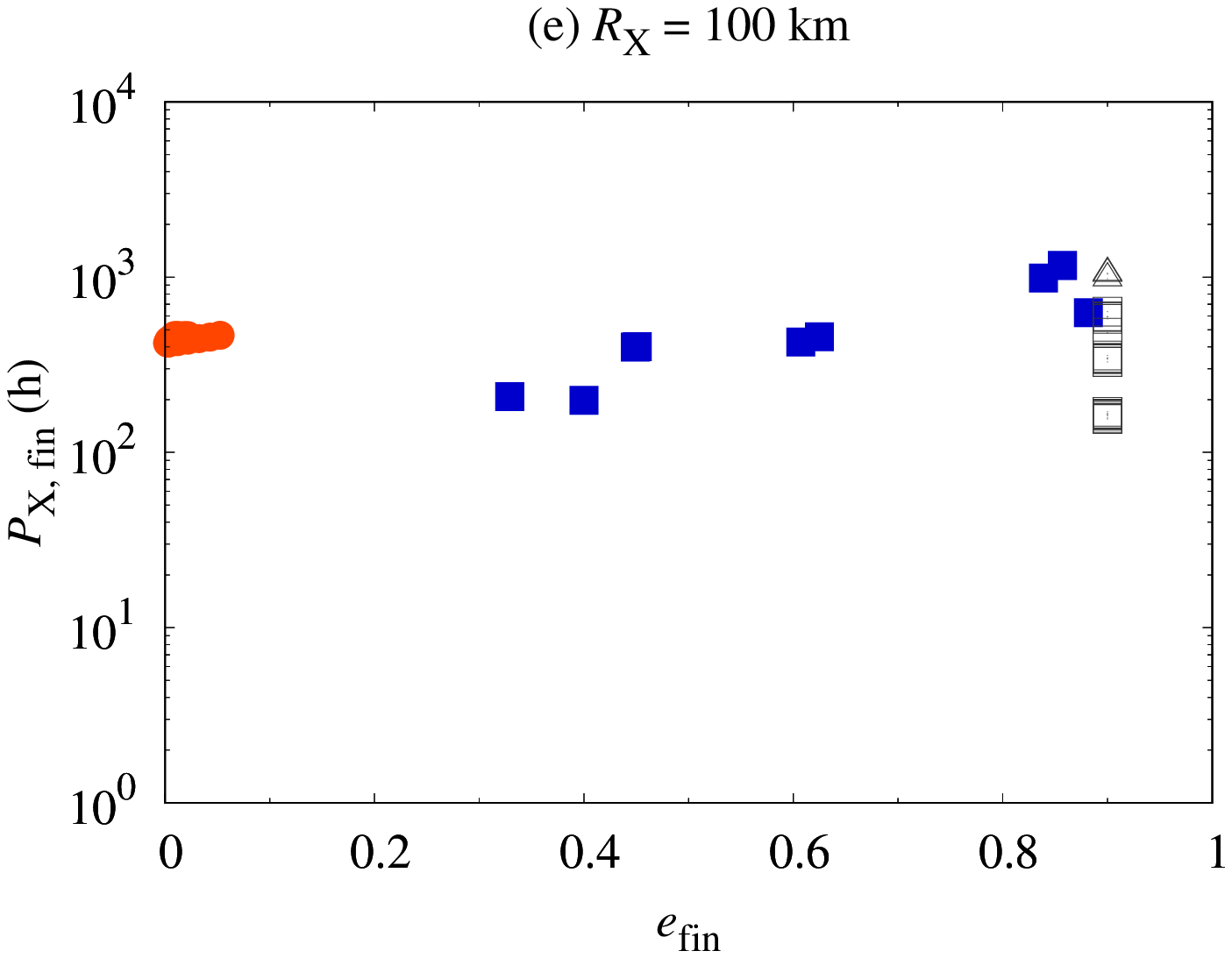}
\includegraphics[width = 0.45\textwidth]{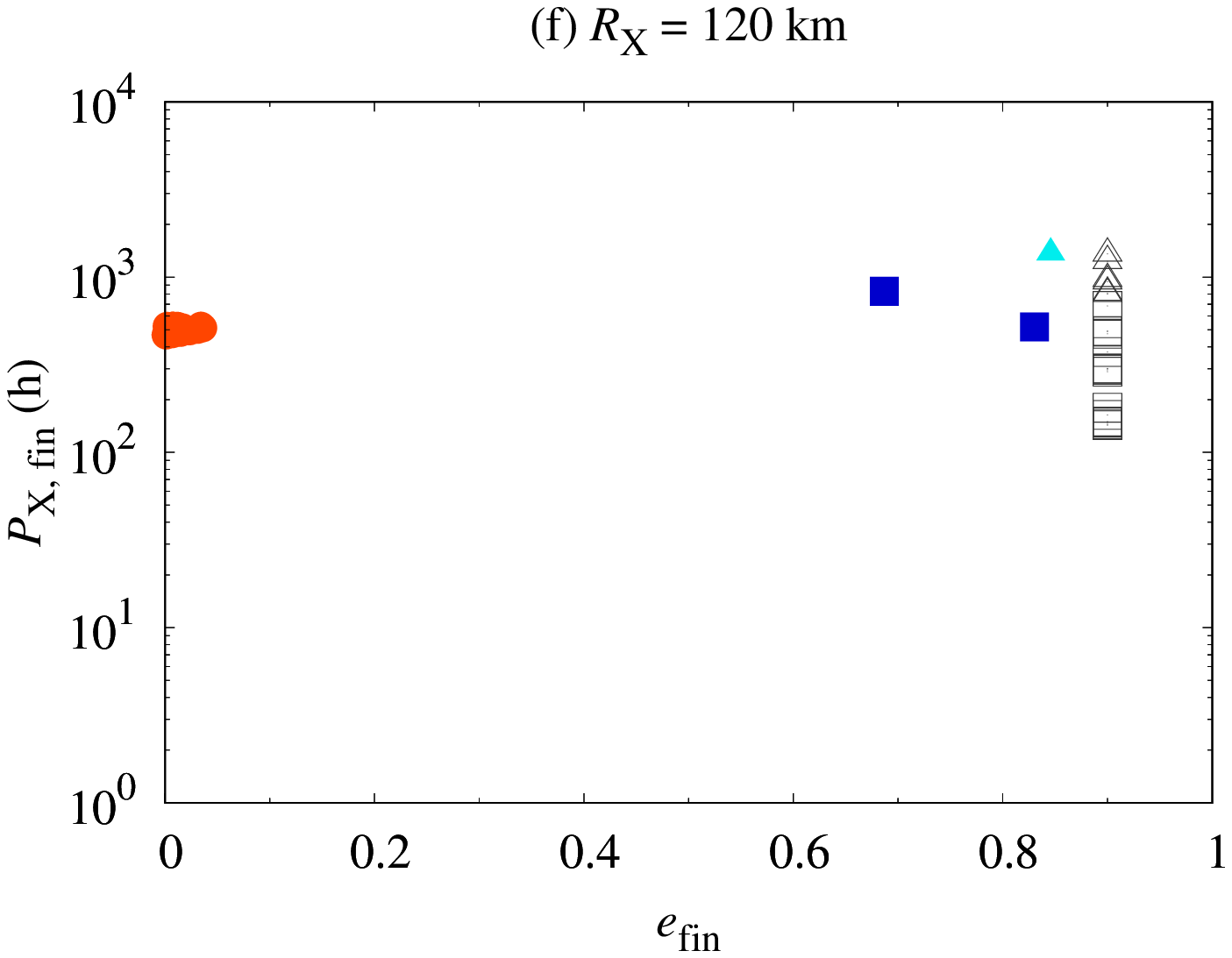}
\caption{
Distribution of the final eccentricity and spin period of the secondary, $e_{\rm fin}$ and $P_{\rm X, fin}$, for the standard runs of our simulation.
The different markers represent the different final states of the tidal evolution classified according to the spin state of Xiangliu and the final semimajor axis and eccentricity.
For the cases of Types Bx/Cx/Z, we stopped numerical integrations when the eccentricity reached $e = 0.9$ or when the periapsis distance reached $q_{\rm orb} = R_{\rm G} + R_{\rm X}$.
}
\label{figp_ecc}
\end{figure*}

Figure \ref{figp_ecc} shows the distribution of the final eccentricity and spin period of the secondary, $e_{\rm fin}$ and $P_{\rm X, fin}$, for standard runs of our simulation.
The spin period of Xiangliu at $t = 4.5\ {\rm Gyr}$ is in the range of $10\ {\rm h} \lesssim P_{\rm X, fin} \lesssim 10^{3}\ {\rm h}$, and the range of $P_{\rm X, fin}$ depends on $R_{\rm X}$.
In the case of $R_{\rm X} = 20\ {\rm km}$, the final spin period of Xiangliu is in the range of $10\ {\rm h} \lesssim P_{\rm X, fin} \lesssim 10^{2}\ {\rm h}$ and is always shorter than the orbital period, $P_{\rm orb, fin}$.
In contrast, for $R_{\rm X} = 100\ {\rm km}$ and $120\ {\rm km}$, the final spin period of Xiangliu is in the range of $10^{2}\ {\rm h} \lesssim P_{\rm X, fin} \lesssim 10^{3}\ {\rm h}$, and $P_{\rm X, fin}$ is significantly longer than the initial spin period, $P_{\rm X, ini} = 12\ {\rm h}$.
Thus, the determination of the spin period of Xiangliu by future observations and light curve analyses is the key to estimating the radius of Xiangliu.

The final spin period of Xiangliu is close to the initial spin period, $P_{\rm X, fin} \simeq P_{\rm X, ini}$, for the case of Type C.
The spin period of satellites formed via a giant impact should provide abundant information regarding the condition of the colliding TNOs, including the mass ratio, the spin state before the impact, and the (un)differentiated state.
Therefore, we want to determine the spin period of satellites from astronomical observations for Gonggong--Xiangliu and for other satellite systems around large TNOs.

\subsection{Final eccentricity and the initial condition}

\begin{figure*}
\centering
\includegraphics[width = 0.45\textwidth]{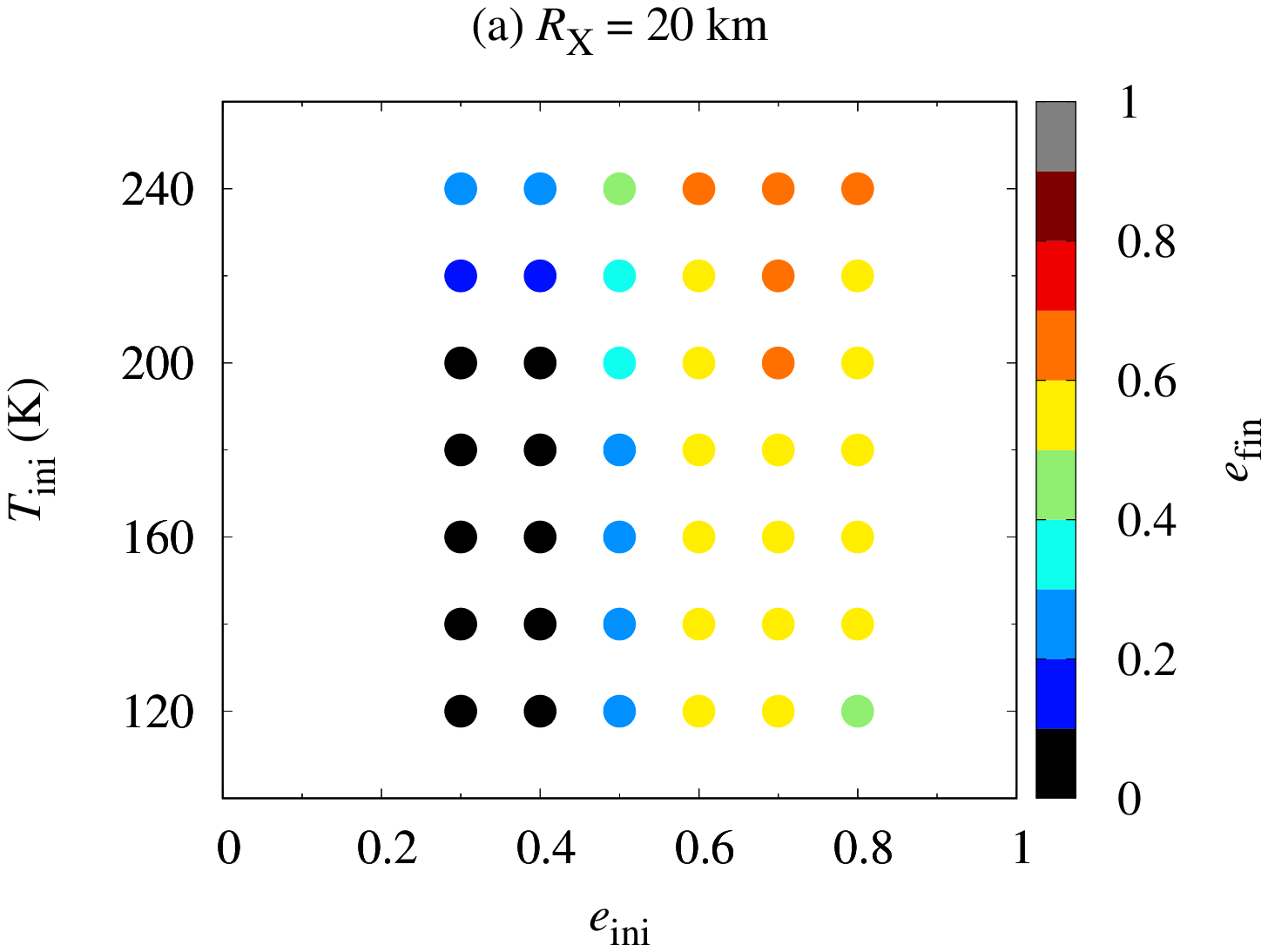}
\includegraphics[width = 0.45\textwidth]{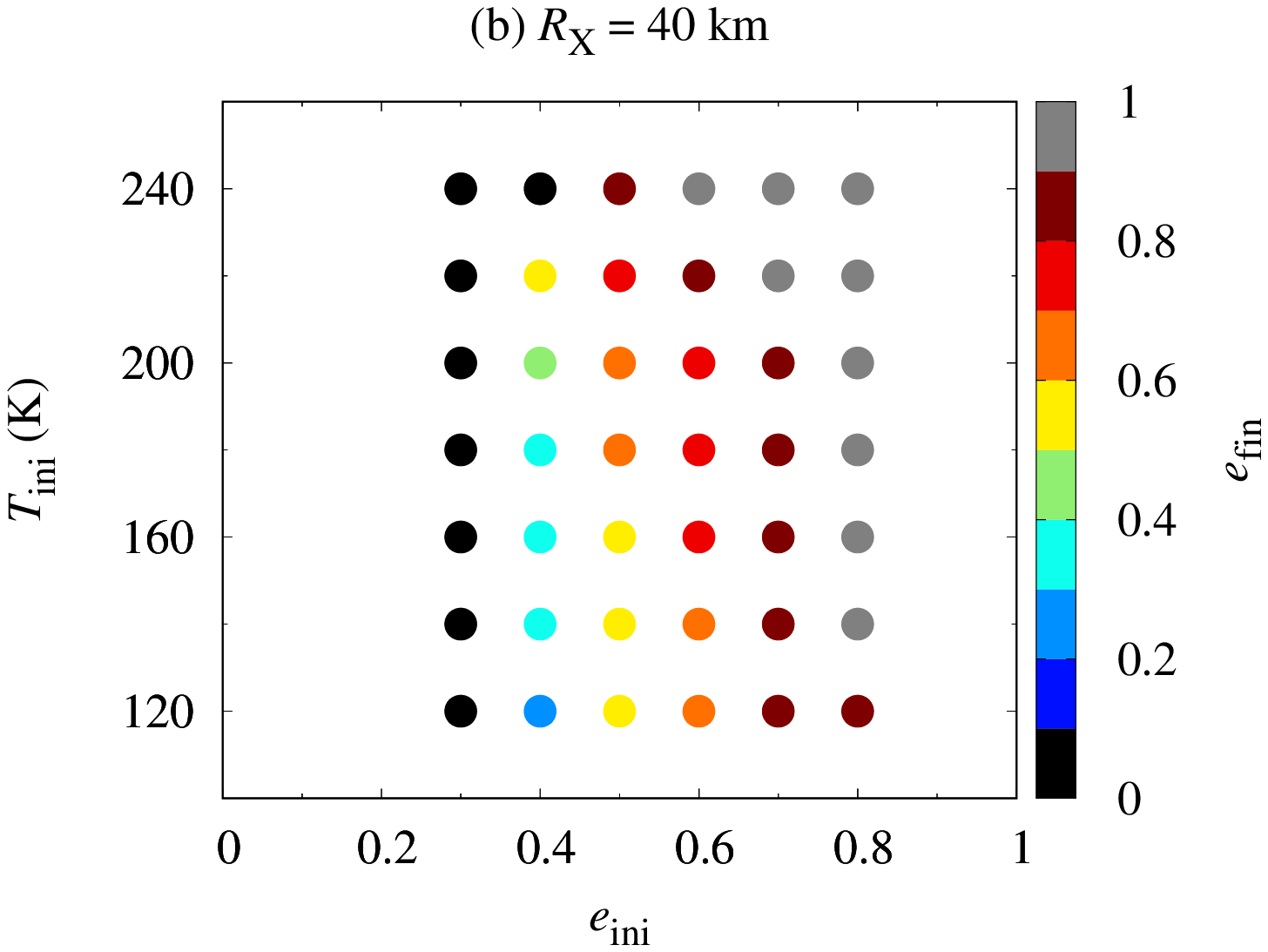}
\includegraphics[width = 0.45\textwidth]{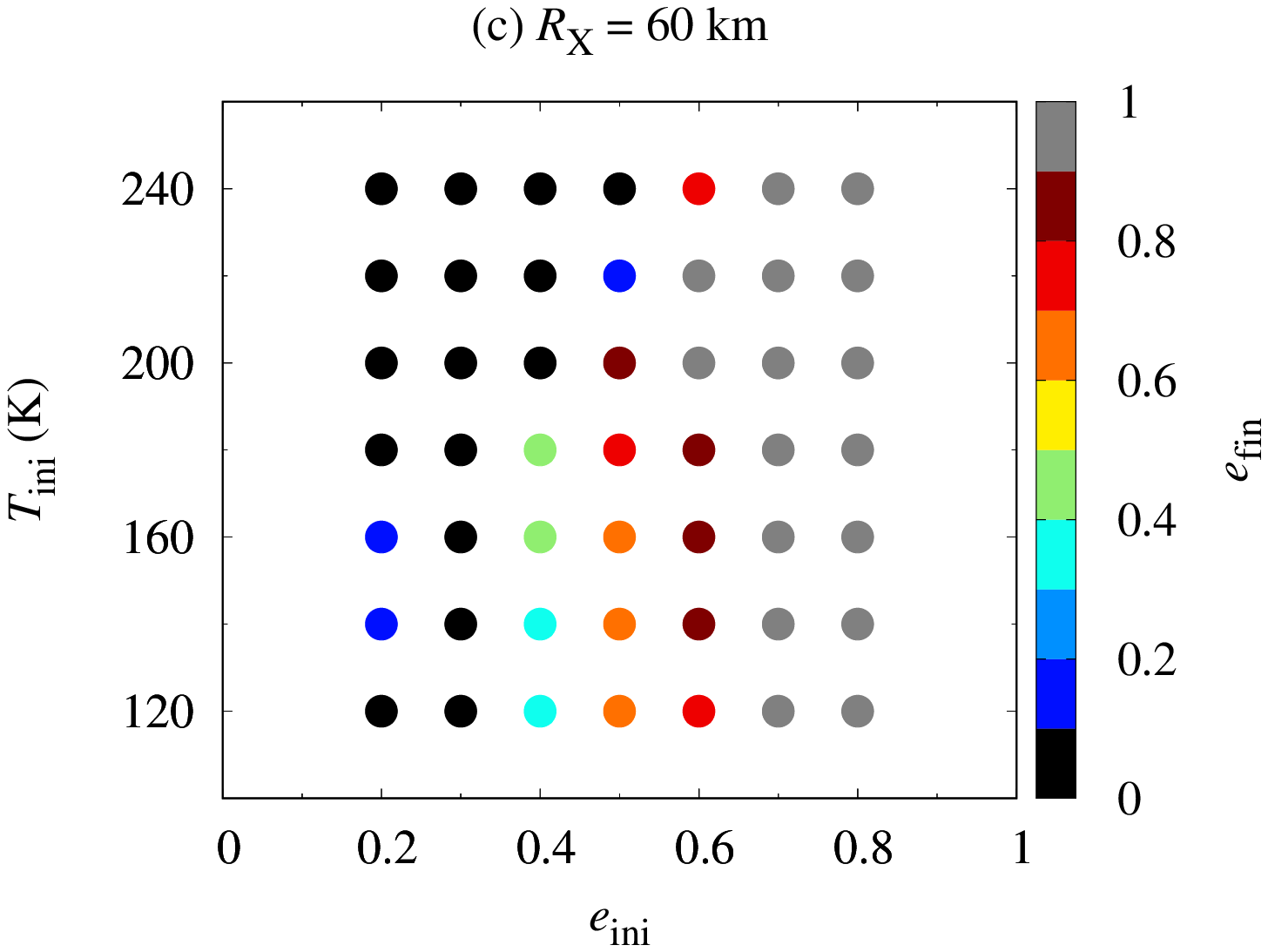}
\includegraphics[width = 0.45\textwidth]{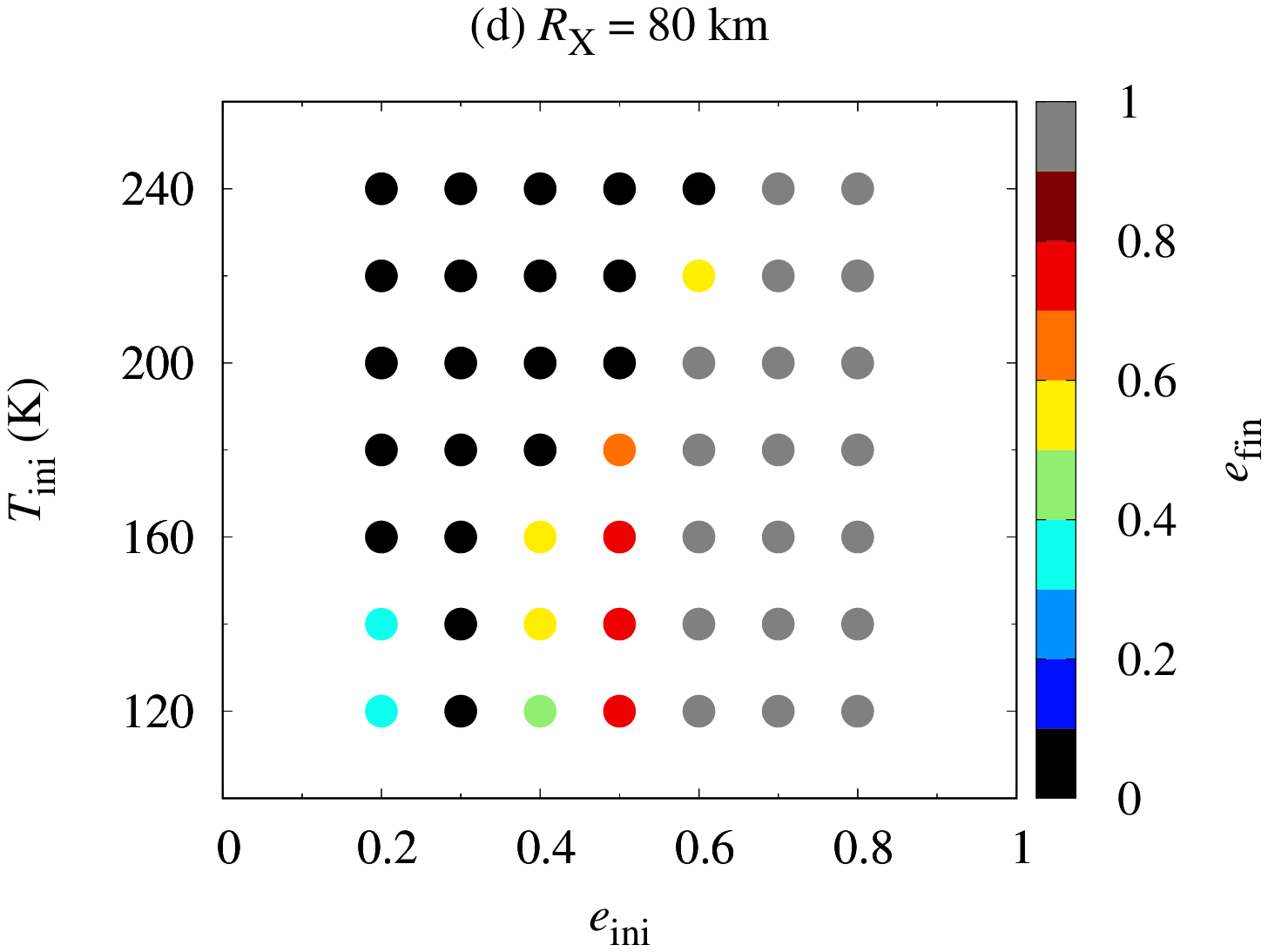}
\includegraphics[width = 0.45\textwidth]{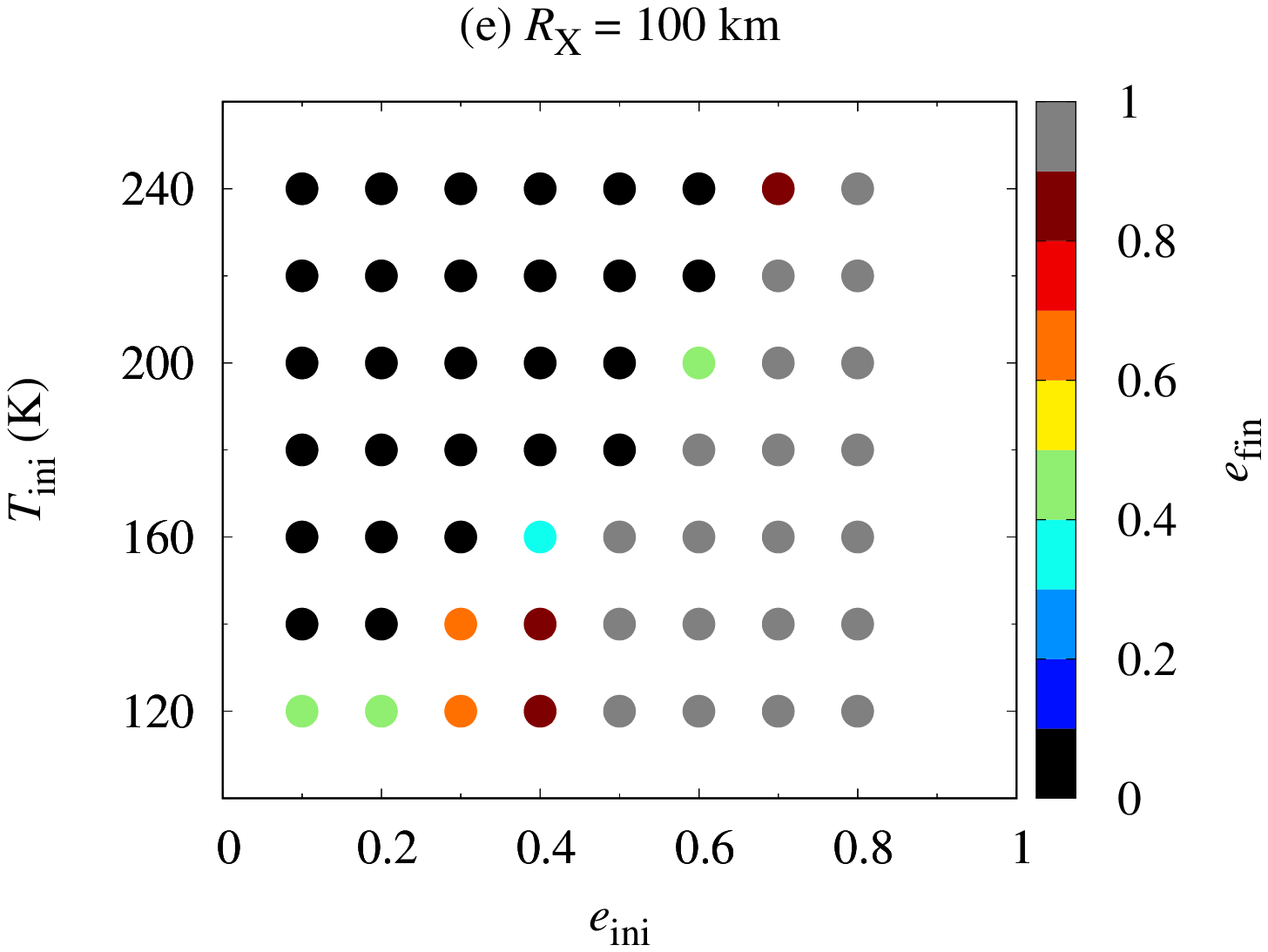}
\includegraphics[width = 0.45\textwidth]{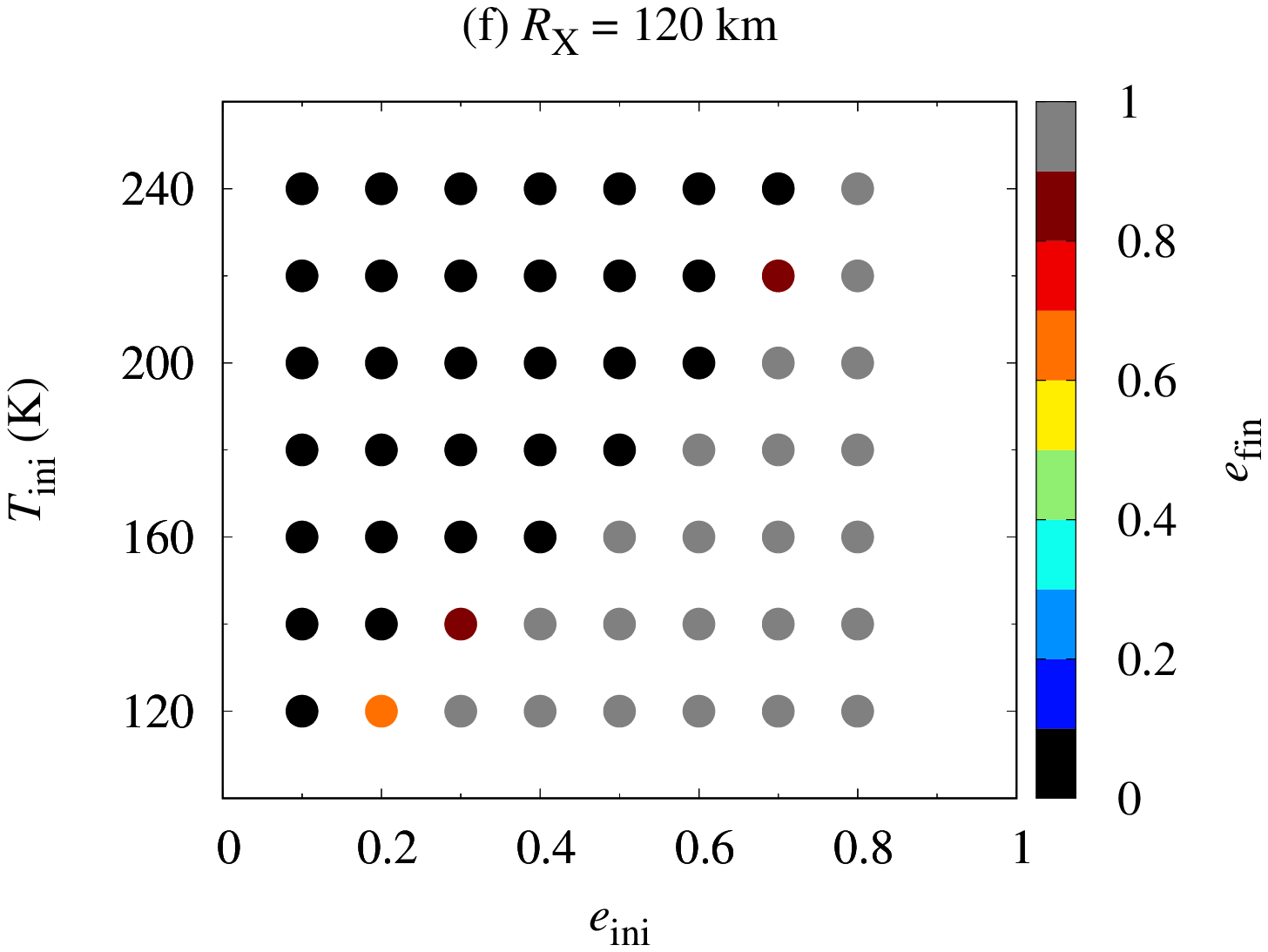}
\caption{
Color maps for the final eccentricity of the system, $e_{\rm fin}$, for standard runs of our simulation.
}
\label{figeccfin}
\end{figure*}

Figure \ref{figeccfin} shows the final eccentricity of the system, $e_{\rm fin}$, for standard runs of our simulation.
We note that we do not plot the results classified into Type Z.

Figure \ref{figeccfin} shows that satellite systems with an initial eccentricity of $0.2 \le e_{\rm ini} \le 0.5$ tend to become a final eccentricity of $0.2 \le e_{\rm fin} \le 0.5$.
When the initial eccentricity is $e_{\rm ini} \le 0.1$, the final eccentricity at $t = 4.5\ {\rm Gyr}$ is typically $e_{\rm fin} \le 0.1$ or the system becomes Type Z.
When the initial eccentricity is $e_{\rm ini} \ge 0.6$, on the other hand, the final eccentricity at $t = 4.5\ {\rm Gyr}$ is typically $e_{\rm fin} \ge 0.5$ for the case of $R_{\rm X} \le 60\ {\rm km}$, and $e_{\rm fin} \ge 0.9$ or $e_{\rm fin} \le 0.1$ for the case of $R_{\rm X} \ge 80\ {\rm km}$.

Therefore, we conclude that the initial eccentricity of Gonggong--Xiangliu system after the moon-forming giant impact may be in the range of $0.2 \lesssim e_{\rm ini} \lesssim 0.5$.
This non-zero $e_{\rm ini}$ is also consistent with the fact that $e_{\rm ini} \gtrsim 0.2$ is needed to avoid the inward migration of Xiangliu (see Figure \ref{figType}).

Debris disks can be formed by a giant impact \citep{Canup2005}, collisional disruption of a pre-existing satellite \citep{Hyodo+2017a}, or by tidal disruption of a passing body \citep{Hyodo+2017b}.
If a single large satellite is formed via accretion of debris disk materials around Gonggong, the initial eccentricity becomes small: $e \lesssim 0.1$ \citep[e.g.,][]{Ida+1997,Kokubo+2000}.
In this case, we cannot reproduce the observed eccentricity of the system.
Thus, Gonggong--Xiangliu system was more likely to be born as an ``intact moon'' rather than ``disk-origin moon'', and their formation process is similar to that of Pluto--Charon system \citep[e.g.,][]{Canup2005,Sekine+2017,Arakawa+2019}.

\section{Discussion}
\label{sec.discussion}

\begin{figure*}
\centering
\includegraphics[width = 0.45\textwidth]{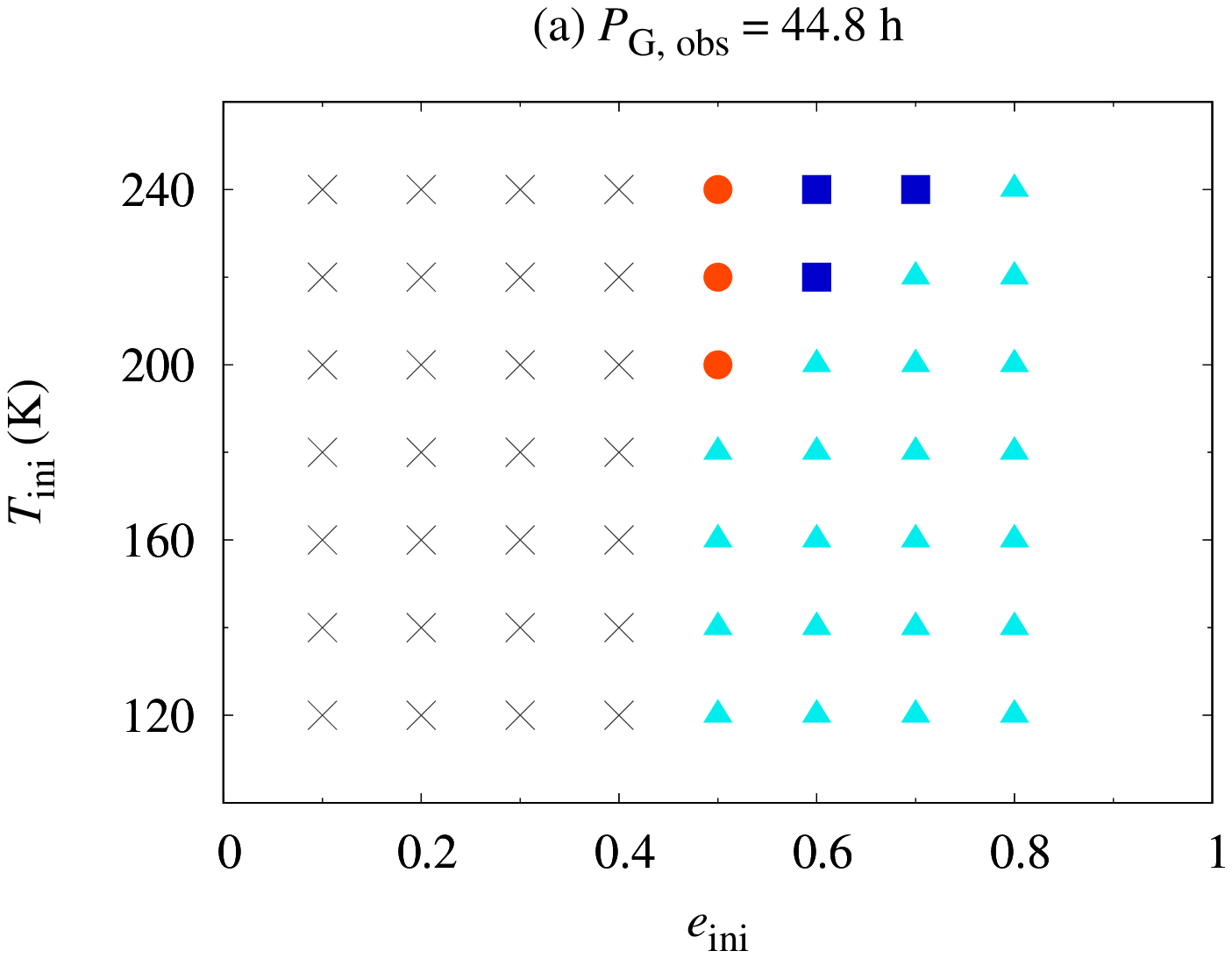}
\includegraphics[width = 0.45\textwidth]{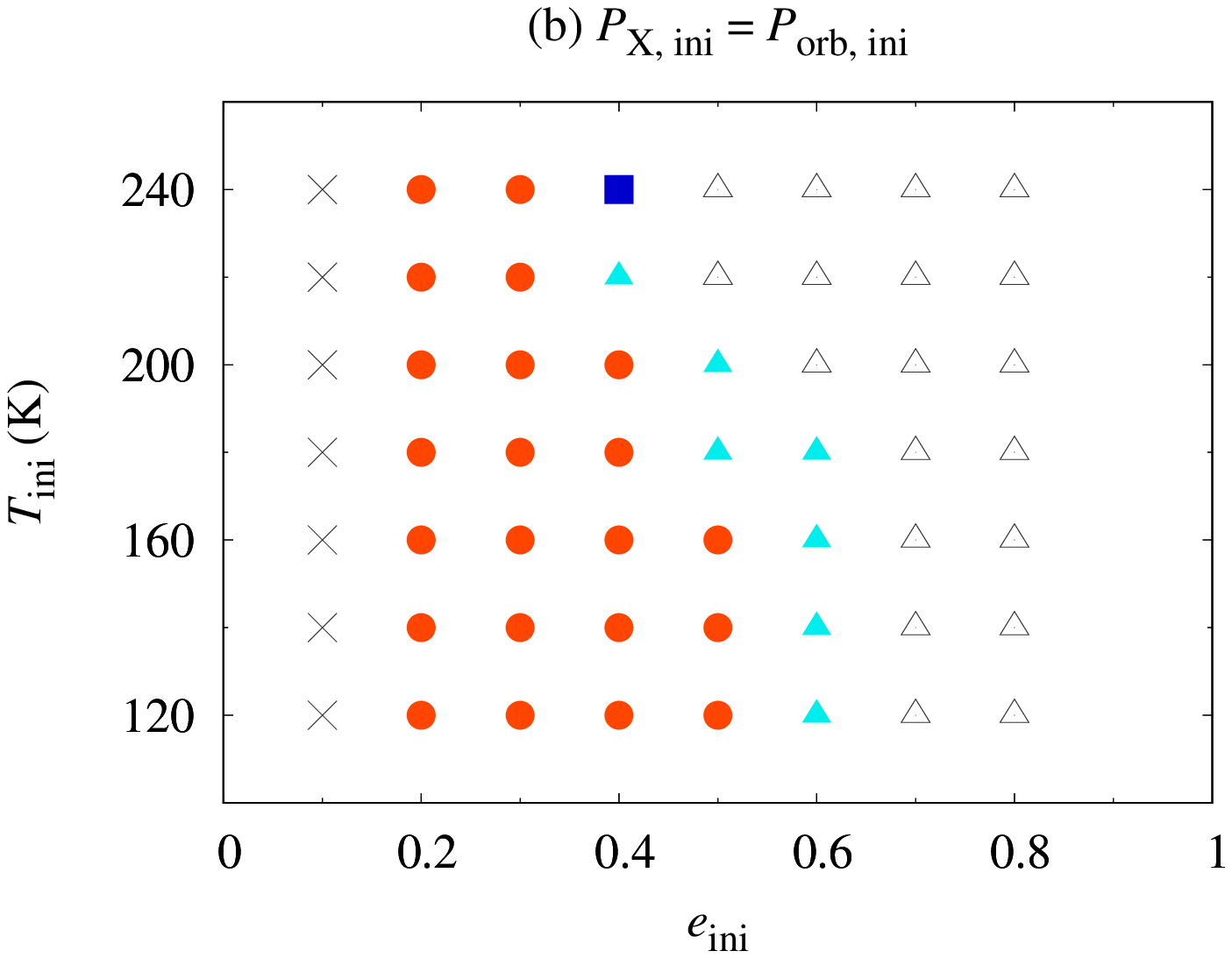}
\includegraphics[width = 0.45\textwidth]{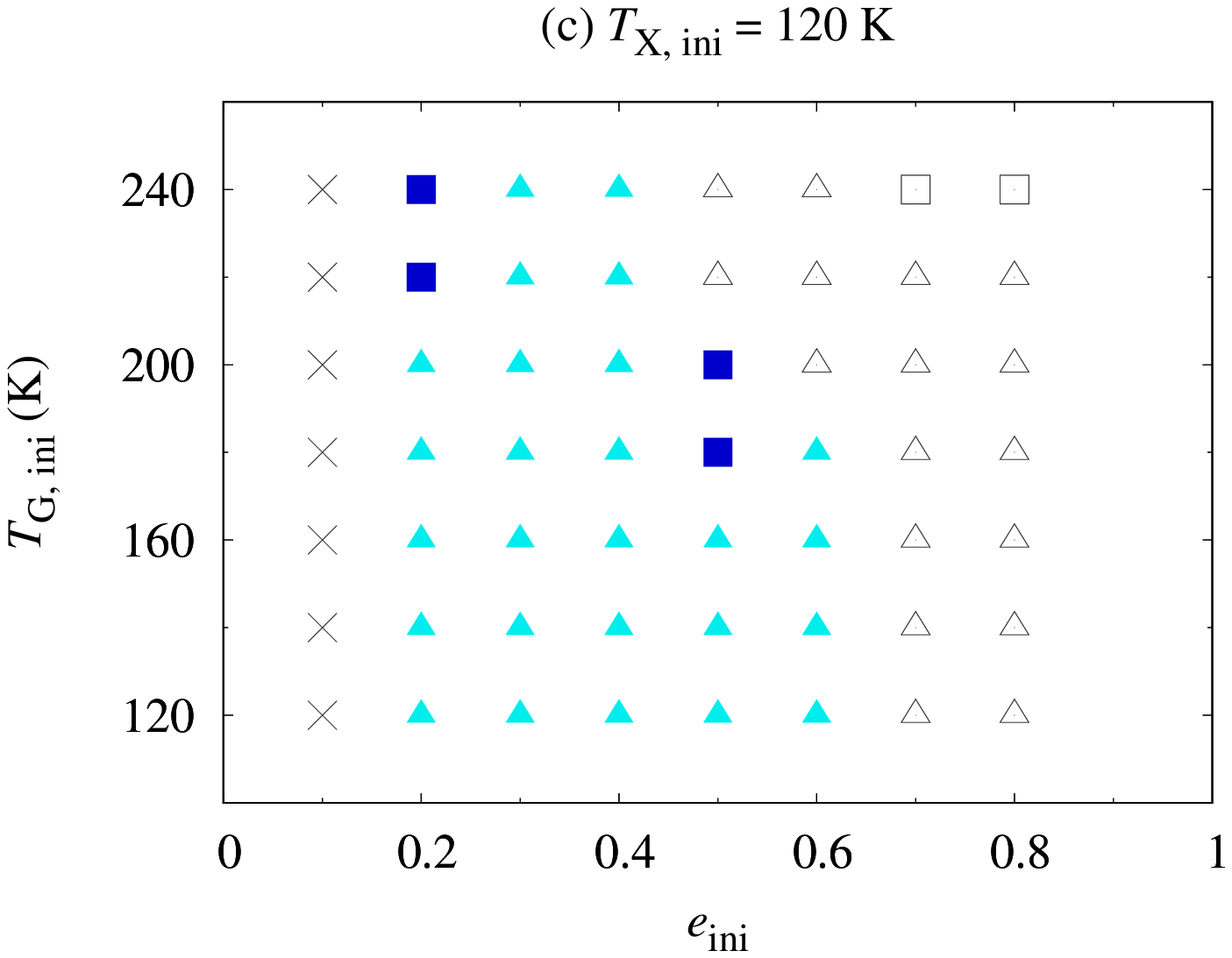}
\includegraphics[width = 0.45\textwidth]{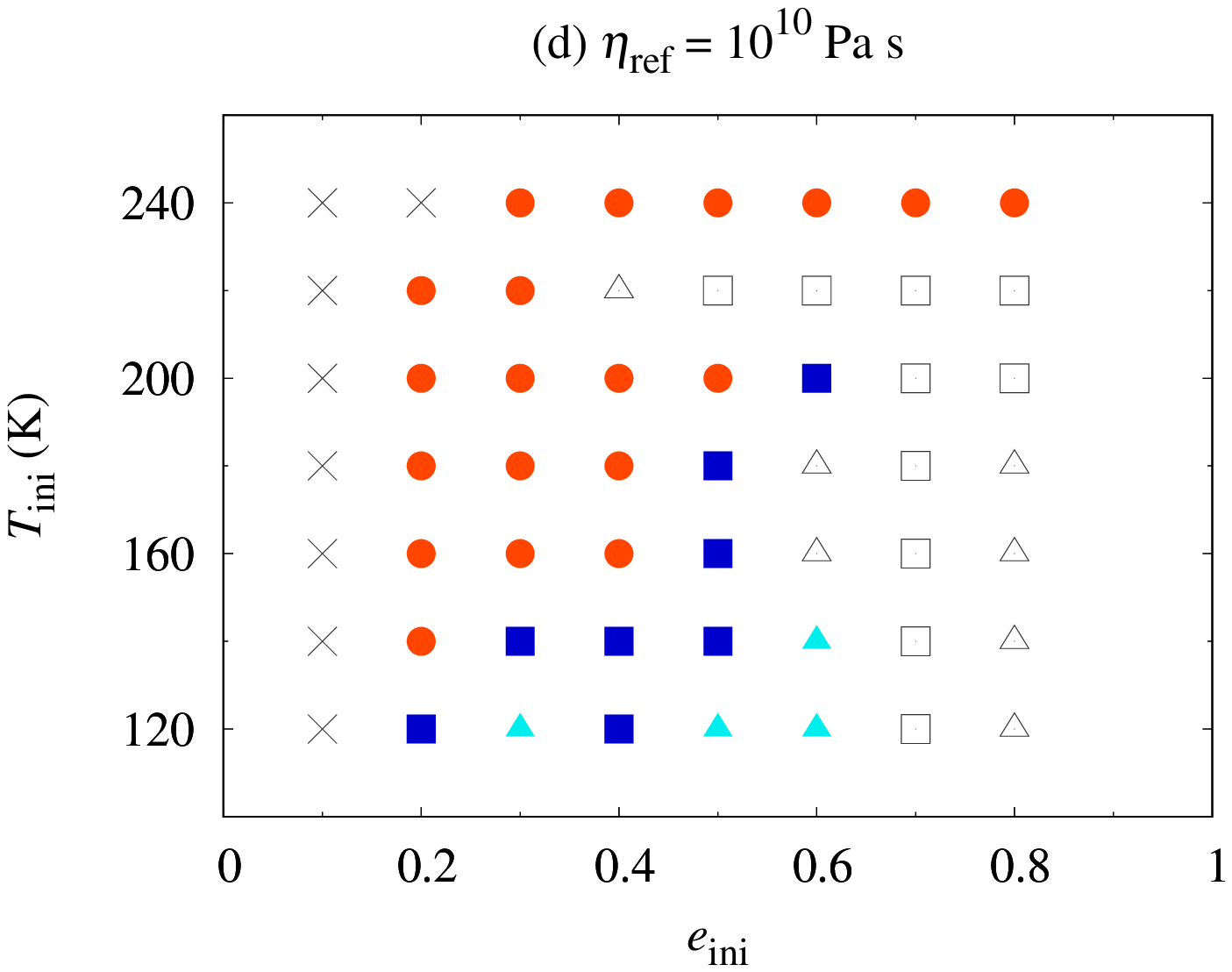}
\includegraphics[width = 0.45\textwidth]{fig_final_cap.eps}
\caption{
Summary of the final state for non-standard runs of our simulation.
We changed the initial eccentricity, $e_{\rm ini}$, and the initial temperature (of Gonggong), $T_{\rm ini}$ ($T_{\rm G, ini}$), as parameters.
The different markers represent the different final states of the tidal evolution classified according to the spin state of Xiangliu and the final semimajor axis and eccentricity.
}
\label{figType2}
\end{figure*}

\begin{figure*}
\centering
\includegraphics[width = 0.45\textwidth]{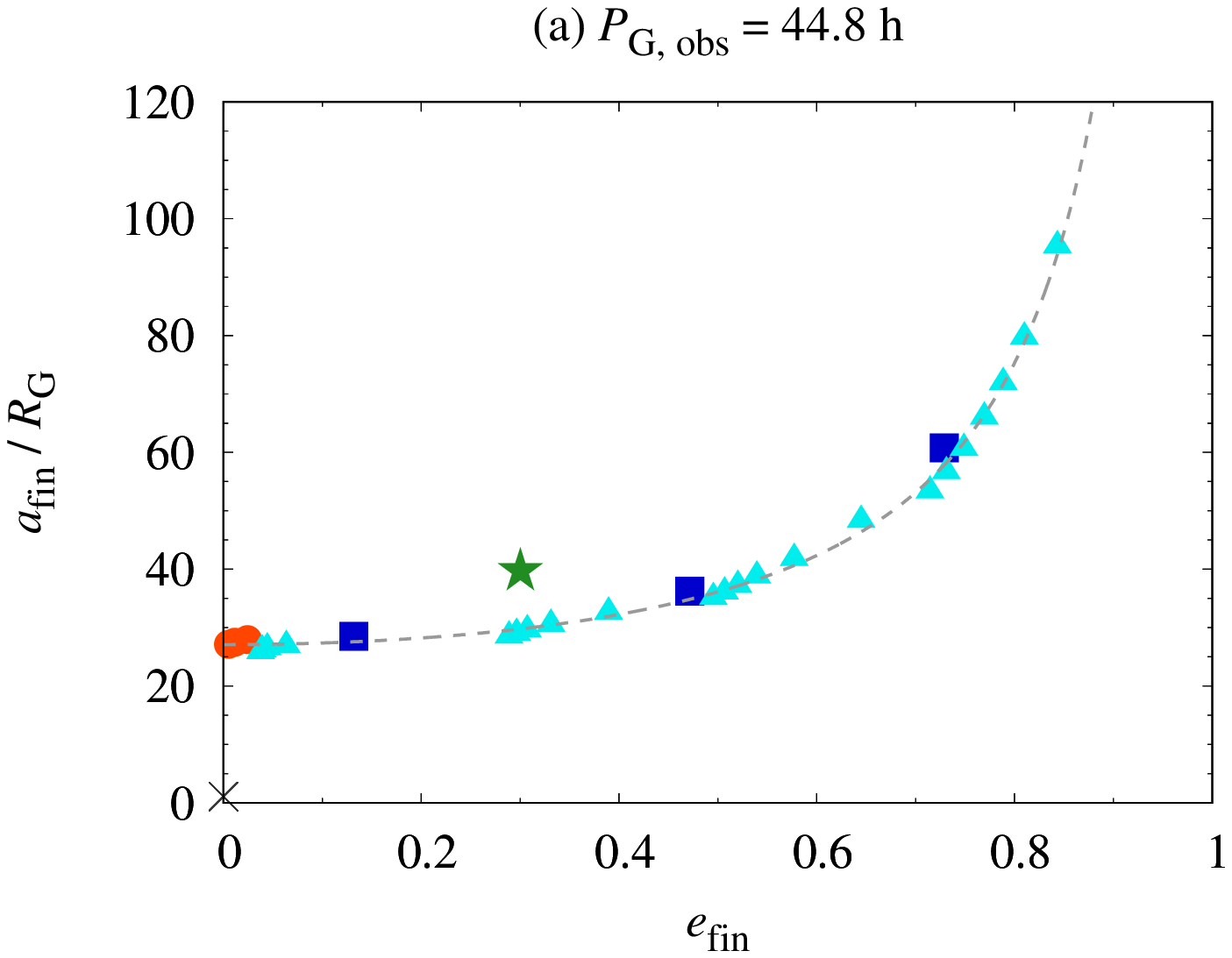}
\includegraphics[width = 0.45\textwidth]{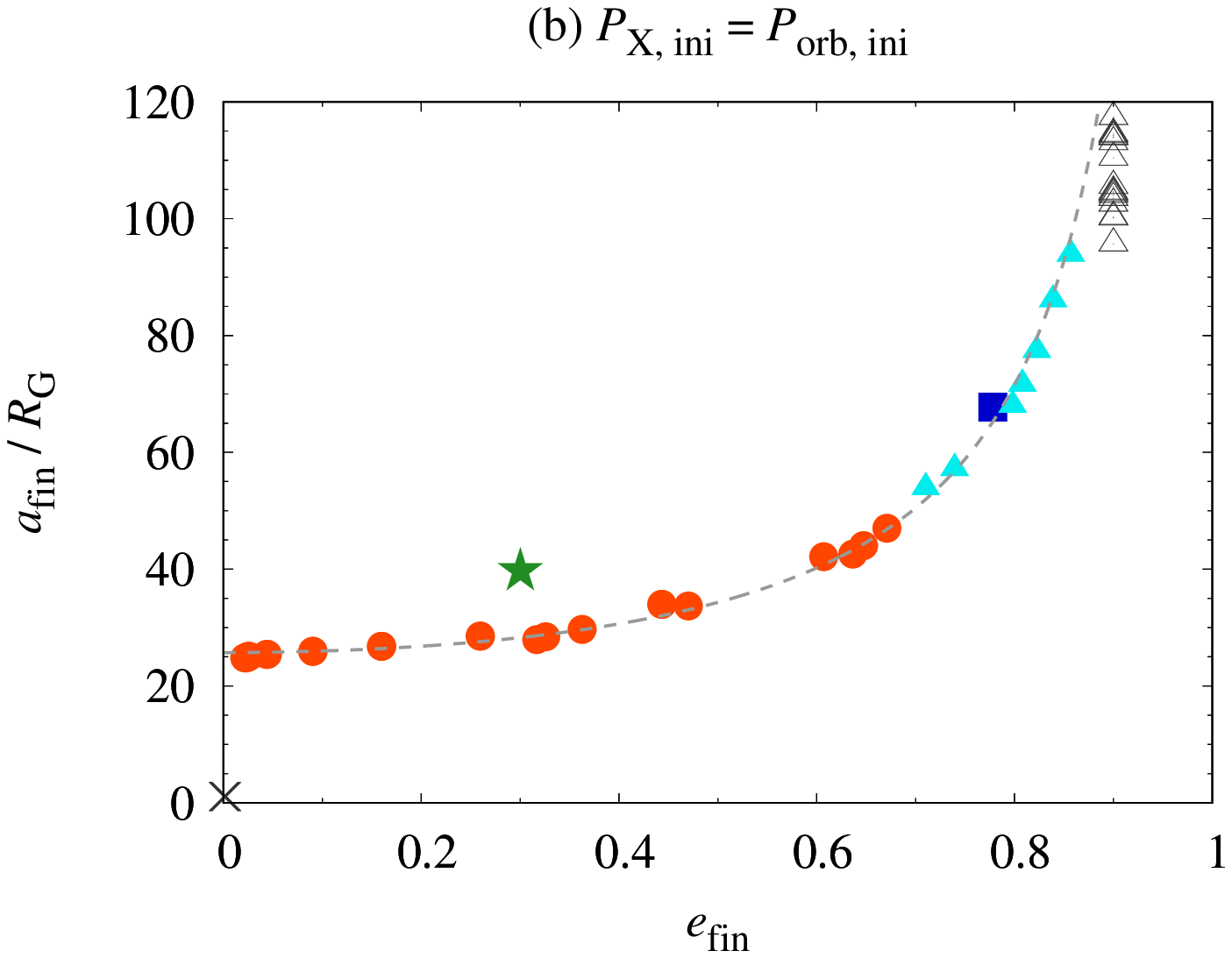}
\includegraphics[width = 0.45\textwidth]{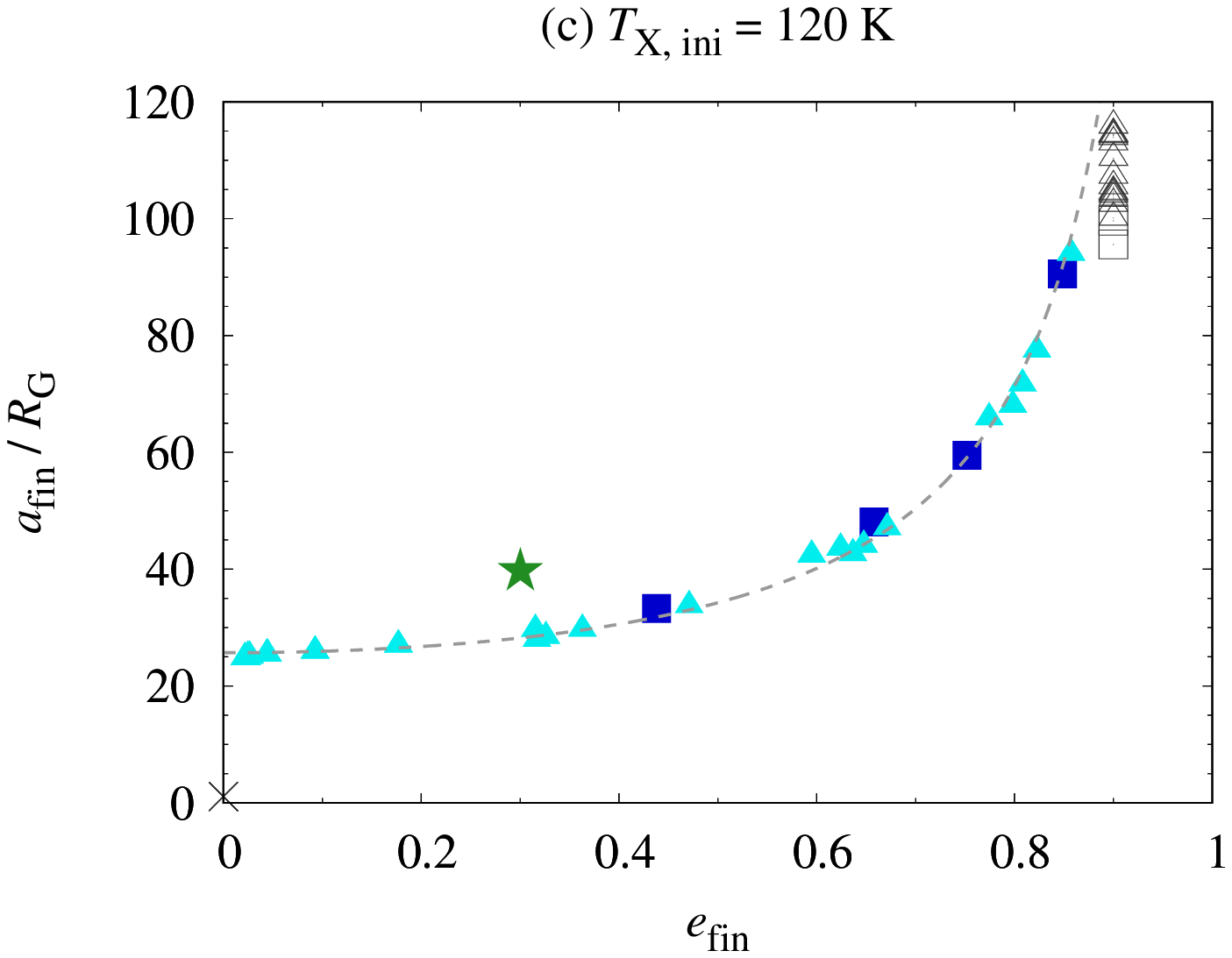}
\includegraphics[width = 0.45\textwidth]{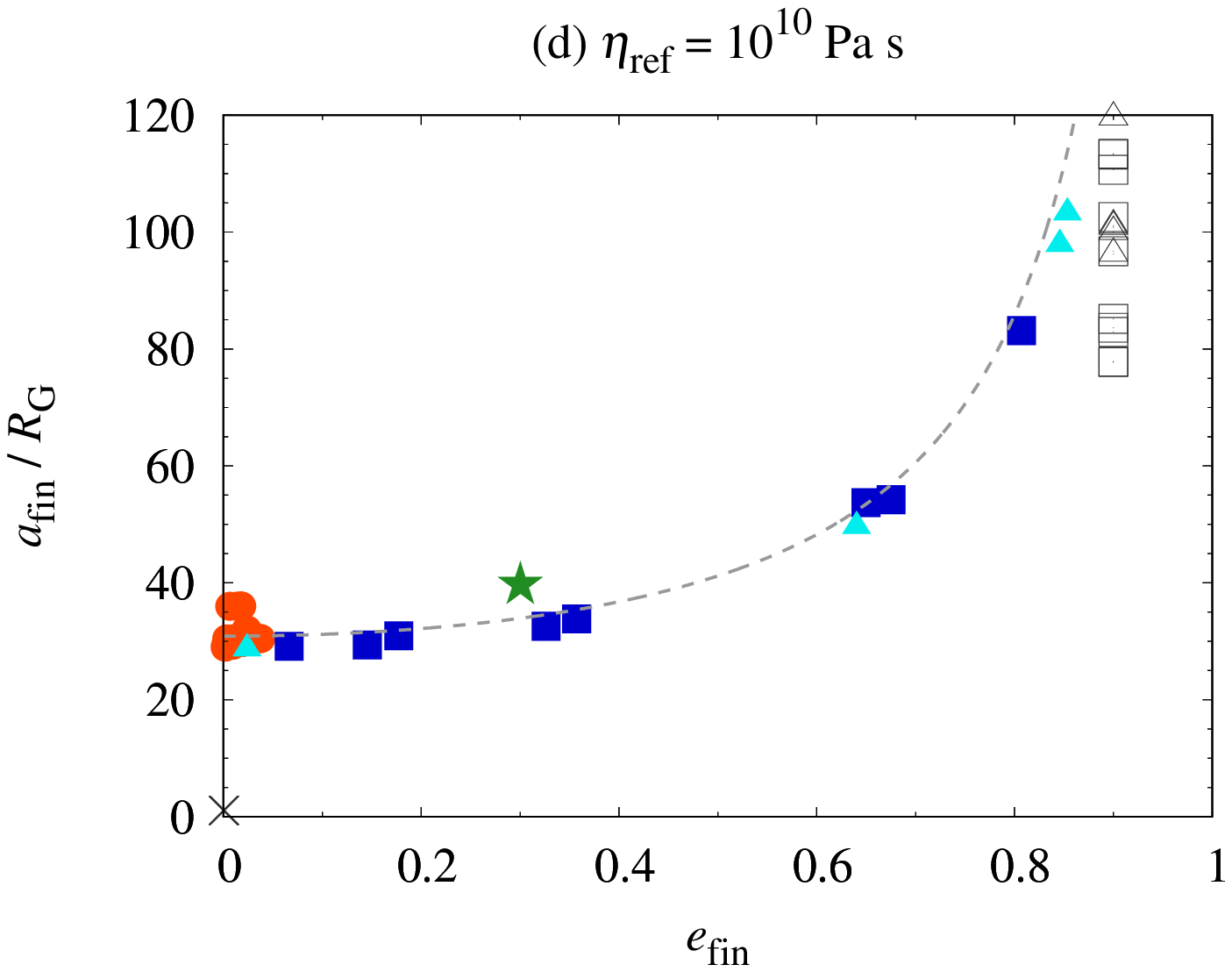}
\caption{
Distribution of the final eccentricity and semimajor axis, $e_{\rm fin}$ and $a_{\rm fin}$, for non-standard runs of our simulation.
}
\label{figa_ecc2}
\end{figure*}

\begin{figure*}
\centering
\includegraphics[width = 0.45\textwidth]{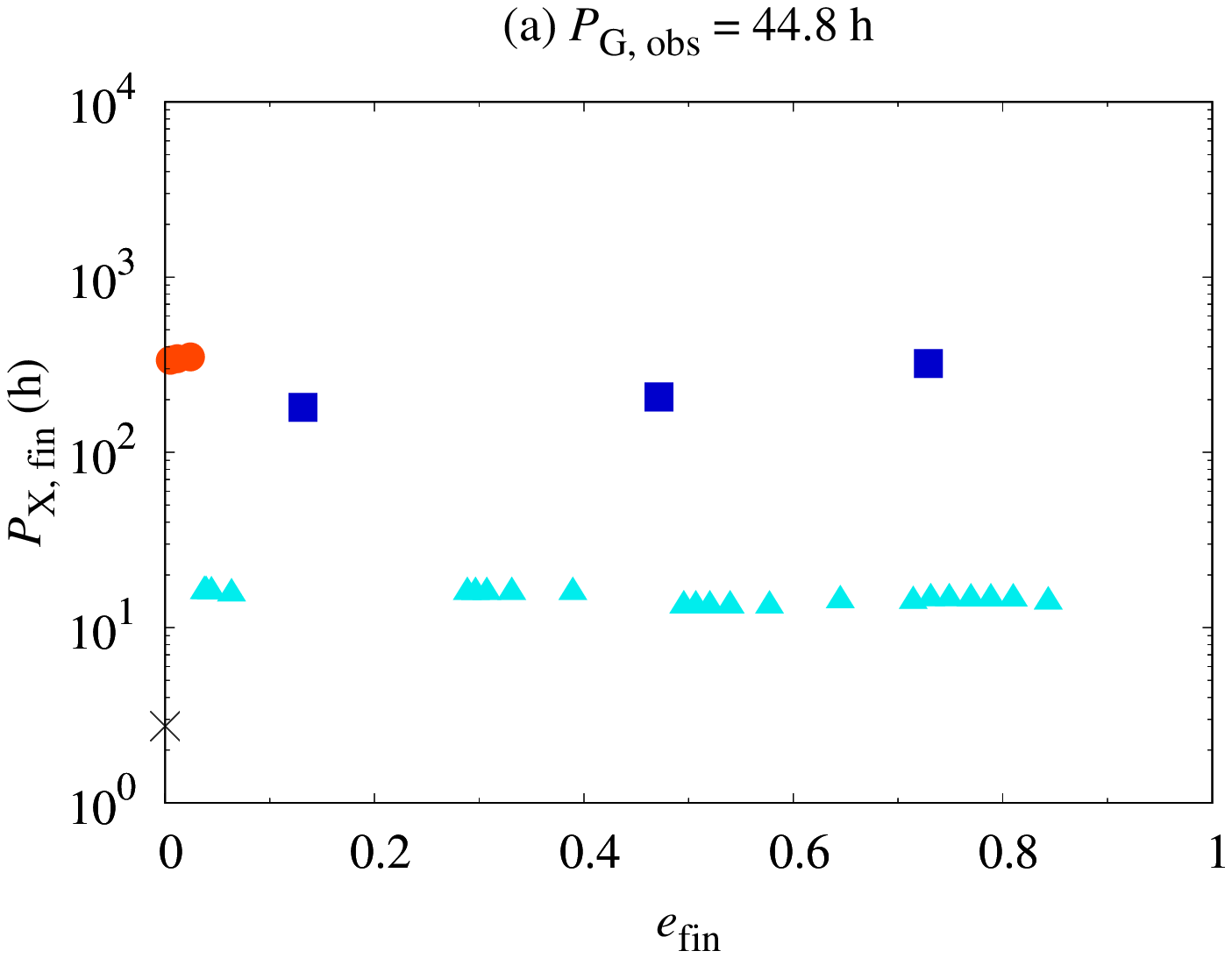}
\includegraphics[width = 0.45\textwidth]{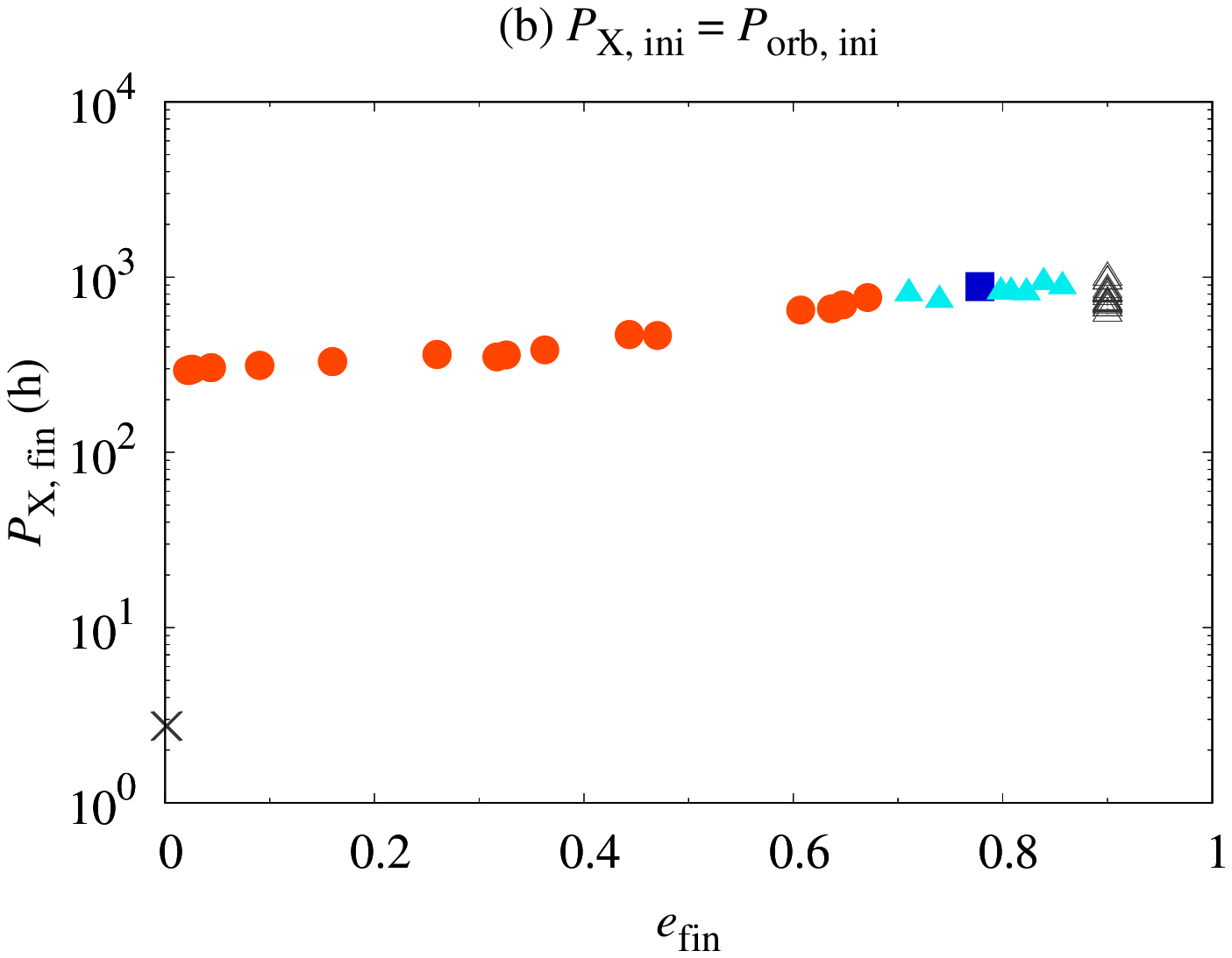}
\includegraphics[width = 0.45\textwidth]{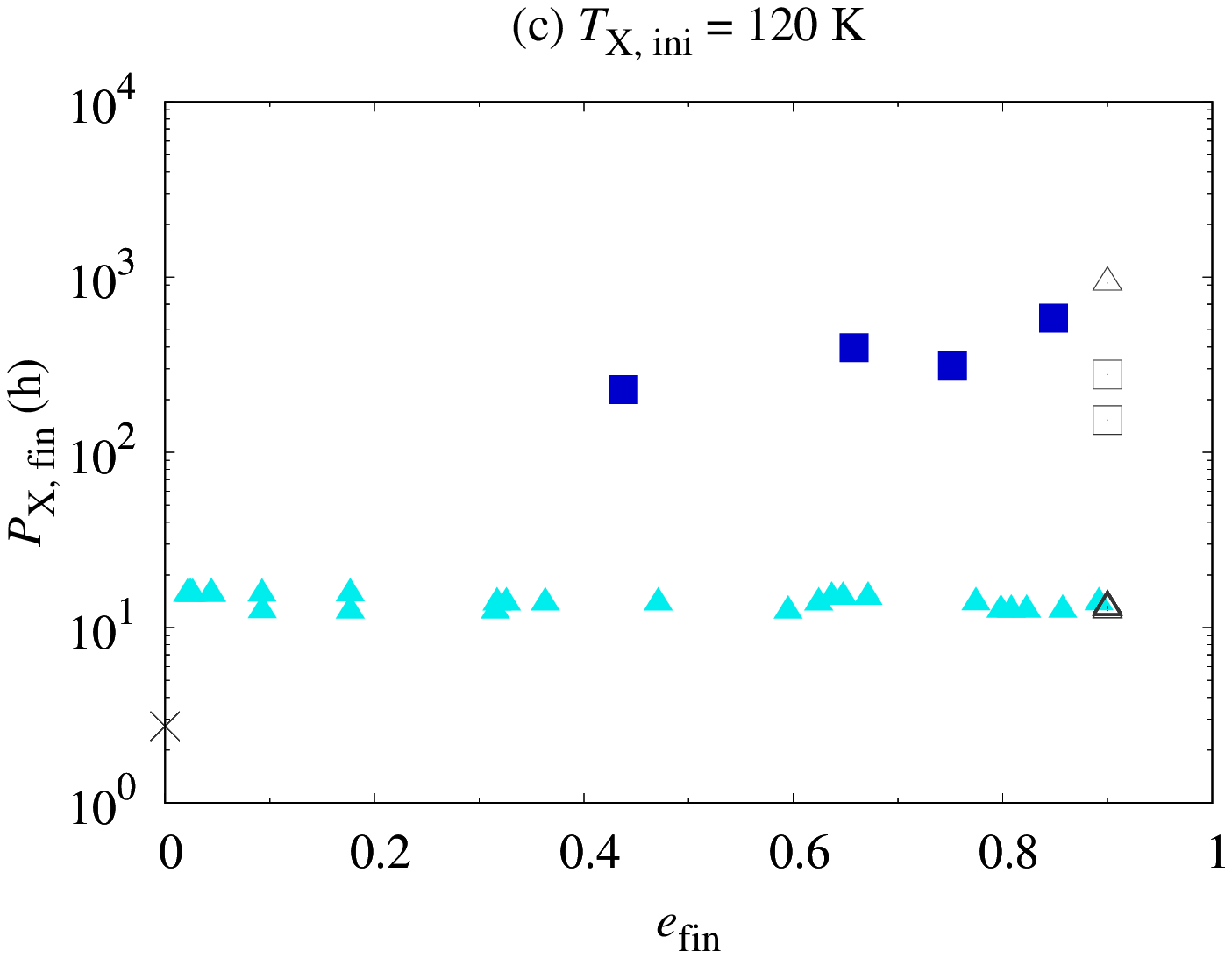}
\includegraphics[width = 0.45\textwidth]{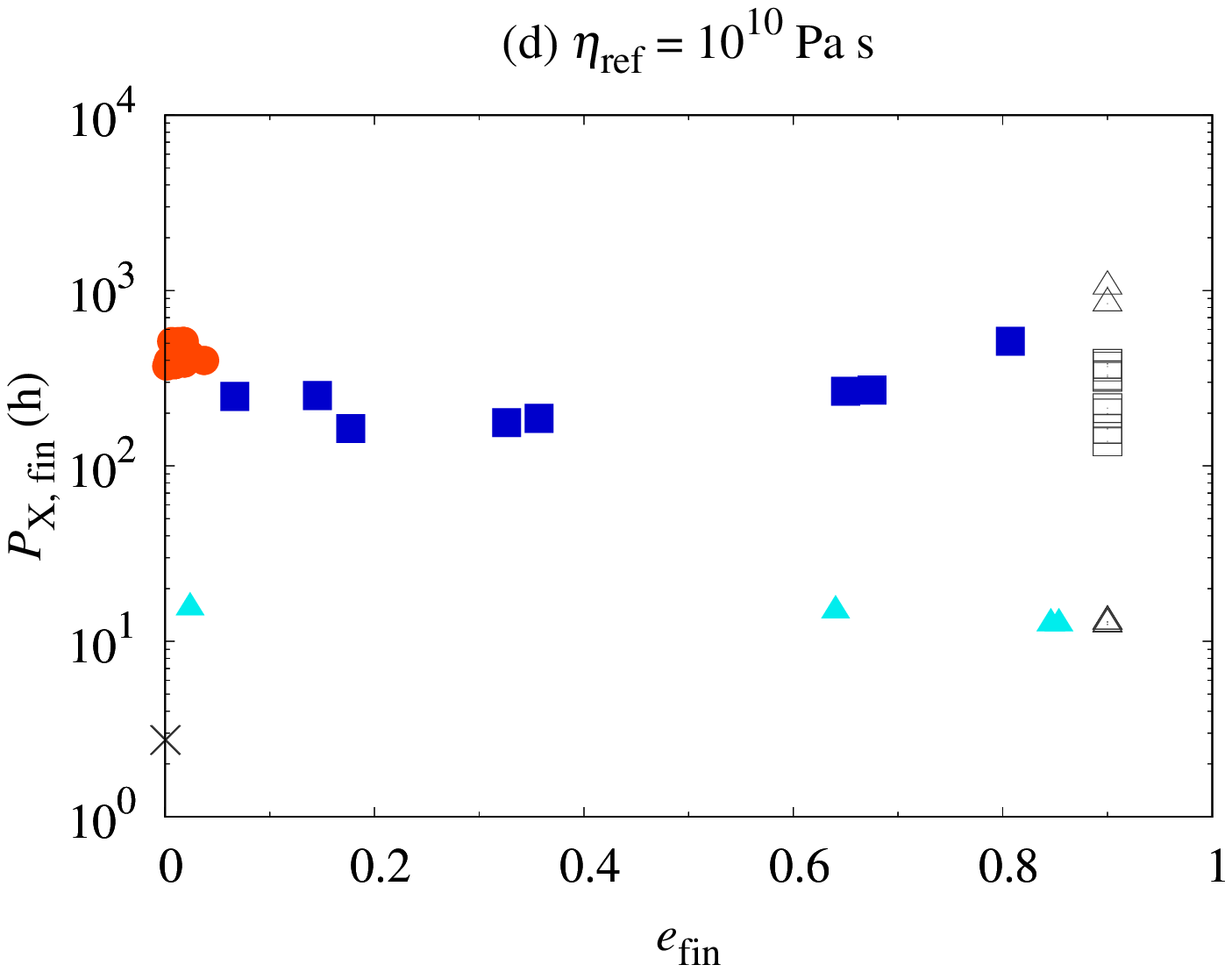}
\caption{
Distribution of the final eccentricity and spin period of the secondary, $e_{\rm fin}$ and $P_{\rm X, fin}$, for non-standard runs of our simulation.
}
\label{figp_ecc2}
\end{figure*}

\begin{figure*}
\centering
\includegraphics[width = 0.45\textwidth]{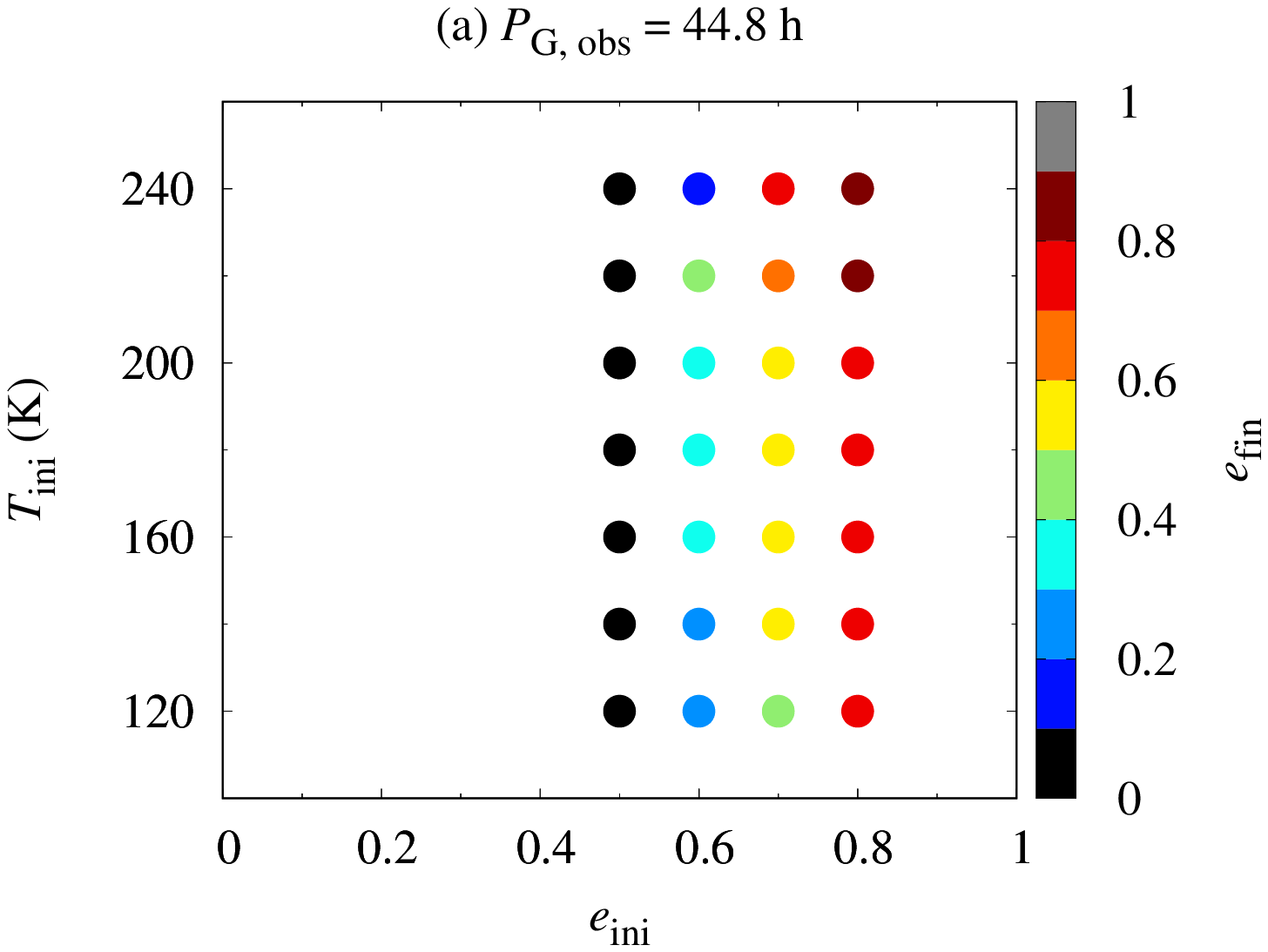}
\includegraphics[width = 0.45\textwidth]{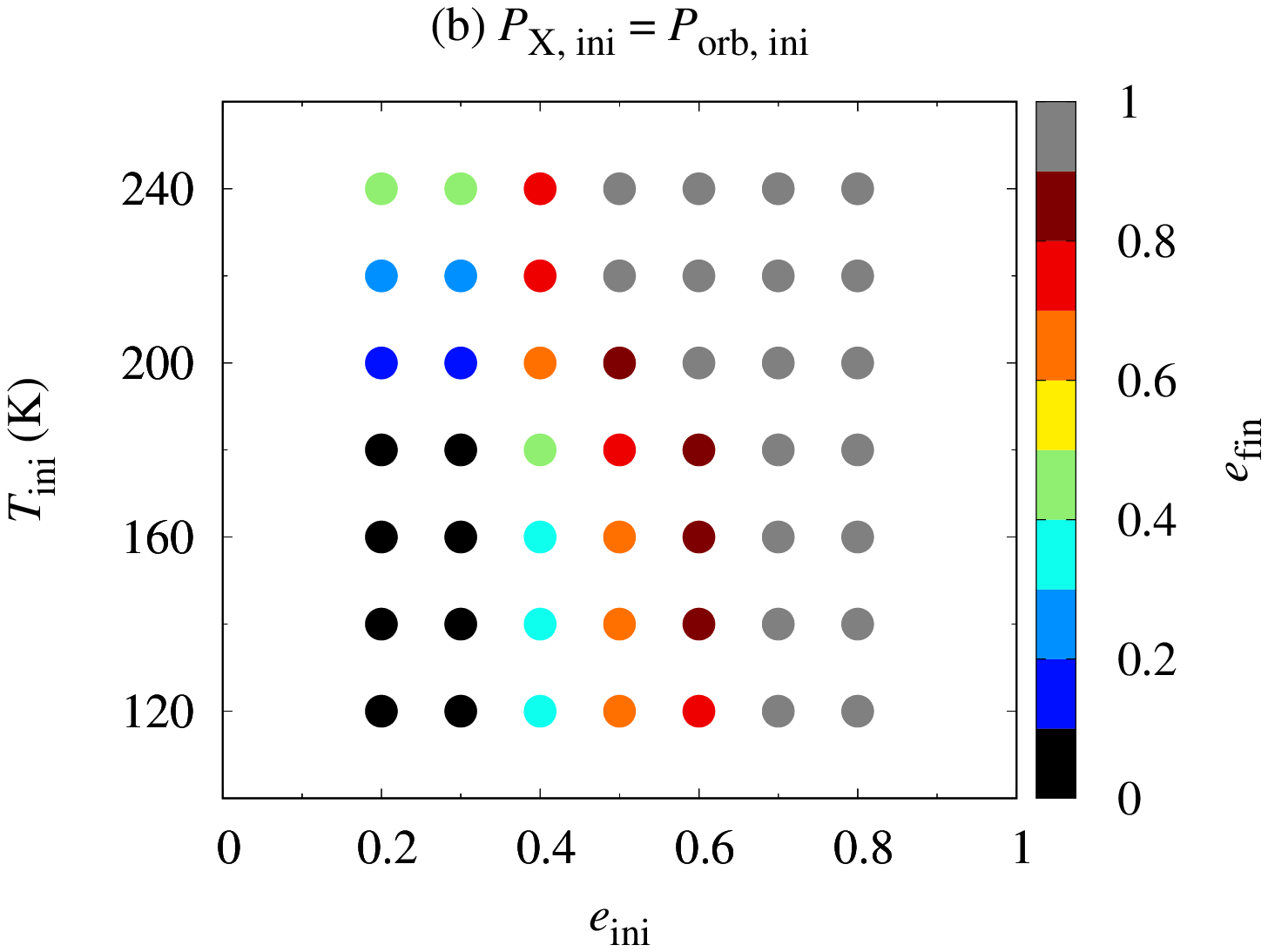}
\includegraphics[width = 0.45\textwidth]{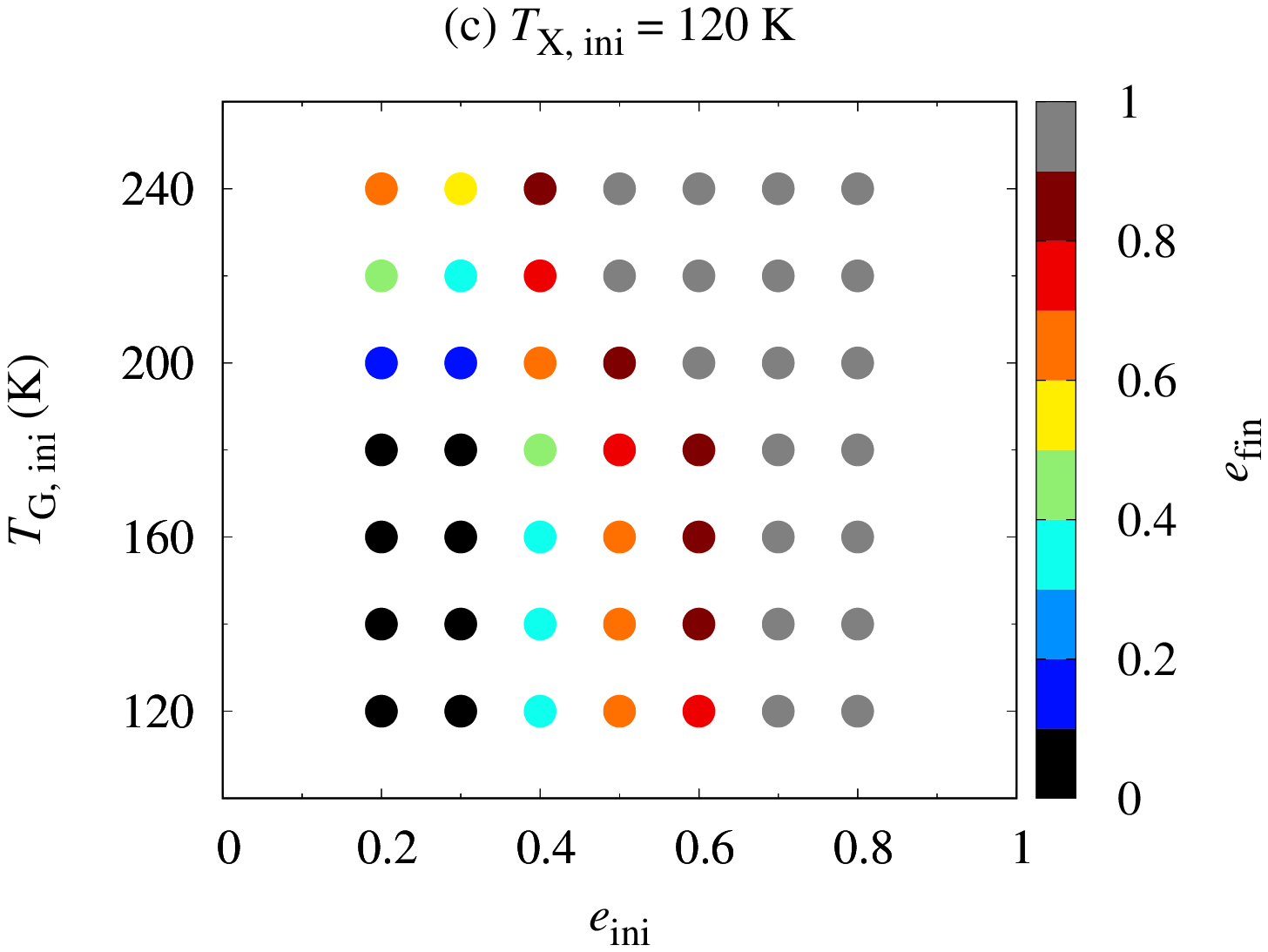}
\includegraphics[width = 0.45\textwidth]{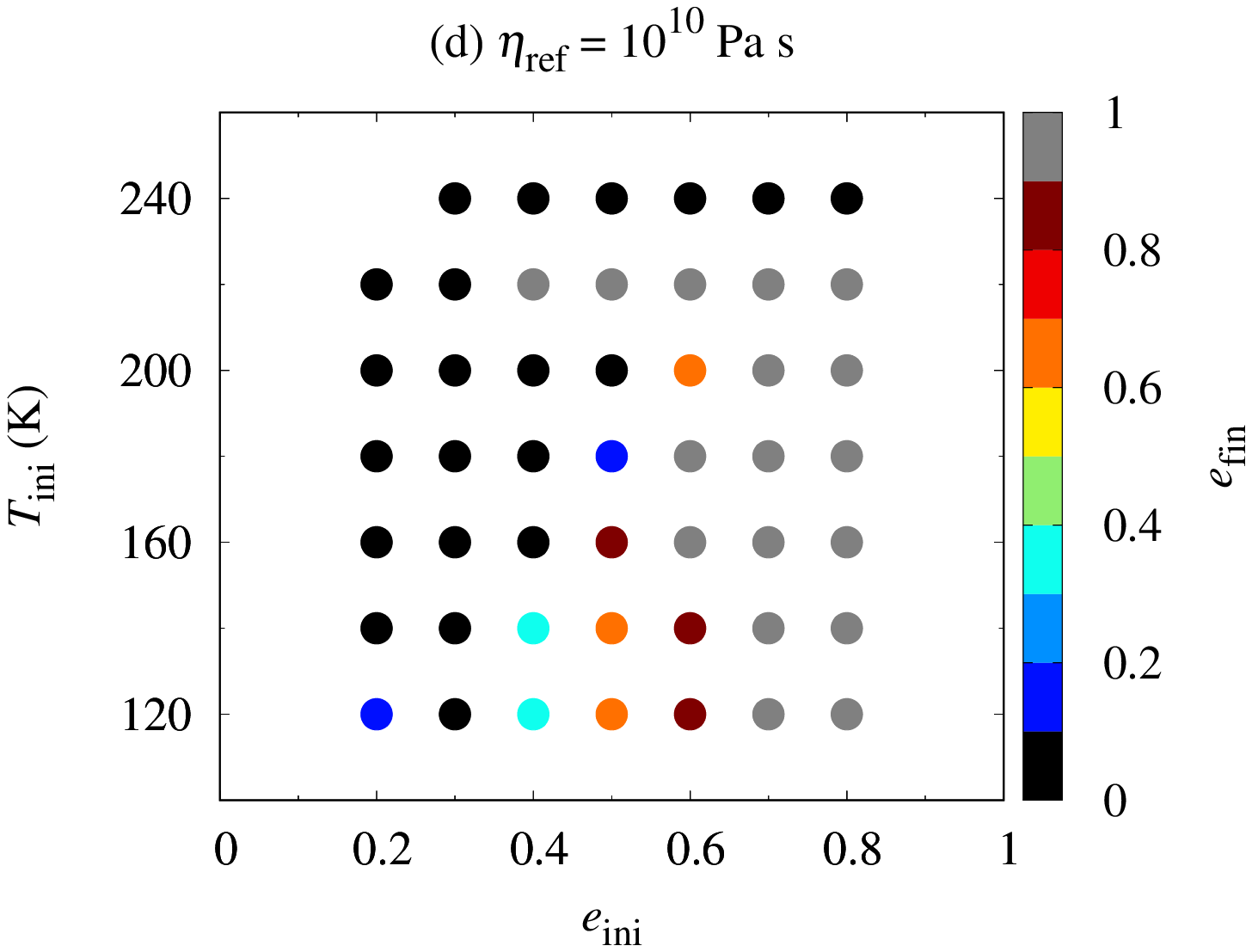}
\caption{
Color maps for the final eccentricity of the system, $e_{\rm fin}$, for non-standard runs of our simulation.
}
\label{figeccfin2}
\end{figure*}

In Section \ref{sec.distribution}, we show the distributions of the final spin/orbital properties of Gonggong--Xiangliu system for our standard setting (i.e., $P_{\rm G, obs} = 22.4\ {\rm h}$, $P_{\rm X, ini} = 12\ {\rm h}$, $T_{\rm X, ini} = T_{\rm G, ini} = T_{\rm ini}$, and $\eta_{\rm ref} = 10^{14}\ {\rm Pa}\ {\rm s}$).
In Section \ref{sec.discussion}, we discuss the effect of the non-standard setting on the results of the tidal evolution.
We investigate four cases in this section: (a) a slow spin of Gonggong ($P_{\rm G, obs} = 44.8\ {\rm h}$ instead of $22.4\ {\rm h}$), (b) an initially tidally-synchronized Xiangliu ($P_{\rm X, ini} = P_{\rm orb, ini}$ instead of $12\ {\rm h}$), (c) a cold start of Xiangliu ($T_{\rm X, ini} = 120\ {\rm K}$ instead of $T_{\rm G, ini}$), and (d) undifferentiated bodies with low reference viscosity ($\eta_{\rm ref} = 10^{10}\ {\rm Pa}\ {\rm s}$ instead of $10^{14}\ {\rm Pa}\ {\rm s}$).
Figures \ref{figType2}, \ref{figa_ecc2}, \ref{figp_ecc2}, and \ref{figeccfin2} are the results of the non-standard runs of our simulations.
In these simulations, we set $R_{\rm X} = 60\ {\rm km}$.

\subsection{Slow-spinning Gonggong}

In our standard models, we set the current spin period of Gonggong as $P_{\rm G, obs} = 22.4\ {\rm h}$.
The spin period of Gonggong was obtained from the light-curve analysis by \citet{Pal+2016} and has two possible values: $P_{\rm G, obs} = 22.4\ {\rm h}$ or $44.8\ {\rm h}$.
In Section \ref{sec.iniGX}, we discuss the impact of $P_{\rm G, obs}$ on the location of the initial corotation radius.
We showed that a large $e_{\rm ini}$ is necessary to avoid the inward migration of Xiangliu when $P_{\rm G, obs} = 44.8\ {\rm h}$; however, a non-zero value of $e_{\rm ini}$ is required even for the case of $P_{\rm G, obs} = 22.4\ {\rm h}$.
As we cannot exclude the possibility of $P_{\rm G, obs} = 44.8\ {\rm h}$ from the light-curve analyses, we discuss the case in this section.

Panels (a) in Figures \ref{figType2}, \ref{figa_ecc2}, \ref{figp_ecc2}, and \ref{figeccfin2} show the results for the case of $P_{\rm G, obs} = 44.8\ {\rm h}$.
Figure \ref{figType2}(a) shows the satellite system results in Type Z for an initial eccentricity of $e_{\rm ini} \le 0.4$.
The critical value of $e_{\rm ini}$ to avoid the inward migration of Xiangliu is significantly larger than that for our standard runs (see Figure \ref{figType}(c)), which is consistent with our prediction from the analytical estimate of the corotation radius in Section \ref{sec.iniGX}.

Figure \ref{figa_ecc2}(a) shows the distribution of $e_{\rm fin}$ and $a_{\rm fin}$.
The final eccentricity of satellite systems classified as Type B and Type C is in the wide range \added{($0 < e_{\rm fin} < 0.9$)}, as in the case shown in Figure \ref{figa_ecc}(c).
The final semilatus rectum was slightly lower than the observed value for Gonggong--Xiangliu system.
Figure \ref{figp_ecc2}(a) shows the distribution of $e_{\rm fin}$ and $P_{\rm X, fin}$.
The general trend is similar to that of the standard shown in Figure \ref{figp_ecc}(c).

Figure \ref{figeccfin2}(a) is the color map for $e_{\rm fin}$.
In the case of $P_{\rm G, obs} = 44.8\ {\rm h}$, satellite systems with an initial eccentricity of $e_{\rm ini} \simeq 0.6$ tend to obtain a final eccentricity of $0.2 \le e_{\rm fin} \le 0.5$.
Systems with large $e_{\rm ini}$ tend to result in a large $e_{\rm fin}$; this trend is similar to that observed in the standard runs (see Figure \ref{figeccfin}(c)); although the suitable range of $e_{\rm ini}$ to reproduce the observed value for Gonggong--Xiangliu system is different.

\subsection{Initially tidally-synchronized Xiangliu}

In our standard models, we set the initial spin period of Xiangliu as $P_{\rm X, ini} = 12\ {\rm h}$.
This assumption is based on numerical simulations of moon-forming giant imapacts of 1000-km-sized TNOs (see Figure \ref{figPini}(b)).
In this case, $P_{\rm X, ini} \ll P_{\rm orb, ini}$ and a large number of runs are trapped in higher-order spin--orbit resonances (see Figure \ref{figType}).
In this section, we instead set $P_{\rm X, ini} = P_{\rm orb, ini}$.
Although the assumption of $P_{\rm X, ini} = P_{\rm orb, ini}$ seems unrealistic, we discuss the effects of the initially synchronized Xiangliu on the tidal evolution of the system to better understand our results.

Panels (b) in Figures \ref{figType2}, \ref{figa_ecc2}, \ref{figp_ecc2}, and \ref{figeccfin2} show the results for the case of $P_{\rm X, ini} = P_{\rm orb, ini}$.
Figure \ref{figType2}(b) shows that the final state of the system is not Type A (i.e., 1:1 spin--orbit resonance) in some cases even if the initial spin period of the secondary is $P_{\rm X, ini} = P_{\rm orb, ini}$.

Figure \ref{figa_ecc2}(b) shows the distribution of $e_{\rm fin}$ and $a_{\rm fin}$.
For the case of $P_{\rm X, ini} = P_{\rm orb, ini}$, the final eccentricity is in the wide range, even for Type A \added{($0 < e_{\rm fin} < 0.7$)}.
We note, however, that the assumption of $P_{\rm X, ini} = P_{\rm orb, ini}$ is unrealistic if they are formed via giant impacts (see Figure \ref{figPini}(b)).
Figure \ref{figp_ecc2}(b) shows the distribution of $e_{\rm fin}$ and $P_{\rm X, fin}$.
In this case, the final spin period of Xiangliu ranges from $10^{2}\ {\rm h} \lesssim P_{\rm X, fin} \lesssim 10^{3}\ {\rm h}$.
In addition, all satellite systems with $e_{\rm fin} \le 0.6$ are classified as Type A.

Figure \ref{figeccfin2}(b) is the color map for $e_{\rm fin}$.
In the case of $P_{\rm X, ini} = P_{\rm orb, ini}$, satellite systems with an initial eccentricity of $0.2 \le e_{\rm ini} \le 0.4$ tend to obtain a final eccentricity of $0.2 \le e_{\rm fin} \le 0.5$.
This result is similar to that of our standard runs (see Figure \ref{figeccfin}).

\subsection{Cold start for Xiangliu}
\label{sec.coldstart}

In our standard models, we assumed that the initial temperatures of Gonggong and Xiangliu are the same: $T_{\rm G, ini} = T_{\rm G, ini} = T_{\rm ini}$.
Although this assumption is somewhat natural, we can imagine a case in which the initial temperature of Xiangliu is lower than that of Gonggong.
The motivation of this setting is as follows.

Giant impact simulations of 1000-km-sized TNOs \citep[e.g.,][]{Canup2005,Sekine+2017,Arakawa+2019} revealed that satellites are directly formed as intact fragments of the impactor (``intact moons'').
As the mass of the impactor is several times smaller than that of the target body, the cooling rate of the impactor would be larger than that of the target, and the temperature of the impactor might be lower than that of the target upon collision.
Although we need to assess the effect of the impact heating on the initial temperatures of the primary and secondary, the setting in which the initial temperature of Xiangliu is lower than that of Gonggong seems to be reasonable.

Panels (c) in Figures \ref{figType2}, \ref{figa_ecc2}, \ref{figp_ecc2}, and \ref{figeccfin2} show the results for the case of $T_{\rm X, ini} = 120\ {\rm K}$.
Figure \ref{figType2}(c) shows that no systems classified as Type A are formed in this case.

Figure \ref{figa_ecc2}(c) shows the distribution of $e_{\rm fin}$ and $a_{\rm fin}$, and Figure \ref{figp_ecc2}(c) shows the distribution of $e_{\rm fin}$ and $P_{\rm X, fin}$.
These distributions are generally similar to those for our standard runs (Figures \ref{figa_ecc}(c) and \ref{figp_ecc}(c)); however, no systems result in 1:1 spin--orbit resonance.

Figure \ref{figeccfin2}(c) is the color map for $e_{\rm fin}$.
In the case of $T_{\rm X, ini} = 120\ {\rm K}$, satellite systems with an initial eccentricity of $0.2 \le e_{\rm ini} \le 0.4$ tend to obtain a final eccentricity of $0.2 \le e_{\rm fin} \le 0.5$.
This result is also similar to those of our standard runs (see Figure \ref{figeccfin}(c)).

\subsection{Undifferentiated bodies made of soft ice}
\label{sec.softice}

In our standard models, we set the reference viscosity of icy bodies as $\eta_{\rm ref} = 10^{14}\ {\rm Pa}\ {\rm s}$.
This value is widely used in studies of differentiated icy bodies, including of the thermal evolution of Pluto \citep[e.g.,][]{Kamata+2019}, the orbital evolution of Pluto--Charon \citep[e.g.,][]{Renaud+2021}, and the tidal heating of Europa and Titan \citep[e.g.,][]{Tobie+2005}.

However, we note that the reference viscosity of icy bodies strongly depends on the grain size of ice crystals \citep[e.g.,][]{Goldsby+2001,Kubo+2006}, and $\eta_{\rm ref}$ for ice mixed with a small volume fraction of dust grains may be orders of magnitude lower than the canonical value for differentiated ice bodies \citep[e.g.,][]{Kubo+2009}.
When the ice viscosity is controlled by diffusion creep, the reference viscosity, $\eta_{\rm ref}$, at the reference temperature, $T_{\rm ref} = 273\ {\rm K}$, is given by \citep{Kalousova+2016}
\begin{equation}
\eta_{\rm ref} = \frac{{d_{\rm ice}}^{2}}{A} \exp{\left( \frac{E_{\rm a}}{R_{\rm gas} T_{\rm ref}} \right)},
\end{equation}
where $d_{\rm ice}$ is the grain size of the ice crystals, $A = 3.3 \times 10^{-10}\ {\rm Pa}^{-1}\ {\rm m}^{2}\ {\rm s}^{-1}$ is the pre-exponential constant, $R_{\rm gas} = 8.31\ {\rm J}\ {\rm K}^{-1}\ {\rm mol}^{-1}$ is the gas constant, and $E_{\rm a}$ is the activation energy.
Assuming $E_{\rm a} = 60\ {\rm kJ}\ {\rm mol}^{-1}$ \citep{Kamata+2019}, we obtain the following relation between $d_{\rm ice}$ and $\eta_{\rm ref}$:
\begin{equation}
\eta_{\rm ref} = 9.3 \times 10^{8}\ {\left( \frac{d_{\rm ice}}{1\ \textmu{\rm m}} \right)}^{2}\ {\rm Pa}\ {\rm s}.
\end{equation}
When undifferentiated icy bodies are mixtures of {\textmu}m-sized dust grains and matrix ices, the grain size of ice crystals would be maintained in the {\textmu}m-size owing to the Zener pinning effect \citep{Kubo+2009}.
Thus, the reference viscosity of $\eta_{\rm ref} \sim 10^{10}\ {\rm Pa}\ {\rm s}$ may be somewhat reasonable for undifferentiated icy bodies.

Panels (d) in Figures \ref{figType2}, \ref{figa_ecc2}, \ref{figp_ecc2}, and \ref{figeccfin2} show the results for the case of $\eta_{\rm ref} = 10^{10}\ {\rm Pa}\ {\rm s}$.
Figure \ref{figType2}(d) shows that runs with a small $e_{\rm ini}$ and high $T_{\rm ini}$ tend to result in Type A.

Figure \ref{figa_ecc2}(d) shows the distribution of $e_{\rm fin}$ and $a_{\rm fin}$, and Figure \ref{figp_ecc2}(d) shows the distribution of $e_{\rm fin}$ and $P_{\rm X, fin}$.
These distributions are similar to that of our standard runs.
The final semilatus rectum, $p_{\rm fin}$, for the case of $\eta_{\rm ref} = 10^{10}\ {\rm Pa}\ {\rm s}$ is not significantly different from that for the standard case of $\eta_{\rm ref} = 10^{14}\ {\rm Pa}\ {\rm s}$.
This is because the temperature of Gonggong, $T_{\rm G}$, also depends on $\eta_{\rm ref}$; and the viscosity of Gonggong, $\eta_{\rm G} = \eta {( T_{\rm G} )}$, is not strongly dependent on $\eta_{\rm ref}$ between the two settings, $\eta_{\rm ref} = 10^{10}\ {\rm Pa}\ {\rm s}$ and $10^{14}\ {\rm Pa}\ {\rm s}$.
Figure \ref{figtypeeta10} shows the thermal and orbital evolutions of Gonggong and Xiangliu for the case of $\eta_{\rm ref} = 10^{10}\ {\rm Pa}\ {\rm s}$, $T_{\rm ini} = 140\ {\rm K}$, and $e_{\rm ini} = 0.4$.

Figure \ref{figeccfin2}(d) is the color map for $e_{\rm fin}$.
In the case of $\eta_{\rm ref} = 10^{10}\ {\rm Pa}\ {\rm s}$, satellite systems with an initial temperature of $T_{\rm ini} \ge 160\ {\rm K}$ tend to result in Type A or Type Bx/Cx  (i.e., the final eccentricity of $e_{\rm fin} \le 0.1$ or $e_{\rm fin} \ge 0.9$).
Therefore, the initial temperature of $T_{\rm ini} \le 140\ {\rm K}$ might be suitable for explaining the formation of moderately eccentric satellite systems when $\eta_{\rm ref} = 10^{10}\ {\rm Pa}\ {\rm s}$.

\section{Summary}
\label{sec.summary}

Recent astronomical observations by \citet{Kiss+2019} revealed that (225088) Gonggong, a 1000-km-sized trans-Neptunian dwarf planet, hosts an eccentric satellite, Xiangliu, whose eccentricity is approximately $e_{\rm obs} = 0.3$.
As the majority of known satellite systems around large TNOs have circular orbits with $e_{\rm obs} \lesssim 0.1$, the observed eccentricity of Gonggong--Xiangliu system may reflect the singular properties of the system.
However, no detailed analysis of the orbital evolution of Gonggong--Xiangliu system has been conducted so far.

In this study, we investigated the secular tidal evolution of Gonggong--Xiangliu system under the simplifying assumption of homogeneous bodies, assuming that their initial orbital properties are those obtained from a giant impact (see Table 2).
We conducted the coupled thermal--orbital evolution simulations using the Andrade viscoelastic model and by including higher-order eccentricity functions up to $| q | \le 200$ for ${\left[ G_{2, 0, q} {( e )} \right]}^{2}$ and ${\left[ G_{2, 1, q} {( e )} \right]}^{2}$. 
Our findings are summarized as follows.

\begin{enumerate}
\item{
A re-evaluation of giant impact simulations by \citet{Arakawa+2019} revealed that the initial semilatus rectum, $p_{\rm ini} = a_{\rm ini} {\left( 1 - {e_{\rm ini}}^{2} \right)}$, ranges from $3 < p_{\rm ini} / R_{\rm G} < 8$ when a giant impact forms a satellite as an intact fragment.
In addition, the initial spin period of secondaries after giant impacts, $P_{\rm X, ini}$, ranges from $3\ {\rm h} \lesssim P_{\rm X, ini} \lesssim 18\ {\rm h}$ (see Figure \ref{figPini}).
}
\item{
In Section \ref{sec.typical}, we reported four typical tidal evolution pathways of satellite systems: 1:1 spin--orbit resonance (Type A), higher-order spin--orbit resonance (Type B), non-resonance (Type C), and collision with the primary (Type Z).
}
\item{
Figure \ref{figType} shows the outcomes of the tidal evolution.
We found that Xiangliu migrates inward and finally collides with Gonggong (i.e., Type Z) when the initial eccentricity is too small.
The critical value of $e_{\rm ini}$ required for the outward migration of Xiangliu is clearly dependent on the radius of Xiangliu, $R_{\rm X}$; smaller satellites have a larger critical value of $e_{\rm ini}$.
In addition, runs with small $e_{\rm ini}$ and high $T_{\rm ini}$ tend to turn into Type A as a consequence of tidal evolution.
Table \ref{tableStat} shows the statistics of the final state for the standard runs of our simulation.
The fraction of each type is clearly dependent on the radius of the secondary.
}
\item{
We found that the final eccentricity is $e_{\rm fin} \lesssim 0.05$ for Type A.
Therefore Xiangliu would not be in 1:1 spin--orbit resonance (i.e., Type A) because it has a moderate eccentricity of $e_{\rm obs} = 0.3$ \citep{Kiss+2019}.
The final eccentricity of satellite systems classified as Type B and Type C is in the wide range \added{($0 < e_{\rm fin} < 0.9$)}.
However, that the fraction of the system whose final eccentricity is between $0.2 \le e_{\rm fin} \le 0.5$ strongly depends on the radius of Xiangliu, and no systems with $0.2 \le e_{\rm fin} \le 0.5$ were formed for the case of $120\ {\rm km}$.
These results suggest that the radius of Xiangliu would not be larger than $100\ {\rm km}$.
}
\item{
We also found a significant relationship between $a_{\rm fin}$ and $e_{\rm fin}$.
In Section \ref{sec.pfin}, we derived a simple formula for $p_{\rm fin}$; the final semilatus rectum is given by Equation (\ref{eqp}).
Figures \ref{figa_ecc} and \ref{figaprime} show that the radius of Xiangliu might be close to $100\ {\rm km}$ from the point of view of $p_{\rm fin}$.
}
\item{
From the aspect of the evolution of the eccentricity and semilatus rectum, our findings suggest that the radius of Xiangliu is approximately $100\ {\rm km}$.
This value is consistent with the estimate of \citet{Kiss+2017} and indicates that Gonggong and Xiangliu have similar albedos.
}
\item{
Figure \ref{figp_ecc} shows the distribution of the final eccentricity and spin period of the secondary, $e_{\rm fin}$ and $P_{\rm X, fin}$, for standard runs of our simulation.
The spin period of Xiangliu at $t = 4.5\ {\rm Gyr}$ is in the range of $10\ {\rm h} \lesssim P_{\rm X, fin} \lesssim 10^{3}\ {\rm h}$; the range of $P_{\rm X, fin}$ may depend on $R_{\rm X}$.
In the case of $R_{\rm X} = 20\ {\rm km}$, the final spin period of Xiangliu is in the range of $10\ {\rm h} \lesssim P_{\rm X, fin} \lesssim 10^{2}\ {\rm h}$ and is always shorter than the orbital period, $P_{\rm orb, fin}$.
In contrast, for $R_{\rm X} = 100\ {\rm km}$ and $120\ {\rm km}$, the final spin period of Xiangliu is in the range of $10^{2}\ {\rm h} \lesssim P_{\rm X, fin} \lesssim 10^{3}\ {\rm h}$, which is significantly longer than the initial spin period, $P_{\rm X, ini} = 12\ {\rm h}$.
Thus, the determination of the spin period of Xiangliu by future observations and by light curve analyses is necessary to evaluate the radius of Xiangliu.
}
\item{
We found that satellite systems with an initial eccentricity of $0.2 \le e_{\rm ini} \le 0.5$ tend to achieve a final eccentricity of $0.2 \le e_{\rm fin} \le 0.5$ (Figure \ref{figeccfin}).
In contrast, for initial eccentricity of $e_{\rm ini} \le 0.1$, the final eccentricity at $t = 4.5\ {\rm Gyr}$ is typically $e_{\rm fin} \le 0.1$, or the system become Type Z.
For initial eccentricity of $e_{\rm ini} \ge 0.6$, on the other hand, the final eccentricity at $t = 4.5\ {\rm Gyr}$ is typically $e_{\rm fin} \ge 0.5$ for the case of $R_{\rm X} \le 60\ {\rm km}$, and $e_{\rm fin} \ge 0.9$ or $e_{\rm fin} \le 0.1$ for the case of $R_{\rm X} \ge 80\ {\rm km}$.
Therefore, the initial eccentricity of Gonggong--Xiangliu system after the moon-forming giant impact may have ranged from $0.2 \lesssim e_{\rm ini} \lesssim 0.5$.
}
\end{enumerate}

Our results highlight the importance of coupled thermal--orbital evolution simulations using a realistic viscoelastic model with higher-order eccentricity functions.
We also note that both the thermal history and the strength of tides strongly depend on the internal structure of the bodies.
We assumed that the primary and secondary are undifferentiated homogeneous bodies for simplicity; however, 1000-km-sized TNOs, including Gonggong, might actually be differentiated.
In future studies, we should calculate the coupled thermal--orbital evolution of the differentiated bodies.
In this case, the effects of subsurface oceans are also of great interest.
The differentiated state of large TNOs is closely associated with their accretion history, which is the key to understanding the planet formation in the outer solar system.
Thus, more detailed analyses of the coupled thermal--orbital evolution of satellite systems around large TNOs are necessary.

\acknowledgments

S.A.\ was supported by JSPS KAKENHI Grants Nos.\ JP17J06861 and JP20J00598.
R.H.\ was supported by JSPS KAKENHI Grants Nos.\ JP17J01269 and JP18K13600.
R.H.\ also acknowledges JAXA's International Top Young program.
H.G.\ acknowledges the financial support of MEXT KAKENHI Grant No.\ JP17H06457.
This work was supported by the Publications Committee of NAOJ.

\appendix
\restartappendixnumbering

\section{Eccentricity Function}
\label{app.G2pq}

The eccentricity functions are the Cayley expansions for the solutions of the Keplerian motion \citep{Cayley1861}.
The eccentricity functions, $G_{2, p, q} {( e )}$, can be calculated as follows \citep[see][for details]{Ferraz-Mello2013}:
\begin{eqnarray}
G_{2, p, q} {( e )} & = & \frac{1}{2 \pi} \int_{0}^{2 \pi} {\rm d}\ell\ {\left( \frac{a}{r} \right)}^{3} \cos{\left[ {\left( 2 - 2 p \right)} \nu - {\left( 2 - 2 p + q \right)} \ell \right]}, \nonumber \\
& = & \frac{1}{2 \pi \sqrt{1 - e^{2}}} \int_{0}^{2 \pi} {\rm d}\nu\ \frac{a}{r} \cos{\left[ {\left( 2 - 2 p \right)} \nu - {\left( 2 - 2 p + q \right)} \ell \right]},
\label{eqG2pq}
\end{eqnarray}
where $r$ is the distance between the primary and secondary, $\nu$ is the true anomaly, and $\ell$ is the mean anomaly.
The relation between $r$ and $\nu$ is
\begin{equation}
r = \frac{a {\left( 1 - e^{2} \right)}}{1 + e \cos{\nu}},
\end{equation}
and the mean anomaly is given by
\begin{equation}
\ell = \int_{0}^{\nu} {\rm d}\nu'\ \frac{\sqrt{1 - e^{2}}}{1 + e \cos{\nu'}}.
\end{equation}

\begin{figure*}
\centering
\includegraphics[width = 0.45\textwidth]{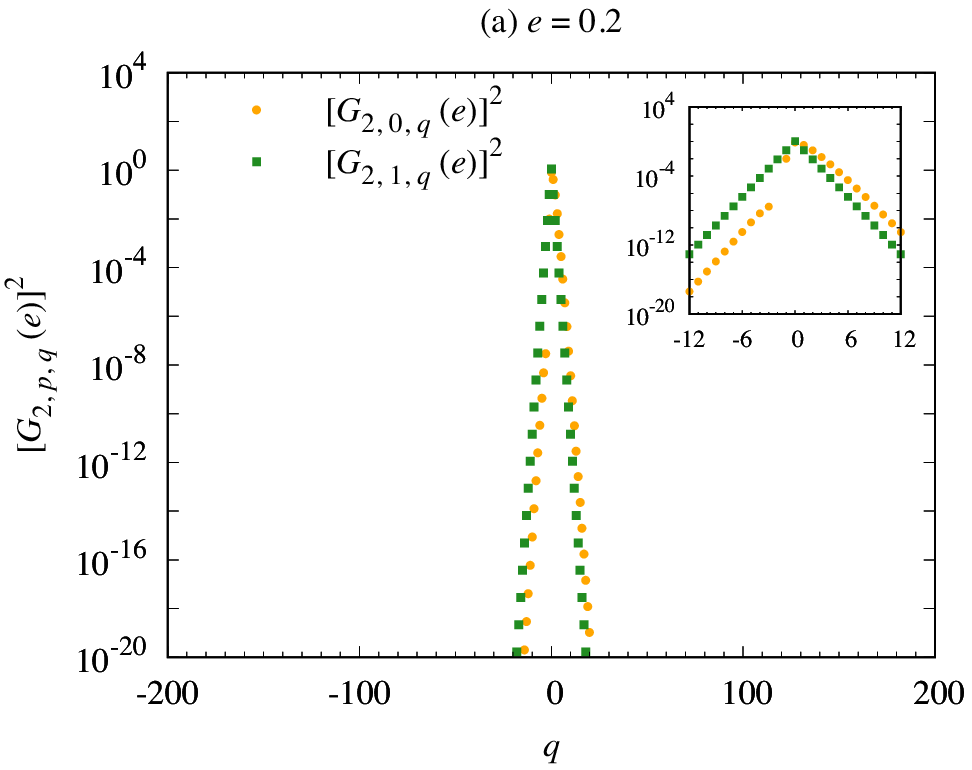}
\includegraphics[width = 0.45\textwidth]{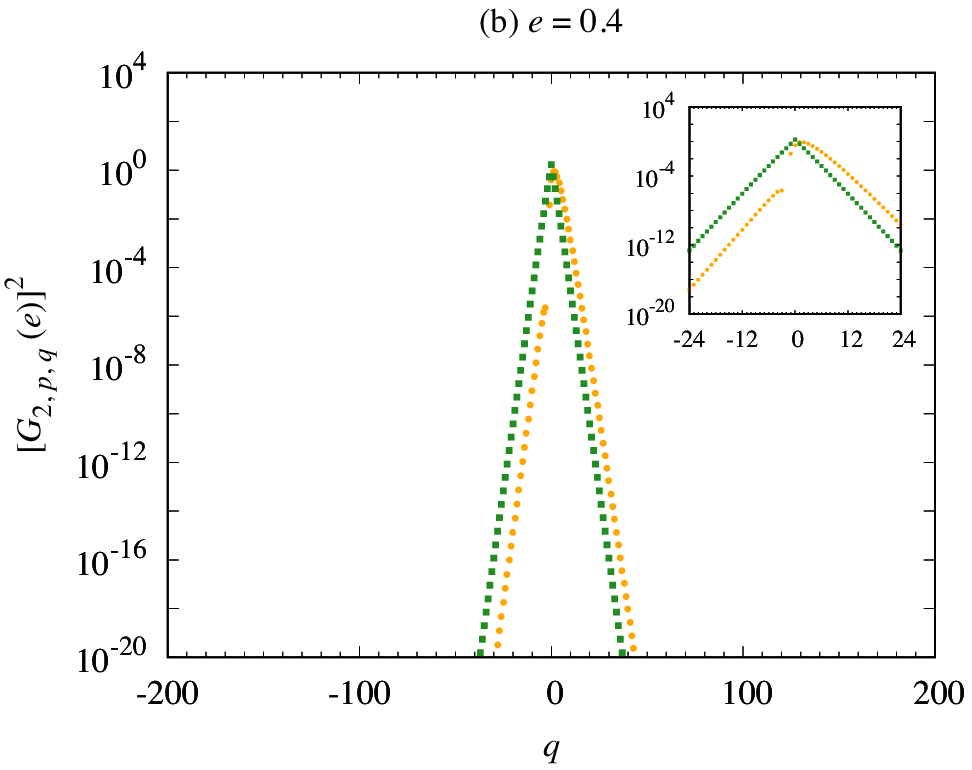}
\includegraphics[width = 0.45\textwidth]{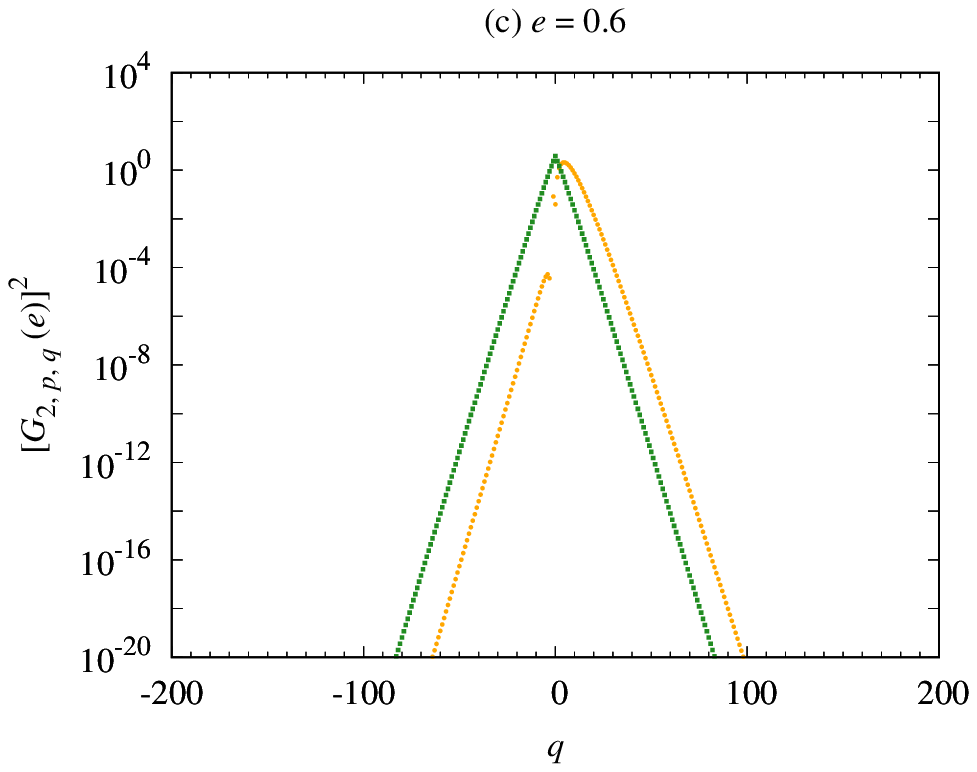}
\includegraphics[width = 0.45\textwidth]{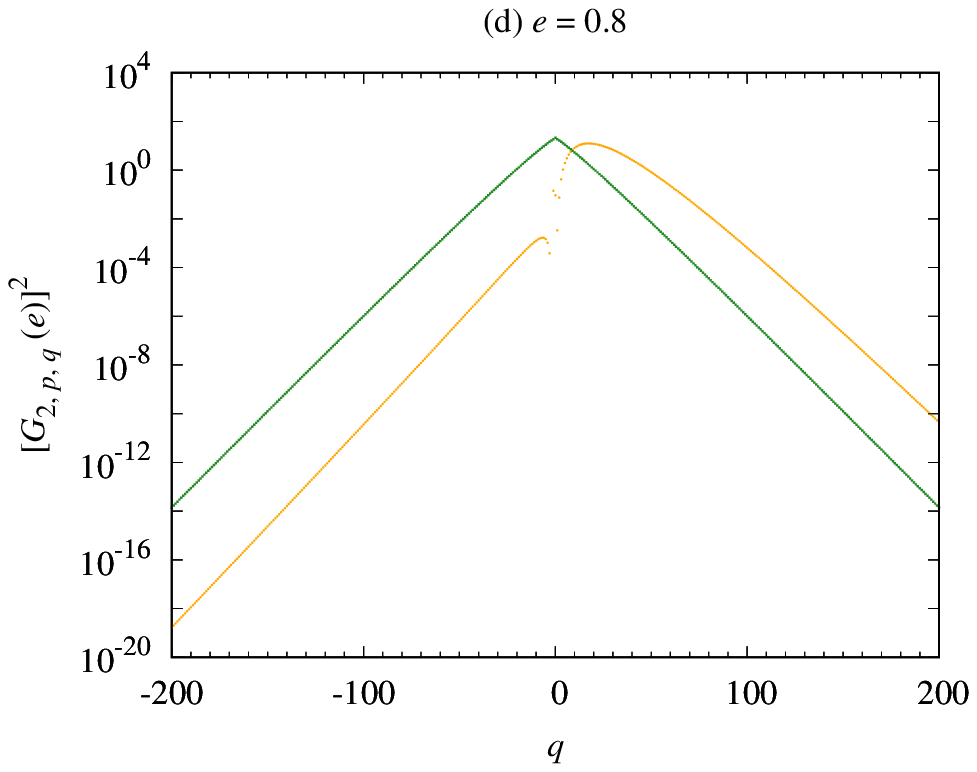}
\caption{Square of the eccentricity functions, ${\left[ G_{2, 0, q} {( e )} \right]}^{2}$ and ${\left[ G_{2, 1, q} {( e )} \right]}^{2}$ for different eccentricities: (a) $e = 0.2$, (b) $e = 0.4$, (c) $e = 0.6$, and (d) $e = 0.8$.
}
\label{figG2pq}
\end{figure*}

\citet{Ferraz-Mello2013} noted that the Fourier series solutions of $G_{2, p, q} {( e )}$ \citep[e.g.,][]{Efroimsky2012b,Renaud+2021} are based on the Taylor expansion of the solution of Kepler's equation, whose convergence radius is $e^{*} = 0.6627434$ \citep[see][]{Wintner1941,Hagihara1970,Murray+1999}.
Therefore we should calculate $G_{2, p, q} {( e )}$ from direct integration of Equation (\ref{eqG2pq}) when we consider the tidal evolution of highly eccentric systems with $e \gg 0.5$.
Figure \ref{figG2pq} shows the square of the eccentricity functions, ${\left[ G_{2, 0, q} {( e )} \right]}^{2}$ and ${\left[ G_{2, 1, q} {( e )} \right]}^{2}$, as functions of the eccentricity $e$ and the order $q$.

Table 2 of \citet{Renaud+2021} shows the series solutions of ${\left[ G_{2, p, q} {( e )} \right]}^{2}$, with eccentricity terms up to and including $e^{10}$.
In the range of $|q| \leq 2$, the series solutions of ${\left[ G_{2, 0, q} {( e )} \right]}^{2}$ and ${\left[ G_{2, 1, q} {( e )} \right]}^{2}$ are given by \citep{Boue+2019}
\begin{eqnarray}
{\left[ G_{2, 0, -2} {( e )} \right]}^{2} & = & 0, \\
{\left[ G_{2, 0, -1} {( e )} \right]}^{2} & = & \frac{1}{4} e^{2} - \frac{1}{16} e^{4} + {\mathcal O} {( e^{6} )}, \\
{\left[ G_{2, 0, 0} {( e )} \right]}^{2} & = & 1 - 5 e^{2} + \frac{63}{8} e^{4} + {\mathcal O} {( e^{6} )}, \\
{\left[ G_{2, 0, 1} {( e )} \right]}^{2} & = & \frac{49}{4} e^{2} - \frac{861}{16} e^{4} + {\mathcal O} {( e^{6} )}, \\
{\left[ G_{2, 0, 2} {( e )} \right]}^{2} & = & \frac{289}{4} e^{4} + {\mathcal O} {( e^{6} )}, \\
{\left[ G_{2, 1, -2} {( e )} \right]}^{2} & = & {\left[ G_{2, 1, 2} {( e )} \right]}^{2} = \frac{81}{16} e^{4} + {\mathcal O} {( e^{6} )}, \\
{\left[ G_{2, 1, -1} {( e )} \right]}^{2} & = & {\left[ G_{2, 1, 1} {( e )} \right]}^{2} = \frac{9}{4} e^{2} + \frac{81}{16} e^{4} + {\mathcal O} {( e^{6} )}, \\
{\left[ G_{2, 1, 0} {( e )} \right]}^{2} & = & {\left( 1 - e^{2} \right)}^{- 3}.
\end{eqnarray}

\begin{figure}
\centering
\includegraphics[width = 0.45\textwidth]{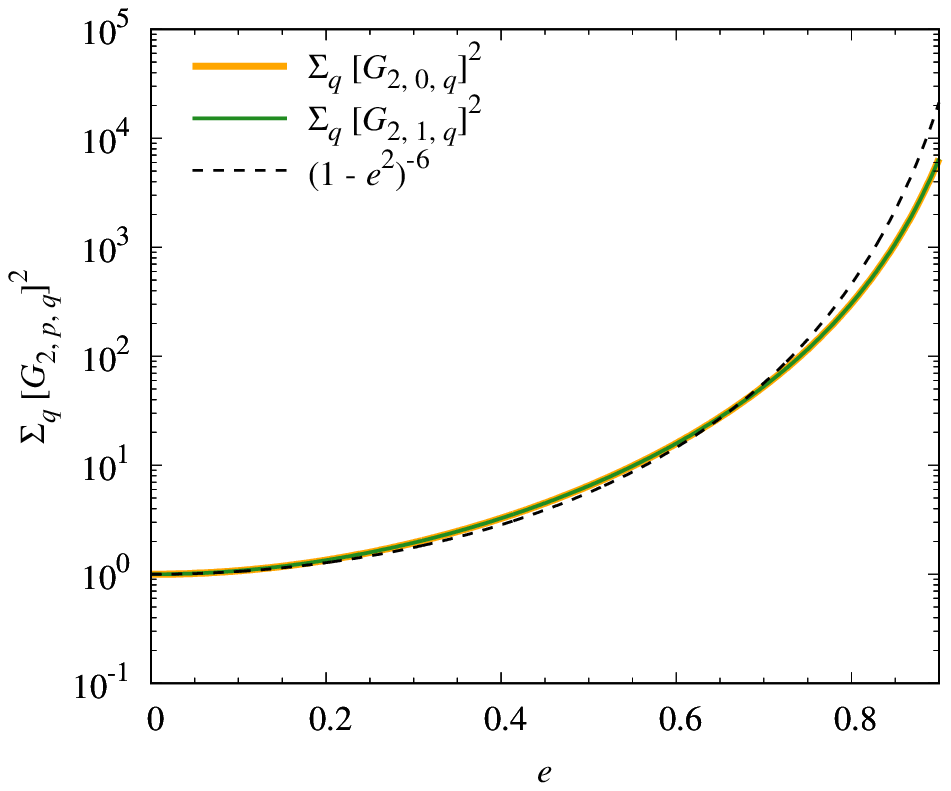}
\caption{
Summations of ${\left[ G_{2, 0, q} {( e )} \right]}^{2}$ and ${\left[ G_{2, 1, q} {( e )} \right]}^{2}$ from $q = - 200$ to $+ 200$.
The dashed line indicates ${\left( 1 - e^{2} \right)}^{-6}$ and the summations of ${\left[ G_{2, 0, q} {( e )} \right]}^{2}$, and ${\left[ G_{2, 1, q} {( e )} \right]}^{2}$ are well approximated by ${\left( 1 - e^{2} \right)}^{-6}$ in the range of $0 \le e \le 0.8$.
}
\label{figG_ecc}
\end{figure}

Figure \ref{figG_ecc} shows the summations of ${\left[ G_{2, 0, q} {( e )} \right]}^{2}$ and ${\left[ G_{2, 1, q} {( e )} \right]}^{2}$ from $q = - 200$ to $+ 200$.
This wide range of $q$ allows sufficient summation convergence.
Both $\sum_{q} {\left[ G_{2, 0, q} {( e )} \right]}^{2}$ and $\sum_{q} {\left[ G_{2, 1, q} {( e )} \right]}^{2}$ are approximately given by
\begin{equation}
\sum_{q = - \infty}^{+ \infty} {\left[ G_{2, 0, q} {( e )} \right]}^{2} = \sum_{q = - \infty}^{+ \infty} {\left[ G_{2, 1, q} {( e )} \right]}^{2} \simeq {\left( 1 - e^{2} \right)}^{-6},
\end{equation}
for the range of $0 \le e \le 0.8$.

\section{Rheological Model}
\label{app.rheology}

\subsection{Complex shear modulus}

Viscoelastic models are widely used to describe the tidal response of small icy bodies.
The Andrade model is an empirical model based on experimental results for the viscous flow of metals and ice \citep[e.g.,][]{Andrade1910,Glen1955}.
The model is widely used to compute tidal dissipation within icy satellites.
For example, \citet{Shoji+2013} calculated the tidal heating of Enceladus using the Andrade model.
\citet{Neveu+2019} also performed numerical simulations of coupling thermal, geophysical, and orbital evolution using the model.

The Andrade viscoelastic model contains an anelastic part, and the effect of elasticity decreases in the high-frequency region.
The complex shear modulus, $\tilde{\mu} {( \omega )}$, is expressed in terms of a creep function as follows:
\begin{eqnarray}
\tilde{\mu} {( \omega )} = \frac{1}{\tilde{J} {( \omega )}},
\end{eqnarray}
where $\tilde{J} {( \omega )}$ is the creep function, and $\omega$ is the angular velocity of the forcing cycle.
The creep function is given by \citep[e.g.,][]{Efroimsky2012a,Efroimsky2012b}
\begin{eqnarray}
\tilde{J} {( \omega )} = {\left[ \frac{1}{\mu} + \frac{1}{\mu {\left( \tau_{\rm A} \omega \right)}^{\alpha}} \cos{\left( \frac{\alpha \pi}{2} \right)} \Gamma {\left( \alpha + 1 \right)} \right]} - i {\left[ \frac{1}{\eta \omega} + \frac{1}{\mu {\left( \tau_{\rm A} \omega \right)}^{\alpha}} \sin{\left( \frac{\alpha \pi}{2} \right)} \Gamma {\left( \alpha + 1 \right)} \right]},
\end{eqnarray}
where $\mu$ is the shear modulus, $\eta$ is the viscosity, and $0 < \alpha < 1$ is the Andrade exponent.
We assume $\alpha = 0.33$ for ice \citep[e.g.,][]{Rambaux+2010}.
The relaxation time of the Andrade model is given by $\tau_{\rm A} = \eta / \mu$.

We note that ${\tilde{\mu}}$ depends on a dimensionless parameter, $\tau_{\rm A} \omega$, and can be divided into two regions.
For the case of $\tau_{\rm A} \omega \gg 1$, the complex shear modulus is approximately given by
\begin{equation}
\tilde{\mu} {( \omega )} \simeq \mu {\left[ 1 + i {\left( \tau_{\rm A} \omega \right)}^{- \alpha} \sin{\left( \frac{\alpha \pi}{2} \right)} \Gamma {\left( \alpha + 1 \right)} \right]},
\end{equation}
and ${| \tilde{\mu} {( \omega )} |} / \mu \simeq 1$, respectively.
On the other hand, when $\tau_{\rm A} \omega \ll 1$, the complex shear modulus is approximately given by
\begin{equation}
\tilde{\mu} {( \omega )} \simeq \eta \omega {\left[ {\left( \tau_{\rm A} \omega \right)}^{1 - \alpha} \cos{\left( \frac{\alpha \pi}{2} \right)} \Gamma {\left( \alpha + 1 \right)} + i \right]},
\end{equation}
and ${| \tilde{\mu} {( \omega )} |} / \mu \simeq \tau_{\rm A} \omega$, respectively.


\subsection{Love number}

For a homogeneous spherical body, the tidal potential Love number, $\tilde{k}_{2} {( \omega )}$, is given by
\begin{eqnarray}
\tilde{k}_{2} {( \omega )} = \frac{3 \rho g R}{2 \rho g R + 19 {\tilde{\mu} {( \omega )}}} = \frac{3 / 2}{1 + \mu_{\rm eff} {\left( {\tilde{\mu}} / {\mu} \right)}},
\end{eqnarray}
where $g = {\mathcal G} M / R^{2}$ is the surface gravity, and ${\mathcal G}$ is the gravitational constant, $\rho$ and $R$ are the density and radius of the body, respectively.
We introduce the (dimensionless) effective rigidity, $\mu_{\rm eff}$, as follows \citep[e.g.,][]{Goldreich+2009,Cheng+2014}:
\begin{equation}
\mu_{\rm eff} \equiv \frac{19 \mu}{2 \rho g R},
\end{equation}
and we found that $\mu_{\rm eff} \sim 10^{2}$ for 1000-km-sized TNOs.
The imaginary part of the complex Love number is negative when $\omega$ is positive.
When the angular velocity of the forcing cycle is negative, the complex Love number is given by
\begin{eqnarray}
\tilde{k}_{2} {( - \omega )} = \overline{\tilde{k}_{2} {( \omega )}},
\end{eqnarray}
where $\overline{\tilde{k}_{2} {( \omega )}}$ is the complex conjugate of $\tilde{k}_{2} {( \omega )}$.

In the case of $\mu_{\rm eff} \gg 1$, the tidal potential Love number is classified into three regions: (i) high-frequency region, (ii) intermediate region, and (iii) low-frequency regions \citep[see also][]{Efroimsky2012a}.
We found that ${\left| {\rm Im}{\left[ \tilde{k}_{2} {( \omega )} \right]} \right|}$ reaches a maximum at $\tau_{\rm A} \omega \simeq \pm {\mu_{\rm eff}}^{-1}$; the maximum value is ${\left| {\rm Im}{\left[ \tilde{k}_{2} {( \omega )} \right]} \right|} \simeq 3/4$.

\subsubsection{High-frequency region}

When $\tau_{\rm A} \omega \gg 1$, the absolute value and the imaginary part of the complex Love number, ${\left| \tilde{k}_{2} {( \omega )} \right|}$ and ${\rm Im}{\left[ \tilde{k}_{2} {( \omega )} \right]}$, are approximately given by
\begin{eqnarray}
{\left| \tilde{k}_{2} {( \omega )} \right|} & \simeq & \frac{3}{2 \mu_{\rm eff}}, \\
{\rm Im}{\left[ \tilde{k}_{2} {( \omega )} \right]} & \simeq & - \frac{3}{2 \mu_{\rm eff}} {\left( \tau_{\rm A} \omega \right)}^{- \alpha} \sin{\left( \frac{\alpha \pi}{2} \right)} \Gamma {\left( \alpha + 1 \right)}.
\end{eqnarray}
The tidal quality factor, $\mathcal{Q} {( \omega )} \equiv {\left| {\rm Im}{\left[ \tilde{k}_{2} {( \omega )} \right]} / \tilde{k}_{2} {( \omega )} \right|}$, is also given by
\begin{equation}
\mathcal{Q} {( \omega )} \simeq {\left( \tau_{\rm A} \omega \right)}^{- \alpha} \sin{\left( \frac{\alpha \pi}{2} \right)} \Gamma {\left( \alpha + 1 \right)}.
\end{equation}

\subsubsection{Intermediate region}

When ${\mu_{\rm eff}}^{-1} \ll \tau_{\rm A} \omega \ll 1$, ${\left| \tilde{k}_{2} {( \omega )} \right|}$, ${\rm Im}{\left[ \tilde{k}_{2} {( \omega )} \right]}$, and $\mathcal{Q} {( \omega )}$ are approximately given by
\begin{eqnarray}
{\left| \tilde{k}_{2} {( \omega )} \right|} & \simeq & \frac{3}{2} {\left( \mu_{\rm eff} \tau_{\rm A} \omega \right)}^{- 1}, \\
{\rm Im}{\left[ \tilde{k}_{2} {( \omega )} \right]} & \simeq & - \frac{3}{2} {\left( \mu_{\rm eff} \tau_{\rm A} \omega \right)}^{- 1},
\end{eqnarray}
and
\begin{equation}
\mathcal{Q} {( \omega )} \simeq 1.
\end{equation}

\subsubsection{Low-frequency region}

When $\tau_{\rm A} \omega \ll {\mu_{\rm eff}}^{-1}$, ${\left| \tilde{k}_{2} {( \omega )} \right|}$, ${\rm Im}{\left[ \tilde{k}_{2} {( \omega )} \right]}$, and $\mathcal{Q} {( \omega )}$ are approximately given by
\begin{eqnarray}
{\left| \tilde{k}_{2} {( \omega )} \right|} & \simeq & \frac{3}{2}, \label{eq_abs_k2_low} \\
{\rm Im}{\left[ \tilde{k}_{2} {( \omega )} \right]} & \simeq & - \frac{3}{2} \mu_{\rm eff} \tau_{\rm A} \omega, \label{eq_im_k2_low}
\end{eqnarray}
and
\begin{equation}
\mathcal{Q} {( \omega )} \simeq \mu_{\rm eff} \tau_{\rm A} \omega. \label{eq_q_low}
\end{equation}

\section{Cooling and heating rates of the primary}
\label{app.coolingrate}

\added{
In this study, we assumed that the interior of the primary is homogeneous for simplicity, then we calculated the time evolution of the internal temperature.
We found that the temperature evolution is approximately given by the balance between the conduction/convection cooling and the decay heating.
Here we show the cooling and heating rates of the primary.}

\added{The cooling rate of the primary due to conduction/convection is given by
\begin{eqnarray}
\frac{{\rm d}T_{\rm con, G}}{{\rm d}t} & = & \min{\left( \frac{{\rm d}T_{\rm cond, G}}{{\rm d}t}, \frac{{\rm d}T_{\rm conv, G}}{{\rm d}t} \right)}, \\
\frac{{\rm d}T_{\rm cond, G}}{{\rm d}t} & = & \frac{Q_{\rm cond, G}}{M_{\rm G} c_{\rm G}}, \\
\frac{{\rm d}T_{\rm conv, G}}{{\rm d}t} & = & \frac{Q_{\rm conv, G}}{M_{\rm G} c_{\rm G}},
\end{eqnarray}
where ${{\rm d}T_{\rm cond, G}} / {{\rm d}t}$ and ${{\rm d}T_{\rm conv, G}} / {{\rm d}t}$ are the cooling rates due to conduction and convection, respectively.
Similarly, the heating rate of the primary due to decay the decay of long-lived radioactive elements is given by
\begin{equation}
\frac{{\rm d}T_{\rm dec, G}}{{\rm d}t} = \frac{Q_{\rm dec, G}}{M_{\rm G} c_{\rm G}}.
\end{equation}
As the specific heat and the thermal conductivity depend on the temperature, the cooling and heating rates also depend on temperature of the primary.
We note that the decay heating rate is also a function of the time.
}

\begin{figure*}
\centering
\includegraphics[width = 0.45\textwidth]{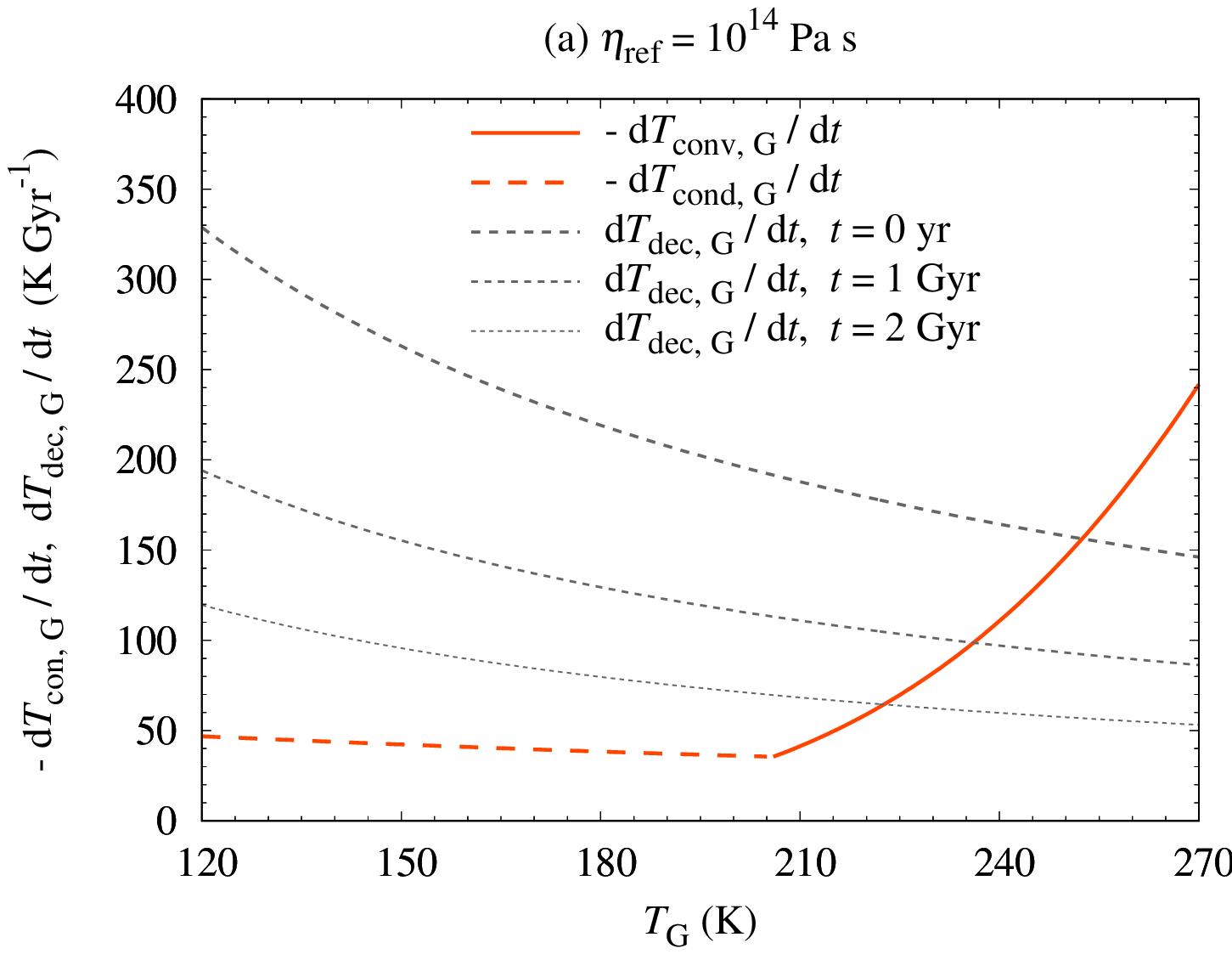}
\includegraphics[width = 0.45\textwidth]{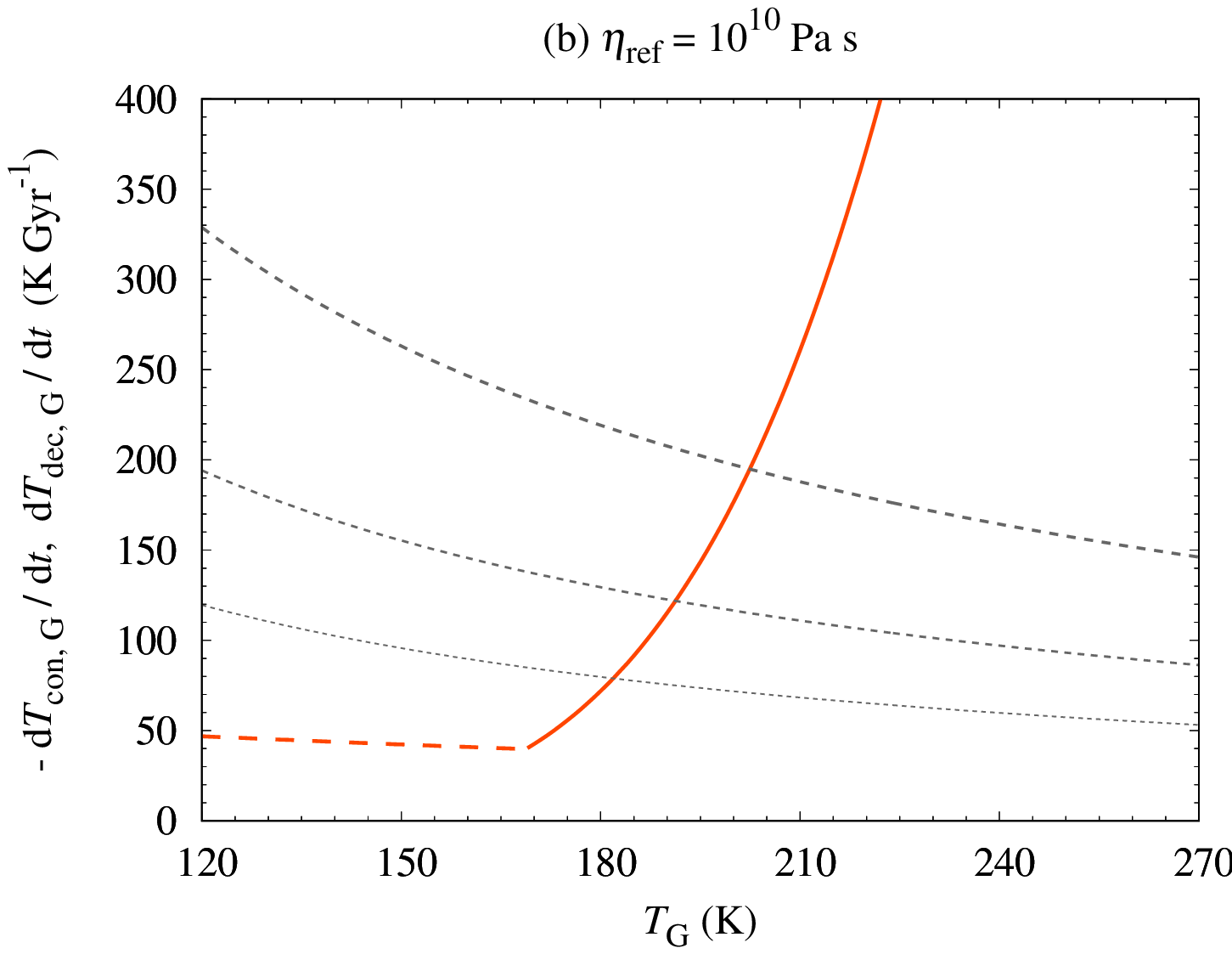}
\caption{
Cooling rate of the primary due to conduction/convection, ${{\rm d}T_{\rm con, G}} / {{\rm d}t}$, and heating rate due to decay of long-lived radioactive elements, ${{\rm d}T_{\rm dec, G}} / {{\rm d}t}$.
}
\label{figcooling}
\end{figure*}

\added{Figure \ref{figcooling} shows the cooling/heating rates of the primary as a function of the temperature, the reference viscosity, and the time.
For the case of $\eta_{\rm ref} = 10^{14}\ {\rm Pa}\ {\rm s}$, the cooling rate is controlled by the convection when $T_{\rm G} > 206\ {\rm K}$.
We found that the cooling and heating terms balance at a certain temperature, and the equilibrium temperature is consistent with the temperature evolution of the primary shown in Appendix \ref{app.typical}.
We also show that the equilibrium temperature strongly depends on $\eta_{\rm ref}$.
For the case of $\eta_{\rm ref} = 10^{10}\ {\rm Pa}\ {\rm s}$, the equilibrium temperature is lower than 210 K even at $t = 0$.
This low equilibrium temperature is also consistent with the temperature evolution shown in Figure \ref{figtypeeta10}(d).
}

\added{We also discuss the validity of the assumption of the fixed surface temperature.
As shown in Appendix \ref{app.typical}, the internal temperature of the primary reaches the equilibrium temperature at $t \lesssim 1\ {\rm Gyr}$.
When we assume that the decay heating is balanced with the radiative cooling at the surface, the surface temperature of the primary, $T_{\rm surf}$, is given by the following equation:
\begin{equation}
4 \pi {R_{\rm G}}^{2} \sigma_{\rm SB} {\left( {T_{\rm surf}}^{4} - {T_{\rm BG}}^{4} \right)} = Q_{\rm dec, G},
\end{equation}
where $\sigma_{\rm SB}$ is the Stefan--Boltzmann constant and $T_{\rm BG}$ is the effective background temperature.
Here we rewrite $T_{\rm surf}$ as $T_{\rm BG} + {\Delta T}$, then the difference between $T_{\rm surf}$ and $T_{\rm BG}$ is approximately given by
\begin{eqnarray}
{\Delta T} & \simeq & \frac{R_{\rm G} \rho_{\rm G}}{12 \sigma_{\rm SB} {T_{\rm BG}}^{3}} f_{\rm rock} \sum_{j} 2^{- t / t_{{\rm HL}, j}} H_{0, j}, \\
           & \simeq & 0.1 {\left( \frac{T_{\rm BG}}{40\ {\rm K}} \right)}^{-3} {\left( \frac{\sum_{j} 2^{- t / t_{{\rm HL}, j}} H_{0, j}}{10^{-11}\ {\rm W}\ {\rm kg}^{-1}} \right)}\ {\rm K}. 
\end{eqnarray}
and ${\Delta T}$ is negligibly smaller than $T_{\rm BG}$.
Thus we can assume $T_{\rm surf} = T_{\rm BG}$ when the decay heating is balanced with the radiative cooling at the surface.
}

\added{We acknowledge that our thermal evolution calculations would not be accurate, especially for the early stage (i.e., $t \ll 1\ {\rm Gyr}$).
In the early stage, the temperature structure inside the primary may not reach the equilibrium, and the cooling rate should be modified when we calculate the internal temperature structure.
As the thermal evolution in the early phase would be important for the eccentricity evolution of satellite systems (see Appendix \ref{app.typical}), we need to conduct the coupled thermal--orbital evolution simulations considering the internal temperature structure in future studies.
}

\section{Typical results for tidal evolution calculations}
\label{app.typical}

In Appendix \ref{app.typical}, we introduce typical tidal evolution pathways of satellite systems.
We classify the outcome of tidal evolution into four types: 1:1 spin--orbit resonance (Type A; Figures \ref{figtypeA} and \ref{figtypeAini}), higher-order spin--orbit resonance (Type B; Figures \ref{figtypeB} and \ref{figtypeBini}), not resonance (Type C; Figures \ref{figtypeC} and \ref{figtypeCini}), and collision with the primary (Type Z; Figure \ref{figtypeZ}).
We also show a typical evolution pathway with $\eta_{\rm ref} = 10^{10}\ {\rm Pa}\ {\rm s}$ (Figure \ref{figtypeeta10}).

\subsection{Type A: 1:1 spin--orbit resonance}

\begin{figure*}
\centering
\includegraphics[width = 0.45\textwidth]{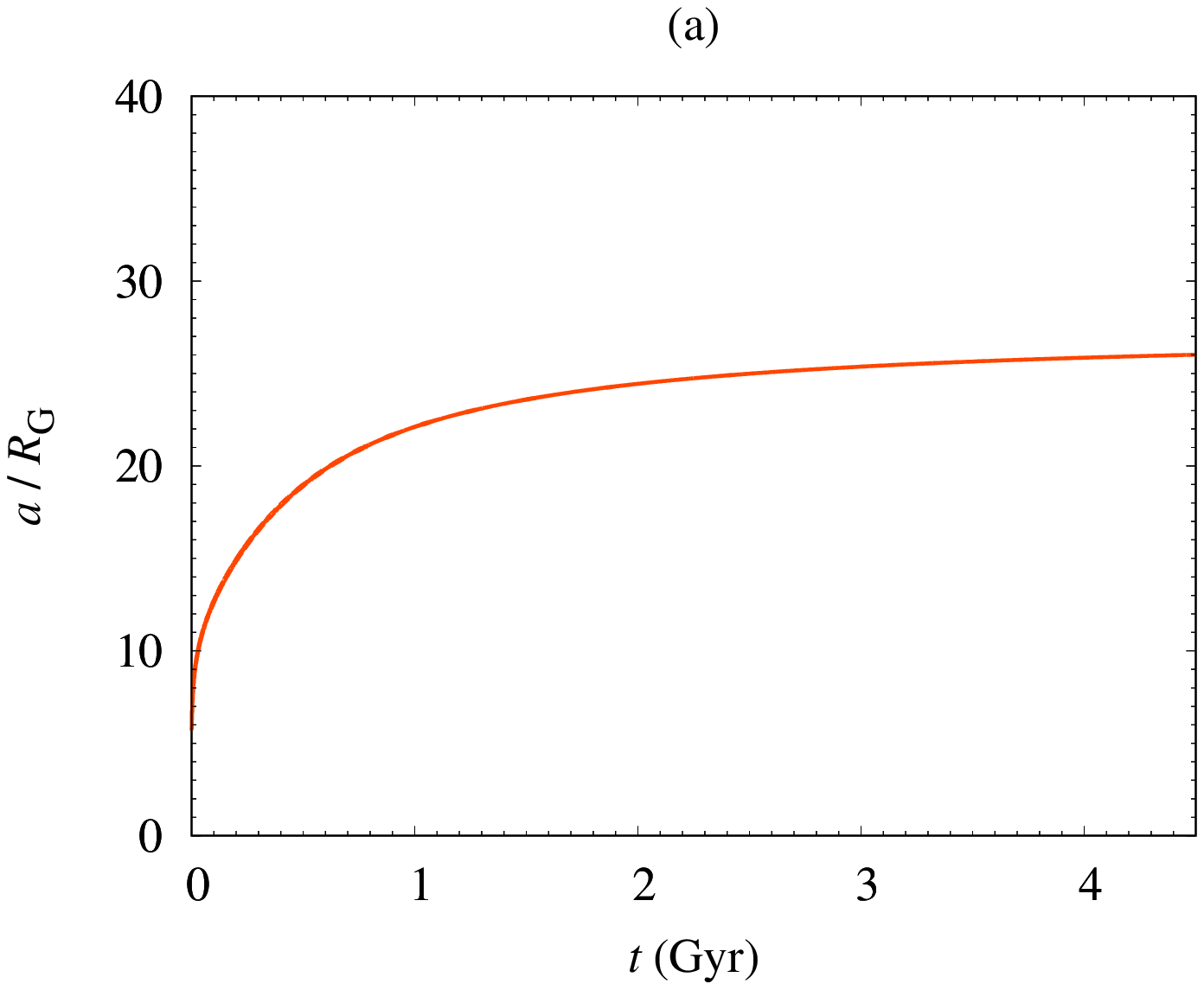}
\includegraphics[width = 0.45\textwidth]{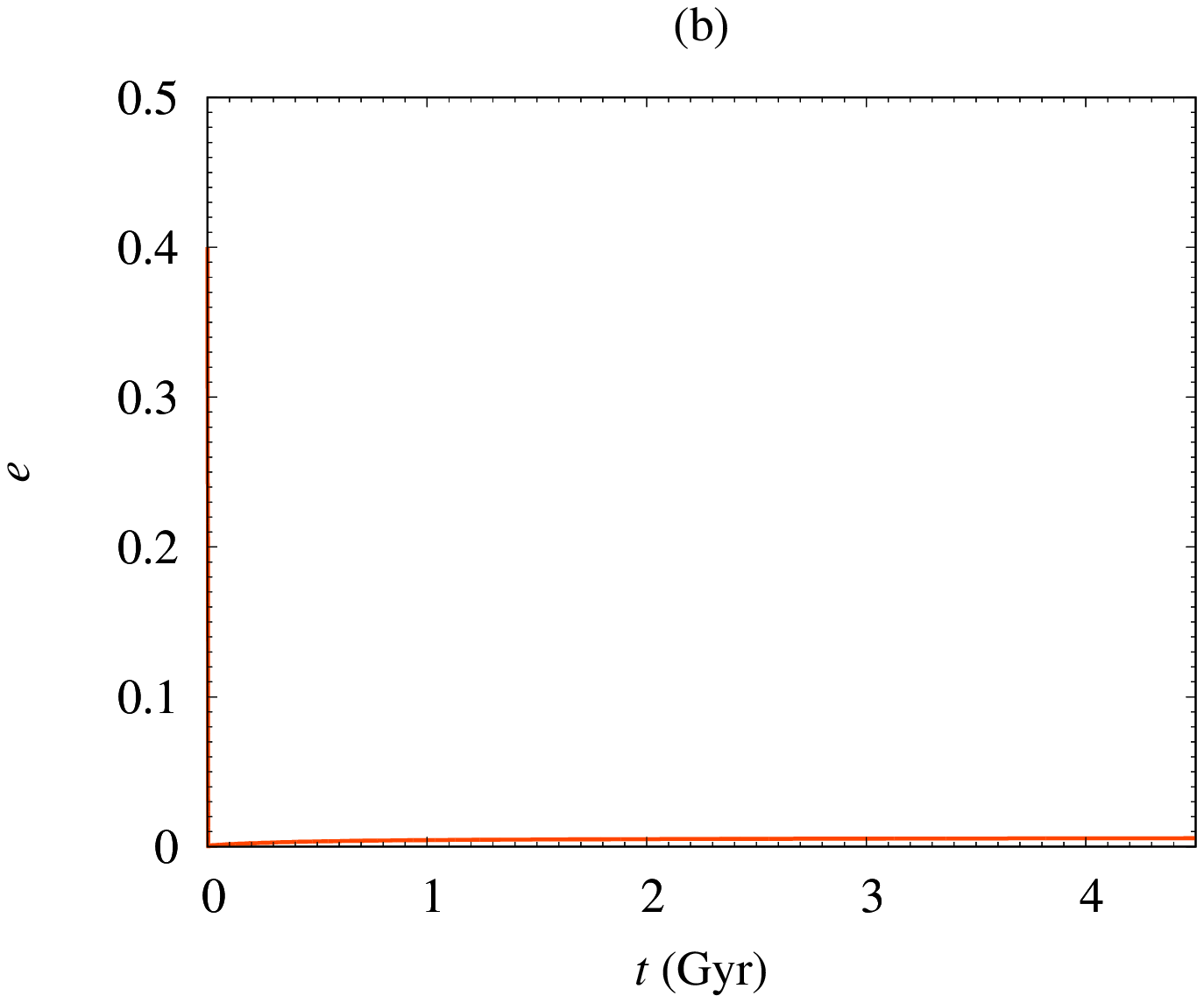}
\includegraphics[width = 0.45\textwidth]{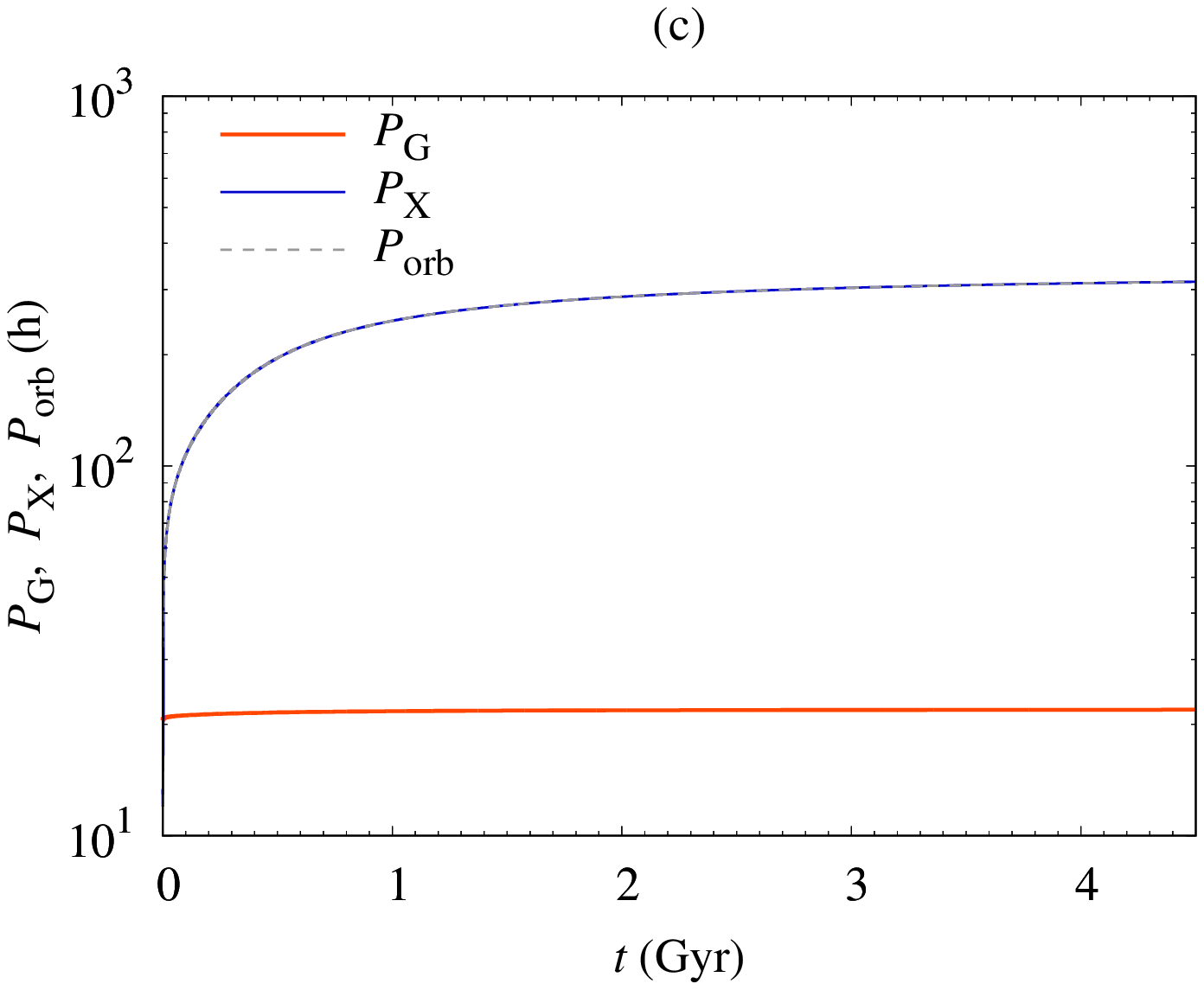}
\includegraphics[width = 0.45\textwidth]{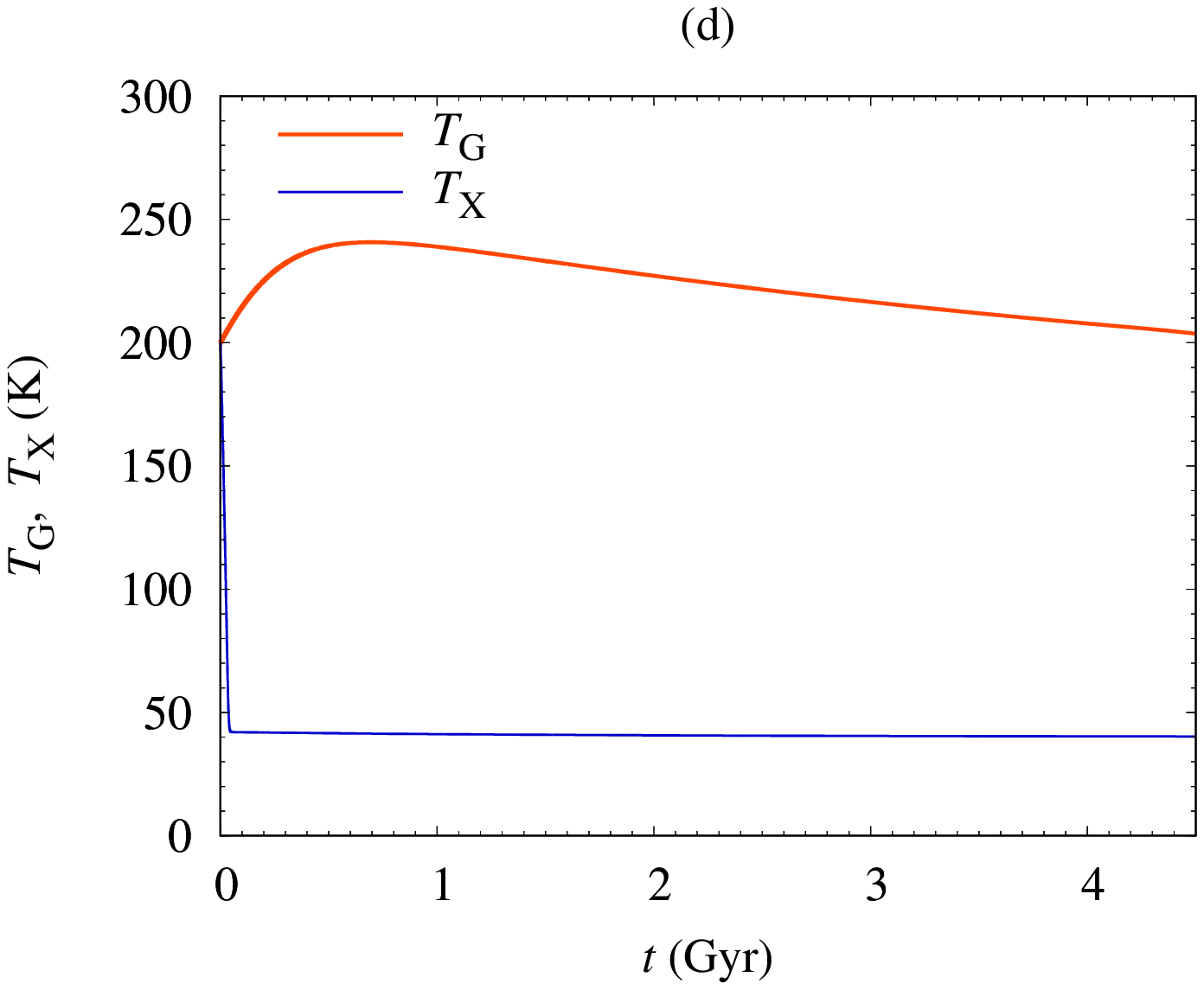}
\caption{
Evolution of the (a): semimajor axis, (b): eccentricity, (c): spin/orbital periods, and (d): temperature of the primary and secondary for Model A.
The initial condition of Model A is $R_{\rm X} = 60\ {\rm km}$, $T_{\rm ini} = 200\ {\rm K}$, and $e_{\rm ini} = 0.4$.
}
\label{figtypeA}
\end{figure*}

Figure \ref{figtypeA} shows a typical tidal evolution pathway of a satellite system resulting in Type A.
The initial condition of this model is $R_{\rm X} = 60\ {\rm km}$, $T_{\rm ini} = 200\ {\rm K}$, and $e_{\rm ini} = 0.4$ (see Figure \ref{figType}(c)). 
The final semimajor axis and eccentricity are $a_{\rm fin} / R_{\rm G} = 26.0$ and $e_{\rm fin} = 5.6 \times 10^{-3}$, respectively; and the spin of the secondary is in 1:1 spin--orbit resonance at $t = 4.5\ {\rm Gyr}$.

The thermal evolution of both the primary and secondary hardly depends on their spin/orbital evolution.
This is because their thermal evolution pathways are determined by the balance between decay heating and conduction/convection cooling.

\begin{figure*}
\centering
\includegraphics[width = 0.45\textwidth]{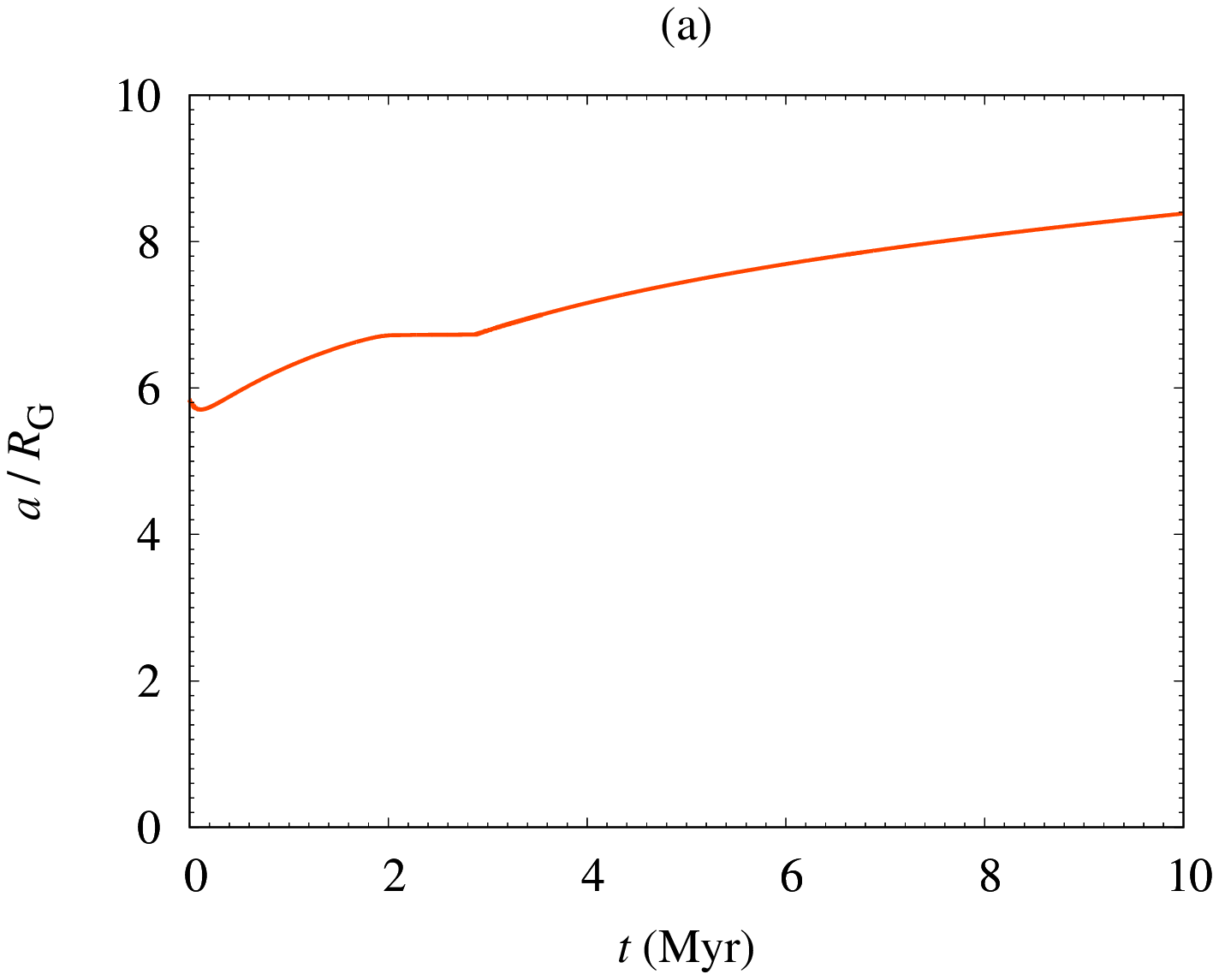}
\includegraphics[width = 0.45\textwidth]{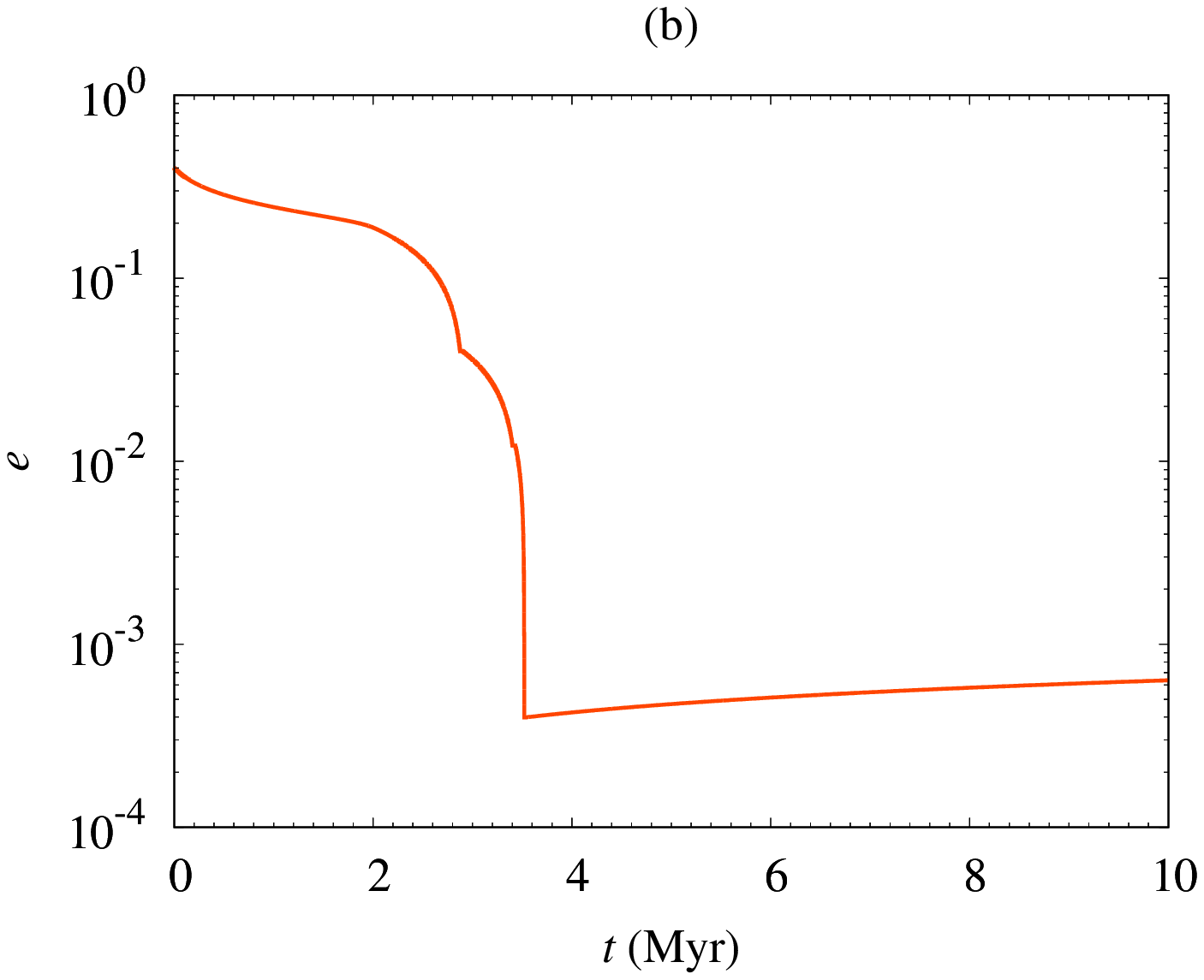}
\includegraphics[width = 0.45\textwidth]{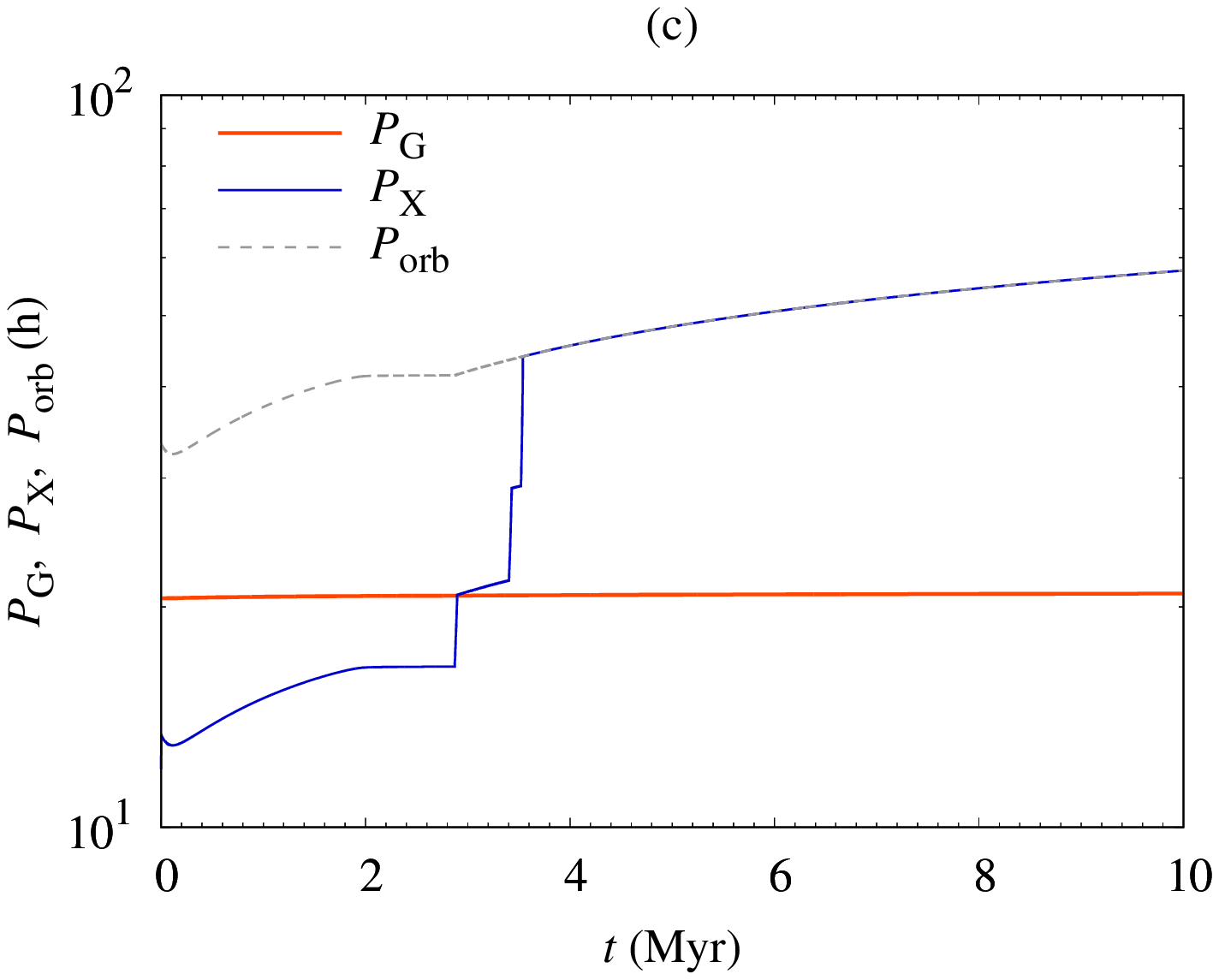}
\includegraphics[width = 0.45\textwidth]{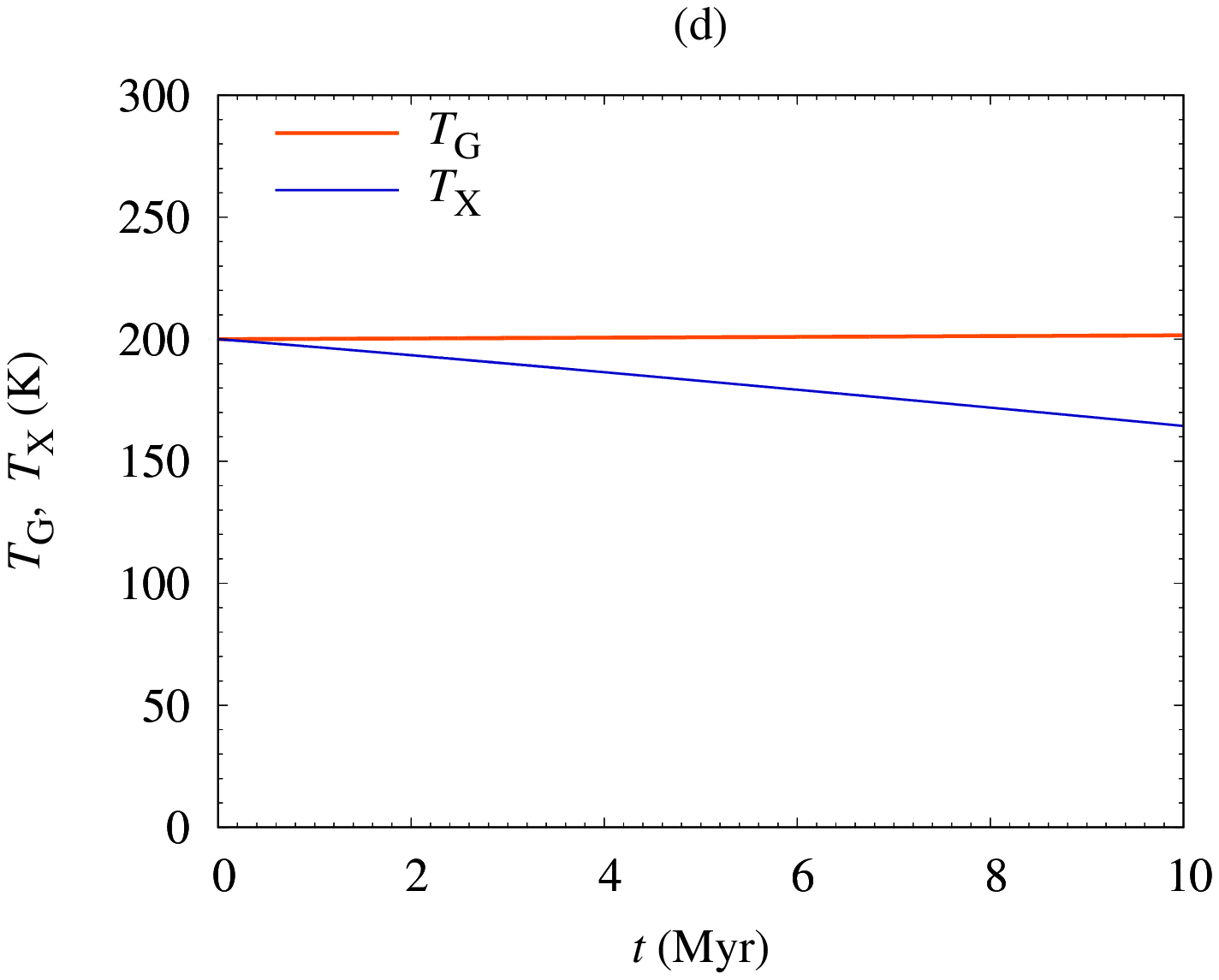}
\includegraphics[width = 0.45\textwidth]{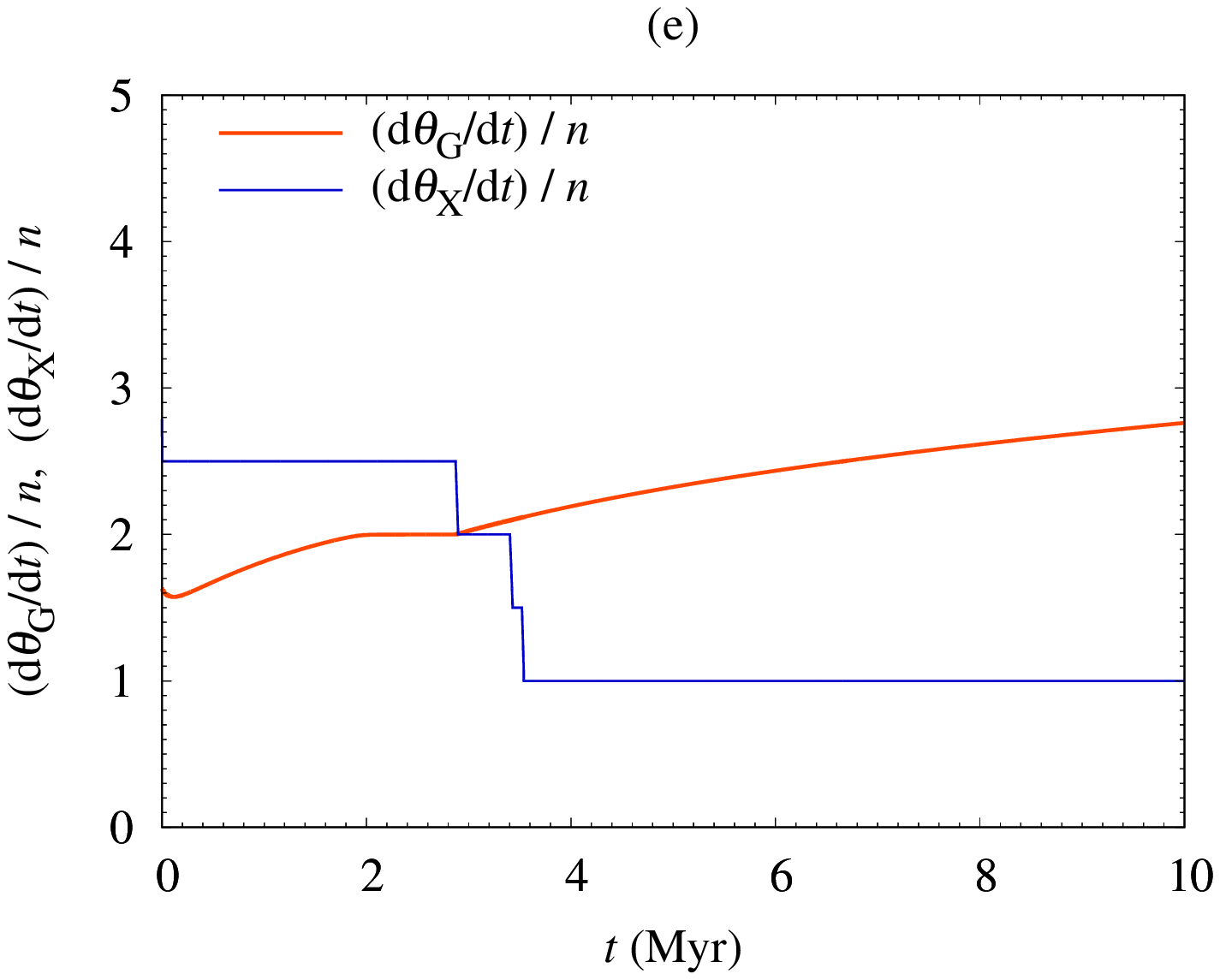}
\caption{
Early phase of the evolution of the (a): semimajor axis, (b): eccentricity, (c): spin/orbital periods, (d): temperature of the primary and secondary, and (e) spin-to-orbit period ratio for Model A.
}
\label{figtypeAini}
\end{figure*}

Figure \ref{figtypeAini} shows the first 10 Myrs of the tidal evolution of this case.
Both the primary and secondary are initially captured into spin--orbit resonances.
At $t = 2.872\ {\rm Myr}$, the eccentricity becomes $e = 4.15 \times 10^{-2}$, and the secondary is released from 5:2 spin--orbit resonance.
Then, the primary is also released from 2:1 spin--orbit resonance at $t = 2.875\ {\rm Myr}$.
The secondary is recaptured into 2:1 spin--orbit resonance at $t = 2.897\ {\rm Myr}$, and finally trapped in 1:1 resonance at $t = 3.54\ {\rm Myr}$.

\subsection{Type B: higher-order spin--orbit resonance}

\begin{figure*}
\centering
\includegraphics[width = 0.45\textwidth]{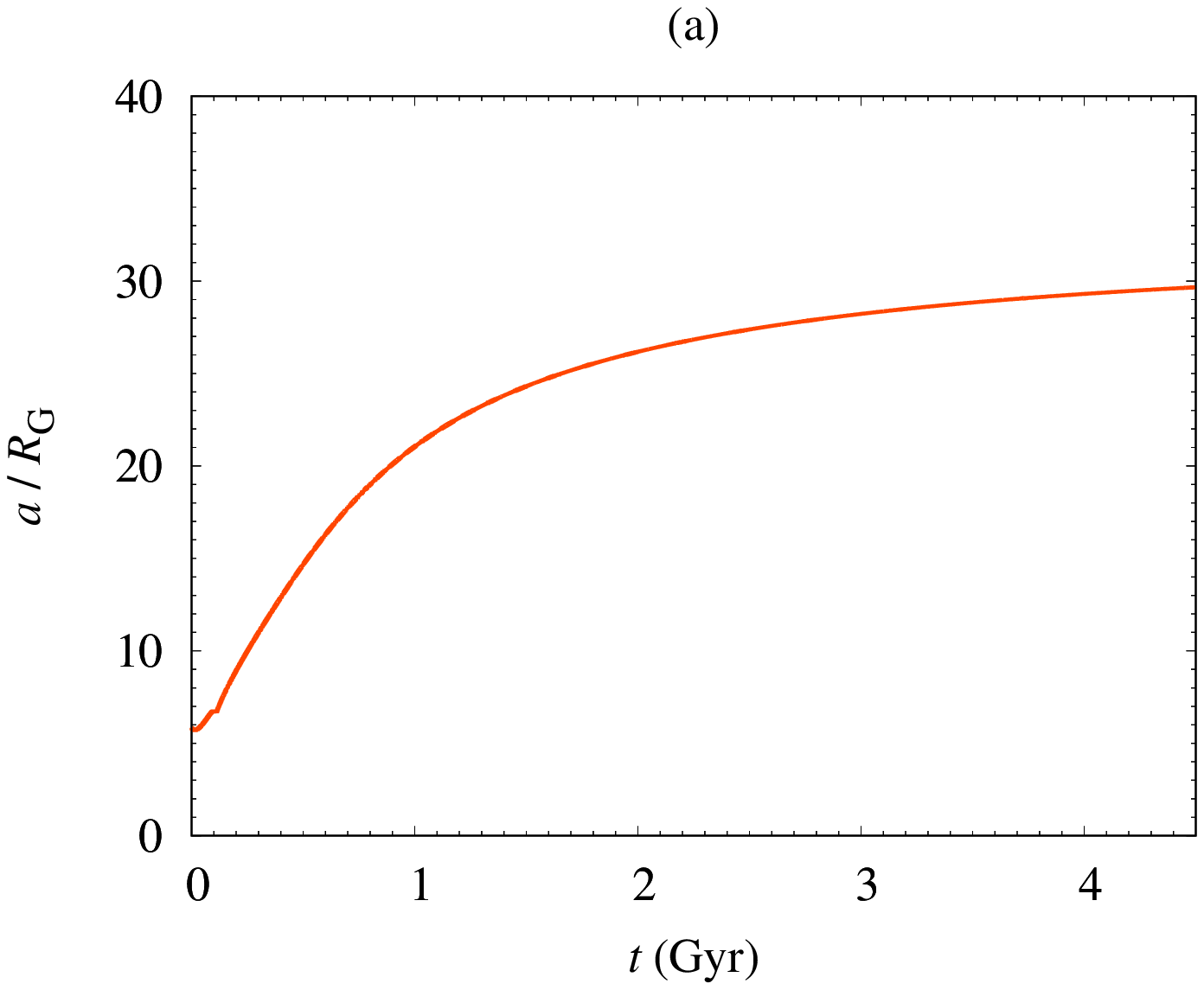}
\includegraphics[width = 0.45\textwidth]{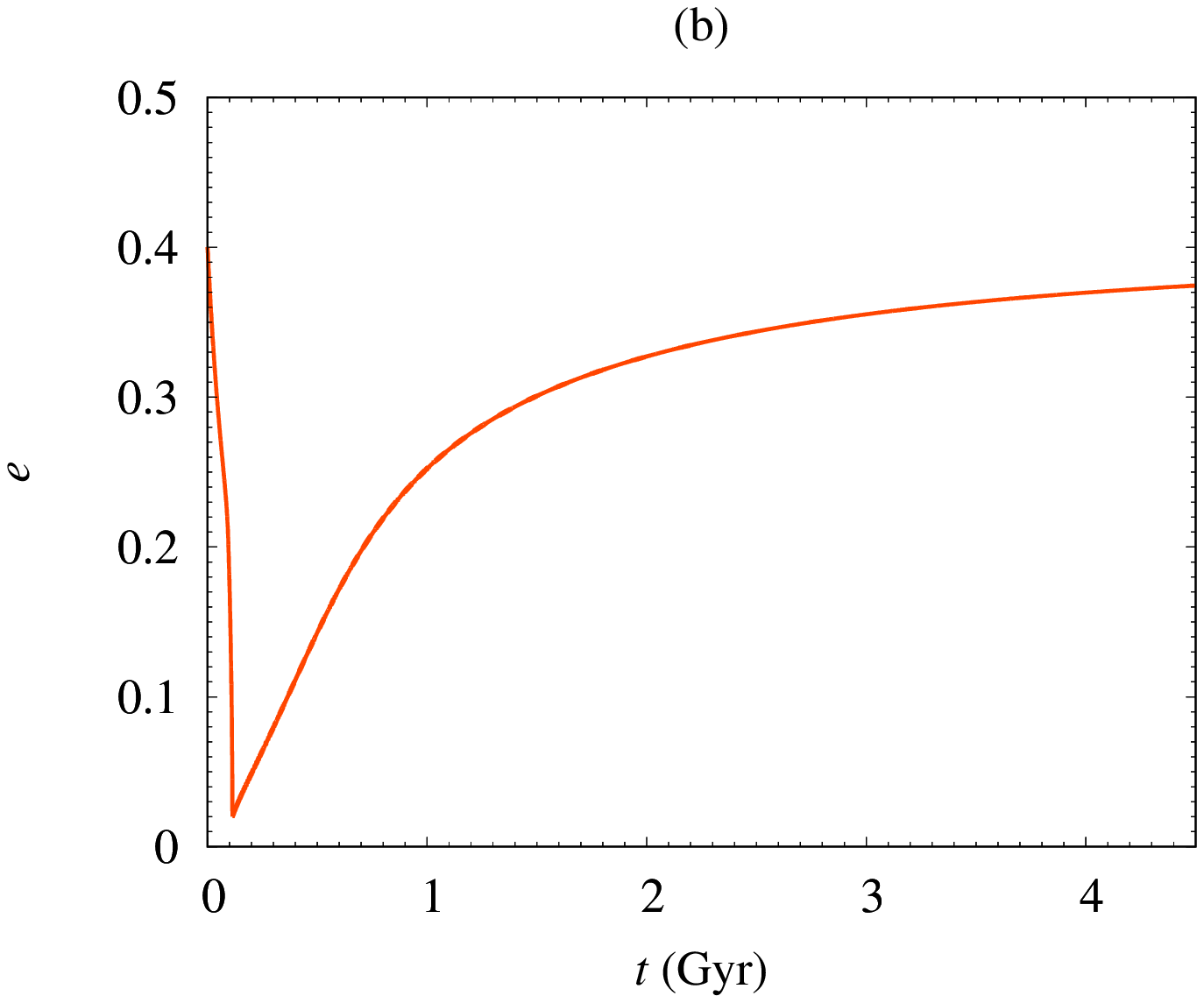}
\includegraphics[width = 0.45\textwidth]{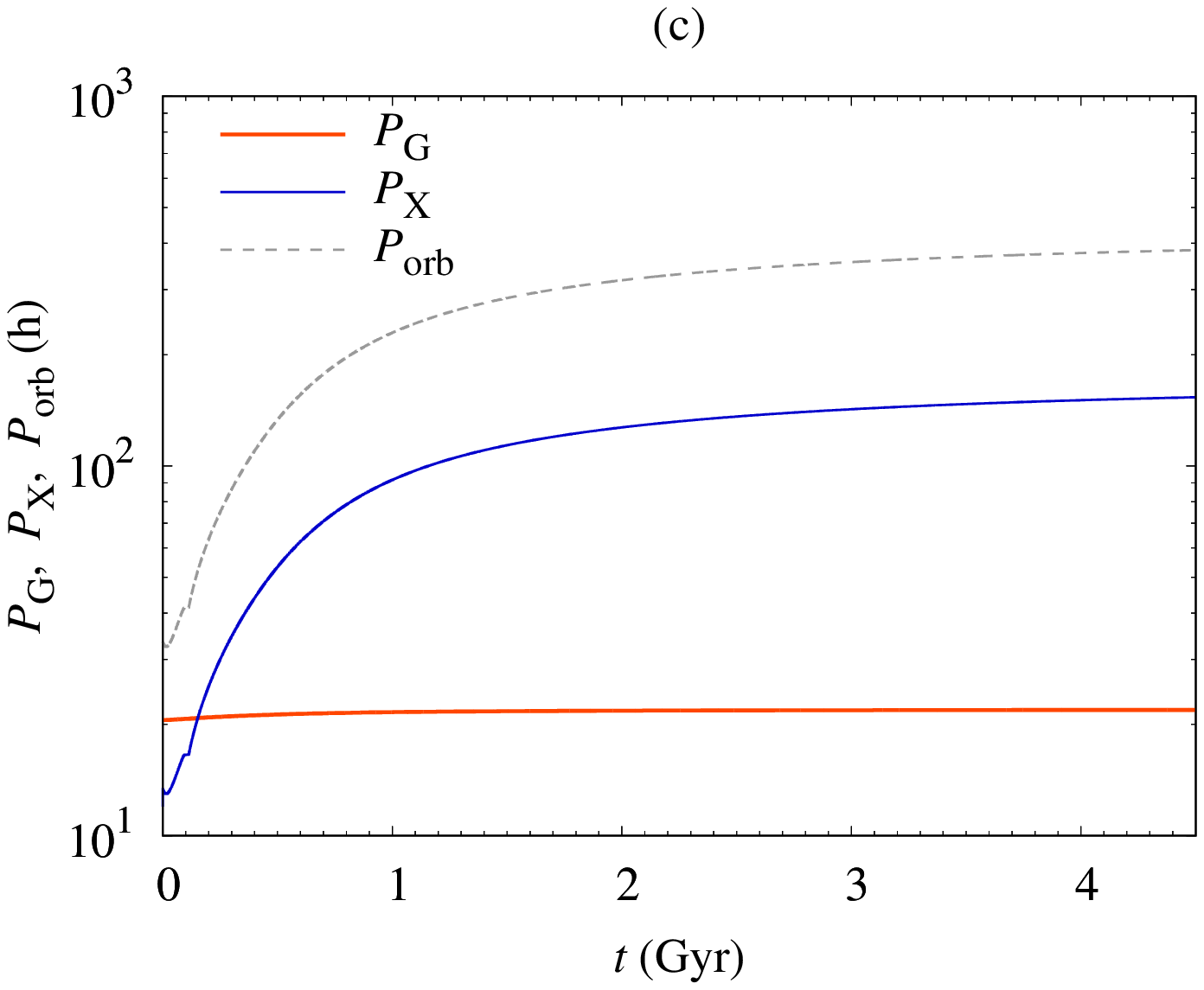}
\includegraphics[width = 0.45\textwidth]{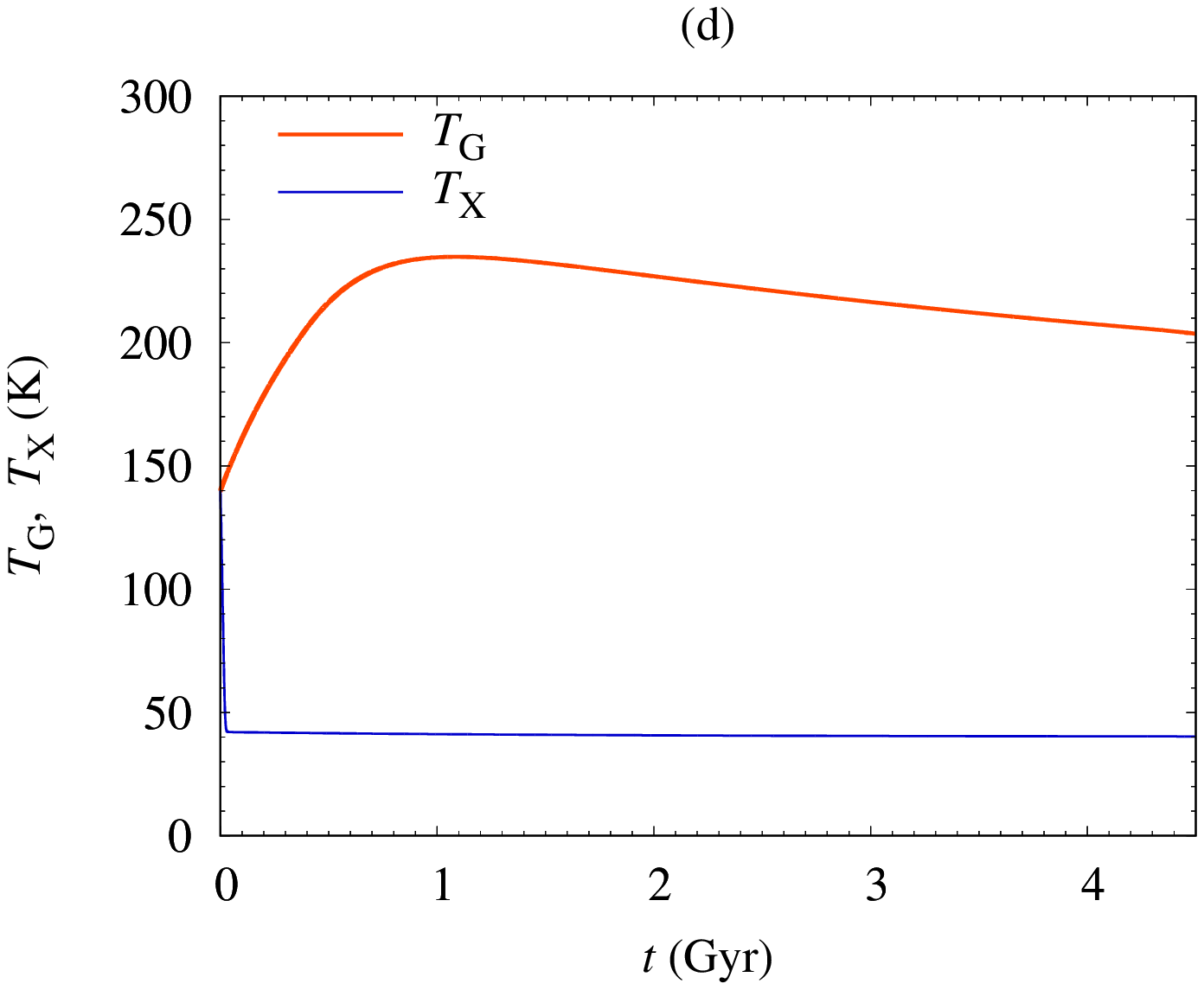}
\caption{
Evolution of the (a): semimajor axis, (b): eccentricity, (c): spin/orbital periods, and (d): temperature of the primary and secondary for Model B.
The initial condition of Model B is $R_{\rm X} = 60\ {\rm km}$, $T_{\rm ini} = 140\ {\rm K}$, and $e_{\rm ini} = 0.4$.
}
\label{figtypeB}
\end{figure*}

Figure \ref{figtypeB} shows a typical tidal evolution pathway of a satellite system resulting in Type B.
The initial condition of this model is $R_{\rm X} = 60\ {\rm km}$, $T_{\rm ini} = 140\ {\rm K}$, and $e_{\rm ini} = 0.4$ (see Figure \ref{figType}(c)).
The final semimajor axis and eccentricity are $a_{\rm fin} / R_{\rm G} = 29.7$ and $e_{\rm fin} = 0.37$, respectively, and the spin of the secondary is in 5:2 spin--orbit resonance at $t = 4.5\ {\rm Gyr}$.

The temperature of the 1000-km-sized primary hardly depends on the initial temperature approximately 1 Gyrs after the start of calculations (see Figures \ref{figtypeA}(d), \ref{figtypeB}(d), and \ref{figtypeC}(d)).
This shows that the temperature of the primary reaches the equilibrium determined by the balance of the decay heating and the conduction/convection cooling at $t \lesssim 1\ {\rm Gyr}$.

\begin{figure*}
\centering
\includegraphics[width = 0.45\textwidth]{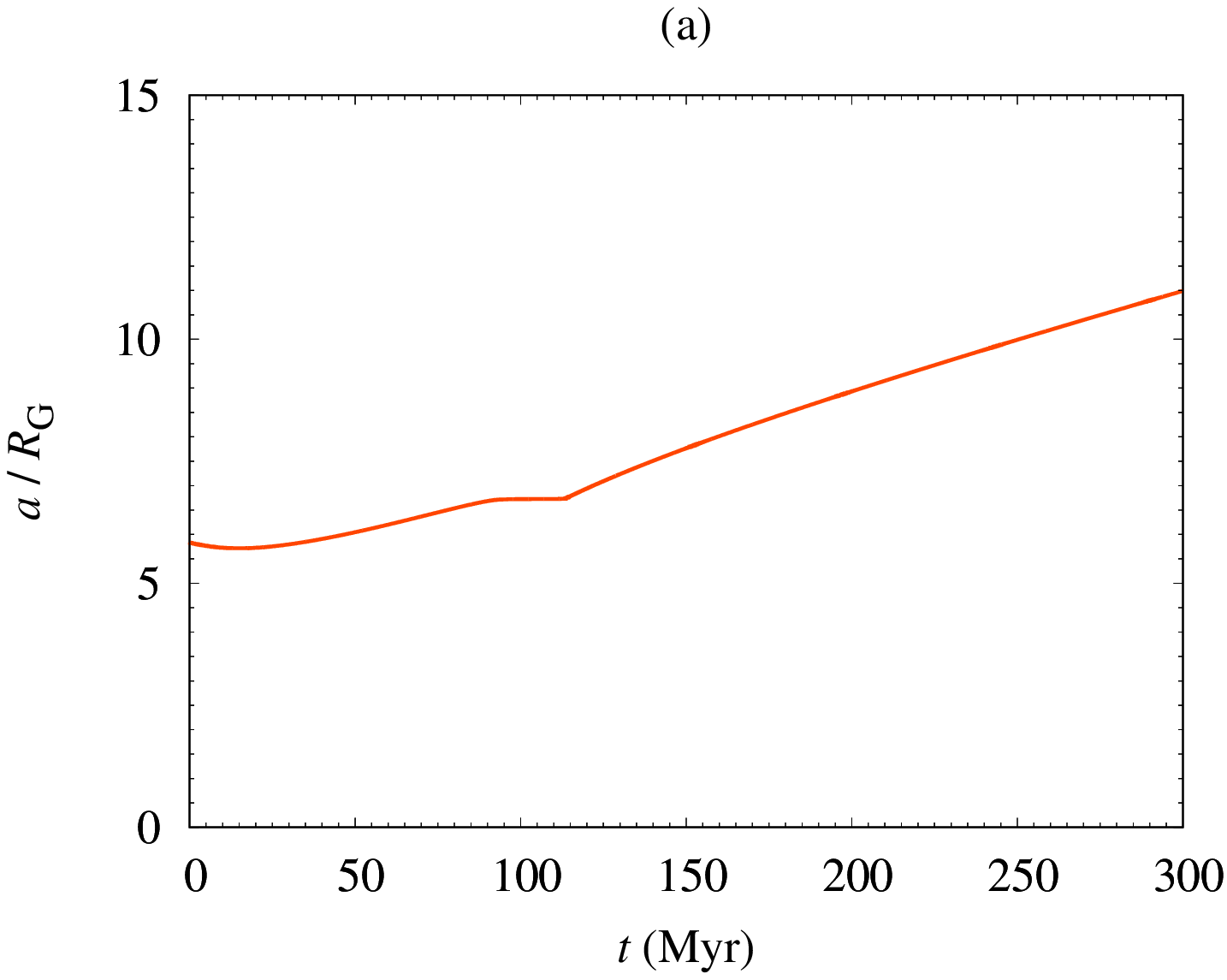}
\includegraphics[width = 0.45\textwidth]{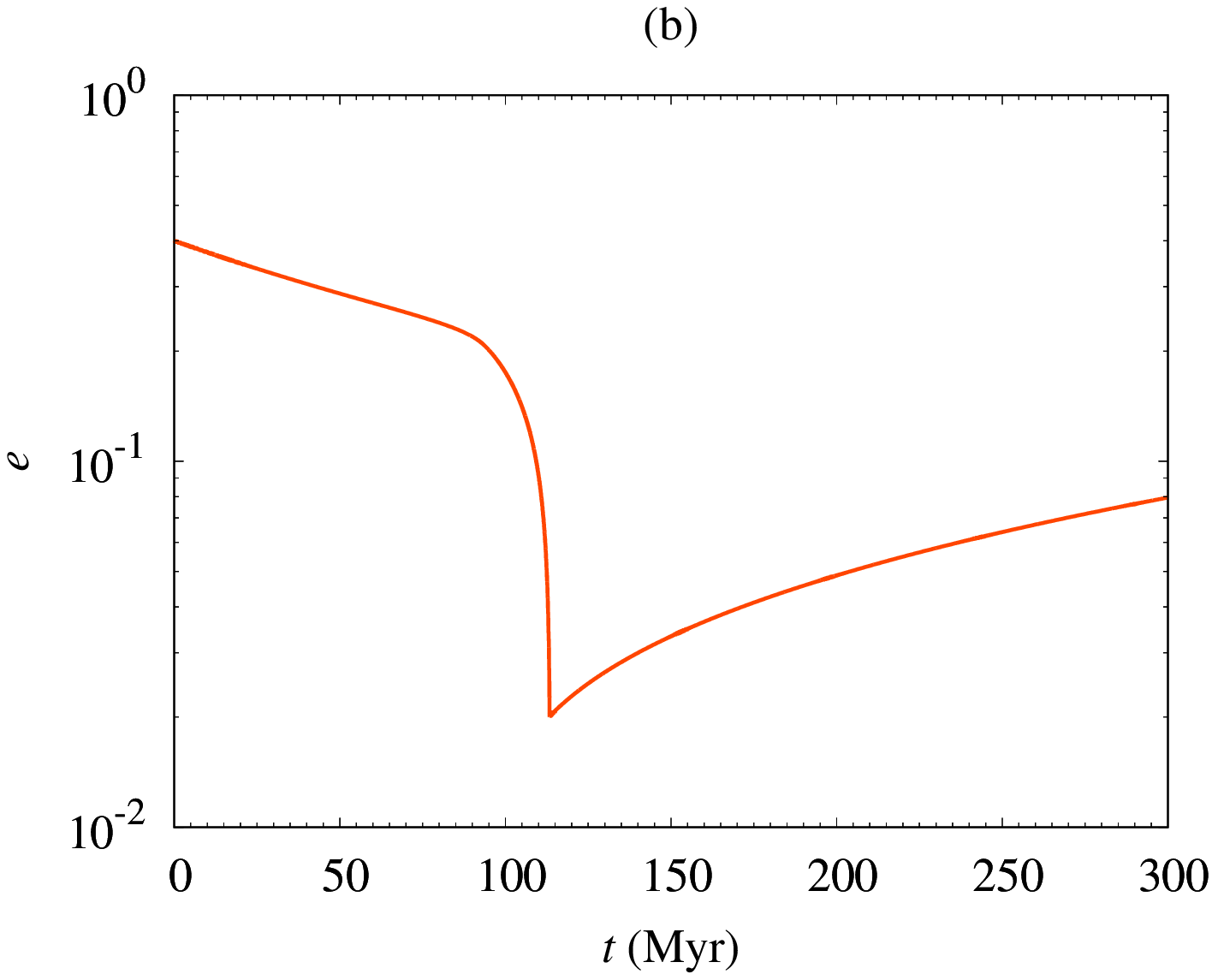}
\includegraphics[width = 0.45\textwidth]{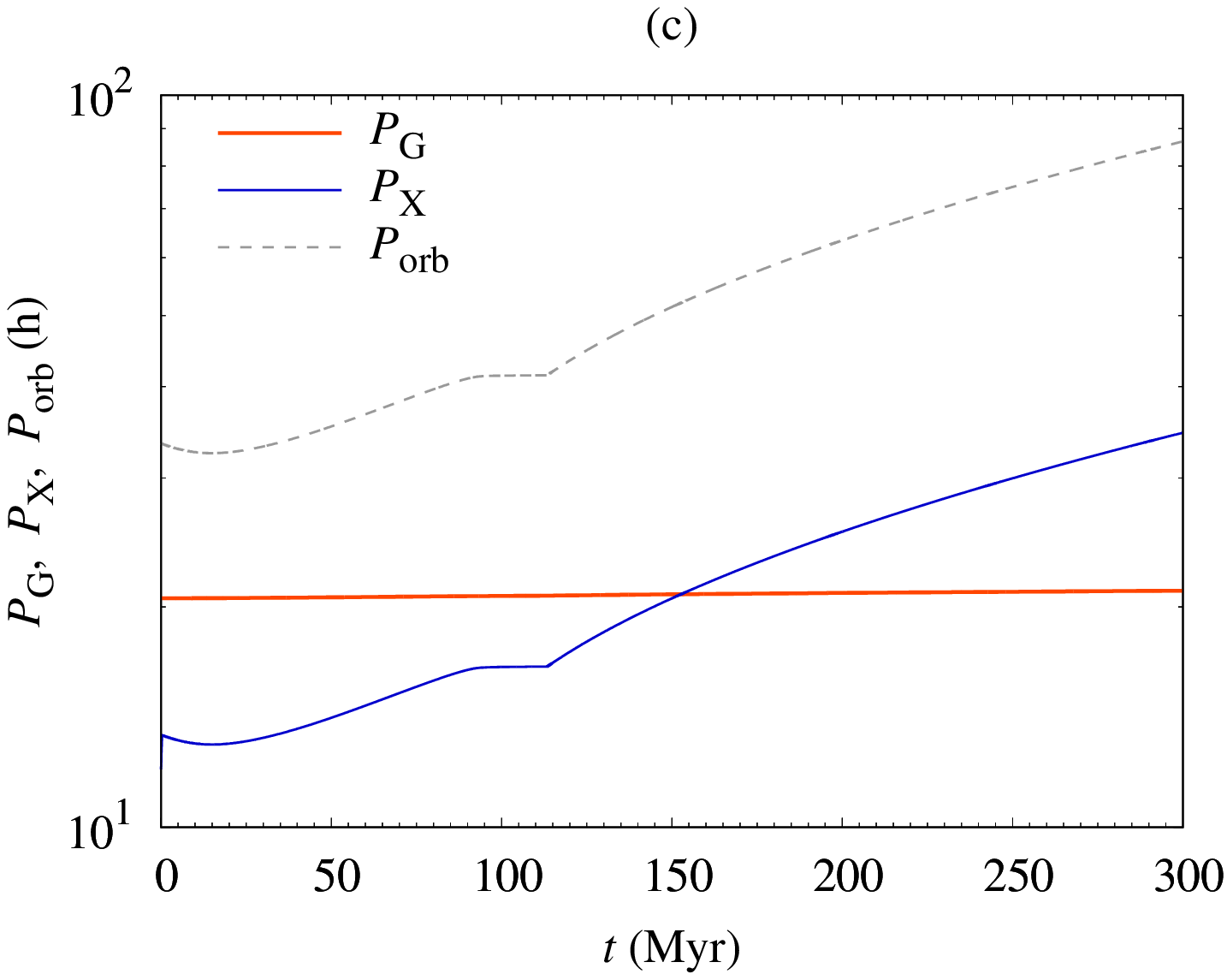}
\includegraphics[width = 0.45\textwidth]{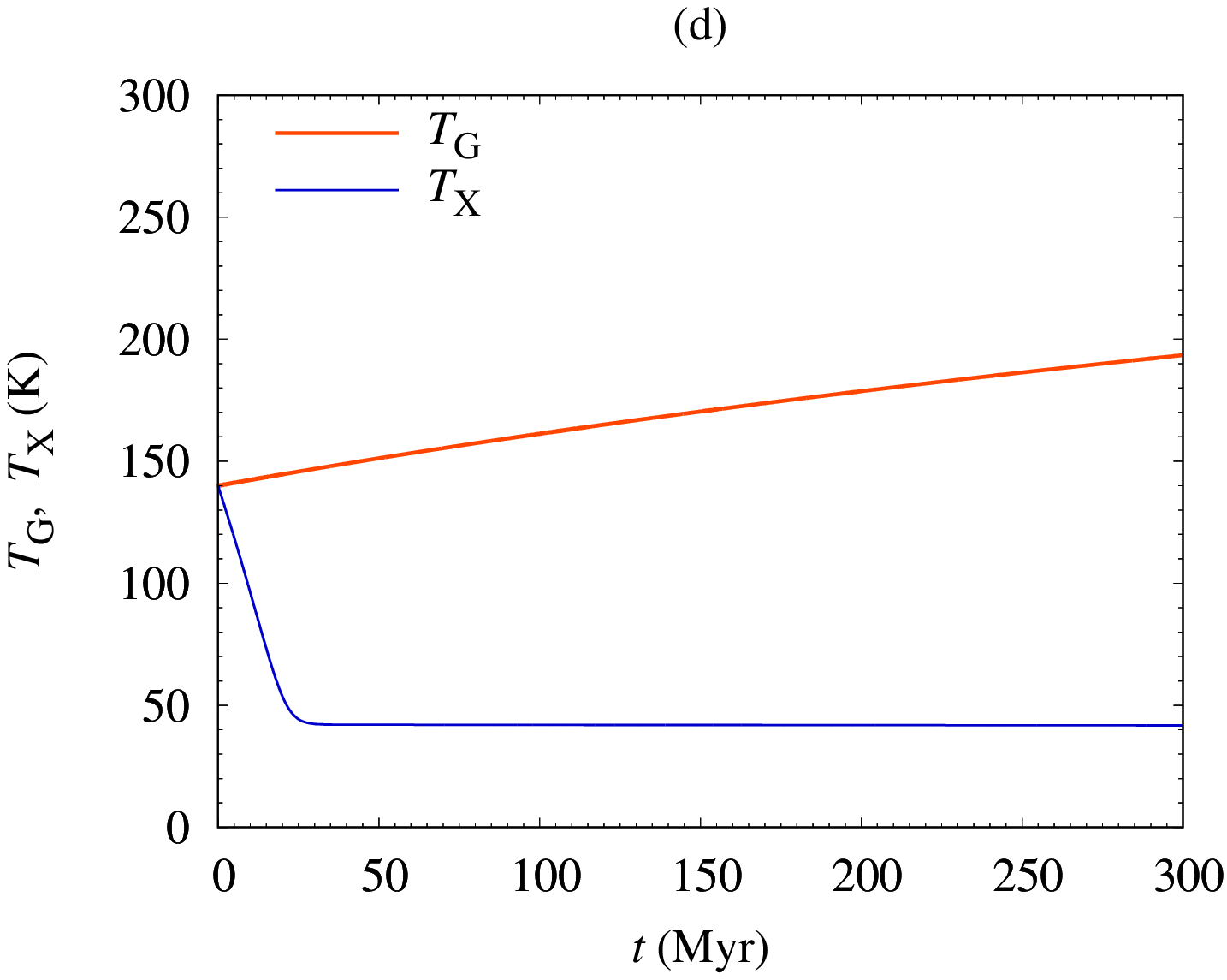}
\includegraphics[width = 0.45\textwidth]{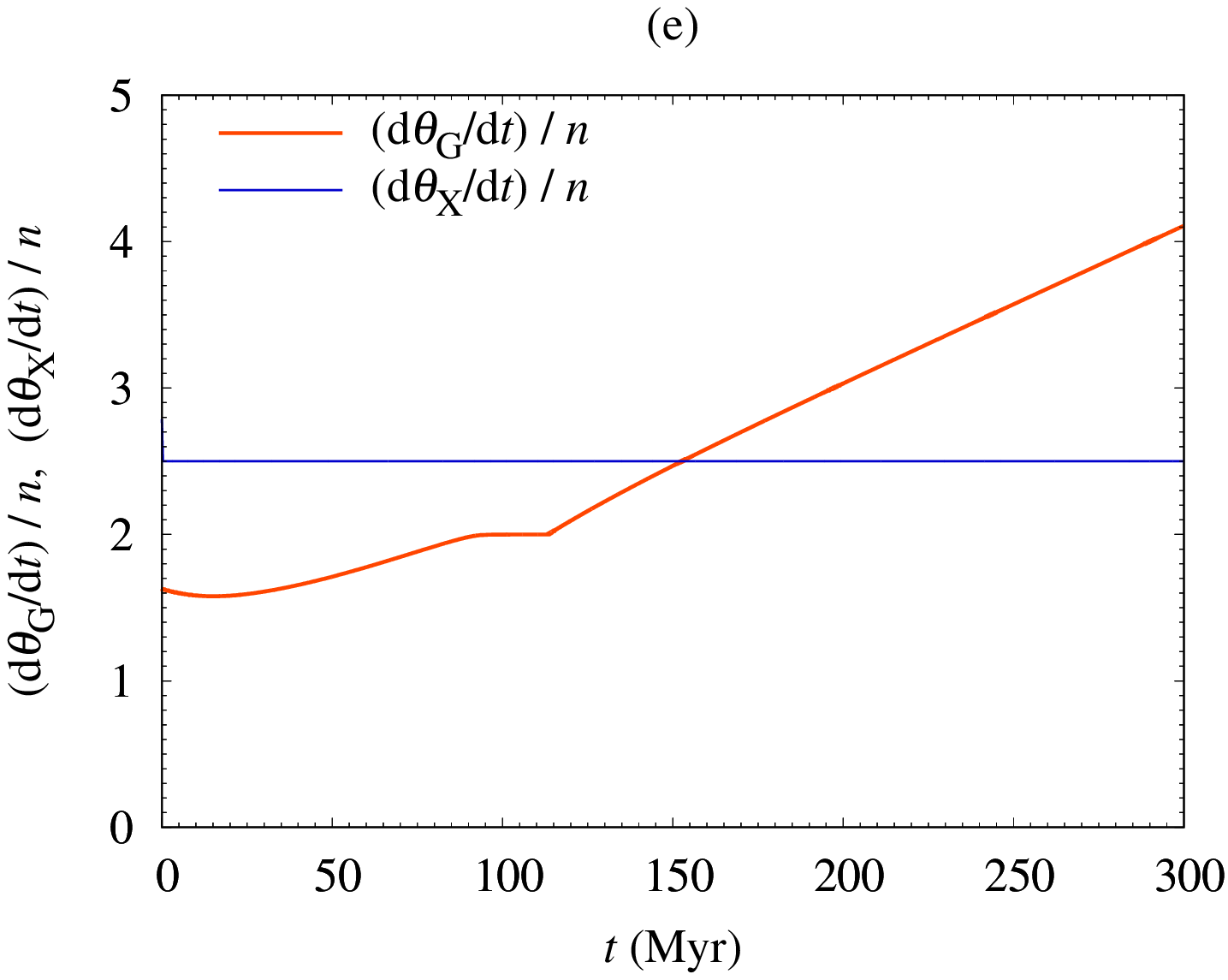}
\caption{
Early phase of the evolution of the (a): semimajor axis, (b): eccentricity, (c): spin/orbital periods, (d): temperature of the primary and secondary, and (e) spin-to-orbit period ratio for Model B.
}
\label{figtypeBini}
\end{figure*}

Figure \ref{figtypeBini} shows the first 300 Myrs of the tidal evolution of this case.
Both the primary and secondary are initially captured into spin--orbit resonances.
The primary is released from 2:1 spin--orbit resonance at $t = 113\ {\rm Myr}$, at which time the eccentricity begins increasing.
Conversely, 5:2 spin--orbit resonance of the secondary is stable in this case.

\subsection{Type C: not resonance}

\begin{figure*}
\centering
\includegraphics[width = 0.45\textwidth]{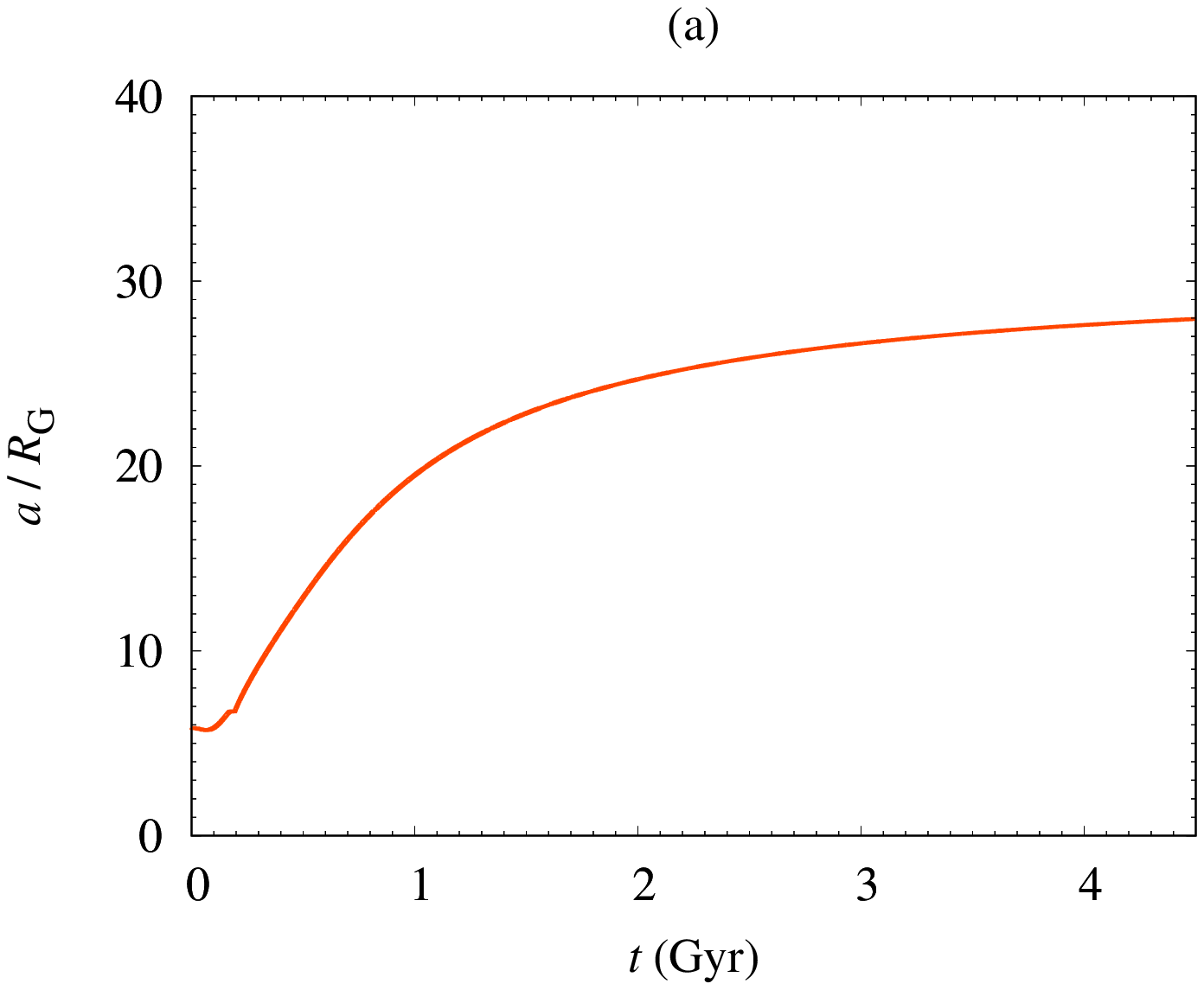}
\includegraphics[width = 0.45\textwidth]{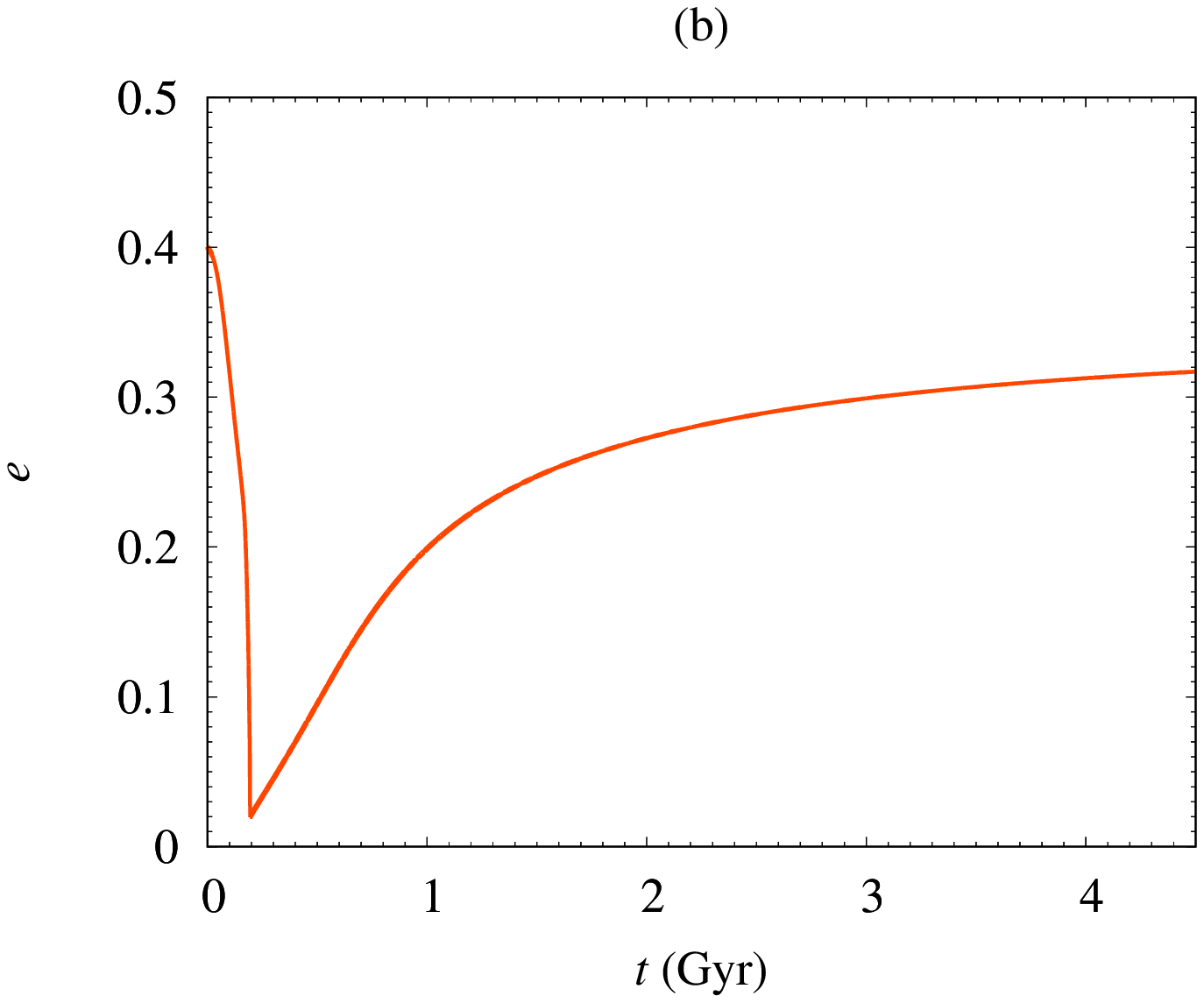}
\includegraphics[width = 0.45\textwidth]{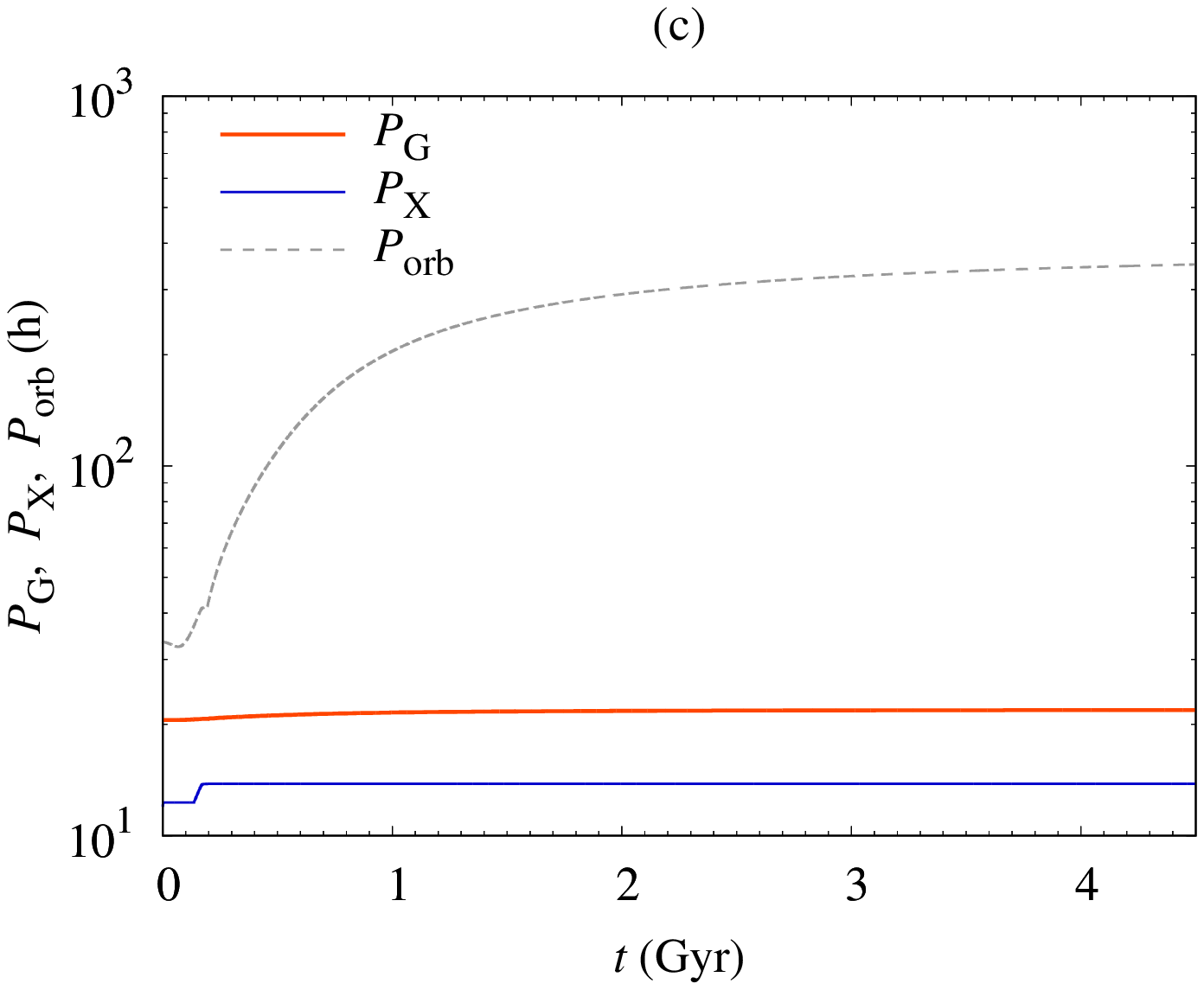}
\includegraphics[width = 0.45\textwidth]{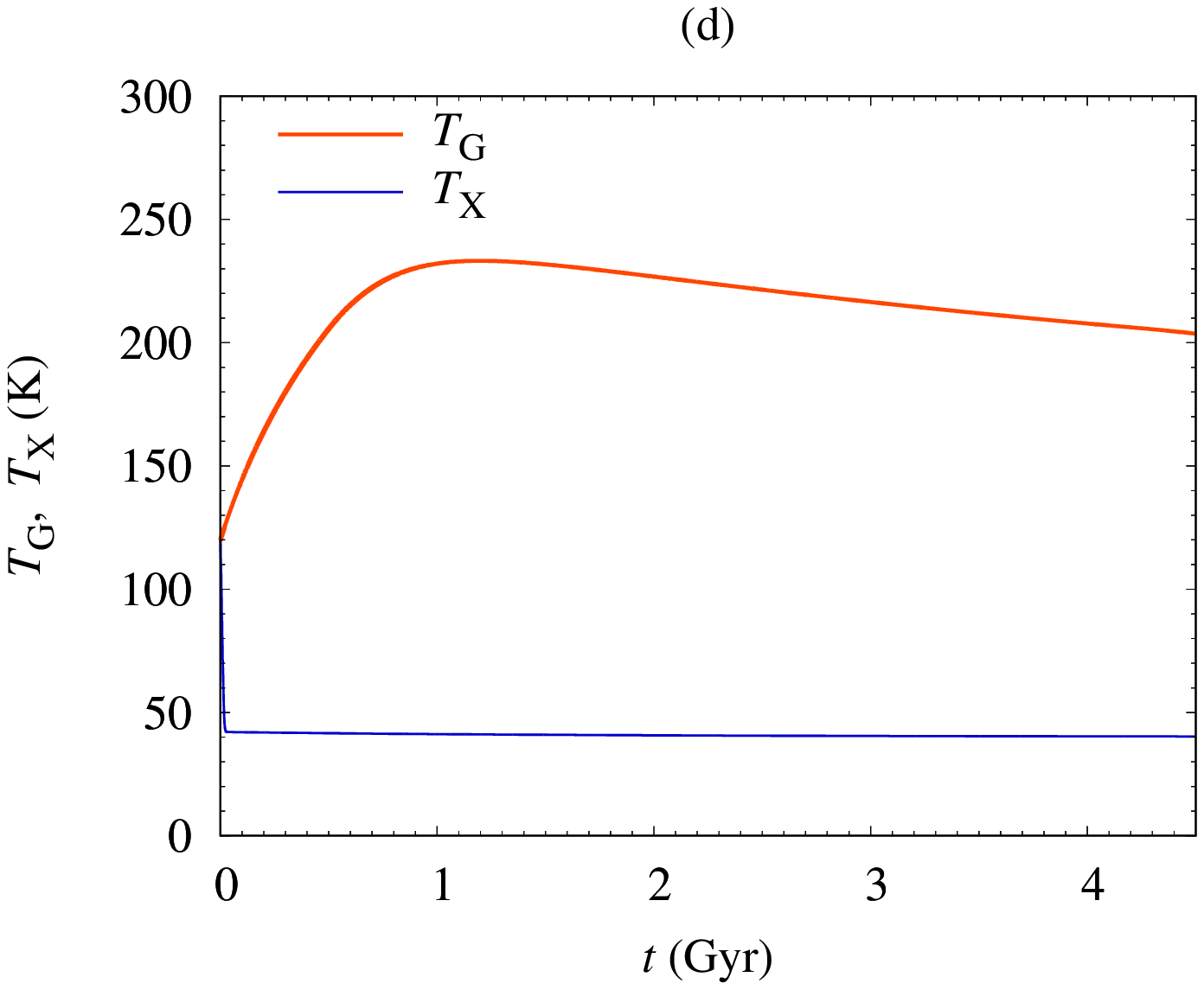}
\caption{
Evolution of the (a): semimajor axis, (b): eccentricity, (c): spin/orbital periods, and (d): temperature of the primary and secondary for Model C.
The initial condition of Model C is $R_{\rm X} = 60\ {\rm km}$, $T_{\rm ini} = 120\ {\rm K}$, and $e_{\rm ini} = 0.4$.
}
\label{figtypeC}
\end{figure*}

Figure \ref{figtypeC} shows a typical tidal evolution pathway of a satellite system resulting in Type C.
The initial condition of this model is $R_{\rm X} = 60\ {\rm km}$, $T_{\rm ini} = 120\ {\rm K}$, and $e_{\rm ini} = 0.4$ (see Figure \ref{figType}(c)).
The final semimajor axis and eccentricity are $a_{\rm fin} / R_{\rm G} = 28.0$ and $e_{\rm fin} = 0.32$, respectively, and the spin of the secondary is not in spin--orbit resonances at $t = 4.5\ {\rm Gyr}$.

\begin{figure*}
\centering
\includegraphics[width = 0.45\textwidth]{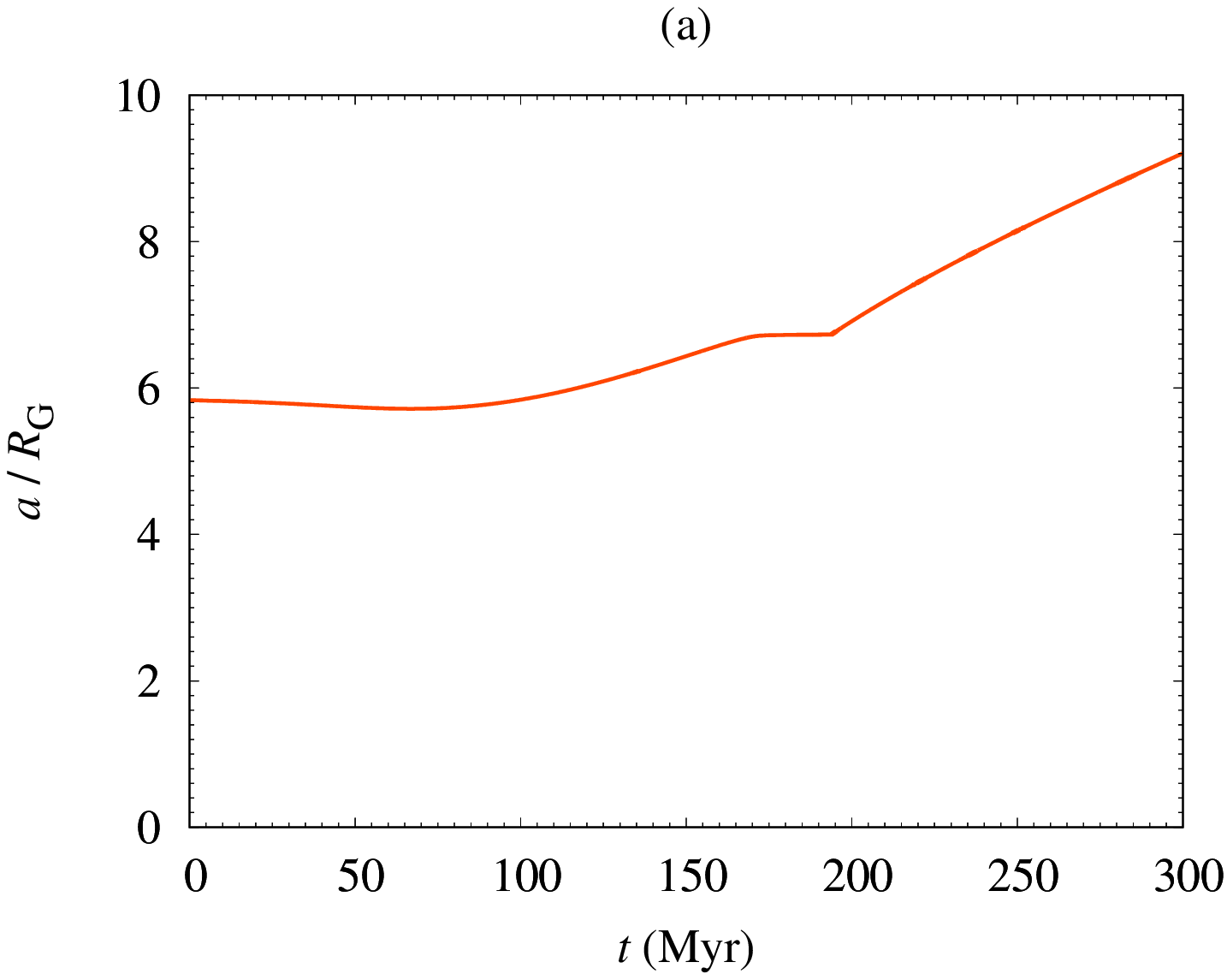}
\includegraphics[width = 0.45\textwidth]{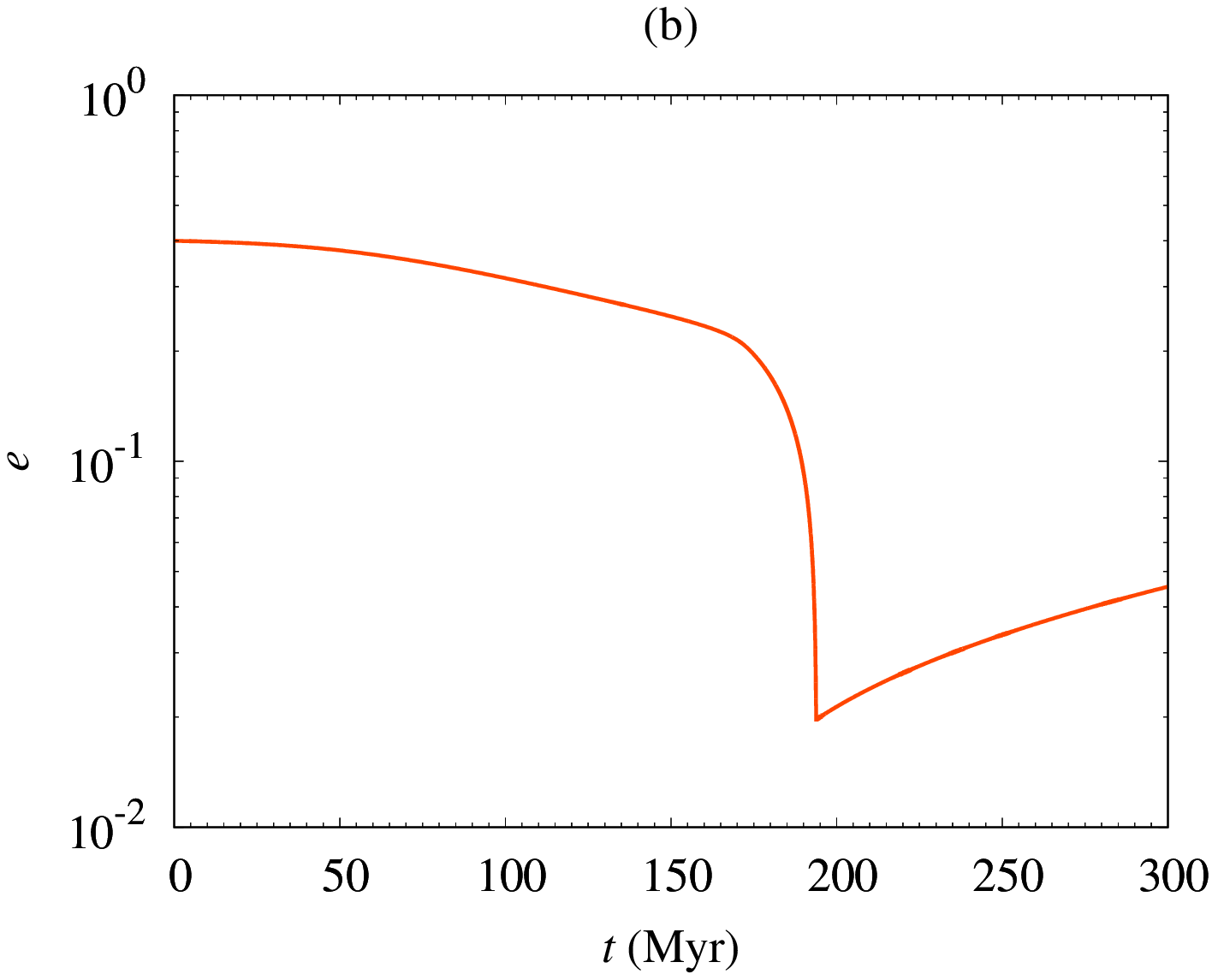}
\includegraphics[width = 0.45\textwidth]{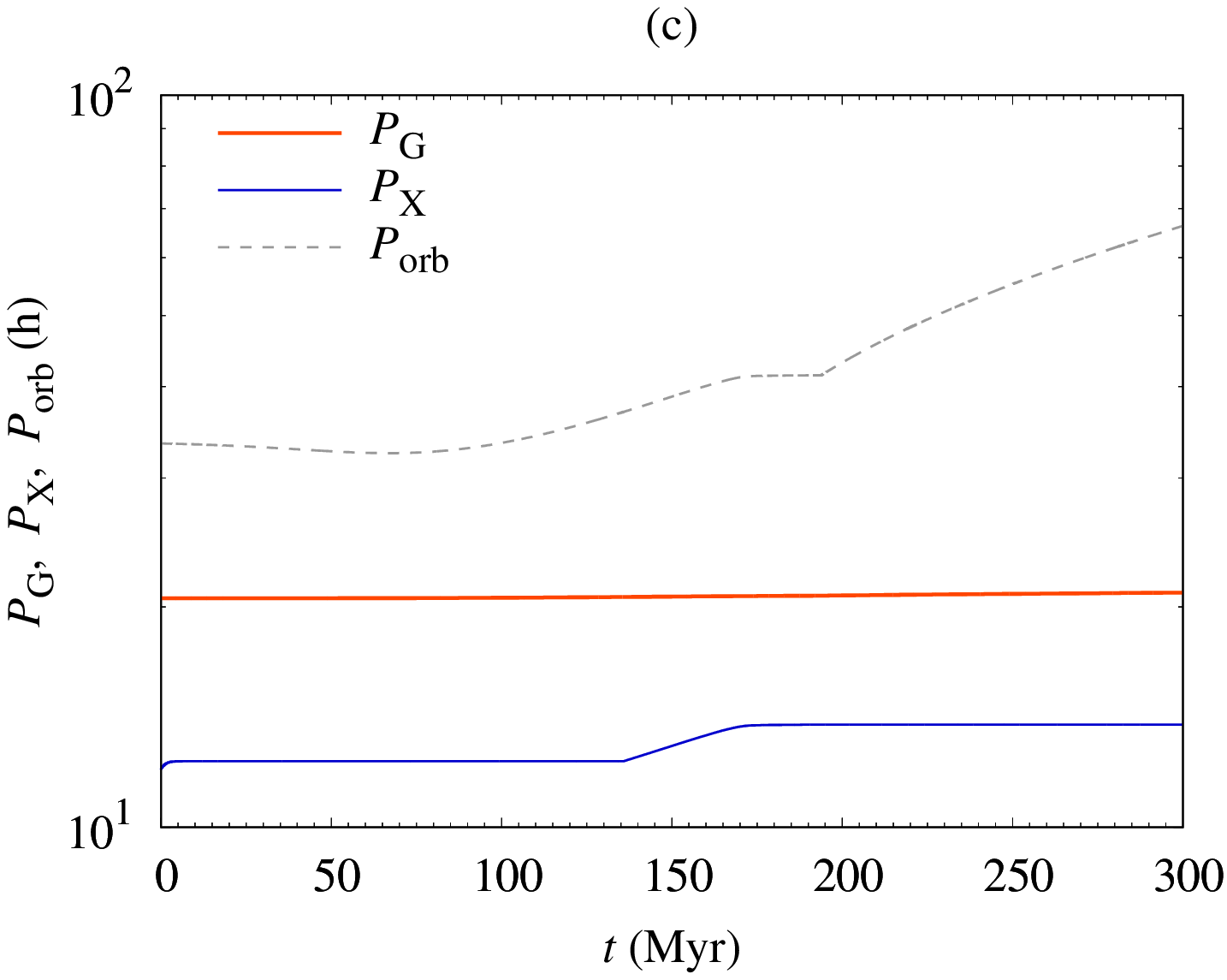}
\includegraphics[width = 0.45\textwidth]{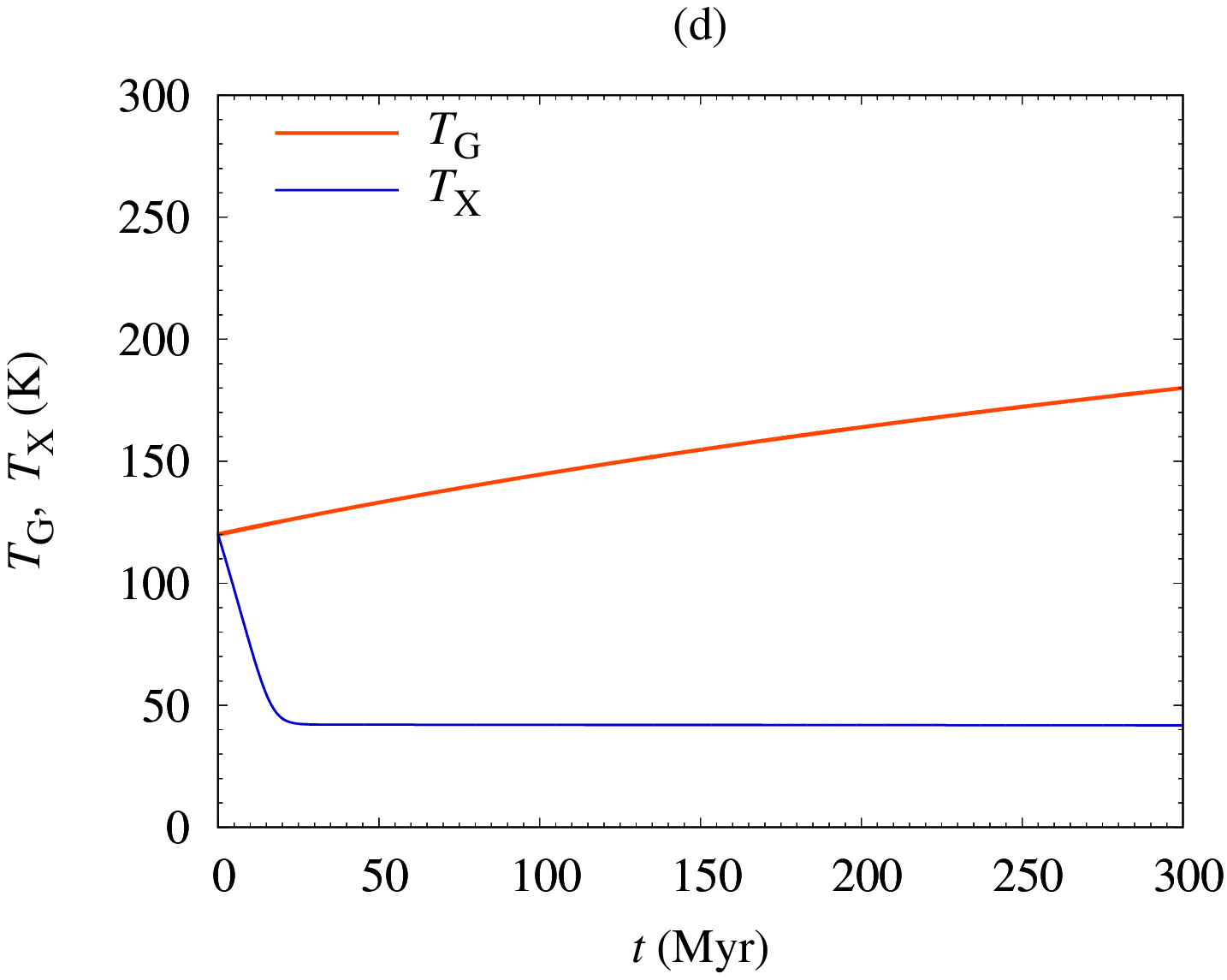}
\includegraphics[width = 0.45\textwidth]{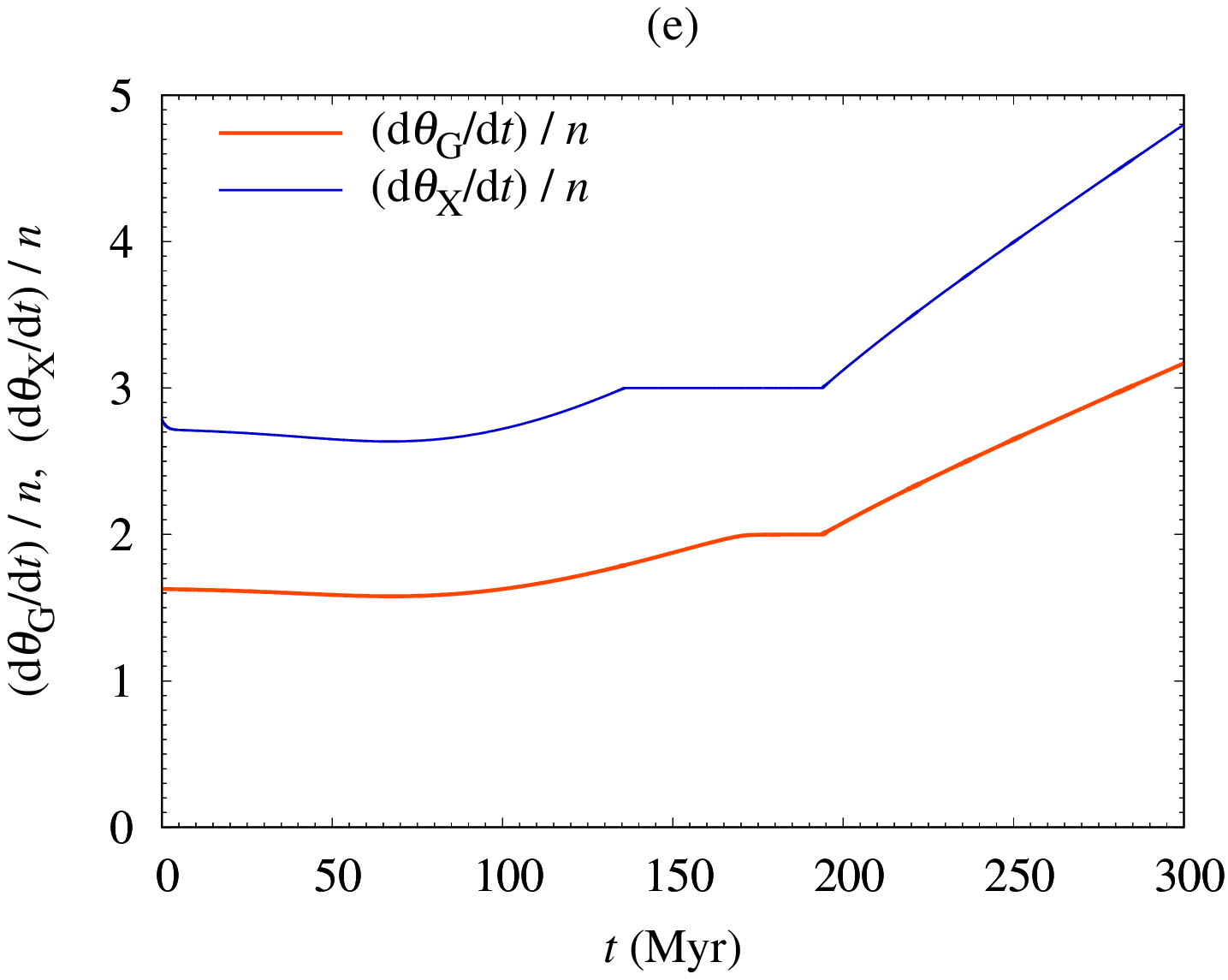}
\caption{
Early phase of the evolution of the (a): semimajor axis, (b): eccentricity, (c): spin/orbital periods, (d): temperature of the primary and secondary, and (e) spin-to-orbit period ratio for Model C.
}
\label{figtypeCini}
\end{figure*}

Figure \ref{figtypeCini} shows the first 300 Myrs of the tidal evolution of this case.
The secondary is captured into 3:1 spin--orbit resonance at first, and then the primary is captured into 2:1 spin--orbit resonance.
Both the primary and secondary are released from the spin--orbit resonance at $t = 193.8\ {\rm Myr}$, and the breaking of the spin--orbit resonance of the primary is prior to that of the secondary.
The eccentricity starts to increase at this time, similar to the case shown in Figure \ref{figtypeBini}.
In Section \ref{sec.analyticalspin}, we show the required conditions to maintain spin--orbit resonances.

\subsection{Type Z: collision with the primary}

\begin{figure*}
\centering
\includegraphics[width = 0.45\textwidth]{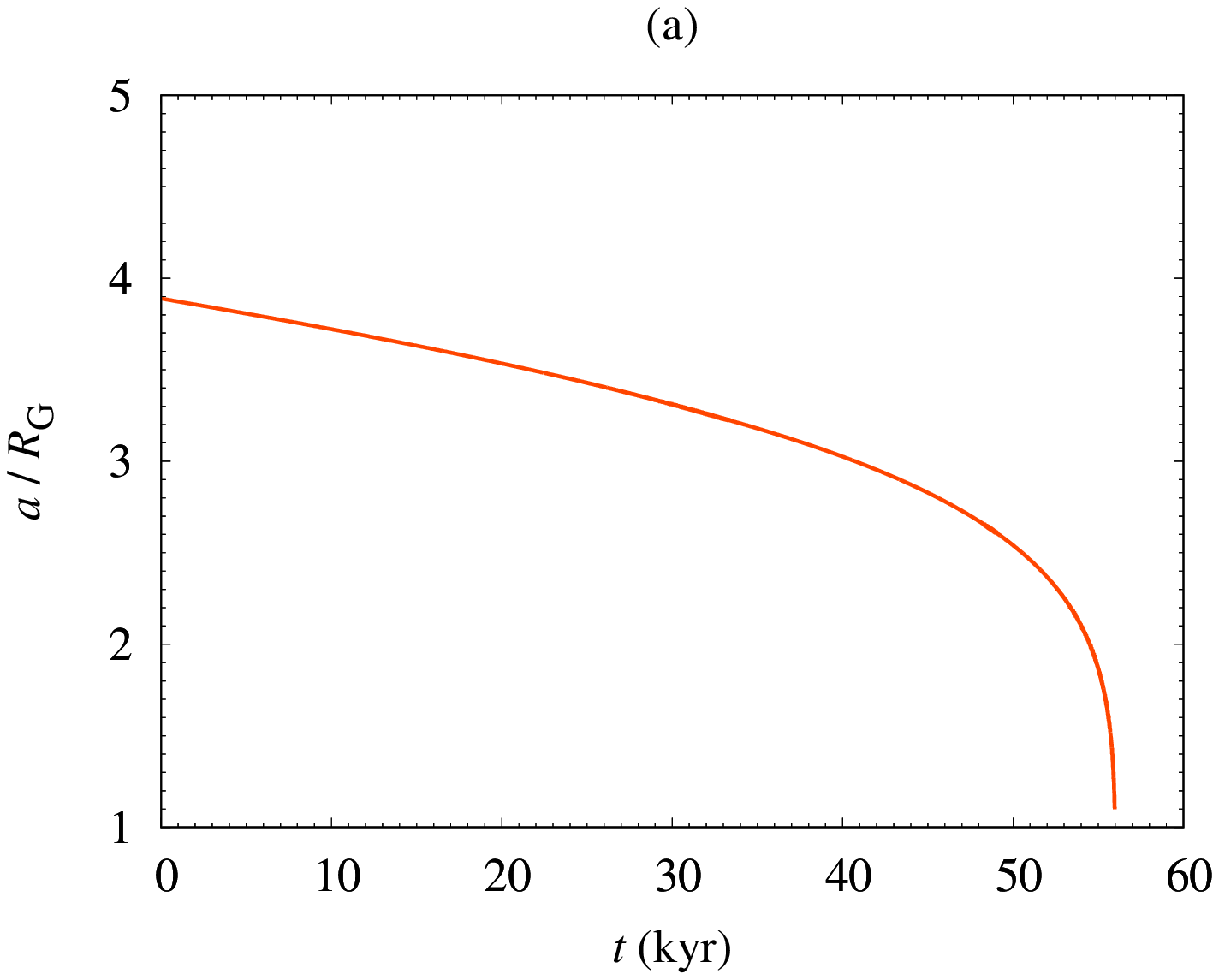}
\includegraphics[width = 0.45\textwidth]{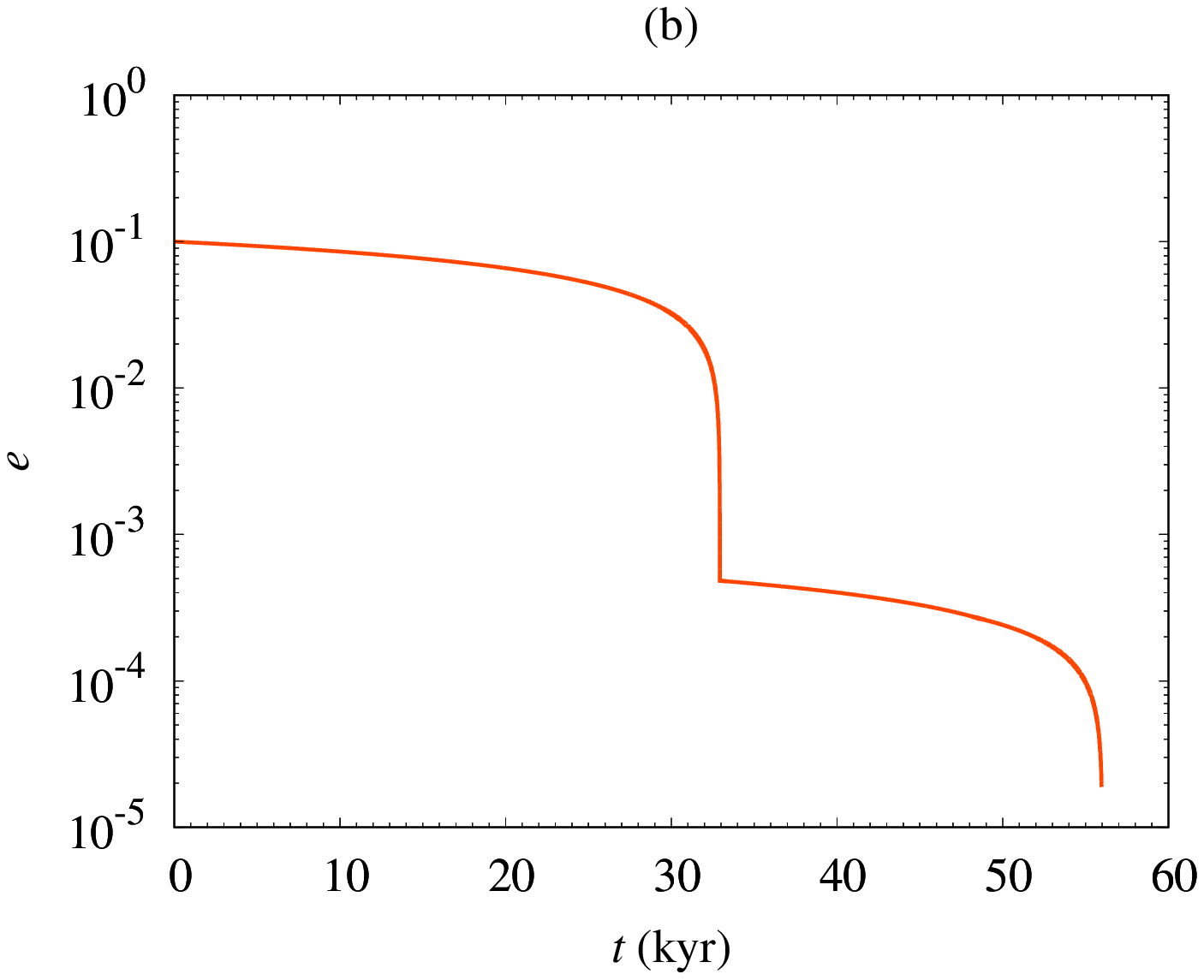}
\includegraphics[width = 0.45\textwidth]{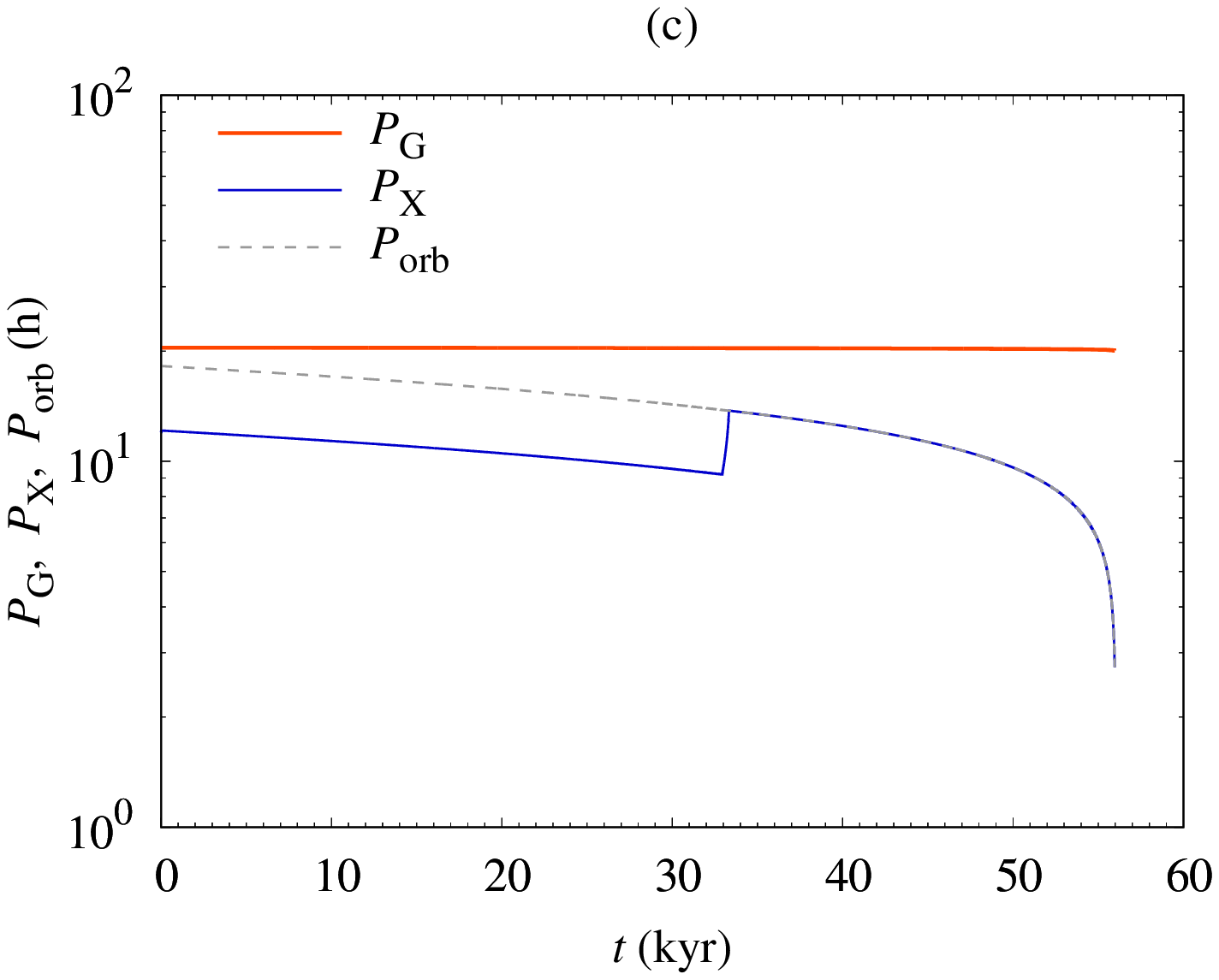}
\includegraphics[width = 0.45\textwidth]{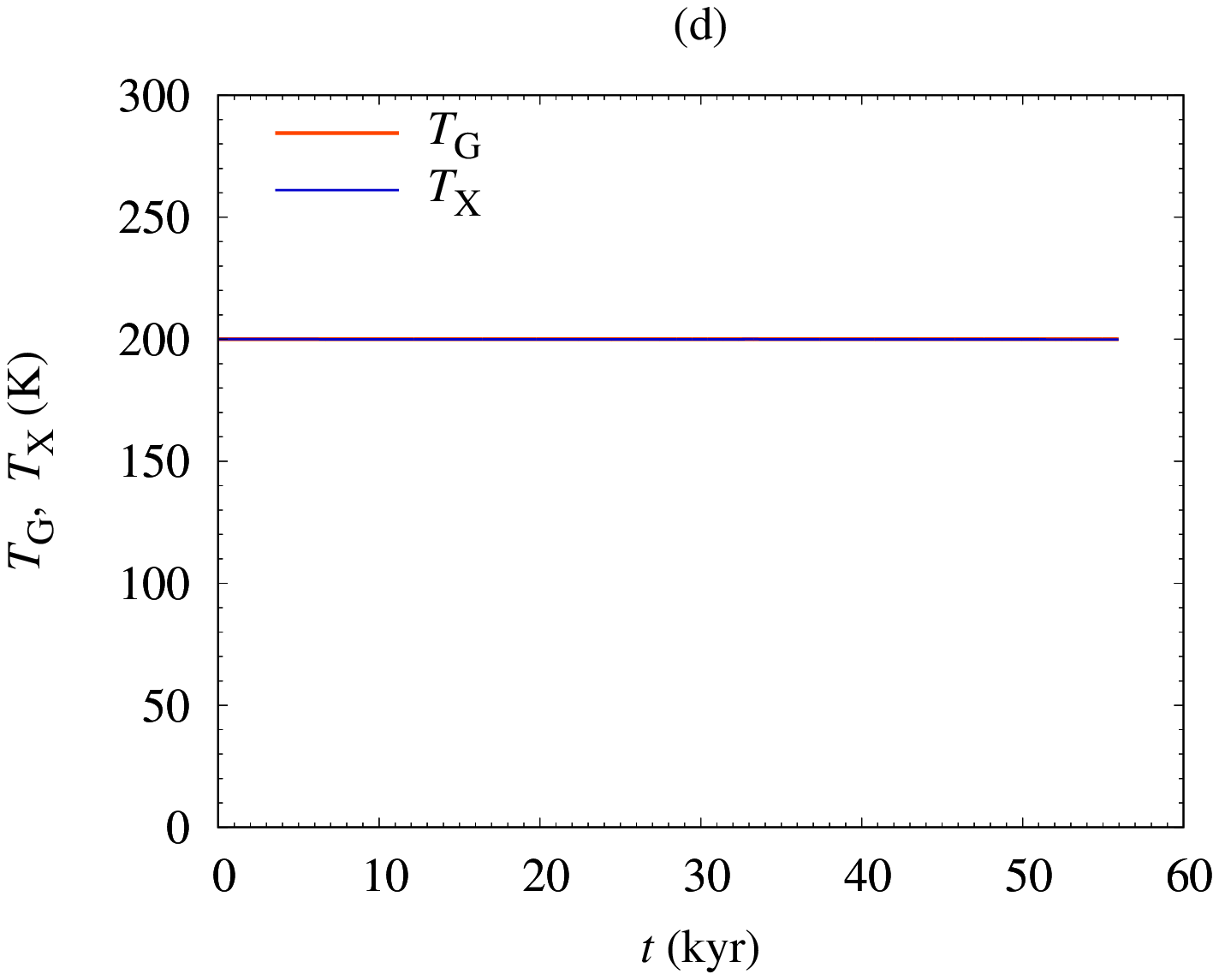}
\caption{
Evolution of the (a): semimajor axis, (b): eccentricity, (c): spin/orbital periods, and (d): temperature of the primary and secondary for Model Z.
The initial condition of Model Z is $R_{\rm X} = 60\ {\rm km}$, $T_{\rm ini} = 200\ {\rm K}$, and $e_{\rm ini} = 0.1$.
}
\label{figtypeZ}
\end{figure*}

Figure \ref{figtypeZ} shows a typical tidal evolution pathway of a satellite system resulting in Type Z.
The initial condition of this model is $R_{\rm X} = 60\ {\rm km}$, $T_{\rm ini} = 200\ {\rm K}$, and $e_{\rm ini} = 0.1$ (see Figure \ref{figType}(c)).
The final semimajor axis and eccentricity are $a_{\rm fin} / R_{\rm G} = 1.1$ and $e_{\rm fin} = 1.9 \times 10^{-5}$, and the spin of the secondary is in 1:1 spin--orbit resonance at $t = 56\ {\rm kyr}$.
In this case, the secondary collides with the primary at the end of the simulation.
Or, in reality, the secondary is tidally disrupted and a second-generation satellites/rings might be formed after the disruption event.

\subsection{Undifferentiated bodies made of soft ice}

\begin{figure*}
\centering
\includegraphics[width = 0.45\textwidth]{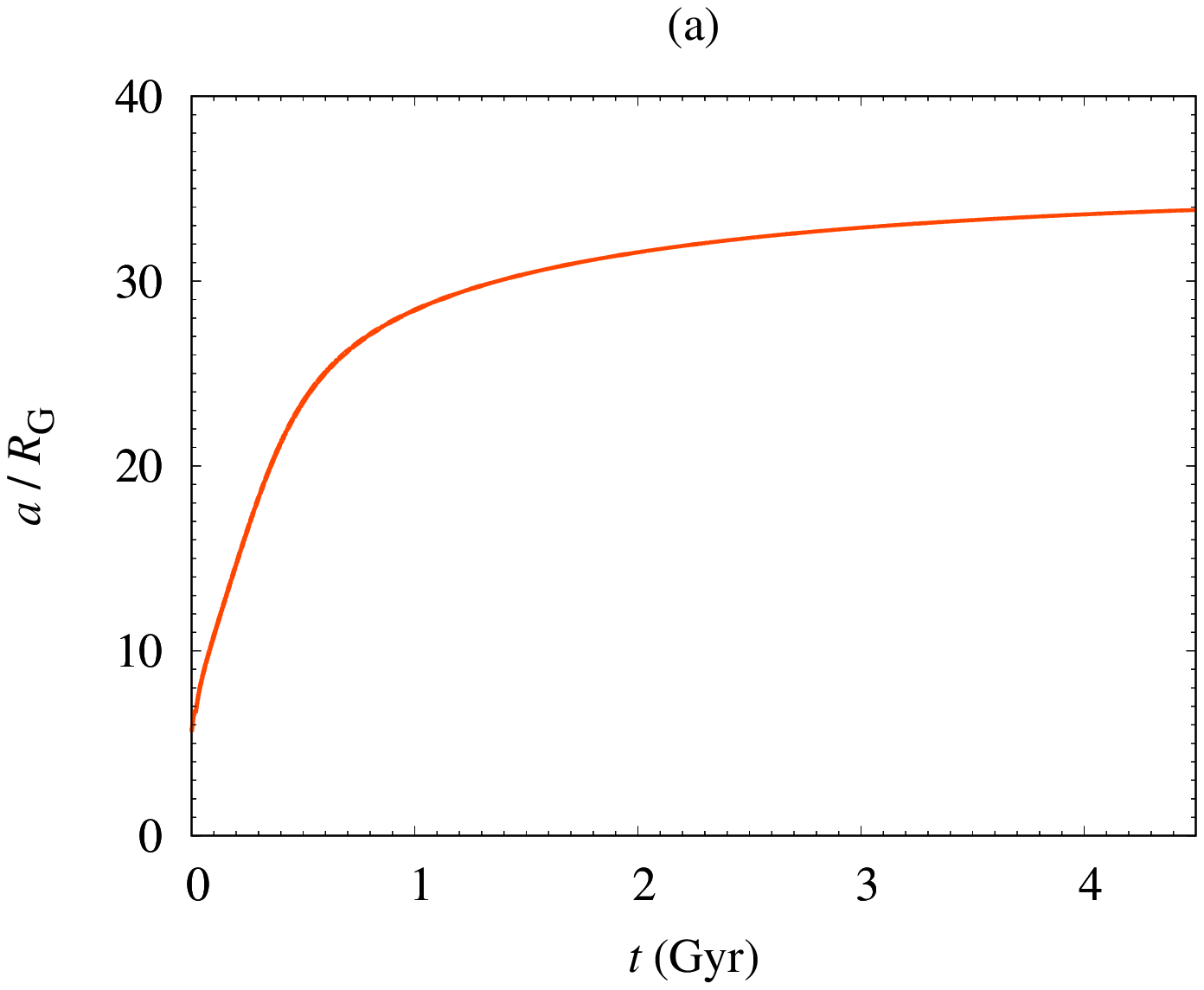}
\includegraphics[width = 0.45\textwidth]{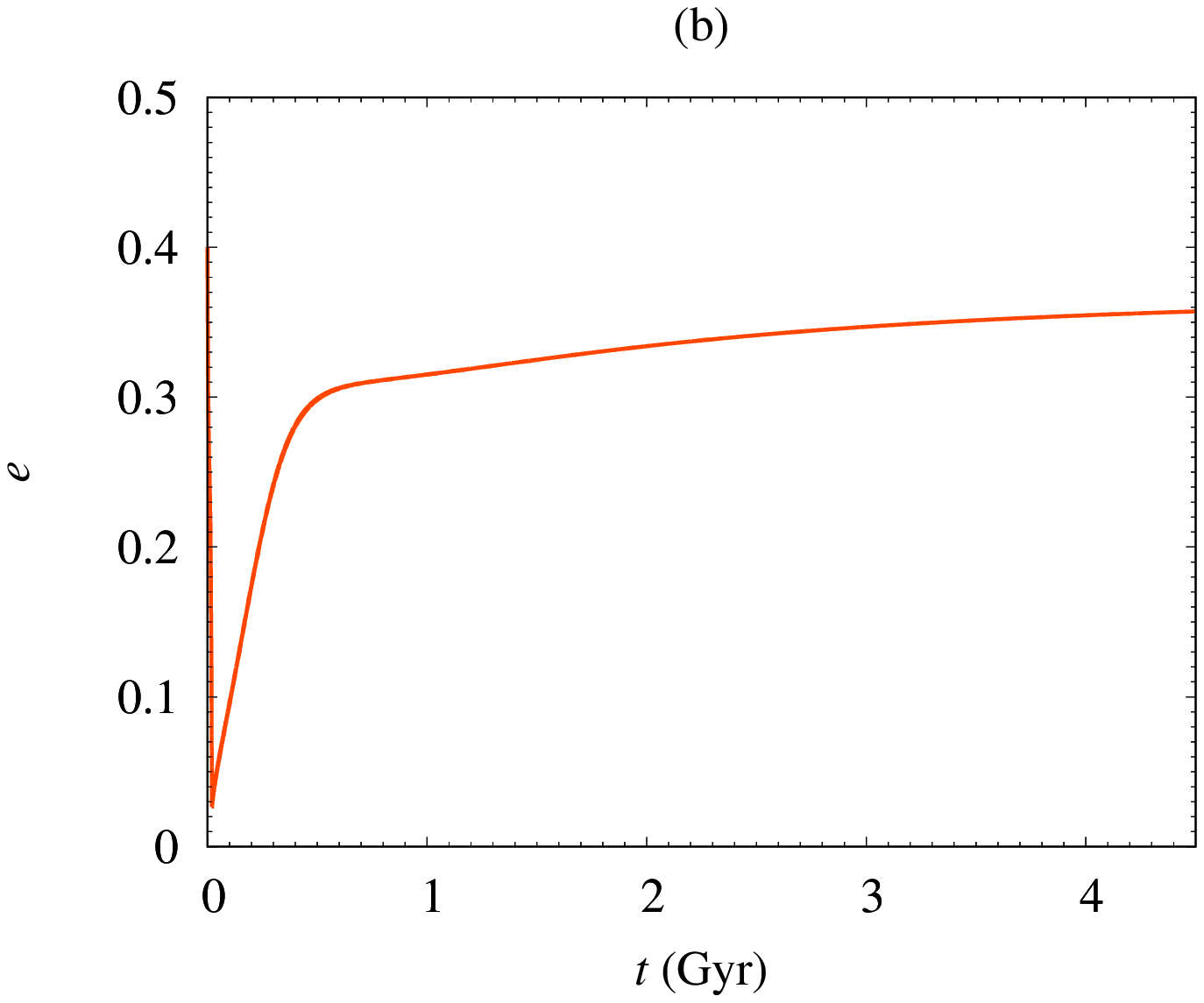}
\includegraphics[width = 0.45\textwidth]{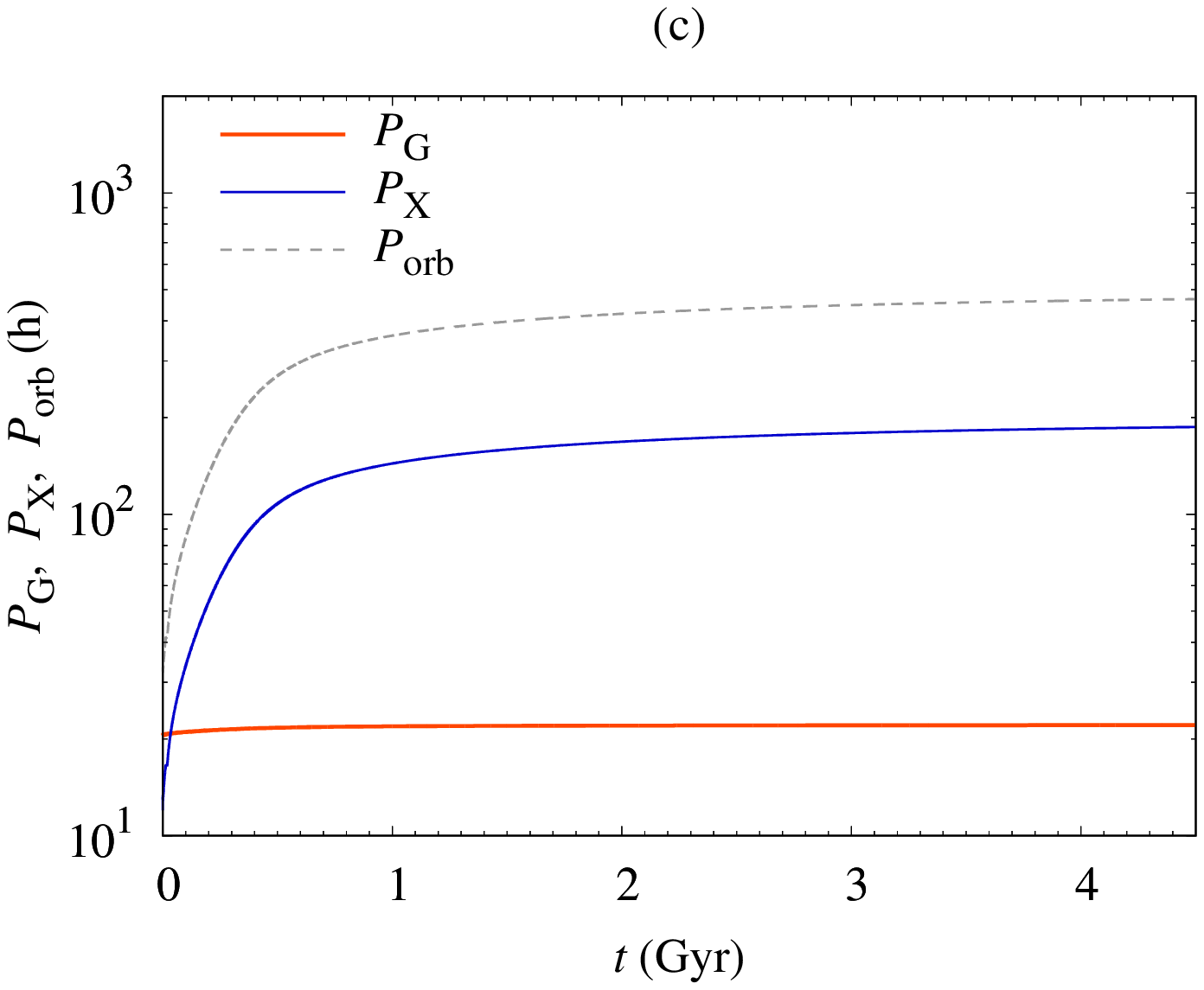}
\includegraphics[width = 0.45\textwidth]{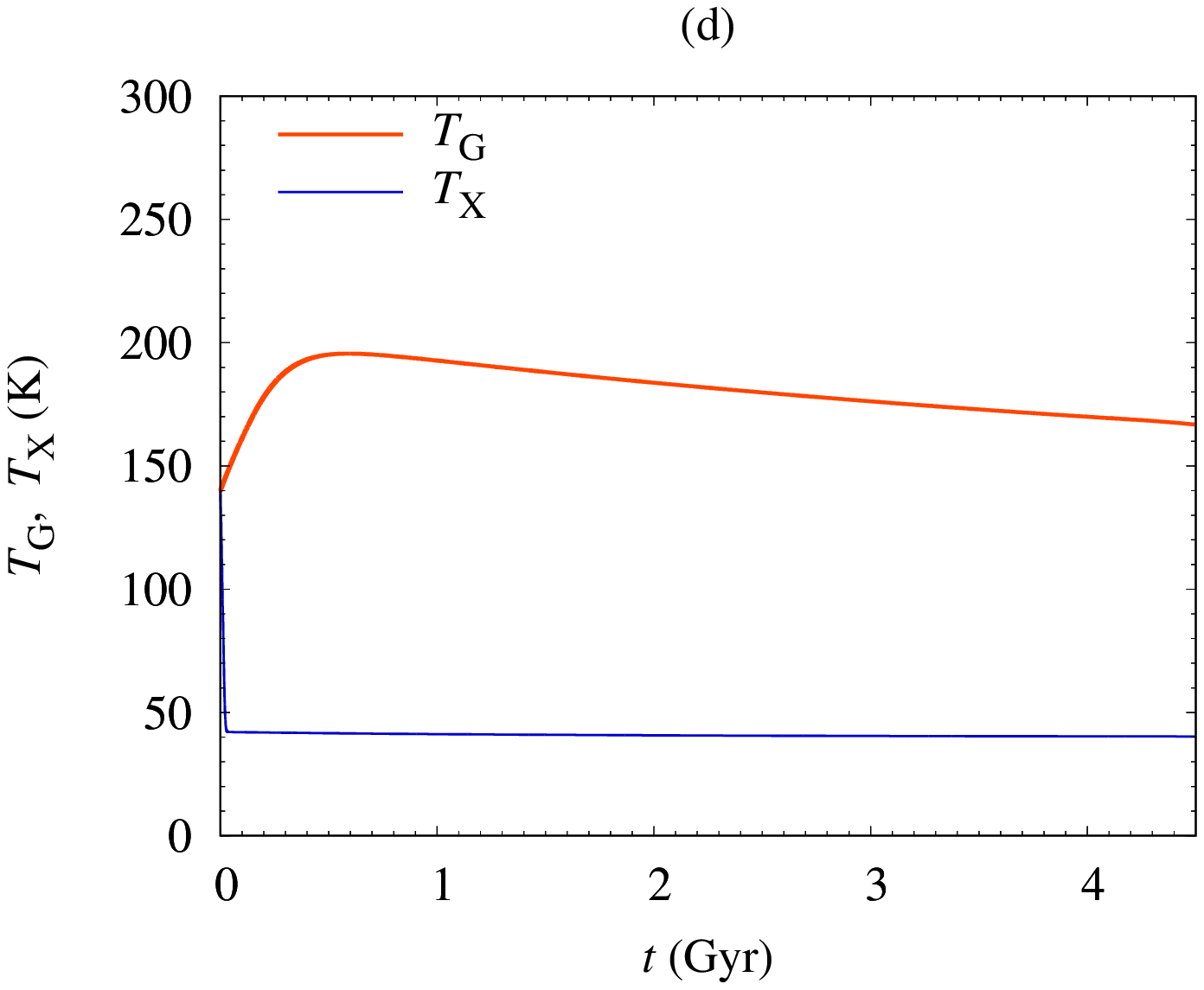}
\caption{
Evolution of the (a): semimajor axis, (b): eccentricity, (c): spin/orbital periods, and (d): temperature of the primary and secondary for a typical run with $\eta_{\rm ref} = 10^{10}\ {\rm Pa}\ {\rm s}$.
The initial condition of this run is $\eta_{\rm ref} = 10^{10}\ {\rm Pa}\ {\rm s}$, $R_{\rm X} = 60\ {\rm km}$, $T_{\rm ini} = 140\ {\rm K}$, and $e_{\rm ini} = 0.4$.
}
\label{figtypeeta10}
\end{figure*}

Figure \ref{figtypeeta10} shows a tidal evolution pathway of a satellite system for the case of $\eta_{\rm ref} = 10^{10}\ {\rm Pa}\ {\rm s}$.
The initial condition of this model is $R_{\rm X} = 60\ {\rm km}$, $T_{\rm ini} = 140\ {\rm K}$, and $e_{\rm ini} = 0.4$ (see Figure \ref{figType2}(d)).
The final semilatus rectum is not very different from that of the standard case of $\eta_{\rm ref} = 10^{14}\ {\rm Pa}\ {\rm s}$ (see Figures \ref{figtypeA}(d), \ref{figtypeB}(d), and \ref{figtypeC}(d)), because the temperature of Gonggong, $T_{\rm G}$, also depends on $\eta_{\rm ref}$.
In the case of $\eta_{\rm ref} = 10^{10}\ {\rm Pa}\ {\rm s}$, $T_{\rm G}$ is lower than that for the case of $\eta_{\rm ref} = 10^{14}\ {\rm Pa}\ {\rm s}$; and the viscosity of Gonggong, $\eta_{\rm G} = \eta {( T_{\rm G} )}$, is not very different between the two settings, $\eta_{\rm ref} = 10^{10}\ {\rm Pa}\ {\rm s}$ and $\eta_{\rm ref} = 10^{14}\ {\rm Pa}\ {\rm s}$.

\clearpage
\bibliography{sample63}{}
\bibliographystyle{aasjournal}



\listofchanges
\end{document}